\documentclass[12pt]{article}
\usepackage{graphicx} % Required for inserting images
\usepackage{natbib}
\usepackage{graphics}
\usepackage[left=1.50cm, right=1.00cm, top=2.00cm, bottom=2.00cm]{geometry}
\usepackage{amsmath}
\usepackage{amsfonts}
\usepackage{latexsym}
\usepackage{setspace}
\usepackage{mathtools}
\usepackage{subcaption}
\usepackage{hyperref}
\usepackage{comment}
\usepackage{float}

\usepackage[ruled]{algorithm2e}
\usepackage{color}

\usepackage{subfiles}
\usepackage{textcomp}
\usepackage{adjustbox}
\usepackage{multicol}
\newcommand{\ctn}{\cite}
\newcommand{\R}{\mathbb{R}}
\newcommand{\cov}{\mbox{cov}}

\newtheorem{lem}{\bf Lemma}[section]
\newtheorem{thm}{\bf Theorem}[section]
\newtheorem{remark}{Remark}
\newtheorem{definition}{Definition}

\newcommand{\bi}[1]{\mbox{\boldmath{$ #1 $}}}
\usepackage{lastpage}
\usepackage{authblk}

\title{\bf{A New Spatio-Temporal Model Exploiting Hamiltonian Equations}}
\author{ }
\author[1]{Satyaki Mazumder\thanks{email: satyaki@iiserkol.ac.in}}
\author[2]{Sayantan Banerjee}
\author[3]{Sourabh Bhattacharya}
\affil[1]{DMS, IISER Kolkata, Mohanpur}
\affil[2]{Operations Management \& Quantitative Techniques Area, IIM Indore}
\affil[3]{Interdisciplinary Statistical Research Unit, ISI Kolkata}
\date{}

\begin{document}

\maketitle
\begin{abstract}%   <- trailing '%' for backward compatibility of .sty file
The solutions of Hamiltonian equations are known to describe the underlying phase space of a mechanical system. %Hamiltonian Monte Carlo is the sole use of the properties of solutions to the Hamiltonian equations in Bayesian statistics.
In this article, we propose a novel spatio-temporal model using a strategic modification of the Hamiltonian equations, incorporating appropriate stochasticity via Gaussian processes. The resultant spatio-temporal process, continuously varying with time, turns out to be nonparametric, non-stationary, non-separable, and non-Gaussian. Additionally, the lagged correlations converge to zero as the spatio-temporal lag goes to infinity. We investigate the theoretical properties of the new spatio-temporal process, including its continuity and smoothness properties. We derive methods for complete Bayesian inference using MCMC techniques in the Bayesian paradigm. The performance of our method has been compared with that of a non-stationary Gaussian process (GP) using two simulation studies, where our method shows a significant improvement over the non-stationary GP. Further, applying our new model to two real data sets revealed encouraging performance. 
\end{abstract}
%\begin{keywords}

  \tableofcontents

\newpage
%\spacingset{1.9} % DON'T change the spacing!
%\setstretch{1.9}
\singlespacing
\section{Introduction}
\label{sec:intro}

	%Modeling spatially and spatio-temporally dependent data has drawn much attention within the statistics community in the last few decades. A wide array of scientific areas produce spatio-temporal data, including meteorology, environment, and ecology. %Furrer and Sain in 2009 (\cite{furrer2009spatial}) fitted a spatial model for climate and microarray data. A study on the summer and winter temperature of  northern hemisphere was carried out using a spatial model by \cite{jun2008spatial}. } 
    The modeling of spatial and spatio-temporal data has gained significant attention in the statistics community over the past few decades. Various scientific fields, including meteorology, environmental science, and ecology, generate spatio-temporal data, necessitating robust analytical methods.
 %A significant amount of work has been done on spatial modeling, including hierarchical Bayesian models with applications to climate data, for example, see \cite{jun2008spatial, furrer2009spatial, tebaldi2009joint, sain2011spatial} among others. 
 Significant research has been carried out on spatial modeling, with a particular focus on hierarchical Bayesian methods for climate applications (e.g., \cite{jun2008spatial, furrer2009spatial, tebaldi2009joint, sain2011spatial}). 
 Results of different climate models concerning the increase in temperature for 22 different regions have been combined by \cite{smith2009bayesian} using Bayesian methods, and \cite{sang2011covariance} built a hierarchical statistical model to integrate multiple climate errors. Numerous spatio-temporal models are available for analyzing climate data. To provide a few examples, spatio-temporal models aiming at analyzing rainfall data have been proposed by \cite{cox1988simple, sanso1999venezuelan}, and \cite{sahu2010fusing} proposed a hierarchical Bayesian space-time model for wet deposition analysis.  
Extensive research has applied spatial and spatio-temporal models to the analysis of environmental data, including ozone, particulate matter, sulfur dioxide, and nitrogen oxide. Notable works include \cite{guttorp1994space} on ground-level ozone, \cite{huerta2004spatiotemporal} on ozone concentration in Mexico City. \cite{bruno2009simple} worked on a non-stationary, non-separable model for ozone data analysis. \cite{dou2010modeling} compared two models for hourly ozone concentration, while \cite{paciorek2009practical} contributed fundamentally to particulate matter analysis using a spatio-temporal model. Studies on sulfur dioxide concentration include \cite{holland2000estimation} and \cite{giannitrapani2006sulphur}. Modeling of nitrogen oxide concentration has been explored by \cite{beelen2013development} and \cite{amini2016annual}, among others. Beyond air pollution, \cite{deng2009spatio} and \cite{chakraborty2010modeling} applied spatio-temporal models to ecological data. These references represent only a fraction of the extensive work in this field.

 Spatio-temporal data analysis often relies on two key assumptions: (i) separability of the covariance function in space and time (see Chapter 11 of \cite{banerjee2014hierarchical}) and (ii) covariance stationarity, meaning the covariance between two observations depends on their separation vector in space and time (Chapters 2 and 11 of \cite{banerjee2014hierarchical}). A common special case, isotropic stationarity, assumes covariance is solely a function of the distance between locations and time points. However, stationarity assumptions can be artificial when local influences affect the correlation structure \cite{sampson1992nonparametric}. Studies have shown nonstationarity in PM10 pollution \cite{paciorek2009practical,  das2023kernel, roy2020bayesian} and sea temperature data \cite{bhattacharya2021bayesian}. Furthermore, \cite{paciorek2003nonstationary} showed that assuming stationarity can lead to over- or under-smoothing. For a review on the importance of nonstationary modeling in spatio-temporal data, see Chapter 9 of \cite{gelfand2010handbook}.

%Recently, several attempts have been made to incorporate nonstationarity in the underlying spatio-temporal process. \cite{sampson1992nonparametric} first significantly contributed to capturing nonstationarity of the covariance function based on the idea of spatial deformation. This idea has been further exploited in the Bayesian paradigm by 	\cite{damian2001bayesian} and \cite{schmidt2003bayesian}. 
Recent efforts to incorporate nonstationarity in spatio-temporal processes have explored various approaches. \cite{sampson1992nonparametric} introduced spatial deformation to model nonstationary covariance, later extended in Bayesian frameworks \cite{damian2001bayesian, schmidt2003bayesian}.     
    %Kernel convolution technique was used by \cite{higdon1998process, higdon1999bayesian, fuentes2001, higdon2002space,  fuentes2002spectral, paciorek2006spatial} to model nonstationarity of the underlying processes. Nonlinear, non-stationary, although parametric spatio-temporal models, were considered in \cite{wikle2010general}. A general framework, called a general nonlinear spatio-temporal framework, has been developed by \cite{wikle2010general}, where a hierarchical Bayesian structure was provided with applications to ecology and oceanology. 
    Kernel convolution techniques were employed by \cite{higdon1998process, higdon1999bayesian, fuentes2001, higdon2002space, fuentes2002spectral, paciorek2006spatial} to model nonstationarity, while \cite{wikle2010general} developed a general nonlinear, non-stationary but parametric hierarchical Bayesian model with an application to ecological and oceanographic data. 
    The stochastic partial differential equation (SPDE) approach, first proposed by \cite{lindgren2011explicit}, represents Gaussian fields (GF) as Gaussian Markov random fields (GMRF), a discretely indexed spatial random field. The key ingredient was to use the fact that a GF with Mat\'{e}rn covariance kernel is the solution of a fractional SPDE (Chapter 6 of \cite{blangiardo2015spatial}). In the SPDE framework, non-stationary covariance is modeled by varying Mat\'{e}rn kernel parameters with respect to location \cite{lindgren2011explicit, bolin2011spatial, ingebrigtsen2014spatial, blangiardo2015spatial}. A review of the SPDE approach for spatial and spatio-temporal modeling can be found in \cite{lindgren2022spde}. A major advantage of SPDE is its implementation in R via the INLA package \cite{lindgren2015bayesian}, which efficiently handles large datasets using sparse matrices \cite{blangiardo2015spatial, gomez2020bayesian}. Also see the website \url{r-inla.org/home} for further details on INLA.  Alternative nonstationary models include a Dirichlet process-based Bayesian approach for conditional nonstationarity \cite{gelfand2005bayesian}, a kernel process mixing method \cite{fuentes2013multivariate}, and a discretized stochastic differential equation model \cite{duan2009modeling}. More recently, \cite{driver2018hierarchical} proposed a hierarchical Bayesian method for dynamic parametric models via stochastic differential equations.

	In the above-mentioned works, the following aspects can be noticed. The underlying process is Gaussian (may be non-stationary) or parametric, or the correlation between the two spatio-temporal points does not converge to zero as the distance between the two points (locations and/or time points) increases to infinity or both. For example, our simulation from the general quadratic nonlinear model \cite{wikle2010general} reveals that the correlation does not vanish as the distance increases indefinitely (see Figures \ref{Spatial correlation plot for GQN model} and \ref{Temporal correlation plot for GQN model} in Appendix \ref{Correlation analysis of GQN model}).
	%(as also borne out by our simulation studies not detailed here for brevity) or both. 
    %It is not very unlikely to realize spatio-temporal data that are not Gaussian; see, for example, \cite{fuentes2013multivariate}, \cite{das2023kernel}, \cite{bhattacharya2021bayesian}. Moreover, in many real-life applications often, it is observed that the sample correlation goes to zero as the distance between two spatial locations and/or two time points goes to infinity; see, for example, \cite{bhattacharya2021bayesian}. Since many stationary spatio-temporal processes are ergodic, this property follows naturally from the structure of the covariance function in such cases. However, this does not hold for non-stationary processes in general. Hence, while proposing a non-stationary spatio-temporal process, it is desirable to to take special care to ensure convergence of the correlation to zero as the distance between two locations and/or time points increases to infinity. 
    Moreover, spatio-temporal data may not always be Gaussian \cite{fuentes2013multivariate, das2023kernel, bhattacharya2021bayesian}. Furthermore, in many real-life cases, sample correlation often approaches zero with increasing spatial or temporal distance \cite{bhattacharya2021bayesian}, a property naturally following from the covariance structure of ergodic stationary processes but not necessarily for non-stationary ones. Thus, when modeling non-stationary spatio-temporal processes, ensuring correlation convergence to zero at large distances is crucial.
	
	%Very recently,	\cite{das2023kernel} proposed a nonparametric, non-separable, non-stationary, and non-Gaussian spatio-temporal model based on order-based dependent Dirichlet process, where they showed that the underlying covariance function goes to zero as the distance between the two locations and/or time points increases to infinity. While modeling spatio-temporal data, \cite{das2023kernel}also assumed that time and space vary continuously in their respective domains. Although \cite{das2023kernel} made a successful attempt at building the nonparametric, non-stationary, non-Gaussian model with the 	desirable properties of covariance, their model did not impart dynamic properties to the temporal part, either directly or via any latent process. It is well known that a statistical model intended for explaining temporal variability should evolve dynamically (Chapter 1 of \cite{cressie2015statistics}). 
    %On the other hand, \cite{Suman2017} introduces a nonparametric spatio-temporal model, which, through a nonparametric dynamic latent process, induces desirable dynamic properties in the temporal part. In addition, the model of \cite{Suman2017} is non-stationary, and time does not vary continuously on the respective domain.
    Recently, \cite{das2023kernel} proposed a nonparametric, non-separable, non-stationary, and non-Gaussian spatio-temporal model based on an order-based dependent Dirichlet process, ensuring the covariance function converges to zero as spatial or temporal distance increases. Their model assumes continuous variation in time and space. However, while achieving desirable covariance properties, it lacks dynamic temporal evolution, a crucial aspect of modeling temporal variability (Chapter 1 of \cite{cressie2015statistics}).
    In contrast, \cite{Suman2017} introduced a nonparametric non-stationary spatio-temporal model that incorporates a dynamic latent process to induce temporal dynamics. However, in their model time does not vary continually.			

%In a nutshell, there does not seem to exist any nonparametric, non-stationary, non-separable, non-Gaussian, dynamic spatio-temporal model, which is continuous in time and space with an underlying structured latent process, and with the property that the correlation between two spatio-temporal variables goes to zero as the spatial/temporal lag goes to infinity. To fill this gap, we propose a dynamic spatio-temporal nonparametric, non-Gaussian model, where the underlying process is non-stationary. A structured latent process is incorporated into the proposed model. The time and space in our proposed spatio-temporal model vary continuously over their respective domains. Further, the underlying process enjoys the property that the covariance goes to zero as the spatial/temporal lag tends to infinity.

In summary, no existing model meets all the criteria of being nonparametric, non-stationary, non-separable, non-Gaussian, and dynamic, while ensuring continuity in time and space, incorporating a structured latent process, and guaranteeing correlation decay to zero at infinite spatial/temporal lags. To address this gap, we propose a dynamic, nonparametric, non-Gaussian spatio-temporal model with a non-stationary underlying process. Our model integrates a structured latent process, allows continuous variation in time and space, and ensures convergence of covariance to zero at large spatial/temporal distances.

We construct our model using Hamiltonian dynamics, a concept from physics also applied in Bayesian statistics for Hamiltonian Monte Carlo \cite{betancourt2017conceptual,cheung2009bayesian}. The solutions of Hamiltonian equations define a mechanical system's phase-space and vary continuously over time, motivating their use in our spatio-temporal model. While deterministic Hamiltonian equations have been generalized to stochastic forms, ordinary or partial (\cite{lazaro2007stochastic, hong2022three} and references therein), no temporal or spatio-temporal modeling has been attempted. These equations remain parametric, assuming known functional forms. Typically, a Brownian motion is added to deterministic equations to create stochastic Hamiltonian differential equations and partial differential equations are generalizations of the stochastic Hamiltonian ordinary differential equations. Since it is difficult to see how these stochastic equations with deterministic functional forms and (finite or infinite-dimensional) Brownian motions may be useful for constructing realistic temporal or spatio-temporal models with desirable properties, we do not pursue these in this work.

The use of Hamiltonian dynamics with the leap-frog algorithm \cite{young2014leapfrog} in spatio-temporal modeling has not been explored earlier. Adaptation and exploitation of
the leap-frog algorithm with appropriate choice of Gaussian process priors make the proposed model non-parametric. Specifically, we propose a modified leap-frog algorithm alongside modified Hamiltonian equations to construct a spatio-temporal model incorporating latent and observed processes that continuously depend on time $t$. Since the leap-frog algorithm leverages derivatives, our model exhibits dynamic behavior through dependence on the previous time point $t-\delta t$. The Bayesian nonparametric nature arises from modeling unknown functions in Hamiltonian equations via Gaussian processes, \textit{ensuring non-stationarity, non-separability, and non-Gaussianity}. Further analysis confirms that the covariance function converges to zero at large spatial/temporal lags. Additionally, under regularity conditions, we establish that the proposed process has continuous paths in both almost sure and mean square senses, along with almost sure and mean square differentiability. Moreover, we prove a lemma providing sufficient conditions for the mean square differentiability of a composition of a random spatial process and a Gaussian spatial process, a result of independent significance.
%.    

The article is structured as follows: Section \ref{process proposal} introduces a new spatio-temporal process using modified Hamiltonian equations and a leap-frog algorithm, along with key properties of the observed and latent processes. Proofs of related lemmas, theorems, and corollaries are in Appendix \ref{theorem proofs}. Section \ref{complete likelihood} derives the likelihood function, first obtaining the joint conditional density of data given latent variables and parameters (Subsection \ref{data model}, Appendix \ref{joint density of observed data}), followed by the joint density of latent variables (Subsection \ref{process model}, Appendix \ref{joint density of latent data}), and then the complete likelihood (Subsection \ref{Jt Dist}). Section \ref{Priors} discusses prior distributions and their justification. Section \ref{full conditional} derives full conditional densities for Gibbs sampling, with detailed calculations in Appendix \ref{Appendix B: full conditional densities}. Section \ref{Sec: Simulation Studies} presents two simulation studies comparing our model with a non-stationary Gaussian process fitted using the SPDE approach \cite{lindgren2011explicit, lindgren2015bayesian}. Section \ref{real data analysis} analyzes two real datasets-one non-stationary and non-Gaussian, the other stationary and non-Gaussian, to evaluate model performance. Finally, Section \ref{sec:conclusion} summarizes contributions and concludes the study.

\section{Modified Hamiltonian equations and the proposed model}
	\label{process proposal}
    %%%%%%%%%%%%%%%%%%%%%%%%%%%%%%%%%%%%%%%%%%%%%%%%%%%%%%%%%%%
	%We first briefly describe the notion of Hamiltonian dynamics and the associated equations. Thereafter, we shall connect the idea of the proposed spatio-temporal model to Hamiltonian dynamics.
    We begin with a brief overview of Hamiltonian dynamics and its equations before linking them to our proposed spatio-temporal model.
	Let $(\mathcal{M},\mathcal{L})$ be a mechanical system, where $\mathcal{M}$ is the configuration space and $\mathcal{L}$ is the smooth Lagrangian. 
	The coordinate system of $\mathcal{M}$ is determined by $(\theta,\dot{\theta})$, where $\theta$ is the position of a particle at time $t$ and the 
	$\dot{\theta}$ is the derivative vector with respect to time, thus representing the velocity. The partial derivative of $\mathcal{L}$ with respect 
	to $\dot{\theta}$, known as momenta, is denoted by $p$, which is a function of time $t$, the position $\theta$, and velocity $\dot{\theta}$. 
	Now the Hamiltonian, a function of ${p}$, $\theta$ and $t$, is defined as     
    \allowdisplaybreaks
	$${H} (p,\theta,t) = \sum_i p_i \dot{\theta}_i - \mathcal{L}(\theta,\dot{\theta},t),$$ which is the energy function of the mechanical system. Here $p_i\text{ and } \dot{\theta}_i$ are the $i$th component of $p$ and $\dot{\theta}$. The pair $(\theta,p)$ corresponds to phase space coordinates.	Now assume that $H(p,\theta, t) = V(\theta,t) + W(p,t)$, where $V(\theta,t)$ is the potential energy, $W(p,t)$ is the kinetic energy given as $W(p,t) = p^{T}M^{-1}p/2$, with $M$ being a chosen matrix (mass),  and ($\theta,p$) are phase space coordinates (depend on time $t$) \cite{cheung2009bayesian}. The Hamiltonian equations are given by	
	\begin{align*}
		%\label{Hamiltonian eqn}
        \frac{d \theta}{d t}  = \frac{\partial H}{\partial p} =  M^{-1} p, \mbox{ and } 
		\frac{d p}{d t} = \frac{- \partial H}{\partial \theta} = -\nabla V(\theta), 
	\end{align*}
where $\nabla V(\theta)$ represents gradient of $V$ with respect to $\theta$. Thus, the phase-space coordinates satisfy the Hamiltonian equations, where the solution $\theta$, representing a particle’s position at time $t$, depends on $p$, and varies continuously with time. Similarly, $p$ depends on $\theta$ and also evolves continuously over time. To get the solution $(p,\theta)$ numerically, leap-frog algorithm is employed and is given by
%Thus, the phase space coordinates are the solution of the Hamiltonian equations. We note here that the solution $\theta$, which describes the position of a particle at time $t$ and depends on $p$, the other coordinate of the phase-space, varies continuously with respect to time. Further, $p$ is also dependent on $\theta$ at time $t$, and varies continuously with respect time $t$. To get the solution $(p,\theta)$ numerically, leap-frog algorithm is used and is given by
   \begin{align*}
	\theta(t+\delta t) & = \theta(t) + \delta t M^{-1}\left[p(t) - \frac{1}{2}\delta t \nabla V(\theta(t))\right] \mbox{ and } \\
	p(t+\delta t) & = p(t) - \frac{1}{2} \delta t \bigg[\nabla V(\theta(t)) + \nabla V(\theta(t+\delta t))\bigg].
\end{align*}
%	In our case, we introduce $p(t)$ as the latent and $\theta(t)$ as the observed spatio-temporal processes with the hope that the solution to Hamilton's equations will best describe the underlying state-space as done in the phase-space of the mechanical space.
    %---------------------------------------------
		\subsection{Modification, the new leap-frog algorithm and the proposed model}
		\label{Modified HM and leap frog}	
        %-------------------------------------------
	We modify the original Hamiltonian equations so that the rate of change of $p$ and $\theta$ with respect to $t$ depends on both $p$ and $\theta$. One way to do so is to incorporate $p$ and $\theta$ linearly in the equation of $\frac{d\theta}{dt}$ and $\frac{dp}{dt}$, respectively. That is, the Hamiltonian equations are modified to 
	\begin{align}
    \label{eqn: modified Hamiltonian eqn}
		\frac{d \theta }{d t}  = \beta^* \theta + M^{-1} p, \mbox{ and } \frac{d p }{d t} = \alpha^* p -\nabla V(\theta). 
	\end{align}
    In our proposal of spatio-temporal process, we introduce $p(\cdot)$ as the latent process, $X$, say, and $\theta(\cdot)$ as the data-generating spatio-temporal process, $Y$, say, in equation $\ref{eqn: modified Hamiltonian eqn}$ with the hope that the solution to the modified Hamilton's equations will best describe the underlying state-space 
   as the original Hamiltonian equations model the phase-space of the mechanical space. That is, at every location, say $s\in S$ with $S$ being a non-empty subset of $\mathbb{R}^2$ containing a rectangle of positive volume, we have 
    \begin{align}
    \label{eq: proposed stochastic differential equations}
    \frac{dY(s,t)}{dt} = \beta^* Y(s,t) + M_s^{-1}X(s,t) \text{ and } \frac{dX(s,t)}{dt} = \alpha^* X(s,t) - \nabla V(Y(s,t)),
    \end{align}
    where by $\nabla V(Y(s,t))$ we mean $\frac{d}{dx}V(x)\bigg|_{x = Y(s,t)}$, and $M_s$ is a constant given a location $s$, equivalent to the mass in the original Hamiltonian equation. %the only difference is now it varies with respect to location $s$. 
    The function $V(\cdot)$ is assumed to be a random function. Stochasticity of $X(\cdot,\cdot)$ depends on the stochasticity of $V(\cdot)$ and stochasticity of $Y(\cdot, \cdot)$ depends on the stochasticity of $X(\cdot, \cdot)$. Therefore, stochasticity of both $Y(\cdot, \cdot)$ and $X(\cdot, \cdot)$, in turn, depend on the stochasticity of $V(\cdot)$. 
    
    The reason for choosing $p$ as latent is that we believe that the latent process evolves compositely with respect to a unknown stochastic function; however, the data-generating process is a simple function of the underlying latent process. This philosophy has been followed in other works as well, see for example, \cite{ghosh2014bayesian, mazumder2016bayesian, mazumder2017nonparametric} and the references therein. the proposed model is obtained by discretizing equations \ref{eq: proposed stochastic differential equations} 
%via the Leap-frog algorithm (\citep{Hockney1970, young2014leapfrog}) as 
via the Leap-frog algorithm (\citealt{Hockney1970, Verlet1967, young2014leapfrog}) as
%\allowdisplaybreaks     
	\begin{align}
		\label{observation equation}
		y(s,t+\delta t) &= \beta y(s,t) + \delta t M_s^{-1} \left(\alpha x(s,t) - \nabla V(y(s,t))\frac{\delta t}{2}\right), \text{ and } \\
		\label{latent equation}
		x(s,t+\delta t) &= \alpha^2 x(s,t) - \frac{\delta t}{2} \left\{\alpha \nabla V(y(s,t)) + \nabla V(y(s,{t+\delta t})) \right\},
	\end{align}
    where $y(s,t)$ and $x(s,t)$ are realizations of the stochastic processes $\displaystyle \{Y(s,t): s\in S, t\geq 0\}$ and $\displaystyle \{X(s,t): s\in S, t\geq 0\}$, respectively. Further, $\alpha \text{ and }\beta$ are defined as $\displaystyle \alpha = 1+ \alpha^* \frac{\delta t}{2}$  and  $\displaystyle \beta = 1+ \beta^* \frac{\delta t}{2},$  respectively. {The derivation of equations (\ref{observation equation}) and (\ref{latent equation}) are provided in Appendix \ref{Modified leap-frog}.} 
    Here we assume that $y(s,t)$ is the potential observation while $x(s,,t)$ is unobserved. 
    
    {Generally, Hamiltonian equations are numerically handled using Leap-frog algorithm, since it preserves the energy. Energy preservation is also important for 
    ensuring good acceptance rates for HMC \citep{cheung2009bayesian}. In our case, the new, modified Leap-frog algorithm is not directly related to energy conservation, 
    rather, the use of Leap-frog algorithm leads to mean and covariance non-stationarity of the derived processes (see Section~\ref{properties of the processes}), which was 
    one of the main goals of our work.}  

Equation (\ref{observation equation}) may be regarded as the observation equation while (\ref{latent equation}) deals with the dynamics of the latent process. The parameters $\alpha$, $\beta$ are such that $|\alpha|<1$ and $|\beta|<1$. %The restriction on the parameter space of $\alpha$ and $\beta$ helps our model in two fold. 
We will see subsequently, the restriction on $\beta$ is necessary for the lagged correlations between the observations to tend to zero as the space-time lag tends to infinity, and the restriction on $\alpha$ led to good MCMC mixing in our Bayesian applications. 
%, and $M_s$ is a function of $s$. The function $V(\cdot)$ is modeled as random function. %The details on the choice of $M_s$ and $V(\cdot)$ are described in Section \ref{properties of the processes}. 
	Note that equation (\ref{latent equation}) is not the latent equation {\it per se}, since it involves the observed variable as well. Integrating the conditional distribution of the latent variable (equation (\ref{latent equation})) over the observed variable would give us the latent distribution. 
	To fully specify our spatio-temporal process, we model the function $V(\cdot)$ as a Gaussian process. Under suitable assumptions, this choice ensures that $V'(\cdot)$ is also Gaussian \cite{Stein99, rusmassen2005gaussian}, allowing the conditional likelihoods to be expressed in closed form (see Section \ref{complete likelihood}). Additionally, specifying the proposed process requires an appropriate form for $M_s$. The next subsection provides further details on these aspects and explores the theoretical properties of our spatio-temporal processes.
	%\subsection{Completion of specification of the proposed spatio-temporal process and investigation of its theoretical properties}
    %--------------------------------------------
    \subsection{Remaining specification and analysis of the proposed model}
	\label{properties of the processes}
%-----------------------------------------------
	Let $s\in S$, a compact subset of $\mathbb{R}^{2}$, and $||\cdot||$ denote the Euclidean norm. 
	We consider the following assumptions on the processes $X(s,0), Y(s,0)$, $s\in S$ and the random function $V(\cdot)$.
	\begin{enumerate}
		\item[A1.] $X(s,0)$ and $Y(s,0)$, $s\in S$, are assumed to be centered Gaussian processes with symmetric, positive definite covariance functions 
			having bounded partial derivatives. 
			In particular, due to reasons of popularity, the covariance functions for $X(s,0)$ and $Y(s,0)$ are taken to be squared exponential of the form
        $\text{cov}(X(s_1,0),X(s_2,0)) = \sigma^2_{p}\exp{\{-\eta_1||s_1-s_2||^2\}}$ and $\text{cov}(Y(s_1,0),Y(s_2,0)) = \sigma^2_{\theta}\exp{\{-\eta_2||s_1-s_2||^2\}}$, for $s_1, s_2\in S$, respectively. 
			However, the Mat\'{e}rn covariance function with $\nu > 1$ also has bounded partial derivatives and 
			could be employed. 
        %$X(s,0),\, s\in S$ is assumed to be a centered Gaussian process with a symmetric, positive definite covariance function having bounded partial derivatives. For instance, the Mat\'{e}rn covariance function with $\nu > 1$ has bounded partial derivatives and could be employed. In particular, we consider the squared exponential covariance between $X(s_1,0)$ and $X(s_2,0)$, of the form $\sigma^2_{p}\exp{\{-\eta_1||s_1-s_2||^2\}}$, $s_1, s_2 \in S$. 
	%	\item[A2.] $Y(s,0), \, s\in S$ is also assumed to be a centered Gaussian process with a symmetric, positive definite covariance function with bounded partial derivatives. Again, we consider the squared exponential covariance function of the form 		$\sigma^2_{\theta}\exp{\{-\eta_2 ||s_1-s_2||^2\}}$. 
		\item[A2.] The function $V(\cdot)$ is assumed to be a Gaussian random function with zero mean and covariance function 
		$c_{v}(x_1,x_2) = \sigma^2 \exp{\{-\eta_3||x_1-x_2||^2\}}$. 
        %As for the other cases, different choices of covariance structure can 	be assumed with the assumption that the covariance function is continuously twice differentiable and the mixed partial derivatives are Lipschitz continuous. 		If the covariance function has third bounded partial derivatives, then the function will be Lipschitz. In our example, 		the covariance function is infinitely differentiable and the derivatives are bounded. Other choices may include rational quadratic covariance 		function, Mat\'{e}rn covariance function with $\nu > 2$ among many others. 
        Other covariance structures can be considered, provided the covariance function is twice continuously differentiable with Lipschitz continuous mixed partial derivatives. If it has bounded third-order partial derivatives, it is inherently Lipschitz. In our case, the chosen covariance function is infinitely differentiable with bounded derivatives. Alternative choices include the rational quadratic covariance function or the Mat\'{e}rn covariance function with $\nu > 2$, among others. 
	\end{enumerate}
%	\begin{remark}
%		\label{R0} 
		The above assumptions need well-behaved covariance functions in the sense of smoothness. For a list of such covariance functions, one may see \ctn{banerjee2014hierarchical}. Smoothness properties along with other important properties of Mat\'{e}rn covariance functions have been covered in \ctn{Stein1999} (Chapter 2).
%	\end{remark}
	\begin{remark}
		\label{R1}
		%The assumptions A1 and A2 imply that the covariance functions of $X(s,0)$ and $Y(s,0)$ are symmetric, positive definite, and Lipschitz continuous, and thus $X(s,0)$ and $Y(s,0)$ will have continuous sample paths with probability 1. If the covariance functions are taken to be the Mat\'{e}rn covariance function with $\nu>1$ or the squared exponential covariance function, then $X(s,0)$ and $Y(s,0)$ will have differentiable sample paths in $s$. 
        Assumption A1 implies that the covariance functions of $X(s,0)$ and $Y(s,0)$ are symmetric, positive definite, and Lipschitz continuous, and thus $X(s,0)$ and $Y(s,0)$ will have continuous sample paths with probability 1. If the covariance functions are taken to be the Mat\'{e}rn covariance function with $\nu>1$ or the squared exponential covariance function, then $X(s,0)$ and $Y(s,0)$ will have differentiable sample paths in $s$. 
	\end{remark}
	\begin{remark}
		\label{R2}
		Assumption A2 implies that the derivative process of $V(\cdot)$ is also a Gaussian process with continuous sample paths almost surely. If the squared exponential covariance function is used, then all the derivatives of $V(\cdot)$ remain Gaussian processes. In particular, the differential of $V(\cdot)$ will have differentiable sample paths. Similarly, for the Mat\'{e}rn covariance function with $\nu > 2$, $V(\cdot)$ is differentiable, and its covariance function corresponds to the mixed partial derivative of the original covariance function. As $\nu>2$, the differential of $V(\cdot)$ also has differentiable sample paths (see \cite{rusmassen2005gaussian} and \cite{Stein1999}). 
        %In fact, if the squared exponential covariance function is assumed, then all the derivatives of $V(\cdot)$ will be Gaussian processes. In particular, the differential of $V(\cdot)$ will have differentiable sample paths. Also note that for Mat\'{e}rn covariance function with $\nu > 2$, the random function $V(\cdot)$ is differentiable and the corresponding covariance function is the mixed partial derivative of the covariance function of $V(\cdot)$. Moreover, the differential of $V(\cdot)$ will have differentiable sample paths as $\nu$ is assumed to be more than 2 for Mat\'{e}rn covariance function (refer to \cite{rusmassen2005gaussian} along with \ctn{Stein1999}). 
	\end{remark}	
	Now we propose a form of $M_s$ which is continuous and infinitely differentiable when defined on a compact set $S\subset \R^2$. 
	\begin{definition}[Definition of $M_s$]
		Let $S$ be a compact subset of $\R^2$. Define $M_{s} = \exp(\max$ $\{||s^2-u^2||^2: u\in S\})$, where by $\bi{v}^2$ we mean $\bi{v}^2 = (v_{1}^2, v_{2}^2)^T$, for $\bi{v}\in S$.
	\end{definition}
\begin{remark}
\label{Remark on M_s goes to infinity}
        {We observe that with the choice of $M_s$, for a fixed $s'$ (or for fixed $s$), as $||s-s'||\rightarrow \infty$, $M_s\rightarrow \infty$ (or $M_{s'} \rightarrow \infty$). %Further, if $s$ and $s'$ be such that $||s||\rightarrow \infty$, $||s'||\rightarrow \infty$ and $||s-s'||\rightarrow \infty$, $M_s, M_{s'} \rightarrow \infty$. 
        This is a technical requirement for showing the lagged covariance to go to 0 (Theorem \ref{lem: covariance}). Essentially, it means that as the spatial distance increases to infinity, the constant $M_s$ also escapes to infinity. 
        The proof can be argued in the following fashion.}
        
		%Note that $S$ is compact with Lebesgue measure $\mbox{Leb}(S)$. 
		When $||s-s'||\rightarrow \infty$, $S$ also grows in the sense $S_1\subset S_2 \subset \ldots$, such that at each stage $i$, $S_i$ remains compact. 
		Under this limiting situation, $M_s = \exp\left(\max\{||s^2-u^2||^2: u\in S\}\right) \geq \exp\left(||s^2-s'^2||^2\right) \rightarrow \infty $ 
		and $M_{s'} = \exp\left(\max\{||s'^2-u^2||^2:\right.$ $\left. u\in S\}\right) \geq \exp\left(||s'^2-s^2||^2\right)\rightarrow \infty.$
	\end{remark}
Further, it can be shown that $M_s$ is infinitely smooth in $S$. The following lemma proves the claim. We provide the proof in Appendix~\ref{theorem proofs}.	
	\begin{lem}
		\label{lemma3: differentiability of Ms}
		$M_s$ is infinitely differentiable in $s\in S$. 
	\end{lem}
The next results show that the covariance function of $Y(s,t)$ goes to 0 as the distance between two time points and two spatial locations increases to infinity. The distances between any two time points and any two spatial locations are measured with respect to their corresponding distance metrics.  
	\begin{thm}
		\label{lem: covariance} 
		Under the assumptions A1 and A2, cov$\left(Y(s,h\delta t), Y(s',h'\delta t)\right)$ converges to 0, 
		as $||s-s'||\rightarrow \infty$ and/or $|h-h'|\rightarrow \infty$.
	\end{thm}
		In Theorem \ref{lem: covariance}, $\delta t$ in $h\delta t$ and $h'\delta t$ make the time points continuous for continuous $\delta t$. However, 
		discrete values of $\delta t$ are also allowed.
		\begin{remark}
		The closed form of the covariance function of our proposed process $Y(s,t)$ is unavailable. However, we can argue in the following way that the covariance is non-stationary. Conditional on the latent variables, the covariance depends upon the latent variable. For simplicity, let us assume that the latent process is supported on a compact set. Then, by the mean value theorem for integrals, the unconditional covariance is a function of a specific value of the latent variables. Moreover, the latent variables are almost surely nonlinear in locations and times, implying that the unconditional covariance is a nonlinear function of space and time. Therefore, it must be non-stationary.  
        In this context, it is important to note that even if we assume stationary covariance kernel for $V(\cdot)$, the proposed spatio-temporal process is rendered mean and covariance non-stationary. 
	\end{remark}
	{To complete the discussion on correlation analysis, we provide an illustration to give us an idea regarding the spatial correlation matrix of our model and how it is different from the squared exponential or Mat\'{e}rn covariance. {In particular, this illustration shows that the spatial correlation of the proposed model can be negative, whereas the correlation of squared exponential or Mat\'{e}rn will always be positive .} The details of these simulation studies and the correlation plots (Figure~\ref{Fig:spatial correlation matrices}) are provided in Appendix \ref{sample correlation plots}}.	
%	{We simulated spatio-temporal observations 1000 times from our model taking $\eta_1 = \eta_2 =\eta_3 =1$, $\sigma^2 = \sigma^2_{\theta} = \sigma_{p}^2 = 1$, $\alpha= \beta = 0.9$. The number of locations and time points are taken to be 10 and 4, respectively. Based on 1000 simulations we have calculated the sample spatial correlation matrix. Similarly, we simulated spatio-temporal observations from a Gaussian process with zero mean and squared exponential kernel with the variance and the decaying parameter to be 1. The corresponding sample spatial correlation matrix is calculated based on the 1000 repetitions. The same exercise has been done by replacing the squared exponential kernel with two Mat\'{e}rn covariance kernels, namely, Mat\'{e}rn(3/2) and Mat\'{e}rn(5/2), respectively. The range parameter for Mat\'{e}rn covariance kernel is taken as 1 for both cases.}	
%		{The sample spatial correlations are provided in Figure~\ref{Fig:spatial correlation matrices}.}
%---------------------------------------------------------
%--------------------------------------------------			
	
Next we show that $Y(s,t)$ and $X(s,t)$ are continuous in $s$ with probability 1 and in the mean square sense.  
	Since the (modified) Hamiltonian equations already imply that $Y(s,t)$ and $X(s,t)$ are path-wise differentiable with respect to $t$, we focus on their smoothness properties with respect to $s$. The following theorems (except Lemma \ref{lemma5: mean square differentiability of composition of GP}) are established under the assumption that $A1$ and $A2$ hold.
	\begin{thm}
		\label{lemma1: almost sure continuity}
		%If the assumptions A1 $\&$ A2 hold true, then 
		$Y(s,h\delta t)$ and $X(s,h\delta t)$ are continuous in $s$, for all $h\geq 1$, with probability 1. 
	\end{thm}
	\begin{thm}
		\label{lemma2:mean square continuity}
		%Under assumptions A1 $\&$ A2, 
		$Y(s,h\delta t)$ and $X(s,h\delta t)$ are continuous in $s$ in the mean square sense, for all $h\geq 1$.
	\end{thm} 
	The next two results deal with differentiability of the processes $Y(s,t)$ and $X(s,t)$. 
	\begin{thm}
		\label{lemma4: almost sure differentiability}
		%Under assumptions A1-A3,, %$M_{s} = \exp(\max\{||s-u||^2: u\in S\})$, 
		$Y(s,h\delta t)$ and $X(s,h\delta t)$ have differentiable sample paths with respect to $s$, almost surely. 
	\end{thm}
	\begin{remark}
		Theorem \ref{lemma4: almost sure differentiability} is about once differentiability, however, it can be extended to $k$ times differentiability depending on the structure of the covariances assumed on the processes $Y(s,0)$, $X(s,0)$ and the random function $V(\cdot)$. For example, if we assume squared exponential covariance functions on each process, then $Y(s,h\delta t)$ and $X(s,h\delta t)$ will have $k$ times differentiable sample paths in $s$, for any $k \in \mathbb{N}$. 
	\end{remark}
	To prove that the processes $Y(s,t)$ and $X(s,t)$ are mean square differentiable in $s$, we need a lemma, stated below, which may be of independent interest.  
	\begin{lem}
		\label{lemma5: mean square differentiability of composition of GP}
		Let $f:\mathbb{R}\rightarrow \mathbb{R}$ be a zero mean Gaussian random function with covariance function $c_{f}(x_1,x_2)$, $x_1, x_2\in \R$, which is four times continuously differentiable. Let $\left\{Z(s):s\in S\right\}$ be a random process with the following properties 
		\begin{enumerate}
			\item $E(Z(s)) = 0$,
			\item The covariance function $c_{Z}(s_1,s_2)$, $s_1, s_2 \in S$,  where $S$ is a compact subspace of $\mathbb{R}^2$, is four times continuously differentiable, and
			\item $\frac{\partial Z(s)}{\partial s_i}$ has finite fourth moment. 
		\end{enumerate} 
		Then the process $\{g(s):s\in S\}$, where $g(s)=f(Z(s))$, is mean square differentiable in $s$.     
	\end{lem}
	\begin{thm}
		\label{Thm: Mean-square diff}
		Let A1 and A2 hold true, with the covariance functions of all the assumed Gaussian processes being squared exponential. Then $Y(s,h\delta t)$ and $X(s,h\delta t)$ are mean square differentiable in $s$, for every $h\geq 1$. 
	\end{thm}
    Therefore, we prove that for our choice of the random function $V(\cdot)$, there exist valid stochastic processes $Y(s,t)$ and $X(s,t)$ with continuity and smoothness properties. It is not difficult to check that the finite dimensional distributions satisfy Kolmogorov's consistency theorem and thus, by Kolmogorov's theorem, there exists a probability measure for our stochastic process. Almost surely all paths of our stochastic processes satisfy the modified Hamiltonian equations.

\subsection{Connection to traditional Hamiltonian systems with Gaussian Process potentials}
\label{subsec:connection_hamiltonian_gp}

Although (\ref{observation equation}) and (\ref{latent equation}) have been proposed as a numerical solution to the equations given by
(\ref{eq: proposed stochastic differential equations}), the modified stochastic Hamiltonian system driven by  Gaussian process potentials, 
the mathematical analysis presented in \citep{Bhatta26a} revealed a deeper and more fundamental connection: the stochastic parameterized leapfrog scheme 
actually solves, in an appropriate mean-square sense, the \emph{traditional} Hamiltonian equations with Gaussian process potentials. This mathematical result has 
significant implications for both the theoretical interpretation and practical application of our method.

From a theoretical perspective, this connection establishes that our parameterized framework does not deviate from the classical Hamiltonian structure that 
underlies phase-space dynamics in physical systems. Rather, it represents an enriched version where the deterministic potential energy function  
is replaced by a Gaussian process potential, introducing stochasticity while preserving the fundamental symplectic geometry. The parameterization through 
$\alpha$ and $\beta$ then emerges naturally as a flexible discretization scheme that accommodates this stochastic extension while maintaining numerical 
stability and convergence properties.

Practically, the Gaussian process potential introduces a remarkable flexibility that aligns with the nonparametric nature of many modern statistical models. 
Instead of assuming a specific parametric form for the potential energy, the Gaussian process formulation allows adaptation to complex, high-dimensional energy 
landscapes, making the approach particularly suitable for problems where the target spatio-temporal process exhibits intricate geometry or unknown structure. 
This nonparametric enhancement, combined with the parameterized discretization, creates a powerful synergy: the Gaussian process handles model complexity 
while the parameterized leapfrog provides computational control over the integration process.

The convergence analysis in \citet{Bhatta26a} establishes that this hybrid approach maintains $\mathcal{O}(\delta t)$ mean-square convergence, 
validating its numerical robustness. This theoretical guarantee, coupled with the empirical demonstrations in Sections \ref{Sec: Simulation Studies} and 
\ref{real data analysis} position  our method not as an arbitrary modification of the standard Hamiltonian system, but as a principled stochastic 
extension that preserves the Hamiltonian foundation while introducing valuable flexibility for practical, nonparametric spatio-temporal modeling.

\subsection{Connection to Modified Parameterized Leapfrog HMC}
\label{subsec:connection_mplhmc}

The parameterized leapfrog scheme introduced in this spatio-temporal context serves as the mathematical progenitor of the Modified Parameterized Leapfrog 
Hamiltonian Monte Carlo (MPL-HMC) method developed in our companion work \cite{Bhatta26b}. Indeed, the deterministic counterpart of 
equations~(\ref{observation equation})--(\ref{latent equation})---obtained by removing the spatial index $s$ and treating $V$ as a fixed potential---yields exactly 
the MPL-HMC integrator:
\[
\begin{cases}
	q_{n+1} = \beta (\delta t) q_n + \delta t M^{-1}\bigl(\alpha p_n - \frac{\delta t}{2} \nabla V(q_n)\bigr), \\
	p_{n+1} = \alpha (\delta t)^2 p_n - \frac{\delta t}{2}\bigl(\alpha \nabla V(q_n) + \nabla V(q_{n+1})\bigr),
\end{cases}
\]
where $\alpha (\delta t)$ and $\beta (\delta t)$ are appropriate parameterized in terms of $\delta t$. The stochastic version studied here thus provides 
a rigorous foundation for the deterministic integrator: the mean-square convergence analysis of \cite{Bhatta26a} ensures that the discrete MPL-HMC map approximates 
the continuous Hamiltonian flow with $\mathcal{O}(\delta t)$ global error. Importantly, the parameters $\alpha$ and $\beta$---originally introduced to induce 
spatial non-stationarity and damping/anti-damping in the spatio-temporal model---translate into tunable knobs for controlling numerical stability, exploration, and 
damping behavior in MPL-HMC. This transfer of ideas exemplifies how innovations in stochastic ordinary differential equations and numerical analysis for spatio-temporal 
processes can directly inspire advances in Monte Carlo methodology, enriching both fields with cross‑fertilized techniques and theoretical insights.

 %%%%%%%%%%%%%%%%%%%%%%%%%%%%%%%%%%%%%%%%%%%%%%%%%%%%
\section{Calculation of likelihood functions}
\label{complete likelihood}
%%%%%%%%%%%%%%%%%%%%%%%%%%%%%%%%%%%%%%%%%%%%%%%%%%%%%%%%%%%%%%
In this section, we will derive the data and process models under assumptions A1-A2. We assume here that the random function $V(\cdot)$ is a Gaussian process with mean 0 and squared exponential function as covariance function for simplicity of calculations (Mat\'{e}rn covariance kernel could be another choice but that will lead to very high computational complexity, so we avoid such a choice). %Other covariance functions (with required properties, see assumption A3) will work in the same manner. 
In particular, we assume that the covariance function of $V(\cdot)$ takes the form $\mbox{cov}(V(x),V(y)) = k(h) = \sigma^2 e^{-\eta_3 h^2}$, where $h= ||x-y||$. Then $V'(\cdot)$ will be a Gaussian random function with mean 0 and covariance function (\ctn{Stein1999}, Chapter 2)
\begin{align}
	\label{eq1}
	\mbox{cov}(V'(x),V'(y)) = 2\eta_3 \sigma^2 e^{-\eta_3 h^2}(1-2\eta_3 h^2).
\end{align} 

Before moving forward, we introduce a few notations that will be used for calculations of different distributions. Suppose $s_1, \ldots, s_n$ denote the $n$ locations, where $s_i\in \mathbb{R}^2$, for $i=1, \ldots n$. 
%Let the observations on $Y(s,t)$ be available for positions $s_1, s_2, \ldots, s_n$ and time points $1, 2, \ldots, T$. That is, 
Let the observed data and the corresponding latent variables be denoted by 
$\mathbb{D} = \left\{\bi{y}_1^T, \bi{y}_2^T, \ldots, \bi{y}_T^T \right\},$ and 
 %and the corresponding latent variables be
$\mathbb{L} = \left\{\bi{x}_1^T, \bi{x}_2^T, \ldots, \bi{x}_T^T\right\},$ respectively, where, for $t=1, \ldots, T$, $\bi{y}_t = (y(s_1,t),y(s_2,t), \ldots, y(s_n,t))^T$ and $\bi{x}_t = (x(s_1,t),x(s_2,t), \ldots, x(s_n,t))^T$. By ${\bi{a}}^T$ and $f'$, we mean transpose of a vector $\bi{a}$ and derivative of a function $f$, respectively. %we mean transpose of a vector $\bi{a}$ and by $f'$ we mean derivative of $f$. 
Assuming $i, k \in \{1,\ldots,n\}$, $m\in\{0,\ldots,T\}$ and $r\in \{1, \ldots, T\}$, we define 
\allowdisplaybreaks
\begin{align}
	& h_{ij}(m)  = |y(s_i,m)-y(s_j,m)|, \, \mu_i(m) = \beta y(s_i,m) + \frac{\alpha x(s_i,m)}{M_{s_i}}, \bi{\mu}_m = (\mu_i(m), 1\leq i\leq n)^T\notag \\
	& \bi{W}_{r}  = (V'(y(s_i,r)), 1\leq i\leq n)^T , 
    \mathbb{W}_{r-1}  = \begin{pmatrix}
		\alpha \bi{W}_{r-1} \\
		\bi{W}_{r}
	\end{pmatrix}, \ell_{ik}(r-1,r)  = |y(s_i,r-1) - y(s_k,r)|.
    \notag
\end{align}
%\begin{align}
%	h_{ij}(m) & = |y(s_i,m)-y(s_j,m)|, \, \mu_i(m) = \beta y(s_i,m) + \frac{\alpha x(s_i,m)}{M_{s_i}}, \bi{\mu}_m = (\mu_i(m), 1\leq i\leq n)^T\notag \\
%	\bi{\mu}_m & = (\mu_1(m), \ldots, \mu_n(m))^T, 	\bi{W}_{r}  = (V'(y(s_1,r)), \ldots, V'(y(s_n,r)))^T , \notag \\	
%	 \mathbb{W}_{r-1} & = \begin{pmatrix}
%		\alpha \bi{W}_{r-1} \\
%		\bi{W}_{r}
%	\end{pmatrix}, 
%	\ell_{ik}(r-1,r)  = |y(s_i,r-1) - y(s_k,r)|,	
%	\notag \\
%	\bi{\Theta} & = (\alpha, \beta, \sigma^2,\sigma_{\theta}^2,\sigma_p^2, \eta_1,\eta_2,\eta_3). \notag
%\end{align}
%By ${\bi{a}}^T$, we mean transpose of a vector $\bi{a}$ and by $f'$ we mean derivative of $f$.
%%%%%%%%%%%%%%%%%%%%%%%%%%%%%%%%%%%%%%%%%%%%%
\subsection{Joint conditional density of the observed data}	
\label{data model}
%%%%%%%%%%%%%%%%%%%%%%%%%%%%%%%%%%%%%%%%%%%%%%%
The joint conditional density of the data given the latent variables $\bi{x}_0, \bi{x}_1, \ldots, \bi{x}_{T}$ and the parameter vector $\bi{\Theta}= (\alpha, \beta, \sigma^2,\sigma_{\theta}^2,\sigma_p^2, \eta_1,\eta_2,\eta_3).$ is given by
\begin{align}
	\label{eq4}
	  [\mathbb{D}\big\vert \bi{x}_0;\ldots; \bi{x}_{T-1};\bi{y}_0;\bi{\Theta}] & \propto [\bi{y}_1\big\vert \bi{y}_0;\bi{x}_0; \bi{\Theta}] \ldots \left[\bi{y}_T\big\vert \bi{y}_{T-1}, \ldots, \bi{y}_0; \bi{x}_{T-1}, \ldots, \bi{x}_0; \bi{\Theta}\right] \notag \\
	& \propto \frac{(\sigma^2)^{-nT/2}}{\prod\limits_{t=1}^{T}|\Sigma_{t-1}|^{1/2}}\, \exp\left\lbrace{-\frac{2}{\sigma^2}\sum\limits_{t=1}^{T}\left(\bi{y}_{t} - \bi{\mu}_{t-1}\right)^T\Sigma_{t-1}^{-1}\left(\bi{y}_{t} - \bi{\mu}_{t-1}\right)}\right\rbrace,
\end{align} 
where, for $j=1, 2, \ldots, T$, the
$(k,\ell)$th element of $\Sigma_{j-1}$ is $2\eta_3 e^{-\eta_3 h_{k\ell}^2(j-1)}\left(1-2\eta_3 h_{k\ell}^2(j-1)\right)/(M_{s_k}M_{s_\ell}).$ It is to be noted that due to the use of the Leap-frog algorithm, the randomness of $\mathbb{D}$ given the latent variables is driven by the randomness of $V(\cdot)$. 
The details of the calculation of the joint density are provided in Appendix \ref{joint density of observed data}. It has to be noted that although the conditional joint density of the data is Gaussian but unconditionally the joint density is not Gaussian and not even tractable. 
%%%%%%%%%%%%%%%%%%%%%%%%%%%%%%%%%%%%%%%%%%%%%%%%%%%
\subsection{Joint conditional density of latent data}
\label{process model}
%%%%%%%%%%%%%%%%%%%%%%%%%%%%%%%%%%%%%%%%%%%%%%%%%%%
		It can be shown that (see Appendix \ref{joint density of latent data} for the detailed calculation) 
		\begin{align}
			\label{eq7: Latent model}
			&\bigg[\mathbb{L}\bigg\vert \bi{y}_0;\ldots; \bi{y}_T;\bi{x}_0;\bi{\Theta}\bigg]
			 \propto [\bi{x}_1\vert \bi{x}_0; \bi{y}_0;\bi{y}_1;\bi{\Theta}] \ldots [\bi{x}_T\vert \bi{x}_{T-1};\ldots;  \bi{x}_0; \bi{y}_0;\ldots; \bi{y}_T;\bi{\Theta}] \notag \\ 
			& \propto \frac{(\sigma^2)^{-nT/2}}{\prod\limits_{t=1}^{T} |\Omega_t|^{1/2}} \exp\left\lbrace{-\frac{2}{\sigma^2}\sum\limits_{t=1}^T (\bi{x}_{t} - \alpha^2 \bi{x}_{t-1})^T\Omega_t^{-1}(\bi{x}_{t} - \alpha^2 \bi{x}_{t-1})}\right\rbrace, 
		\end{align}
		where, for $m\in\{1,2,\ldots, T\}$, 
		$\Omega_t = \alpha^2\Sigma_{t-1,t-1} +\alpha \Sigma_{t-1,t} + \alpha \Sigma_{t,t-1} + \Sigma_{t,t},$ where the $(i,k)$th element of $\Sigma_{jj}$, for $j=t-1,t$, is $2\eta_3 e^{-\eta_3 h_{ik}^2(j)}\left(1-2\eta_3 h_{ik}^2(j)\right),$ and the $(i,k)$th element of $\Sigma_{t-1,t} = \Sigma_{t,t-1}^T$ is 
		$2\eta_3 e^{-\eta_3 \ell_{ik}^2(t-1,t)}$ $ 
		\left(1-2\eta_3 \ell_{ik}^2(t-1,t)\right).$
%%%%%%%%%%%%%%%%%%%%%%%%%%%%%%%%%%%%%%%%%%%%%%%%%%
\subsection{Complete likelihood combining observed and latent data}
\label{Jt Dist}
%%%%%%%%%%%%%%%%%%%%%%%%%%%%%%%%%%%%%%%%%%%%%%%%%
		Next we will find the joint distribution of Data and Latent observations given $\bi{x}_0$, $\bi{y}_0$ and $\bi{\Theta}$. Finally, using the prior distributions on $Y(s,0)$ and $X(s,0)$ as mentioned in A1-A2, we shall obtain the complete joint distribution of $(\bi{y}_T, \ldots, \bi{y}_1, \bi{y}_0)$ and $(\bi{x}_T, \ldots, \bi{x}_1, \bi{x}_0)$, given the parameter $\bi{\Theta}$. 
		The joint distribution of $(\bi{y}_T, \ldots, \bi{y}_1)$ and $(\bi{x}_T, \ldots, \bi{x}_1)$ given $(\bi{x}_0, \bi{y}_0, \bi{\Theta})$, using equations 
		(\ref{eq4}) and  (\ref{eq7: Latent model}), is given by 				
\begin{align}
\label{eq8: jt conditional density}
			&\bigg[\mathbb{D},\mathbb{L}\bigg\vert \bi{y}_0, \bi{x}_0, \bi{\Theta}\bigg] = \prod_{t=1}^{T} \bigg[\bi{y}_t\bigg\vert\bi{x}_{t-1},\bi{y}_{t-1},\bi{\Theta}\bigg] \bigg[\bi{x}_t\bigg\vert\bi{x}_{t-1},\bi{y}_{t-1},\bi{y}_{t},\bi{\Theta}\bigg] \propto \frac{(\sigma^2)^{-nT} }{\prod\limits_{t=1}^{T}|\Sigma_{t-1}|^{1/2} |\Omega_t|^{1/2}} \times \notag \\
			&  \exp\left\lbrace{-\frac{2}{\sigma^2}\sum\limits_{t=1}^{T} \left(\left(\bi{y}_{t} - \bi{\mu}_{t-1}\right)^T\Sigma_{t-1}^{-1}\left(\bi{y}_{t} - \bi{\mu}_{t-1}\right)+ (\bi{x}_{t} - \alpha^2 \bi{x}_{t-1})^T\Omega_t^{-1}(\bi{x}_{t} - \alpha^2\bi{x}_{t-1})\right)}\right\rbrace. 
		\end{align}		
%		Now we will find the full joint distribution of $[\bi{y}_T, \ldots, \bi{y}_1,\bi{y}_0;\bi{x}_T, \ldots, \bi{x}_1, \bi{x}_0\vert \bi{\Theta}]$ using the priors on $Y(s,0)$ and $X(s,0)$. %Again for simplicity we will assume that the  $Y(s,0)$ \& $X(s,0)$ are zero mean Gaussian processes with squared exponential functions as their covariance functions. 
Let $\cov(Y(s_1,0),Y(s_2,0)) = \sigma_{\theta}^2\exp\left\{-\eta_2 k^2\right\}$ (see assumption A2), and $\cov(X(s_1,0),X(s_2,$ $0))$ = 
		$\sigma_p^2 \exp\left\{-\eta_1 k^2\right\}$ (see assumption A1), where $k= ||s_1-s_2||$. Therefore, $[\bi{y}_0\vert \bi{\Theta}]\sim N_{n}(\bi{0},\sigma_{\theta}^2 \Delta_0)$ and $[\bi{x}_0\vert \bi{\Theta}]\sim N_{n}(\bi{0},\sigma_p^2 \Omega_0)$, where the $(i,j)$th element of $\Delta_0$ and $\Omega_0$ are $\exp\left\{-\eta_2 k_{ij}^2\right\}$ and $ \exp\left\{-\eta_1 k_{ij}^2\right\}$, respectively, with $k_{ij} = ||s_i-s_j||.$ Thus,
\allowdisplaybreaks		
		\begin{align}
			\label{eq9: complete joint}
			\bigg[\bi{y}_T,\bi{x}_T,\ldots, \bi{y}_1,\bi{x}_1, \bi{y}_0, \bi{x}_0\bigg \vert \bi{\Theta}\bigg]	&\propto	\bigg[\mathbb{D},\mathbb{L}\bigg\vert \bi{y}_0, \bi{x}_0, \bi{\Theta}\bigg] \frac{(\sigma_{\theta}^2)^{-n/2}}{|\Delta_0|^{1/2}} \frac{(\sigma_{p}^2)^{-n/2}}{|\Omega_0|^{1/2}} \exp\left\{-\frac{1}{2\sigma_{\theta}^2}\bi{y}_0^T\Delta_0^{-1}\bi{y}_0 \right\} \notag \\
			&  \quad  \exp\left\{-\frac{1}{2\sigma_p^2}\bi{x}_0^T\Omega_0^{-1}\bi{x}_0 \right\}, 
		\end{align}			 								
where $\bigg[\mathbb{D},\mathbb{L}\bigg\vert \bi{y}_0, \bi{x}_0, \bi{\Theta}\bigg]$ is obtained from (\ref{eq8: jt conditional density}). 

It is important to note that, thanks to our Gaussian process assumption for the random function $V(\cdot)$, we could obtain the conditional likelihoods as Gaussian; 
however, the unconditional distributions are non-Gaussian.
%%%%%%%%%%%%%%%%%%%%%%%%%%%%%%%%%%%%%%%%%%%%%%%%%%%%
%
\section{Prior distributions}
\label{Priors}
%%%%%%%%%%%%%%%%%%%%%%%%%%%%%%%%%%%%%
%%%%%%%%%%%%%%%%%%%%%%%%%%%%%%%%%%%%%%%%
In this section, we will specify the prior distributions of the components of $\bi{\Theta} = (\alpha, \beta,\sigma^2,$ $\sigma_{\theta}^2, \sigma_p^2, \eta_1,\eta_2,\eta_3)^T$. The parameter spaces of each component of $\bi{\Theta}$ are the following:
$|\alpha|<1$, $|\beta|<1$, and $\sigma^2, \sigma^2_{\theta}, \sigma^2_{p}, \eta_{i}>0$, $i=1, 2, 3$. We make the following transformations on $\alpha$, $\beta$, $\eta_{i},$ for $i=1,2,3$ for better MCMC mixing. 
Define $\alpha_* = \log\left(\frac{1+\alpha}{1-\alpha}\right)$, $\beta_* = \log\left(\frac{1+\beta}{1-\beta}\right)$, and $\eta_i^* = \log(\eta_i),$ for $i=1,2,3$, so that 
$\alpha_*, \beta_*, \eta_i^* \in \mathbb{R}$, for $i=1,2,3$. This implies  $\alpha = 1- \frac{2e^{\alpha_*}}{1+e^{\alpha_*}}$,  $\beta = 1- \frac{2e^{\beta_*}}{1+e^{\beta_*}}$, $\eta_i = e^{\eta_i^*}$, $i=1,2,3$, respectively. 
We assume that the prior distributions are independent. Since the parameter spaces of $\alpha_*$ and $\beta_*$ are $\mathbb{R}$, and they are involved in 
the mean function of our proposed model, the prior for $\alpha_*$ and $\beta_*$ are taken as normal with mean 0 and large variances (of the order 100). 
Particular choices of the prior variances are discussed in Sections \ref{Sec: Simulation Studies} and \ref{real data analysis}. Moreover, the parameter spaces 
of $\eta_i^*$, $i=1,2,3$ are also $\mathbb{R}$ and they are involved in the covariance structure of our proposed model in the sense that they determine 
the amount of correlations between spatial and temporal points. So, we take the priors for $\eta^*_i$, $i=1,2,3$, as normal with means $\mu_{\eta_{i}}$ and variances 1. 
Larger variance of the $\eta^*_i$ made the variances of their posterior distributions unreasonably large due to huge data variability.  
%Nevertheless, the prior variance for the $\eta_i$ turned out to be 4.671, which is not too small. It also turned out that in all the simulation and real data analyses, the choice of variance 1 (for $\eta^*_i$) rendered good mixing properties to our MCMC sampler. 
The exact values of the hyper-prior means 
depend upon the data under consideration, and is discussed in Sections \ref{Sec: Simulation Studies} and \ref{real data analysis}. Finally, the prior distributions 
of variance parameters are taken to be inverse-gamma, as they are the conjugate priors, conditionally. It is expected that the variability of the spatio-temporal 
data is very large, which might possibly render the posterior means and variances of $\sigma^2_{\theta}$, $\sigma^2_{p}$ and $\sigma^2$ very large (especially for $\sigma^2$). 
So we decided to choose the hyper-parameters in such a way that the prior means (with exceptions in a few cases) and the variances are both close to zero. 
The exact choices of the hyper-parameters of priors of $\sigma^2_p$, $\sigma^2_{\theta}$ and $\sigma^2$ are mentioned in 
Sections \ref{Sec: Simulation Studies} and \ref{real data analysis}.

The general forms of the prior distributions of $\alpha_*, \beta_*, \sigma^2, \sigma^2_{\theta}, \sigma_p^2, \eta_{i}^*,$ $i=1,2,3$, are taken as follows:
\allowdisplaybreaks
\begin{align*}
	&[\alpha_*]  \propto N(0, \sigma_{\alpha}),  [\beta_*] \propto N(0,\sigma_{\beta}^2), [\sigma^2]  \propto IG(\alpha_v,\gamma_v/2), [\sigma^2_{\theta}]  \propto IG (\alpha_{\theta},\gamma_{\theta}/2) \\
	&[\sigma^2_{p}] \propto IG (\alpha_{p},\gamma_{p}/2), [\eta_{1}^*]  \propto N (\mu_{\eta_{1}},1), 	[\eta_{2}^*]  \propto N (\mu_{\eta_{2}},1),
	[\eta_{3}^*] \propto N (\mu_{\eta_{3}},1),
\end{align*}
where IG stands for inverse gamma distribution.
%%%%%%%%%%%%%%%%%%%%%%%%%%%%%%%%%%%%%%%%%%%%%%%%%%%%%%%%%5
%	\subfile{Full_conditional.tex}
\section{Full conditional distributions}
\label{full conditional}
%%%%%%%%%%%%%%%%%%%%%%%%%%%%%%%%%%%%%
%%%%%%%%%%%%%%%%%%%%%%%%%%%%%%%%%%%%%%%%
In this section, we provide the full conditional distributions of the parameters, which will be used for generating samples from posterior distributions 
of the parameters using Gibbs or Metropolis-Hastings steps. 
The detailed calculations are provided in Appendix \ref{Appendix B: full conditional densities}.
%%%%%%%%%%%%%%%%%%%%%%%%%%%%%%%%%%%%%%%%%%%%%%%%%%
\\[2mm]
\textbf{\underline{Full conditional distribution of $\beta_*$}}
%%%%%%%%%%%%%%%%%%%%%%%%%%%%%%%%%%%%%%%%%%%%%%%%%%%%

\vspace*{0.25cm}
The full conditional density of $\beta_*$, is given by
$[\beta_*\vert \ldots] \propto \exp\lbrace-\beta_*^{2}/2\sigma_{\beta}^2\rbrace g_1(\beta_*),$
where

\[
g_1(\beta_*) = \exp\left\lbrace{-\frac{2\tilde\beta^2}{\sigma^2} \sum\limits_{t=1}^{T} \bi{y}_{t-1}^T\Sigma_{t-1}^{-1}\bi{y}_{t-1} + \frac{4\tilde\beta}{\sigma^2} \sum\limits_{t=1}^{T}  \bi{y}_{t}^T\Sigma_{t-1}^{-1}\bi{y}_{t-1}}\right\rbrace,
\]
with $\tilde\beta = 1- 2e^{\beta_*}/(1+e^{\beta_*})$. 
%The closed form of the full conditional density for $\beta^*$ is not available and thus we shall update it using a random walk Metropolis step. 
%%%%%%%%%%%%%%%%%%%%%%%%%%%%%%%%%%%%%%%%%%%%%%%%%%%%%
\\[2mm]
\noindent
\textbf{\underline{Full conditional distribution of $\alpha_*$}}
%%%%%%%%%%%%%%%%%%%%%%%%%%%%%%%%%%%%%%%%%%%%%%%%%%%%%%%

\vspace*{0.25cm}
The full conditional density of $\alpha_*$ is given by 
$[\alpha_*\vert \ldots] \propto \exp\lbrace{-{\alpha_*}^2}/{2\sigma_{\alpha}^2}\rbrace g_2(\alpha_*),$
where 
\begin{align*}
g_2(\alpha_*) &= \frac{1}{\prod\limits_{t=1}^T |\Omega_{t}|^{1/2}} \exp\left\lbrace -\frac{2}{\sigma^2}\sum\limits_{t=1}^T \left[ (\bi{x}_t-\tilde\alpha^2\bi{x}_{t-1})^T\Omega_{t}^{-1}(\bi{x}_t-\tilde\alpha^2\bi{x}_{t-1})\right]- \right.\\
& \qquad \left. \frac{2\tilde\alpha}{\sigma^2}\sum\limits_{t=1}^T \left[\tilde\alpha \bi{x}_{t-1}^TD\Sigma_{t-1}^{-1}D\bi{x}_{t-1}-2\bi{y}_t^T\Sigma_{t-1}^{-1}D\bi{x}_{t-1}\right]\right\rbrace,
\end{align*}
%$g_2(\alpha^*) = \frac{1}{\prod\limits_{t=1}^T |\Omega_{t}|^{1/2}} e^{-\frac{2}{\sigma^2}\sum\limits_{t=1}^T \left[ (\bi{x}_t-\alpha^2\bi{x}_{t-1})^T\Omega_{t}^{-1}(\bi{x}_t-\alpha^2\bi{x}_{t-1})\right]} \times e^{-\frac{2\alpha}{\sigma^2}\sum\limits_{t=1}^T \left[\alpha \bi{x}_{t-1}^TD\Sigma_{t-1}^{-1}D\bi{x}_{t-1}-2\bi{y}_t^T\Sigma_{t-1}^{-1}D\bi{x}_{t-1}\right]},$ 
and $\tilde\alpha = 1- 2e^{\alpha_*}/(1+e^{\alpha_*})$. 
The closed forms of the full conditionals of $\alpha_*$ and $\beta_*$ are not available, and hence will be updated using random walk Metropolis.
%%%%%%%%%%%%%%%%%%%%%%%%%%%%%%%%%%%%%%%%%%%%55
\\[2mm]
\noindent
\textbf{\underline{Full conditional distribution of $\sigma_{\theta}^2$, $\sigma^2_p$ and $\sigma^2$}}
%%%%%%%%%%%%%%%%%%%%%%%%%%%%%%%%%%%%%%%%%%%%%%%%%%%%

\vspace*{0.25cm}
The full conditional distribution of $\sigma_{\theta}^2$, $\sigma^2_p$ and $\sigma^2$ are IG$\left(\alpha_{\theta}+n/2, \gamma_{\theta}+\bi{y}_0^T\Delta_0^{-1}\bi{y}_0/2\right)$, 
IG$\left(\alpha_{p}+n/2\right.$, $ \left. \gamma_{p}+\bi{x}_0^T\Delta_0^{-1}\bi{x}_0/2\right)$ and IG($\alpha_v+Tn, \gamma_v/2 + 2\zeta$), respectively, where 
$$\zeta = \sum\limits_{t=1}^T \left[(\bi{y}_t - \bi{\mu}_{t})^T\Sigma_{t-1}^{-1} (\bi{y}_t - \bi{\mu}_{t}) + (\bi{x}_t - \alpha^2\bi{x}_{t-1})^T\Omega_{t}^{-1} (\bi{x}_t - 
\alpha^2\bi{x}_{t-1})\right].$$ %Hence it is updated using a Gibbs sampling step. 
Thus, $\sigma^2_{\theta}, \sigma^2_{p}$ and $\sigma^2$ updated using a Gibbs sampling step. 
%%%%%%%%%%%%%%%%%%%%%%%%%%%%%%%%%%%%%%%%%%
%\\[2mm]
%\noindent
%\textbf{\underline{Full conditional distribution of $\sigma_{p}^2$}}

%%%%%%%%%%%%%%%%%%%%%%%%%%%%%%%%%%%%%%%%%%%
%The full conditional distribution of $\sigma_p^2$ is IG$\left(\alpha_{p}+n/2, \gamma_{p}+\bi{x}_0^T\Delta_0^{-1}\bi{x}_0/2\right)$. %Therefore, $\sigma^2_p$ is updated using a Gibbs sampling step. 
%%%%%%%%%%%%%%%%%%%%%%%%%%%%%%%%%%%%%
%\\[2mm]
%\noindent
%\textbf{\underline{Full conditional distribution of $\sigma^2$}}

%%%%%%%%%%%%%%%%%%%%%%%%%%%%%%%%%%%%%%%%%%%%%%
%The full conditional distribution of $\sigma^2$ is inverse-Gamma with parameters $\alpha_v+Tn$ and $\gamma_v/2 + 2\zeta$, where 
%$\zeta = \sum\limits_{t=1}^T \left[(\bi{y}_t - \bi{\mu}_{t})^T\Sigma_{t-1}^{-1} (\bi{y}_t - \bi{\mu}_{t}) + (\bi{x}_t - \alpha^2\bi{x}_{t-1})^T\Omega_{t}^{-1} (\bi{x}_t - \alpha^2\bi{x}_{t-1})\right].$ 

%Thus, $\sigma^2_{\theta}, \sigma^2_{p}$ and $\sigma^2$ updated using a Gibbs sampling step. 
%
%%%%%%%%%%%%%%%%%%%%%%%%%%%%%%%%%%%%%%%%%%%%5
%\\[2mm]

\vspace{0.15cm}
\noindent
\textbf{\underline{Full conditional distributions of $\eta_i^*$, $i=1,2,3$}}
%%%%%%%%%%%%%%%%%%%%%%%%%%%%%%%%%%%%%%%%%%%%%%%

\vspace*{0.25cm}
The full conditional distributions of $\eta_1^*$, $\eta_2^*$ and $\eta_3^*$ are 
$[\eta_1^*\vert \ldots] \propto \pi(\eta_1^*) g_{3}(\eta_1^*)$, 
$[\eta_2^*\vert \ldots] \propto \pi(\eta_2^*) g_{4}(\eta_2^*)$, and 
$[\eta_3^*\vert \ldots] \propto \pi(\eta_3^*) g_{5}(\eta_3^*)$,
where $\eta_j = \exp\{\eta_j^*\}$, $\pi(\eta_j^*) = \exp\{\eta_j^{*2}/2\}$, for $j=1, 2, 3$. Here $g_{3}(\cdot)$, $g_4(\cdot)$ and $g_{5}(\cdot)$ are defined as the following:
\begin{align*}
g_3(\eta_1^*) &= \frac{1}{|\Omega_0|^{1/2}} \exp\left\{-\frac{1}{2\sigma^2_p}\bi{x}_0^T \Omega_{0}^{-1}\bi{x}_0\right\}, g_4(\eta_2^*) = \frac{1}{|\Delta_0|^{1/2}} \exp\left\{-\frac{1}{2\sigma^2_{\theta}}\bi{y}_0^T \Delta_{0}^{-1}\bi{y}_0\right\}, \text{ and } \\
g_5(\eta_3^*) &= \frac{1}{\prod\limits_{t=1}^{T}|\Sigma_{t-1}|^{1/2}|\Omega_t|^{1/2}} \exp\left\lbrace -\frac{2}{\sigma^2}\sum\limits_{t=1}^T \left[(\bi{y}_t - \bi{\mu}_{t})^T\Sigma_{t-1}^{-1} (\bi{y}_t - \bi{\mu}_{t}) + (\bi{x}_t - \alpha^2\bi{x}_{t-1})^T\Omega_{t}^{-1} (\bi{x}_t - \alpha^2\bi{x}_{t-1})\right]\right\rbrace.
\end{align*}
%\noindent $\quad \displaystyle g_3(\eta_1^*) = \frac{1}{|\Omega_0|^{1/2}} \exp\left\{-\frac{1}{2\sigma^2_p}\bi{x}_0^T \Omega_{0}^{-1}\bi{x}_0\right\}$, $\displaystyle g_4(\eta_2^*) = \frac{1}{|\Delta_0|^{1/2}} \exp\left\{-\frac{1}{2\sigma^2_{\theta}}\bi{y}_0^T \Delta_{0}^{-1}\bi{y}_0\right\},$ and 
%$$g_5(\eta_3^*) = \frac{1}{\prod\limits_{t=1}^{T}|\Sigma_{t-1}|^{1/2}|\Omega_t|^{1/2}} e^{-\frac{2}{\sigma^2}\sum\limits_{t=1}^T \left[(\bi{y}_t - \bi{\mu}_{t})^T\Sigma_{t-1}^{-1} (\bi{y}_t - \bi{\mu}_{t}) + (\bi{x}_t - \alpha^2\bi{x}_{t-1})^T\Omega_{t}^{-1} (\bi{x}_t - \alpha^2\bi{x}_{t-1})\right]}.$$
None of the full conditional distributions of $\eta_1^*$, $\eta_2^*$ or $\eta_3^*$ has closed form and hence they will be updated using random walk Metropolis. 
The detailed calculations of the full conditionals for $\eta_i^*$, for $i=1,2,3$, are provided in Appendix \ref{Appendix B: full conditional densities}.
%%%%%%%%%%%%%%%%%%%%%%%%%%%%%%%%%%%%%%%%%%%5
\\[2mm]
\noindent
\textbf{\underline{Full conditional distributions of $\bi{x}_t$, $t = 1,2,\ldots, T$ and $\bi{x}_0$}}
%%%%%%%%%%%%%%%%%%%%%%%%%%%%%%%%%%%%%%%%%%%%%%

\vspace*{0.25cm}
From equation (\ref{eq7: Latent model}), we immediately see that $[\bi{x}_t\vert \ldots] \sim N_n\left(\alpha^2\bi{x}_{t-1}, \frac{\sigma^2}{4}\Omega_{t}\right)$, for $t=1, \ldots, T$. 
%
%%%%%%%%%%%%%%%%%%%%%%%%%%%%%%%%%%%%%%%%%%%5
%\\[2mm]
%\noindent
%\textbf{\underline{Full conditional distribution of $\bi{x}_0$}}
%
%%%%%%%%%%%%%%%%%%%%%%%%%%%%%%%%%%%%%%%%%%%%%%
With $A = \Omega_0^{-1}+ 4\sigma_p^2\alpha^4 \sigma^{-2} \Omega_1^{-1}+ 4\sigma_p^2 \alpha^2\sigma^{-2} D\Sigma_{0}^{-1}D$, $B = 4\sigma_p^2\alpha^2 \sigma^{-2}\Omega_1^{-1}$ and $C = 4\sigma_p^2 \alpha\sigma^{-2}D\Sigma_0^{-1}$, the full conditional density of $\bi{x}_0$ is found to be a $n-$variate normal with mean  $A^{-1}(B\bi{x}_1+C(\bi{y}_1 - \beta \bi{y}_0))$ and the variance-covariance matrix $\sigma_p^2A^{-1}$. Hence, $\bi{x}_t$, for $t=0, 1, \ldots, T$ are updated using a Gibbs sampling step. 
%%%%%%%%%%%%%%%%%%%%%%%%%%%%%%%%%%%%%%%%%%%%%%%
\begin{comment}
\section{Predictive Density}
\label{predictive densities}
%%%%%%%%%%%%%%%%%%%%%%%%%%%%%%%%%%%%%%%%%%%%%%%
This section outlines how the predictive density is estimated at a new location $s^*$ and a time point $t^*$ for the response as well as latent variable. That is, we aim to find $[y(s^*,t^*),x(s^*,t^*)|\mathbb{D}]$. %We first find $[y(s^*,t^*),x(s^*,t^*)|\mathbb{D}]$ and then using the technique of Gibss sampling we obtain the samples from $[y(s^*,t^*)\vert \mathbb{D}]$ and $[x(s^*,t^*)\vert \mathbb{D}]$. 
 Thanks to Gibbs sampling, we simulate samples from $[x(s^*,t^*)]\vert \mathbb{D}, y(s^*,t^*)]$ and then from $[y(s^*,t^*)]\vert \mathbb{D}, x(s^*,t^*)]$ sequentially to get observations from $[y(s^*,t^*), x(s^*,t^*)]\vert \mathbb{D}]$. We write $[x(s^*,t^*)]\vert \mathbb{D}, y(s^*,t^*)]$ and $[y(s^*,t^*)]\vert \mathbb{D}, x(s^*,t^*)]$, respectively, as 
 $$[x(s^*,t^*)\vert \mathbb{D},y(s^*,t^*)] = \int \left[x(s^*,t^*)\vert \Theta, \bi{x}_0, \mathbb{L},\mathbb{D},y(s^*,t^*)\right]\, [\Theta, \bi{x}_0, \mathbb{L}\vert \mathbb{D}, y(s^*,t^*)] d\Theta\, d\bi{x}_0\, d\mathbb{L}$$ and
$$[y(s^*,t^*)\vert \mathbb{D}, x(s^*,t^*)] = \int [y(s^*,t^*)\vert \Theta, \bi{x}_0, \mathbb{L},x(s^*,t^*)] [\Theta, \bi{x}_0, \mathbb{L}\vert \mathbb{D}, x(s^*,t^*)] d\Theta\, d\bi{x}_0\, d\mathbb{L}.$$
\\ 
\newline 
\underline{\textbf{Case 1: $t^*$ is a future time point:}}

\noindent
The following
\end{comment}
\begin{remark}
    We add ridge parameters to the diagonals of all the dispersion matrices while performing the MCMC runs to make the matrices non-singular. Note that the 
	ridge parameter for $\Sigma_{t-1}$, for $t=1, \ldots T$ works as nugget in the model. The ridge parameter is fixed to a small value, which is chosen based on cross-validation with several pilot runs of MCMC.
\end{remark}
\begin{remark}
    Within the Bayesian framework, missing observations can be addressed by initializing the MCMC chain with pseudo-values at the missing locations and time points, and then updating them using predictive densities within MCMC runs. After burn-in, we can obtain the posterior predictive densities at those locations and time points. This approach was applied in one of our real data analysis (see \ref{Alaska}). If needed, any measure of central tendency can be used for imputation. 
\end{remark}
%=======================================
\section{Simulation studies}
\label{Sec: Simulation Studies}
%=======================================
%In order to assess the performance of our model and to compare our findings with those of a well-known non-stationary model based on Gaussian process (described below), we conduct two  extensive simulation studies.The first data set is generated from a mixture of three Gaussian process models, and the other from a mixture of two general quadratic nonlinear (GQN) models. The data generation algorithms are described in the relevant subsections (Sections \ref{mixture of 3-comp GPs} and \ref{Mixture_GQN}). First, we describe the model, whose findings are compared with the results of our model applied to the simulated data sets. The model is as follows:

To evaluate the performance of our model and compare our results with those of a well-known non-stationary Gaussian process model (described below), we conduct two extensive simulation studies. The first data set is generated from a mixture of three Gaussian process models, while the second is generated from a mixture of two general quadratic nonlinear (GQN) models. The data generation algorithms are outlined in Sections \ref{mixture of 3-comp GPs} and \ref{Mixture_GQN}. We begin by describing the model, whose results are compared with the findings from our model applied to the simulated data sets. The model is as follows:

Letting $s$ and $s'$ denote the locations, and $t$, $t'$ denote the time points, we have
%\allowdisplaybreaks
\begin{align*}
	y_{s, t}\vert \omega_{s,t} &\sim N(\eta_{s,t}, \sigma^2_{\epsilon}), \, \, \eta_{s,t} = b_0 + \omega_{s,t}, \\ 
	%\eta_{s,t} &= b_0 + \omega_{s,t}, \\
	\omega_{s,t} &= a \omega_{s,(t-1)} + \xi_{s,t} , \text{ with } |a|<1, \text{ and } \xi_{s,t} \sim NGP(0,C(\cdot,\cdot)),  
    %\\ \xi_{s,t} &\sim NGP(0,C(\cdot,\cdot)),  
\end{align*} 
where $NGP$ stands for non-stationary Gaussian process and 
$$ C(\xi_{s,t},\xi_{s',t'}) = \begin{cases}
	0 & \text{ if } t\neq t' \\
	\mbox{cov}(\xi_{s}, \xi_{s'}), &\text{ if } t= t',
\end{cases}$$
with $\mbox{cov}(\xi_{s}, \xi_{s'})$ being a Mat\'{e}rn covariance kernel with smoothness parameter $2$. 
%The other parameters of the Mat\'{e}rn covariance kernel depend on location $s$, which makes the model non-stationary (\cite{lindgren2011explicit,bolin2011spatial,ingebrigtsen2014spatial} and \cite{blangiardo2015spatial}). For future reference in the manuscript, we shall denote the above mentioned model as the NGP model. The NGP model is fitted using the SPDE approach as described in Chapter 6 of \cite{blangiardo2015spatial} (also see \cite{lindgren2022spde} for a review on the SPDE approach).  
The other parameters of the Matérn covariance kernel vary with location $s$, making the model non-stationary (\cite{lindgren2011explicit, bolin2011spatial, ingebrigtsen2014spatial}, and \cite{blangiardo2015spatial}). For future reference, we will denote this model as the NGP model. The NGP model is fitted using the SPDE approach, as outlined in Chapter 6 of \cite{blangiardo2015spatial} (also see \cite{lindgren2022spde} for a review on the SPDE approach).

%In all the simulation experiments, we chose the number of locations to be $n = 50$ and the number of time points to be $T= 20$.  The data vector corresponding to the last time point, which contains observations at 50 locations, is kept aside for the purpose of evaluating model performance (the details of predictive density at a future time point is given in Appendix \ref{temporal prediction}). Hence, the data set that we observe is $\mathbb{D} = \{y(s_1,1), \ldots, y(s_{50},1), \ldots, y(s_1,19),$ $\ldots, y(s_{50},19)\}$ and the corresponding latent data is $\mathbb{L}=\{x(s_1,1), \ldots, x(s_{50},1), \ldots, x(s_1,19),$ $\ldots, x(s_{50},19)\}$. The validation set contains $\bi{y}_{20} = (y(s_1,20), \ldots, y(s_{50},20))^T$ and $\bi{x}_{20} = (x(s_1,20), \ldots, x(s_{50},20))^T$. 

In all simulation experiments, we set the number of locations to $n=50$ and the number of time points to $T=20$. The data vector corresponding to the last time point, which contains observations at the 50 locations, is kept aside for model performance evaluation (algorithm to generate from predictive density at a future time point is provided in Appendix \ref{temporal prediction}). Therefore, the observed data set is 
$\mathbb{D} = \{y(s_1,1), \ldots, y(s_{50},1), \ldots, y(s_1,19),$ $\ldots, y(s_{50},19)\}$ and the corresponding latent data is $\mathbb{L}=\{x(s_1,1), \ldots, x(s_{50},1), \ldots, x(s_1,19),$ $\ldots, x(s_{50},19)\}$. The validation set contains 
$\bi{y}_{20} = (y(s_1,20), \ldots,$ $ y(s_{50},20))^T$ and $\bi{x}_{20} = (x(s_1,20), \ldots, x(s_{50},20))^T$.

We run our MCMC sampler for 1,75,000 iterations with a burn-in of 1,50,000 for simulation examples. The MCMC computations were carried out in MATLAB R2018a. For each model-fitting it took about 3 hours 49 minutes %and 15.24 seconds 
on a desktop computer having 8GB RAM, 1 TB Hard Drive and 3.8 GHz core\_ i5.
%==============================
\subsection{Mixture of three Gaussian Processes}
\label{mixture of 3-comp GPs}
%========================
\subsubsection{\small Data generation}
\label{data generation for 3-comp GPs}
%==========================
A data set is simulated from a mixture of three Gaussian process models. Given the number of locations $n$ and the number of time points $T$, we obeyed the following steps to simulate the data set. 
\begin{enumerate}
	\item[a.] Generate $n$ locations, $s_1, \ldots, s_n$, uniformly from $[0,1]\times[0,1]$. 
	\item[b.] For $j=1, 2,3$, simulate an $n\times T$ matrix $\mathbb{\xi}_j$, where each column is independently simulated from $n$-variate normal with mean $\bi{0}$ and covariance $C_j$.
	The $n\times n$ matrix $C_j$ is determined by Mat\'{e}rn covariance kernel ($2$) with parameters $\kappa_j$ and $\sigma^2_{j}$. %Hence for each $j$, we obtain an $m\times n$ matrix, with 
	\item[c.] %Calculate $\omega_{j}(s,t)$, for $t=1, \ldots, T$, and $s=s_1, \ldots, s_n$,   ${\omega}_j(s,t) = a_j {\omega}_{j}(s,t-1) + {\xi}_j(s,t)$, for $j=1,2,3$. Here we set $\omega(s,0) = 0$, for $s=s_1, \ldots,s_n$. 
    Calculate $\omega_{j}(s,t) = a_j {\omega}_{j}(s,t-1) + {\xi}_j(s,t)$, for $t=1, \ldots, T$, $s=s_1, \ldots, s_n$, and $j=1,2,3$. Here we set $\omega(s,0) = 0$, for $s=s_1, \ldots,s_n$. 
	\item[d.] Set $\bi{p}=(1/3,1/3,1/3)$, which defines a partition of $[0,1]$ as $\mathbb{P}_1 = [0,1/3)$, $\mathbb{P}_2 = [1/3,2/3)$ and $\mathbb{P}_3 = [2/3,1]$. 
	\item[e.] For $s=s_1,\ldots,s_n$, simulate $u\sim $ $(0,1)$.  If $u\in \mathbb{P}_j$, generate $T$-variate observation vector $\bi{y}(s)= (y(s,1), \ldots, y(s,T))^T$ from a $T$-variate normal distribution with mean $b_{0_{j}} \bi{1}_T + \bi{\omega}_{j}(s)$ and covariance $\sigma^2_{\epsilon_{j}}\mathbb{I}_T$, where %$\bi{\omega}_j(s) = (\omega_j(s,1), \ldots, \omega_{j}(s,T))^T$. 
    $\bi{\omega}_j(s)^T = (\omega_j(s,r), 1\leq r\leq T)$.
    Finally, set $\omega_j(s,t) = x(s,t)$, for $t=1, \ldots,T$. 
\end{enumerate}
The following values of parameters are provided for the purpose of simulation: 
$b_{0_{1}} = 0$, $b_{0_{2}} = 10$, $b_{0_{3}} = 20$, $\sigma^2_{\epsilon_{1}} = 1.0$, $\sigma^2_{\epsilon_{2}} = 0.01$, $\sigma^2_{\epsilon_{2}} = 2.0$, $a_1=-0.75$, $a_2 = 0.75$, $a_3 = 0.25$
$\kappa_1 = 1$, $\kappa_2 = 1.5$, $\kappa_3 = 2$, $\sigma^2_1 = 1$, $\sigma^2_2 = 2$ and $\sigma^2_3 = 0.2$.
%=============================
\subsubsection{Results from our model}
\label{Results from our model: 3 comp GP mix}
%=================================
With the observed data set $\mathbb{D}$, we run MCMC iterations for the proposed model. Based on MCMC iterations, observations from the predictive densities of $\bi{y}_{20}$ and $\bi{x}_{20}$ given the data set $\mathbb{D}$ are simulated using the algorithm given in Appendix \ref{temporal prediction}. 
%Monte Carlo averages. 
For running MCMC iterations, one needs to provide the prior parameters. 
Although we have chosen a finite prior variance for $\eta_3^*$, the posterior variance of $\eta_3$ turned out to be very large, which is probably due to huge variability present 
in the spatio-temporal data. This large posterior variance of $\eta_3$ makes the complete system unstable, so, we fix the value of $\eta_3$ at 
its maximum likelihood estimate (MLE) $2.6889$ computed by simulated annealing.
Note that it is not very uncommon to fix the decay parameter in Bayesian inference of spatial data analysis, 
see for example \cite{zhang2004inconsistent} and \cite{banerjee2020modeling}. 
We give a complete prior specification for the other parameters in the following.
\newline 
%=====================================
\\[2mm]
\noindent
\underline{\textbf{Choice of prior parameters:}}
\newline
%\label{Choice of priors for 3 comp GPs}
%============================
The particular choices of the hyper-prior parameters have been chosen using the leave-one-out cross-validation technique. 
Specifically, for $t=1,\ldots,20$, we leave $\bi{y}_t$ in turn and compute its posterior predictive distribution using relatively short MCMC runs. We then selected
those hyperparameters that yielded the minimum average length of the 95\% prediction intervals. 
The final specifications are as follows:
\begin{align*}
	& \alpha_* \sim N(0,\sqrt{500}), \beta_* \sim N(0,\sqrt{300}), [\sigma^2] \propto IG(170000,2/2), [\sigma^2_{\theta}] \propto IG (5500,780/2), \\
	&  
	[\sigma^2_{p}] \propto IG (800,20/2),	[\eta_{1}^*] \propto N (-3,1), [\eta_{2}^*] \propto N (-5,1). 
\end{align*}
Large values of the shape parameters of the variance parameters are used, otherwise the posterior variance explodes and makes the complete model unstable. The plausible reason for such  behaviour is the inherent large data variability. To control such large posterior variance, we restrict the prior of variance parameters very close to 0, with a very narrow variance (see \cite{bhattacharya2021bayesian} also for similar choices).
%=======================================
\newline
\\[2mm]
\underline{\textbf{Trace plots and posterior analysis:}}
\newline
%\label{posterior predictive density 3 comp mixture}
%==========================
The trace plots of the parameters are given in Figure \ref{fig:trace plot 3 comp GP mixtures} of Appendix \ref{Trace plt and posterior plts for 3 comp mixture GP}. The plots do not exhibit lack of convergence of the MCMC runs. 
The posterior probability densities of the latent variables %at 50 locations for 19 time points are  
at randomly selected 6 locations for 19 time points are
shown in Figures~\ref{fig:posterior density of latent variables 3-comp GP mixtures: first 25 locations} %and \ref{fig:posterior density of latent variables 3-comp GP mixtures: last 25 locations} 
of Appendix \ref{Trace plt and posterior plts for 3 comp mixture GP}, where higher probability densities are depicted by progressively intense colors. 
We observe that true latent time series at all the 50 locations always lie in the high probability density regions.

\begin{figure}[!h]
    \begin{subfigure}{0.5\textwidth}
    \centering
     \includegraphics[width=\linewidth]{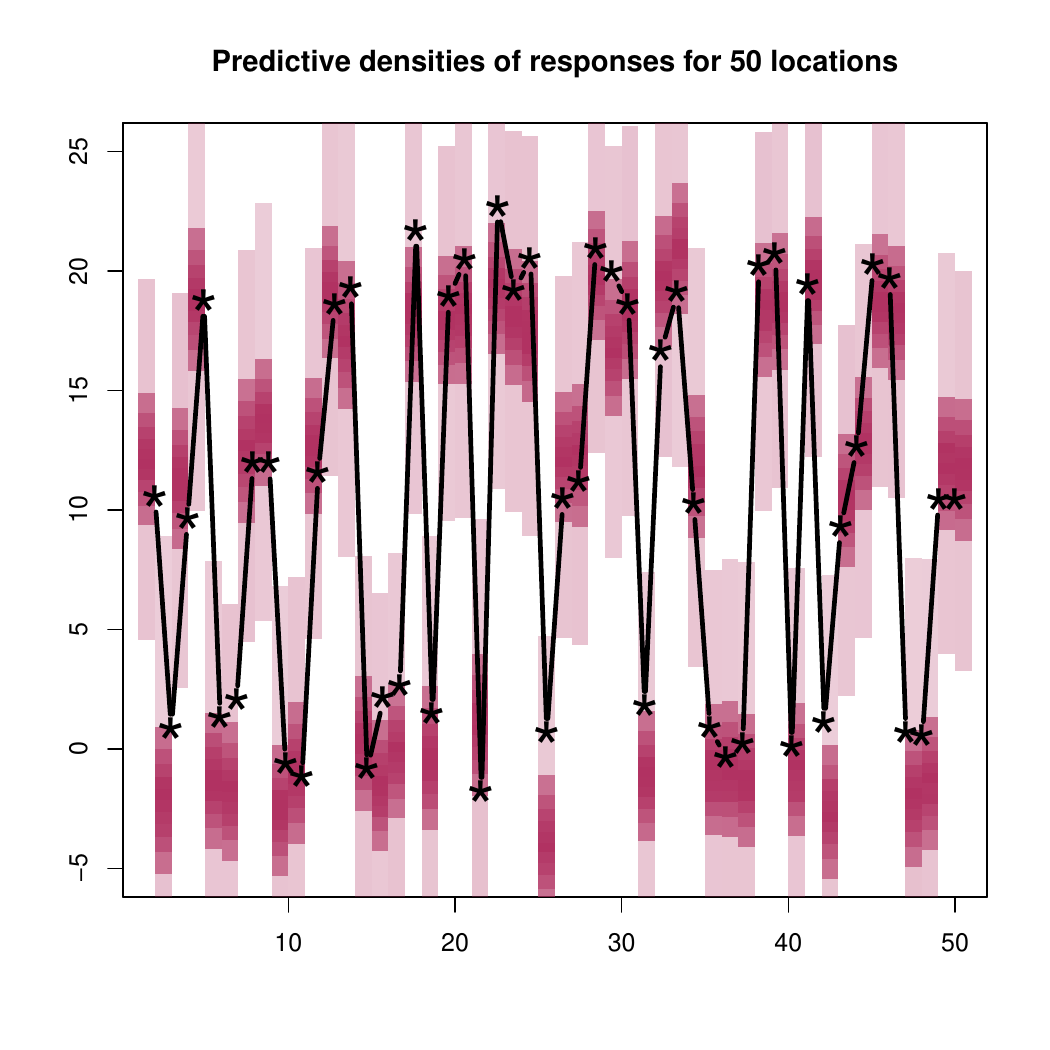}
     \caption{HPD color plot of predictive densities of the elements of $\bi{y}_{20}$.}
     \label{fig:hpd of tin_comp_mixture_GP_response}
    \end{subfigure}
    \begin{subfigure}{0.5\textwidth}
        \includegraphics[width=\linewidth]{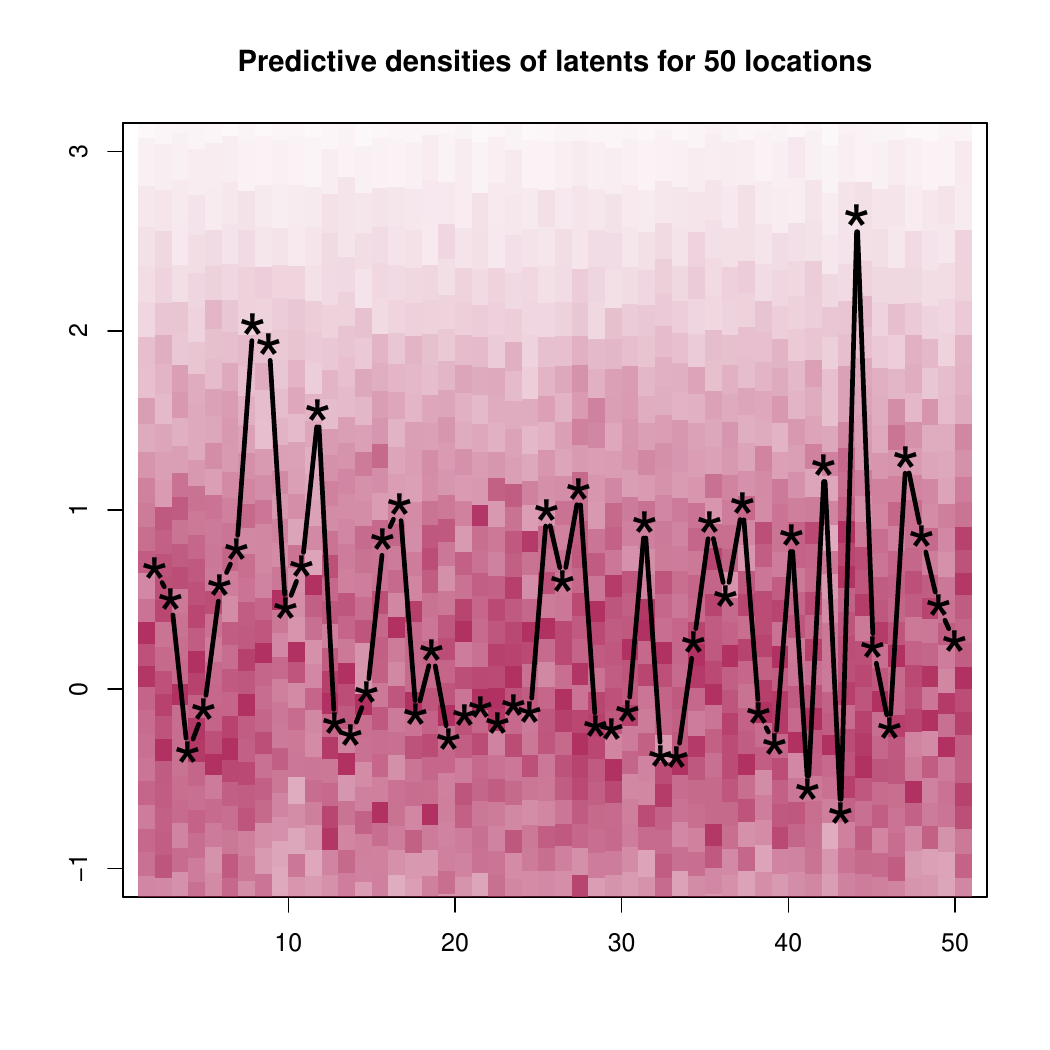}
    \caption{HPD plot of the posterior predictive of the elements of $\bi{x}_{20}$.}
    \label{fig:hpd plot of tin comp mixture GPs latent}
    \end{subfigure}
    \caption{Higher the intensity of the color higher is the density. Black stars denote the true values. Most of the true values lie in the high density regions.}
    \label{HPD color plot for endtime point predictive density of 3 comp GP mixtures}
 \end{figure}

We provide a high probability density (HPD) plot for the predictive densities of response vector $\bi{y}_{20}$ and latent vector $\bi{x}_{20}$ based on our model in Figure~\ref{HPD color plot for endtime point predictive density of 3 comp GP mixtures}.  Figure~\ref{fig:hpd of tin_comp_mixture_GP_response} shows the HPD plot for the predictive density of $\bi{y}_{20}$. It can be seen that the majority of the observations fall in high density regions, which, in turn, follow the pattern of the observed data values. The corresponding HPD plot for the latent vector $\bi{x}_{20}$ is provided in Figure~\ref{fig:hpd plot of tin comp mixture GPs latent}. We have not provided the individual posterior predictive densities for the response vector $\bi{y}_{20}$ and the latent vector $\bi{x}_{20}$ for brevity. %in Figure~\ref{fig: predictive density for 50 locations 3-comp GP mixtures} (first 25 locations in Figure \ref{fig:Predictive densities for first 25 locations 3-comp GP mixtures} and next 25 locations in Figure \ref{fig:Predictive densities for last 25 locations 3-comp GP mixtures}) and Figure~\ref{fig: predictive density of latent variable for 50 locations 3-comp GP mixtures} (Figure \ref{fig:Predictive densities of latent variable for first 25 locations 3-comp GP mixtures} for first 25 locations and Figure  \ref{fig:Predictive densities of latent variable for last 25 locations 3-comp GP mixtures} for next 25 locations), respectively, of Appendix~\ref{predictive densities of response and latent vectors}. 
 However, it is observed that the predictive densities for the responses capture the true values within the 95\% predictive intervals in all but one location (25th location). The lengths of the predictive intervals vary between 7.49 to 8.83 with mean being 8.0219 and standard deviation 0.304879. For the latent vector $\bi{x}_{20}$, all the true values fall within the 95\% predictive intervals of the predictive densities. The lengths of the intervals, in this case, range from 9.44 to 9.69 with mean 9.55808 and standard deviation (sd) 0.0582219.
\subsection{Comparative study with nonstationary GP model}
\label{nonstationary GP model results: 3 comp GP mixtures}
%====================================
%With the data set as generated above, we fit the NGP model using the SPDE technique. %of stochastic partial differential equation (SPDE). 
%It is implemented  using the package INLA in statistical software R. 
Using the generated data set, we fit the NGP model with the SPDE technique, implemented through the INLA package in the statistical software R.
%The posterior predictive densities of $\bi{y}_{20}$ and $\bi{x}_{20}$, given $\mathbb{D}$, are calculated for purpose of comparison with the results of our model. It is seen that the 95\% predictive intervals of each component of $\bi{y}_{20}$ contains the true values of $\bi{y}_{20}$. However, they vary from 1894 to 3364 with mean 2378.483 and standard deviation (sd) = 259.3669 (Figure~\ref{fig: predictive density for 50 locations 3-comp GP mixtures with nonstationary GP model} of Appendix \ref{posterior predictive density plots using INLA for 3 comp GP mixtures}). A similar result is seen for the latent vector $\bi{x}_{20}$ as well (Figure~\ref{fig: predictive density of latent variables for 50 locations 3-comp GP mixtures with nonstationary GP model} of Appendix \ref{posterior predictive density plots using INLA for 3 comp GP mixtures}). All the true latent values are contained within the respective 95\% predictive intervals, and the range of the length of the 95\% intervals is from  1764 to 3053 with mean 2371.798 and standard deviation (sd) = 217.7212. Clearly, the performance of our model, in terms of the length of predictive intervals for both response $\bi{y}_{20}$ and latent $\bi{x}_{20}$ outperformed the NGP model (Table \ref{table of comparison for 3comp GP mixtures}).
The posterior predictive densities of $\bi{y}_{20}$ and $\bi{x}_{20}$, given $\mathbb{D}$, are computed for comparison with the performance of our model. The 95\% predictive intervals for each component of $\bi{y}_{20}$ contain the true values, with lengths ranging from 1894 to 3364, a mean of 2378.483, and a standard deviation of 259.3669. %(Figure~\ref{fig: predictive density for 50 locations 3-comp GP mixtures with nonstationary GP model} of Appendix \ref{posterior predictive density plots using INLA for 3 comp GP mixtures}).  
A similar pattern is observed for the latent vector $\bi{x}_{20}$, %(Figure~\ref{fig: predictive density of latent variables for 50 locations 3-comp GP mixtures with nonstationary GP model}, Appendix \ref{posterior predictive density plots using INLA for 3 comp GP mixtures}), 
with interval lengths ranging from 1764 to 3053, a mean of 2371.798, and a standard deviation of 217.7212. Our model outperforms the NGP model in terms of shorter predictive intervals for both $\bi{y}_{20}$ and $\bi{x}_{20}$ (see Table \ref{table of comparison for 3comp GP mixtures}).

%    \begin{center}
    \begin{table}[!h]
    \centering
\begin{tabular}{| *{9}{c|} }
    \hline
    & \multicolumn{2}{c|}{Mean}
            & \multicolumn{2}{c|}{Std}
                    & \multicolumn{2}{c|}{$\%$ of locations misses}             \\
    \hline
   &   INLA  &   Prop Model  &   INLA  &   Prop Model  &   INLA  &   Prop Model  \\
    \hline
$\bi{y}_{20}$   &  2378.483   &   8.021  &   259.3669  &   0.30487  &   0  &   2  \\
$\bi{x}_{20}$   &   2371.798  &   9.55808  &   217.7212  &   0.0582219  &   0  &   0   \\
\hline    
\end{tabular}
\caption{The table contains the summary of the lengths of 95\% predictive intervals for $\bi{y}_{20}$ and $\bi{x}_{20}$ obtained using INLA and the proposed model for the 3-component mixture of GPs. The columns named ``Mean'' and ``Std'' contain the mean length and the standard deviation of the 95\% posterior predictive intervals across the locations.}
\label{table of comparison for 3comp GP mixtures}
\end{table}
\subsection{Mixture of two GQNs}
\label{Mixture_GQN}
%===================================
For a second comparison study, we choose to simulate from a mixture of two nonlinear models, called general quadratic nonlinear (GQN) models.  The GQN model has been widely used in the literature of spatial and spatio-temporal modeling, see, for example, \cite{wikle2010general}, \cite{cressie2015statistics}, and \cite{bhattacharya2021bayesian} for a significant overview of the GQN model. We define the GQN model as  
\begin{align}
	Y(s_i,t_k) & = \phi_{1}(t_{k},s_i) + \phi_{2}(t_{k},s_i)\, \tan(X(s_i,t_k)) + \epsilon(s_i,t_k), \label{obs eqn for GQN}\\
	X(s_i,t_k) & =  \sum_{j=1}^{n} a_{ij} X(s_j,t_{k-1}) + \sum_{j=1}^{n}\sum_{l=1}^{n} b_{ijl} X(s_j,t_{k-1}) [X(s_l,t_{k-1})]^2 +\eta(s_i,t_k) \label{latent eqn for GQN},
\end{align}
where $i\in\{1, \ldots,n\}$ and $k\in\{1, \ldots, T\}$. The coefficients $\phi_{1}(t_{k},\cdot)$, $\phi_{2}(t_{k},\cdot)$, random errors $\epsilon(t_k,\cdot)$, $\eta(t_k,\cdot)$ and the initial latent variable $X(t_0,\cdot)$ are assumed to be independent zero-mean Gaussian processes having covariance structure 
$c(\bi{s}_1, \bi{s}_2) = \exp\left(-\|\bi{s}_1-\bi{s}_2\|\right)$, for $\bi{s}_1, \bi{s}_2 \in \mathbb{R}^2$, with $\|\cdot\|$ being the Euclidean norm. 
Moreover, it is assumed that for $i, j, l\in  \{1, \ldots, n\}$, $a_{ij}$ and $b_{ijl}$ have independent univariate 0 mean normal distributions with variance $0.001^2$. 
%=================================
\subsubsection{Data generation}
\label{Mixture GQN data generation}
%===================================
Here, a step-by-step description is provided for generating observations from the mixture of two GQNs. Let $n$ and $T$ denote the number of locations and time points. 
\begin{enumerate}
	\item[a.] Generate $n$ locations $s_1, \ldots, s_n$ from $[0,1]\times [0,1]$.
	\item[b.]  Generate $n\times T$ matrix $X$ using equation (\ref{latent eqn for GQN}) by taking the coefficients as described above.
	\item[c.] $\bi{\phi}_1(t)$, $\bi{\phi}_2(t)$ and $\bi{\epsilon}(t)$ are generated independently for $t=1, \ldots, T$ from zero-mean $m-$variate Gaussian with variance-covariance matrix $\mathbb{S}$, whose $(i,j)$th element is $\exp(-||s_i-s_j||)$. Let the $i$th element of $\bi{\phi}_j(t)$, $j=1,2$, and $\bi{\epsilon}(t)$ be $\phi_j(t,s_i)$ and $\epsilon(t,s_i)$, respectively. 
	\item[d.] Let $u\sim \text{unif}(0,1)$. Given all the parameters and latent variables, generate $Y(s_i,t)$, for $i=1, \ldots, n$ and $t=1, \ldots, T$, as	 
%	\allowdisplaybreaks
	$$ Y(s_i,t) = \begin{cases}
		\phi_{1}(t,s_i) + \phi_2(t,s_i) \tan(X(s_i,t)) + \epsilon(t,s_i) &\text{ if } u<0.6, \\
		5+ \phi_{1}(t,s_i) + \phi_2(t,s_i) \tan(X(s_i,t)) + \epsilon(t,s_i) &\text{ if } u\geq 0.6.
	\end{cases} $$ 
\end{enumerate}
%====================================
\subsubsection{Results from our model}
\label{Mixture GQN our results}
%===================================
For this data set as well, we run MCMC iterations for simulating observations from the posterior densities of the parameters involved in our model. These MCMC iterations and Monte Carlo averages provide simulated observations from the posterior densities of $\bi{y}_{20}$ and $\bi{x}_{20}$, given the data set $\mathbb{D}$ (for the algorithm, see Appendix \ref{temporal prediction}). To complete the model, one needs to specify the prior parameters. The specific choices of the hyperparameters are mentioned below, except for $\eta_3$, which is kept fixed at its MLE 5.5042, obtained by simulated annealing.
%======================================
\newline
\\[2mm]
\underline{\textbf{Choice of prior parameters:}}
%\label{Mixture GQN choice of priors}
\newline
%===================================
Using the same cross-validation technique as before, we obtain the following prior specifications:
%
%\allowdisplaybreaks
\begin{align*}
	&\alpha_* \sim N(0,\sqrt{500}), \beta_* \sim N(0,\sqrt{300}), [\sigma^2] \propto IG(780000,2/2), 	[\sigma^2_{\theta}] \propto IG (58900,770/2), \\
	&[\sigma^2_{p}] \propto IG (385,20/2), [\eta_{1}^*] \propto N (-3,1), 	[\eta_{2}^*] \propto N (-5,1).
\end{align*} 
As earlier, here also the shape parameters of the variance parameters are taken large because of the huge inherent data-variability. Unless we restrict the prior mean and variance very close to 0, the posterior variances of the variance parameters become extremely high, making the complete system unstable.
%======================================
\newline
\\[2mm]
\underline{\textbf{Trace plots and posterior analysis:}}
%\label{Mixtuer GQN MCMC convergence}
%====================================
\newline
The trace plots of the parameters are provided in Figure~\ref{fig:trace plot mixture GQNs}, and the posterior density plot of $\bi{x}_t, t=1, \ldots, 19$ are 
displayed in Figures \ref{fig:posterior density of latent variables mixture of GQNs: first 25 locations} and \ref{fig:posterior density of latent variables mixture of GQNs: first 25 locations} of Appendix \ref{Trace plt and posterior plts for 2 comp mixture GQN}. 
%The trace plots provide strong evidence of MCMC convergence. 
The trace plots do not show any evidence of non-convergence of MCMC runs.  
%Further, the posterior density plots show that the posteriors of the latent variables contain the true latent variable time series at all the 50 locations in the high probability density regions successfully. 

\begin{figure}[!h]
    \begin{subfigure}{0.5\textwidth}
    \centering
     \includegraphics[width=\linewidth]{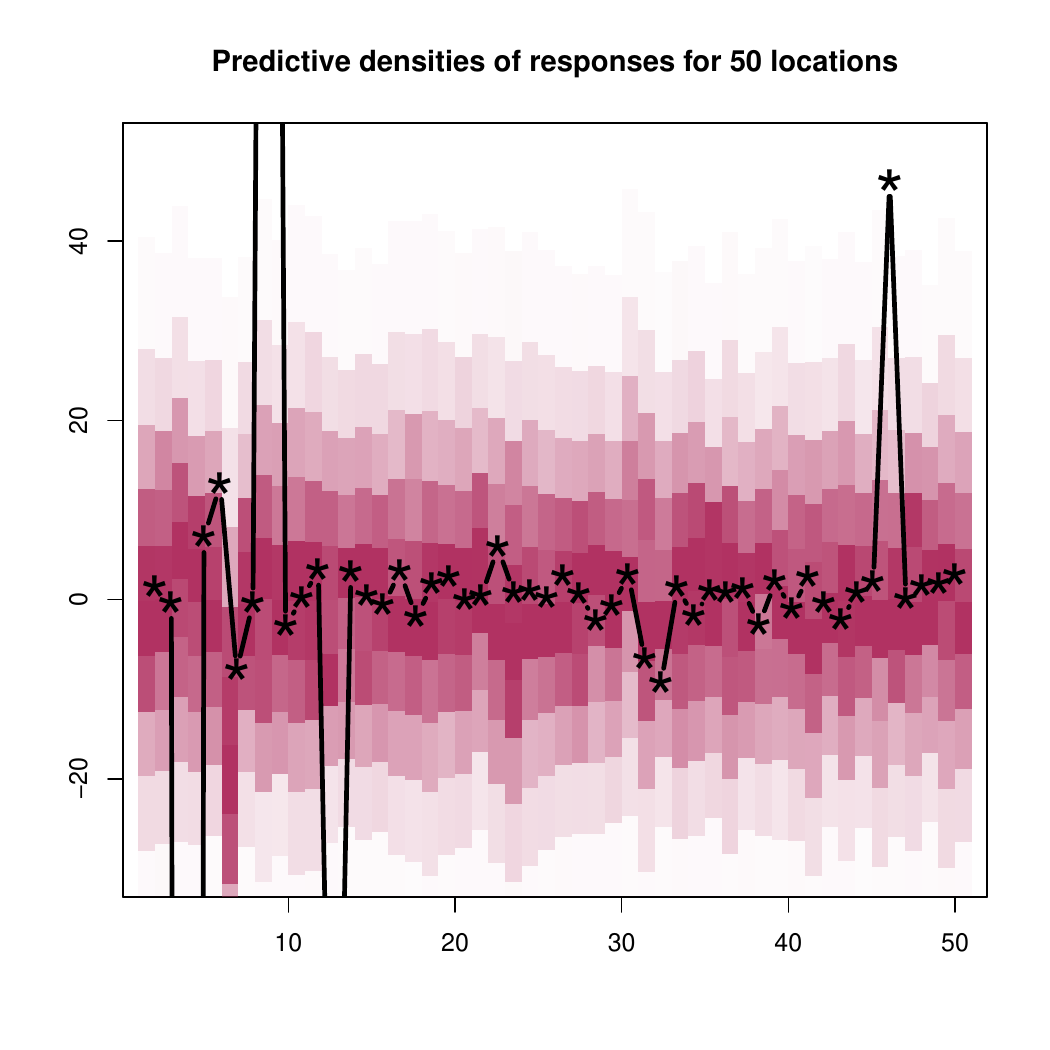}
     \caption{HPD color plot of predictive densities of the elements of $\bi{y}_{20}$.}
     \label{fig:hpd of two_comp_mixture_GQN_response}
    \end{subfigure}
    \begin{subfigure}{0.5\textwidth}
        \includegraphics[width=\linewidth]{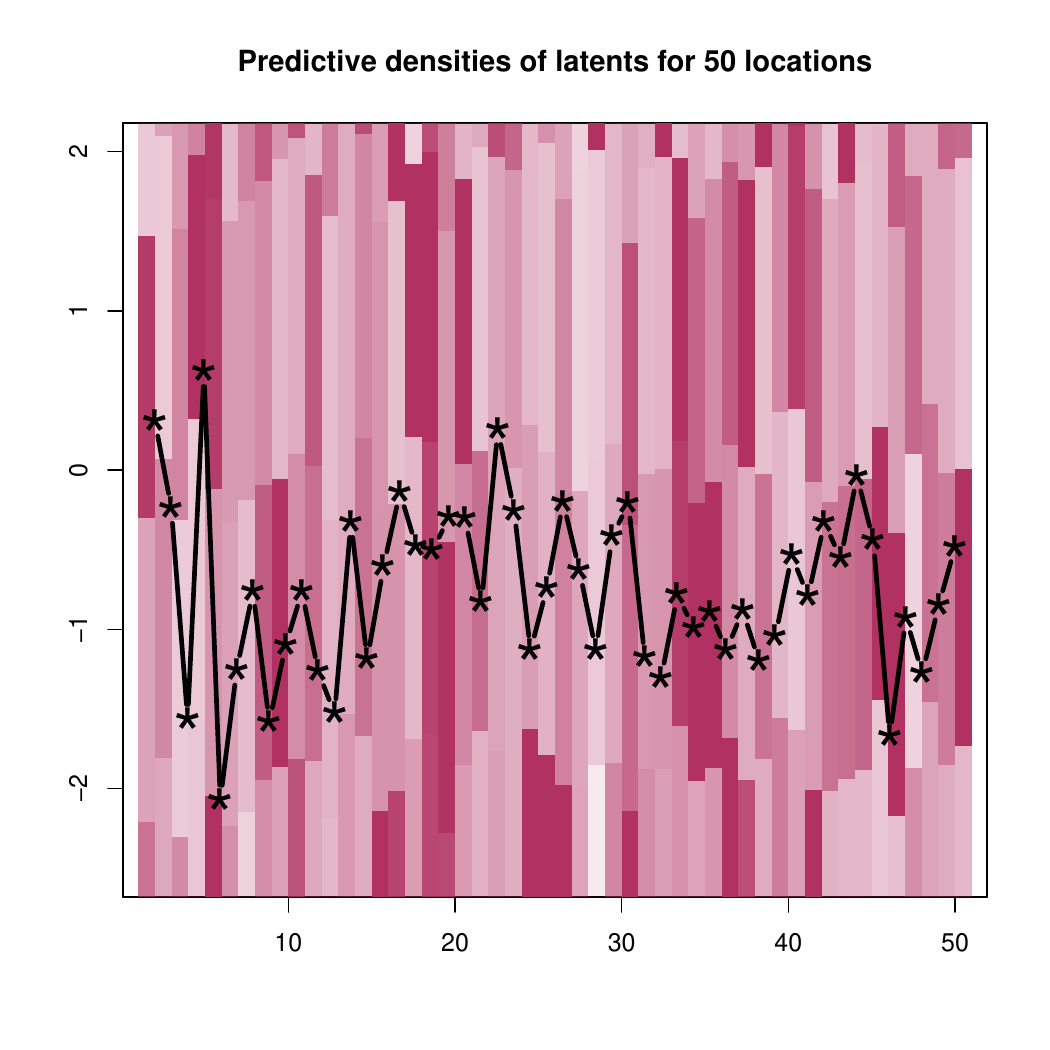}
    \caption{HPD plot of the posterior predictive of the elements of $\bi{x}_{20}$.}
    \label{fig:hpd plot of two_comp_mixture_GQN_latent}
    \end{subfigure}
    \caption{Higher the intensity of the color higher is the density. Black stars denote the true values. Most of the true values lie in the high density regions.}
    \label{HPD color plot for endtime point predictive density of 2 comp GQN mixtures}
 \end{figure}

%Additionally, the posterior density plots indicate that the latent variable posteriors consistently encompass the true latent time series at all 50 locations within high-probability density regions.
The HPD plot of the posterior predictive densities of the elements of latent vector $\bi{x}_{20}$ and response vector $\bi{y}_{20}$ are provided in Figure \ref{HPD color plot for endtime point predictive density of 2 comp GQN mixtures}. Apart from a few true values of $\bi{y}_{20}$, the rest lie well within the high density region (Figure~\ref{fig:hpd of two_comp_mixture_GQN_response}). 
%and almost all the true values of latent vector $\bi{x}_{20}$ fall in the high density region (Figure \ref{fig:hpd plot of two_comp_mixture_GQN_latent}). 
The high probability region of the predictive densities of $\bi{x}_{20}$ could not satisfactorily capture the true values, as seen in Figure \ref{fig:hpd plot of two_comp_mixture_GQN_latent}. {This may possibly be attributed to the complicated structure of the simulation model, making the latent variables  non-identifiable, and also possibly 
to large variability of the data, rendering the densities of the latent variables  wide, leading to much sparser HPD regions.} 
%Along with these, the posterior predictive densities of 50 components of $\bi{y}_{20}$ and $\bi{x}_{20}$ are given in Figures ~\ref{fig: predictive density for 50 locations mixture GQNs} and \ref{fig: predictive density of latent variable for 50 locations mixture GQNs} of Appendix \ref{predictive densities of response and latent vectors 2comp GQN mixtures}, respectively. For the first 25 components of $\bi{y}_{20}$ Figure ~\ref{fig:Predictive densities for first 25 locations mixture GQNs} provides the density plots along with the 95\% predictive intervals.
For the sake of brevity, we have not included component-wise density plots of $\bi{x}_{20}$ and $\bi{y}_{20}$. 
%They failed to capture true values at locations $s_3$, $s_8$ and $s_{13}$. The posterior predictive densities along with the 95\% predictive intervals for the remaining 25 locations are provided in Figure \ref{fig:Predictive densities for last 25 locations mixture GQNs}. 
%It is seen that all the true values fall within the 95\% predictive intervals. 
Out of 50 locations, except for 3 places (at locations 3, 8 and 13), the true response values are captured by the 95\% predictive intervals of the response variable $\bi{y}_{20}$. The lengths of the intervals vary from 99 to 141 units. The mean length of intervals turned out to be 112 with standard deviation 7.9680. In the case of the latent vector, $\bi{x}_{20}$, all the true values are captured by the respective 95\% predictive intervals. The lengths of the intervals vary from 138 to 142.5 with mean 140.4856 and standard deviation 0.8524. %the posterior predictive density plots of the first 25 components of $\bi{x}_{20}$ are given in Figure \ref{fig:Predictive densities of latent variable for first 25 locations mixture GQNs} and that of the next 25 components are provided in Figure \ref{fig:Predictive densities of latent variable for last 25 locations mixture GQNs}, respectively, in Appendix \ref{predictive densities of response and latent vectors 2comp GQN mixtures}. 
%All the true values are captured by the respective predictive intervals. The lengths of the intervals vary from 138 to 142.5 units with mean 140.4856 and standard deviation (sd) 0.8524.
\subsubsection{Comparative study with non-stationary GP model}
\label{Mixture GQN comparison with NSGP model}
%==================================
Similar to the analysis of the mixture of GP models, we fitted the NGP model to the current data set using INLA in R. The same SPDE technique was applied for model fitting. %Figures \ref{fig: predictive density for 50 locations mixtures of GQNs using NSGP} and \ref{fig: predictive density for 50 latent locations mixtures of GQNs using NSGP} in Appendix \ref{posterior predictive density plots using INLA for mixture of GQNs} present the posterior predictive densities for the components of $\bi{y}_{20}$ and $\bi{x}_{20}$, respectively. 
The 95\% predictive intervals of the components of $\bi{y}_{20}$ capture the true values except for the 3rd, 8th and 12th components.  The interval lengths range from 172 to 233.5 units, with a mean of 200.79 and a standard deviation of 10.1872. Notably, the mean predictive interval length exceeds that of our proposed model. 
%It is interesting to see that none of the true values were captured by the 95\% predictive intervals of the posterior predictive densities of the components of $\bi{x}_{20}$ (Figure \ref{fig: predictive density for 50 latent locations mixtures of GQNs using NSGP} of Appendix \ref{posterior predictive density plots using INLA for mixture of GQNs}). Interestingly, the lengths of the intervals turned out to be too small, in the order of $10^{-4}$. Hence from the analysis of this simulated data set, we observe that the performance of our model is significantly better than that of the NGP model for the mixture of GQN data set. The comparison results are summarized in the Table \ref{table of comparison for mixture of 2 GQNs}. 
In contrast, none of the true values fall within the 95\% predictive intervals for $\bi{x}_{20}$, 
%(Figure \ref{fig: predictive density for 50 latent locations mixtures of GQNs using NSGP} in Appendix \ref{posterior predictive density plots using INLA for mixture of GQNs}), 
and the intervals are extremely narrow, in the order of $10^{-4}$. This analysis indicates that our model significantly outperforms the NGP model for the mixture of GQN data. A summary of the comparison is provided in Table \ref{table of comparison for mixture of 2 GQNs}.
    \begin{table}[!h]
    \centering
\begin{tabular}{| *{9}{c|} }
    \hline
    & \multicolumn{2}{c|}{Mean}
            & \multicolumn{2}{c|}{Std}
                    & \multicolumn{2}{c|}{$\%$ of locations misses}             \\
    \hline
   &   INLA  &   Prop Model  &   INLA  &   Prop Model  &   INLA  &   Prop Model  \\
    \hline
$\bi{y}_{20}$   &  200.79  &   112  &   10.1872  &   7.9680  &   6  &   6  \\
$\bi{x}_{20}$   &   $\sim 10^{-4}$  &   140.4856  &   $\sim 10^{-4}$  &   0.8524  &   100  &   0   \\
\hline    
\end{tabular}
\caption{The table contains the summary of the lengths of the 95\% predictive intervals for $\bi{y}_{20}$ and $\bi{x}_{20}$ obtained using INLA and the proposed model for mixture of two GQNs. The columns named ``Mean'' and ``Std'' contain the mean length and the standard deviation of the 95\% posterior predictive intervals across the locations.}
\label{table of comparison for mixture of 2 GQNs}
\end{table}
\section{Real data analyses}
\label{real data analysis}
%%%%%%%%%%%%%%%%%%%%%%%%%%%%%%%%%%%%%
%We evaluate our model performance on two real data sets on temperatures. The first spatio-temporal data is the temperature values taken around Alaska recorded for 65-70 years. The second data set consists of sea surface temperatures over a wide range of areas, recorded for 100 months. The details of the data sets are provided in Sections \ref{Alaska} and \ref{sea temp}, respectively. 
We evaluate our model's performance on two real temperature datasets. The first is a spatio-temporal dataset of temperature measurements recorded in Alaska over 65–70 years. The second consists of sea surface temperatures collected over a wide range of areas for 100 months. Details of these datasets are provided in Sections \ref{Alaska} and \ref{sea temp}, respectively.

%Before applying our model on these data sets, we first make the following transformation (Lambert projection, see \cite{Suman2017}) of the locations so that the Euclidean distance make more sense.Let $ \phi$ be longitude and $\psi$ be latitude in radian. Then the following transformation is made:
Before applying our model to these datasets, we first transform the locations using the Lambert projection (\cite{Suman2017}) to ensure that Euclidean distances are meaningful. Let $ \phi$ be longitude and $\psi$ be latitude in radian. Then the following transformation is made:
\begin{align}
	s_1  = 2 \sin\left( \frac{\pi}{4} - \frac{\psi}{2}\right)\sin \phi, \text{ and }
	s_2  = -2 \sin\left( \frac{\pi}{4} - \frac{\psi}{2}\right)\cos \phi.
\label{Lambert}
\end{align}
%\begin{equation}
%\label{Lambert}
%\begin{split}
%	s_1 & = 2 \sin\left( \frac{\pi}{4} - \frac{\psi}{2}\right)\sin \phi;\text{ and }
%	s_2 & = -2 \sin\left( \frac{\pi}{4} - \frac{\psi}{2}\right)\cos \phi. 
%\end{split}
%\end{equation}
%%%%%%%%%%%%%%%%%%%%%%%%%%%%%%%%%%%%%
\subsection{Alaska temperature data}
\label{Alaska}
%%%%%%%%%%%%%%%%%%%%%%%%%%%%%%%%%%%%%%%%
%A real data analysis is done on the temperature data of Alaska and its surroundings. The data set is collected from \url{https://www.metoffice.gov.uk/hadobs/crutem4/data/download.html} by clicking the link \url{CRUTEM.4.6.0.0.station_files.zip} given under the heading \textbf{Station data}. The details of the data set can be read from \url{https://crudata.uea.ac.uk/cru/data/temperature/crutem4/station-data.htm}. 
%A total of 30 locations are considered for the analysis. Annual average temperature data for the years 1950 to 2015 are taken after detrending. Of these 30 locations, at four locations many data were missing. So, we decided to construct the complete time series for these four locations. Among these 30 locations, data till 2021 were available for 16 positions. We thus have made multiple time predictions for these 16 locations. The 26 spatial points are indicated in Figure~\ref{fig:Alaska} in red and 4 locations (for which the complete time series is reconstructed) are indicted in blue in the same graph. The latitudes and longitudes are provided in Table~\ref{locations of Alaska temp data}.		
We conduct a real data analysis on temperature data from Alaska and its surroundings. The dataset is collected from \url{https://www.metoffice.gov.uk/hadobs/crutem4/data/download.html} by selecting the \url{CRUTEM.4.6.0.0.station_files.zip} link under the \textbf{Station data} section. The details on the dataset can be found at \url{https://crudata.uea.ac.uk/cru/data/temperature/crutem4/station-data.htm}.

\begin{figure}[!ht]
	\centering
	\includegraphics[trim={2cm 2cm 2cm 2cm},width=0.75\textwidth,height=12cm,clip]{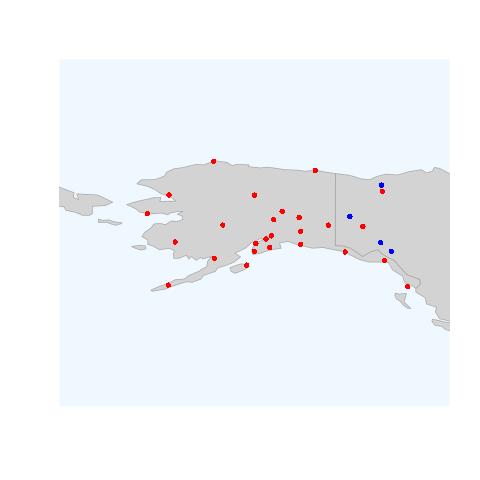}
	\caption{The map of Alaska and its surroundings. The red dots indicate the locations at which the data for the year 1950-2015 are considered. The positions indicated 
	by the blue dots are those at which spatial predictions are made for the complete time series.}
	\label{fig:Alaska}
\end{figure}

For the analysis, we consider 30 locations, using annual average temperature data from 1950 to 2015 after detrending. At four locations, a significant amount of data was missing, so we reconstructed the complete time series for these locations. Among the 30 locations, data were available up to 2021 for 16 locations, allowing us to make multiple time predictions for these sites. Figure~\ref{fig:Alaska} shows the 26 spatial points in red, while the four locations with reconstructed time series are marked in blue. The latitudes and longitudes are listed in Table~\ref{locations of Alaska temp data}.
\begin{table}[!ht]
\centering																	
\begin{tabular}{|c|rr|c|rr|c|rr|c|rr|c|}														
\hline																				
Sl.	No.	&	Lat.	&	Long.	&	Sl.	No.	&	Lat.	&	Long. &	Sl.	No.	&	Lat.	&	Long.\\					
\hline																		
1	&	71.3N	&	156.8W	&	11	&	64.8N	&	147.9W	& 21	&	62.2N	&	145.5W	\\
2	&	60.6N	&	151.3W	&	12	&	60.1N	&	149.5W	& 22	&	66.9N	&	162.6W	\\
3	&	59.5N	&	139.7W	&	13	&	63N	&	155.6W	& 23	&	58.4N	&	134.6W	\\
4	&	55.2N	&	162.7W	&	14	&	63N	&	141.9W	& 24	&	63.7N	&	149W	\\
5	&	61.2N	&	150W	&	15	&	55N	&	131.6W	& {\textcolor{blue}{25}}	&	68.2N	&	135W	\\
6	&	70.1N	&	143.6W	&	16	&	61.6N	&	149.3W	& {\textcolor{blue}{26}}	&	59.6N	&	133.7W	\\
7	&	64.5N	&	165.4W	&	17	&	64N	&	145.7W	& 27	&	62.8N	&	137.4W	\\
8	&	66.9N	&	151.5W	&	18	&	60.8N	&	161.8W	& {\textcolor{blue}{28}}	&	64.1N	&	139.1W	\\
9	&	59.6N	&	151.5W	&	19	&	57.8N	&	152.5W	& 29	&	67.4N	&	134.9W	\\
10	&	60.5N	&	145.5W	&	20	&	58.7N	&	156.7W	& {\textcolor{blue}{30}}	&	60.7N	&	135.1W	\\
\hline																			\end{tabular}	
\caption{Latitude	and	Longitude	(in	degrees)	of	30	locations	in	and	around	Alaska.	The	locations	corresponding	to	the	serial	numbers,	indicated	in	bold,	are	used	for	multiple	predictions.	The	serial	numbers	which	are	denoted	in	blue,	for	the	corresponding	spatial	locations,	complete	time	series	are	reconstructed.	These	blue	serial	numbers	are	indicated	by	$L_{25},	L_{26},	L_{28}$	and	$L_{30}$	for	future	references.}
\label{locations	of	Alaska	temp	data}										
\end{table}		

Thus, the dataset considered here consists of 26 locations and 65 time points from 1950 to 2014, with the final time point set aside for single-time-point prediction. We denote the dataset as
$\mathbb{D} = \{\bi{y}_1, \ldots, \bi{y}_{65}\}$, where $\bi{y}_{t}$, for $t \in \{1, \ldots, 65\}$, is a 26-dimensional vector. In this analysis, we perform both temporal and spatial predictions after detrending the data. At these 26 locations, we obtain predictive densities for the year 2015.

For 16 locations (indicated by $L_1, \ldots, L_7, L_{11}, \ldots, L_{13}, L_{18}, \ldots, L_{20}, L_{22}, L_{27}, L_{29}$ for references hereafter), we make multiple time-point predictions (see Appendix \ref{temporal prediction}). Additionally, we construct 95\% predictive intervals for the complete time series at four locations with missing data (the spatial prediction algorithm is detailed in Appendix \ref{spatial prediction}). To achieve this, we augment the dataset $\mathbb{D}$ with initial values for the 65 time points at these four locations and iteratively update them by computing conditional densities given $\mathbb{D}$. That is, we define an augmented dataset $\mathbb{D}^* = \{\bi{y}_{1}^*, \ldots, \bi{y}_{65}^*\}$, where each $\bi{y}_{t}^{*}$ is a 30-dimensional vector, with the last four values initialized.% and subsequently updated based on the model parameters and latent variables. This process continues throughout the entire MCMC run.
The parameters of our model, including the latent variables, are updated given $\mathbb{D}^*$, and then the last four values of $\bi{y}^*_{t}$ are updated given the parameter values and latent variables. This process continues throughout the entire MCMC run. In Appendix \ref{appendix:alaska}, we validate that the underlying spatio-temporal process generating the Alaska data is non-Gaussian, strictly stationary, and exhibits lagged correlations that converge to zero as lag increases. Moreover, there is no justification for assuming separability in the spatio-temporal covariance structure. While our spatio-temporal process is non-stationary, it retains other desirable properties, and the analysis results demonstrate that it is well-suited for modeling this dataset.
%%%%%%%%%%%%%%%%%%%%%%%%%%%%%%%%%%%%%%%%%%%%%%%%%%%%%%
\subsubsection{\small Prior choices for the Alaska temperature data}
%%%%%%%%%%%%%%%%%%%%%%%%%%%%%%%%%%%%%%%%%%%%%%%%%%%%%%%
\label{choice of prior param for Alaska temp data}
With the same cross-validation technique as in the simulation experiments, the complete specification of the priors are obtained as follows:
\begin{align*}
	& \alpha^* \sim N(0,\sqrt{500}), \beta^* \sim N(0,\sqrt{300}),[\sigma^2]  \propto IG(550000,2/2), [\sigma^2_{\theta}] \propto IG (700,780/2),\\
	& 
	[\sigma^2_{p}] \propto IG (1000,100/2),[\eta_{1}^*]  \propto N (-3,1), [\eta_{2}^*] \propto N (-5,1).
\end{align*}
%As before, we fix $\eta_3$ at its maximum likelihood estimate 4.5579. As in the cases of simulation studies, the shape parameters for the prior distribution of the variance parameters are kept large to make sure that the posterior variances of the variance parameters remain finite; see \ctn{bhattacharya2021bayesian} for a similar strategy.
As before, we fix $\eta_3$ at its MLE of 4.5579. Similar to the simulation studies, the shape parameters for the prior distribution of the variance parameters are set to large values to ensure that the posterior variances of the variance parameters remain finite; see \cite{bhattacharya2021bayesian} for a similar strategy.
%%%%%%%%%%%%%%%%%%%%%%%%%%%%%%%%%%%%%%%%%%%%%%%%%%%%%%
%===================================
\subsubsection{\small Results of the Alaska temperature data}
\label{results: Alaska temp data}
%==================================
%For the current study, we run 2,50,000 MCMC iterations with the first 1,25,000 iterations as burn-in. Moreover, we make a thinning of width 5 for better mixing of the MCMC samples. The time taken was about 8 hours 12 minutes on our desktop computer.The MCMC trace plots for the parameters, except $\eta_3$, are given in Figure \ref{fig:trace plot of Alaska temp data} of Appendix \ref{Alaska: trace plot}. The trace plots do not showsigns of non-convergence. 
%which bear clear evidence of convergence in each case. The posterior densities of the latent variables are displayed in Figure \ref{fig:posterior density of latent variables first 13 Alaska temp data} of Appendix \ref{Alaska: posterior densities of the latent variables}.
For this study, we run 250,000 MCMC iterations, discarding the first 125,000 iterations as burn-in. Additionally, we apply a thinning of width 5 to improve mixing of the MCMC samples. The total computation time was approximately 8 hours and 12 minutes on our desktop computer. 

\begin{figure}[!h]
	\centering
	\includegraphics[trim={0cm 0cm 0cm 0cm},width=0.75\textwidth,height=12cm,keepaspectratio,clip]{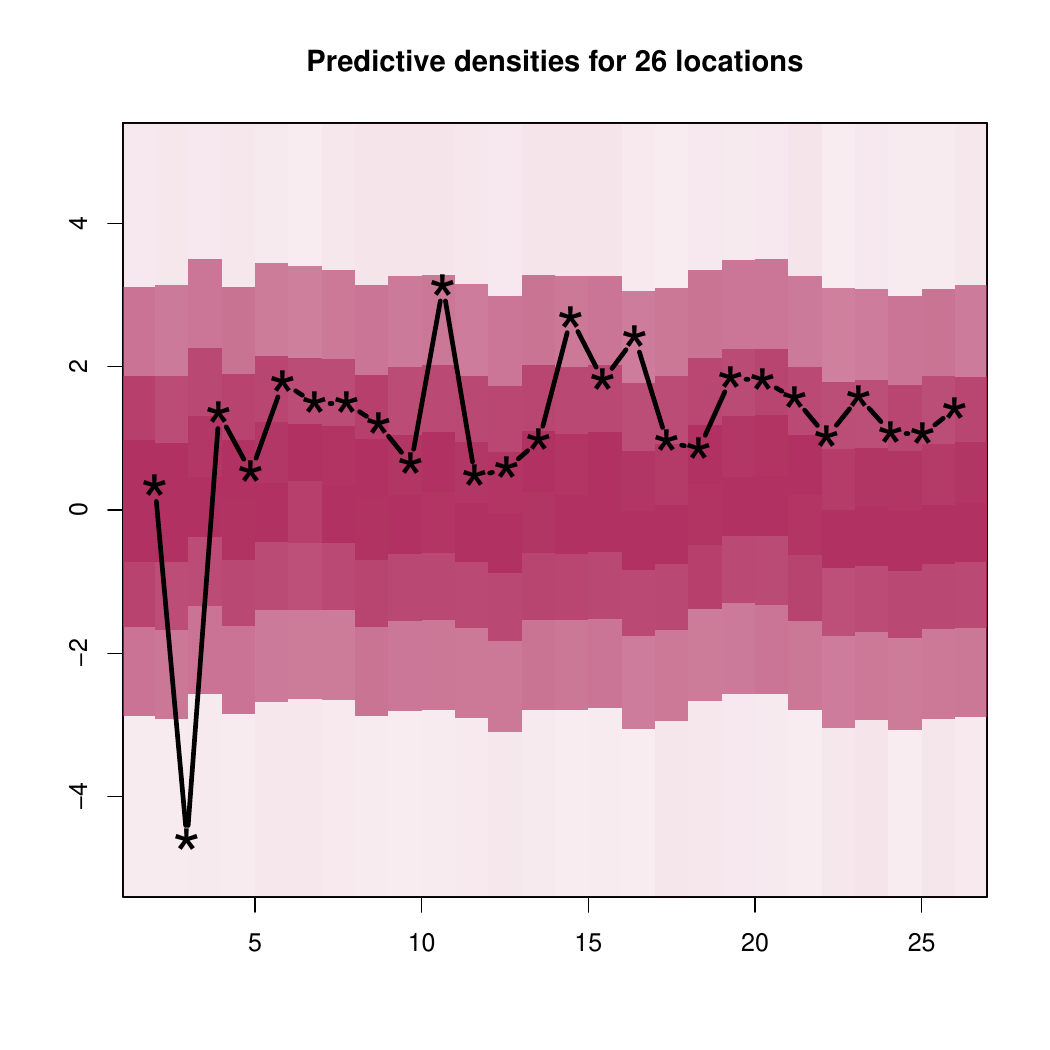}
	\caption{The highest probability density regions of the predictive densities and the true values of detrended temperatures (Alaska temperature data) for the year 2015 at 26 locations are depicted in this figure. The x-axis of of the figure indicates the locations, and at each location, color-coded densities are plotted in the vertical axis. The intensity of the color is proportional to the density. The true values are indicated by the black stars. Majority of the true values fall in the high density regions.}
	\label{fig: hpd predictive density for 26 locations Alaska temp data}
\end{figure}

The trace plots for all parameters except $\eta_3$ are shown in Figure \ref{fig:trace plot of Alaska temp data} in Appendix \ref{Alaska: trace plot}. These plots do not indicate any sign of non-convergence.  
%The posterior densities of the latent variables are presented in Figure \ref{fig:posterior density of latent variables first 13 Alaska temp data} in Appendix \ref{Alaska: posterior densities of the latent variables}.
%
%
%The predictive densities for $\bi{y}_{66}$ (detrended temperatures of the Alaska temperature data for the year 2015) at 26 locations along with the highest density regions are depicted in Figure~\ref{fig: hpd predictive density for 26 locations Alaska temp data}. The intensity of the color is proportional to the density of the predictive densities of $\bi{y}_{66}$. The plot shows that most of the true values fall in the highest density regions. The individual predictive densities with 95\% prediction intervals and the true values are provided in Figure~\ref{fig: predictive density for 15 locations Alaska temp data} in Appendix \ref{Alaska: posterior predictive density for end time point} for further reference. Except one, all the true values lie well within the 95\% predictive intervals.
The HPD plots of the predictive densities for $\bi{y}_{66}$ (detrended temperatures of the Alaska temperature data for the year 2015) at 26 locations are shown in Figure~\ref{fig: hpd predictive density for 26 locations Alaska temp data}. The color intensity is proportional to the density of the predictive distributions of $\bi{y}_{66}$. The plot indicates that most true values fall within the highest density regions.  
%
%For further reference, individual predictive densities with 95\% prediction intervals and true values are provided in Figure~\ref{fig: predictive density for 15 locations Alaska temp data} in Appendix \ref{Alaska: posterior predictive density for end time point}. 
We have not provided the predictive density plots of the individual components of $\bi{y}_{66}$ for brevity, however, it is observed that except for one location (at 2nd location), all true values lie well within the 95\% predictive intervals. The mean length of the predictive interval is 10.6387 (standard deviation=0.1). 

%The predictive densities together with highest density regions (the higher the intensity of color, higher the density) for the years 2016-2021 at the 16 locations, $L_1, \ldots, L_7$, $L_{11}, L_{12}, L_{13}$, $L_{18}, L_{19}, L_{20}$, $L_{22}$, $L_{27}$, $L_{28}$, are shown in Figure~\ref{fig: hpd predictive densities for 16 locations Alaska temp data for multiple time points}. The two figures (Figure~\ref{fig: predictive density for first 8 locations Alaska temp data for multiple time points} and Figure~\ref{fig: predictive density for last 8 locations Alaska temp data for multiple time points}) in Appendix \ref{Alaska: posterior predictive densities multiple time points}, each containing 8 locations, show the predictive densities with 95\% prediction intervals. Except for one location, at one time point, in all other scenarios, the true values are captured by the 95\% predictive intervals associated with our model.

\begin{figure}[!h]
	%\centering
	\includegraphics[trim={0cm 0cm 0cm 0cm},width=0.5\textwidth,height=6.5cm,clip]{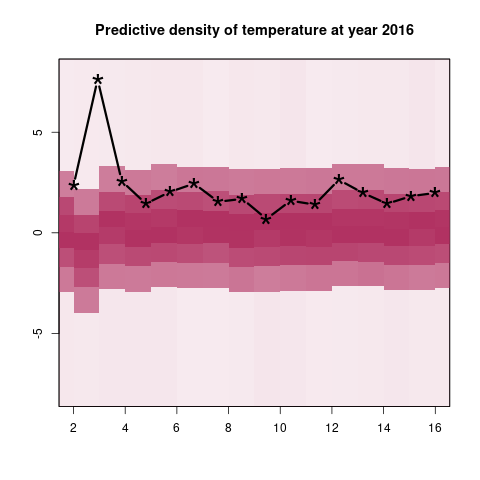}
    \includegraphics[trim={0cm 0cm 0cm 0cm},width=0.5\textwidth,height=6.5cm,clip]{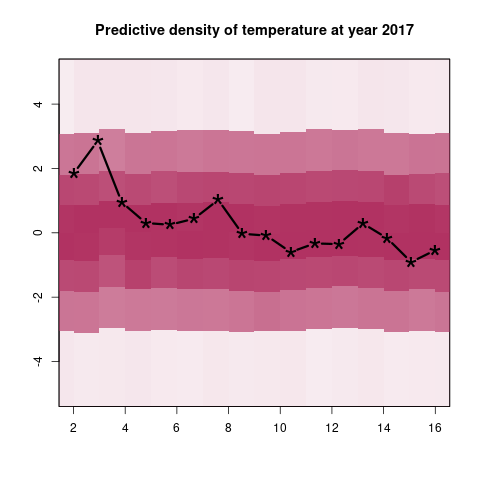}
    \includegraphics[trim={0cm 0cm 0cm 0cm},width=0.5\textwidth,height=6.5cm,clip]{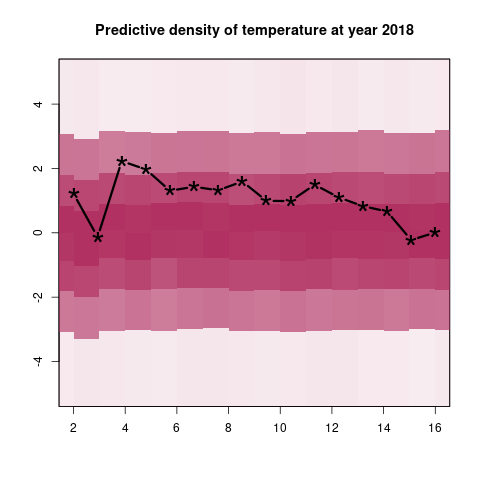}
    \includegraphics[trim={0cm 0cm 0cm 0cm},width=0.5\textwidth,height=6.5cm,clip]{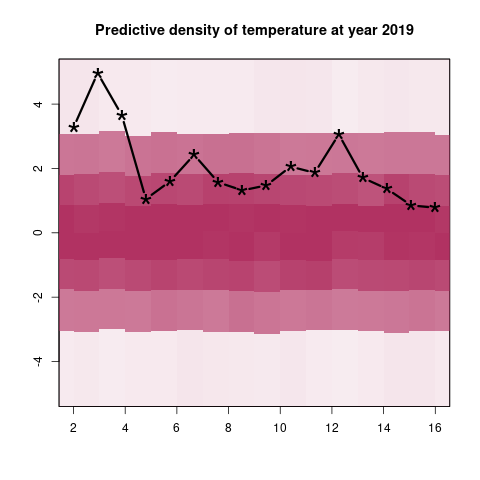}
    \includegraphics[trim={0cm 0cm 0cm 0cm},width=0.5\textwidth,height=6.5cm,clip]{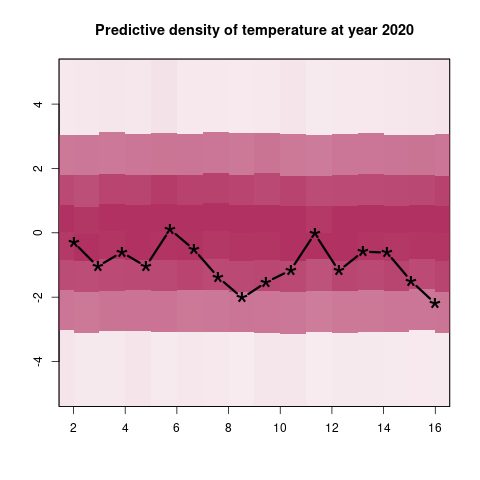}
    \includegraphics[trim={0cm 0cm 0cm 0cm},width=0.5\textwidth,height=6.5cm,clip]{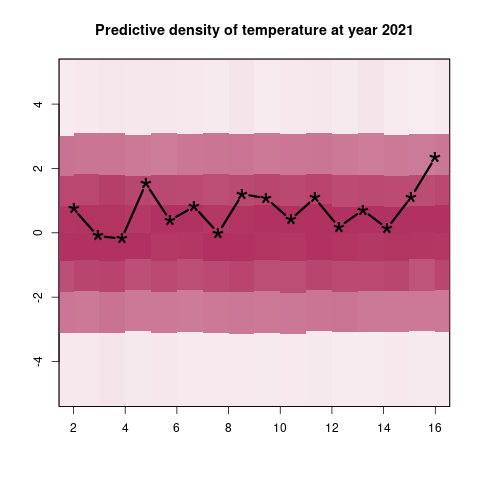}
	\caption{The highest probability density regions of the predictive densities and the true values of detrended temperatures (Alaska temperature data) for the years 2016 to 2021 at 16 locations are depicted in this figure. The x-axis of the figure indicates the locations, and at each location, color-coded densities are plotted in the vertical axis. The intensity of the color is proportional to the density. The true values are indicated by the black stars. Majority of the true values for all the 6 years fall in the high density regions.}
	\label{fig: hpd predictive densities for 16 locations Alaska temp data for multiple time points}
\end{figure}

The predictive densities, along with the highest density regions (where higher color intensity indicates higher density), for the years 2016–2021 at the 16 locations $L_1, \ldots, L_7$, $L_{11}, L_{12}, L_{13}$, $L_{18}, L_{19}, L_{20}$, $L_{22}$, $L_{27}$, and $L_{28}$ are shown in Figure~\ref{fig: hpd predictive densities for 16 locations Alaska temp data for multiple time points}. Apart from a few observations, the true values fall within high density regions.  
%Additionally, Figure~\ref{fig: predictive density for first 8 locations Alaska temp data for multiple time points} and Figure~\ref{fig: predictive density for last 8 locations Alaska temp data for multiple time points} in Appendix \ref{Alaska: posterior predictive densities multiple time points} display the predictive densities with 95\% prediction intervals for these locations, grouped into two sets of eight. Except for one location at a single time point, all true values fall within the 95\% predictive intervals of our model.

\begin{figure}[!ht]
	%	\begin{subfigure}{.5\textwidth}
		\centering
		\includegraphics[trim={0cm 0cm 0cm 0cm}, width=0.45\textwidth,height=8.0cm, keepaspectratio,clip]{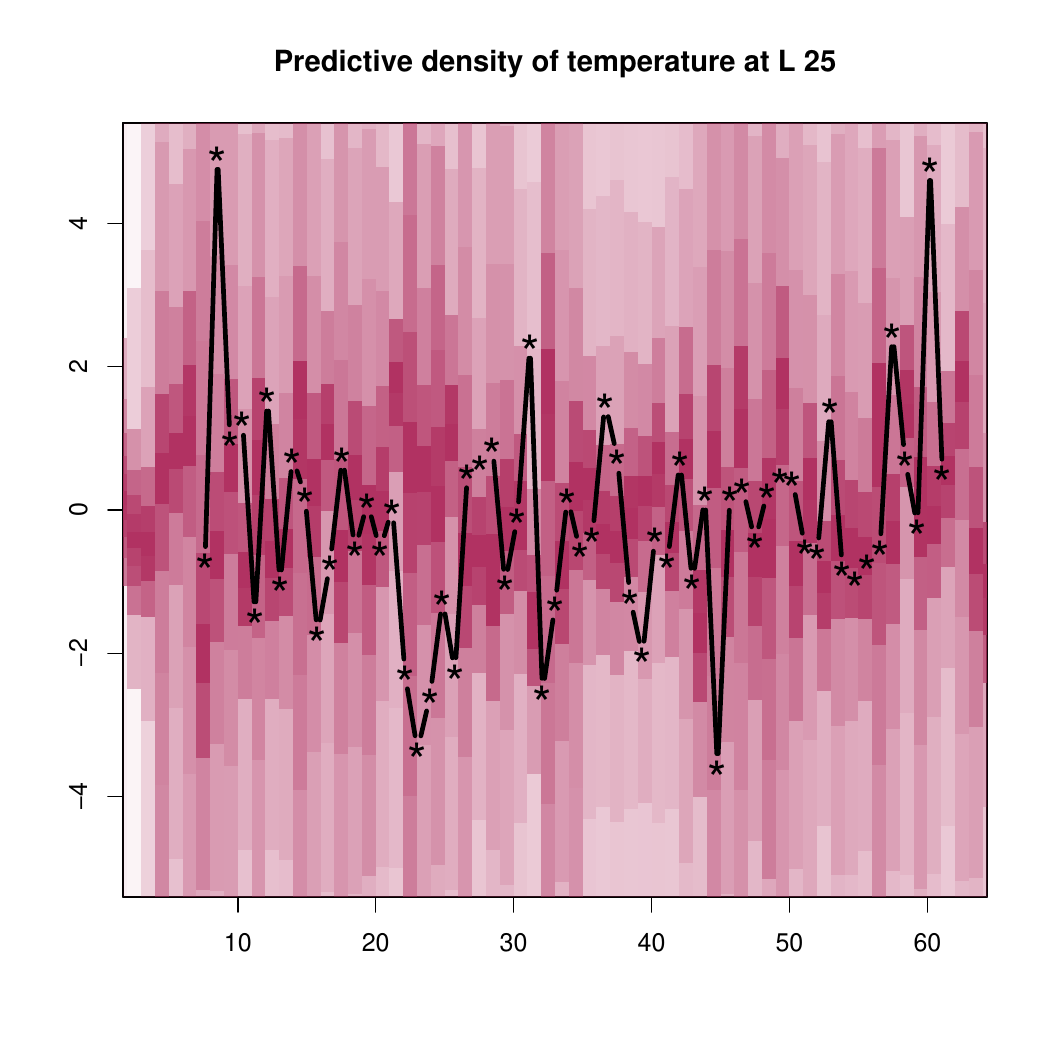}
		\includegraphics[trim={0cm 0cm 0cm 0cm}, width=0.45\textwidth,height=8.0cm, keepaspectratio,clip]{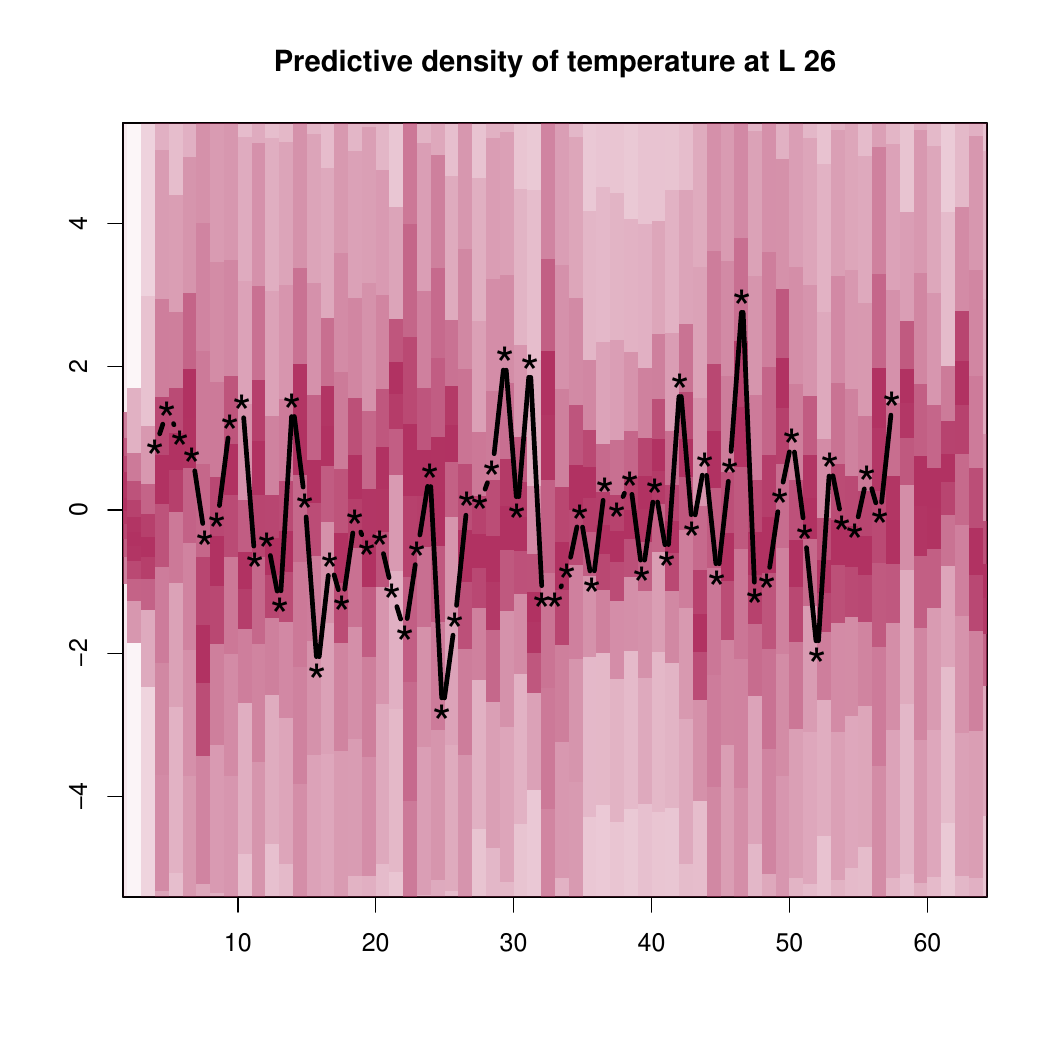}
		\includegraphics[trim={0cm 0cm 0cm 0cm}, width=0.45\textwidth,height=8.0cm, keepaspectratio,clip]{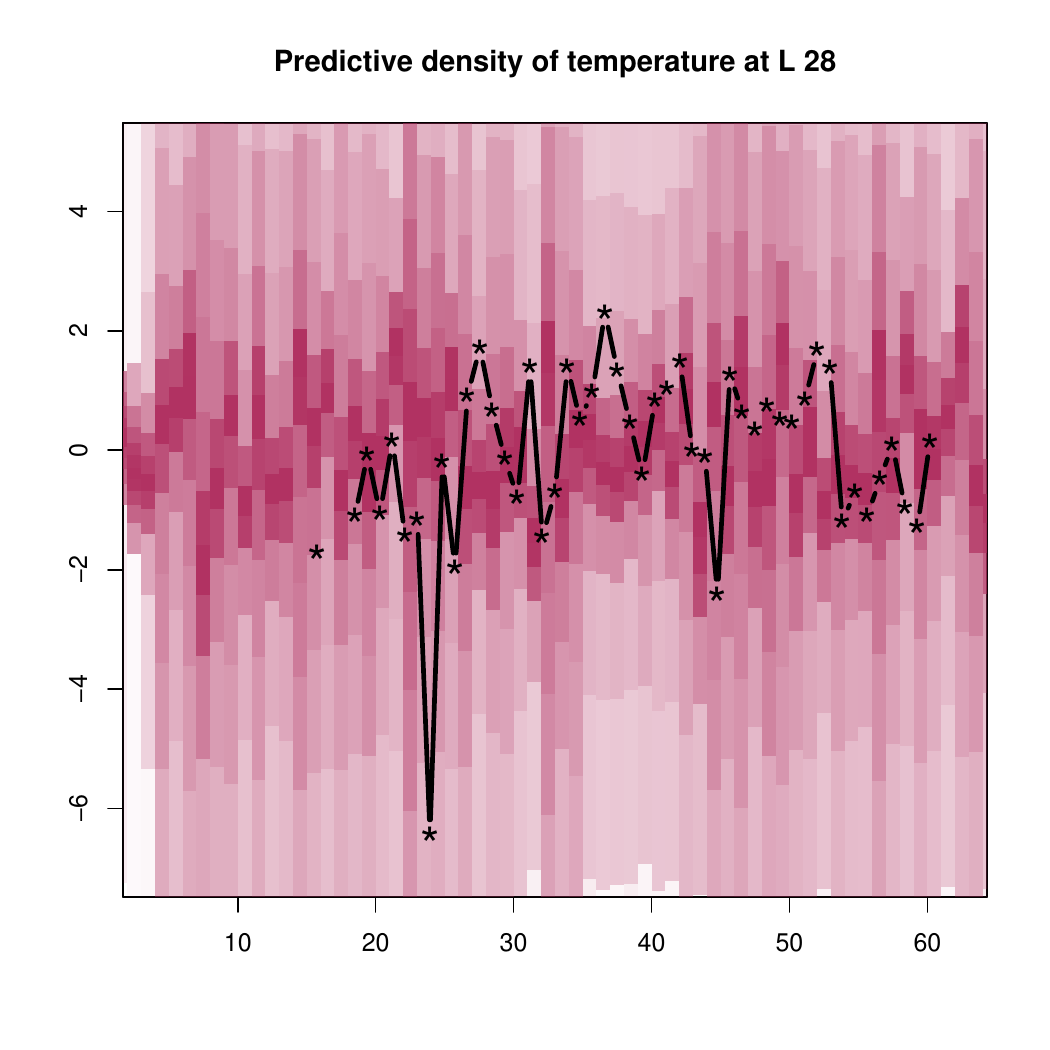}
		\includegraphics[trim={0cm 0cm 0cm 0cm}, width=0.45\textwidth,height=8.0cm, keepaspectratio,clip]{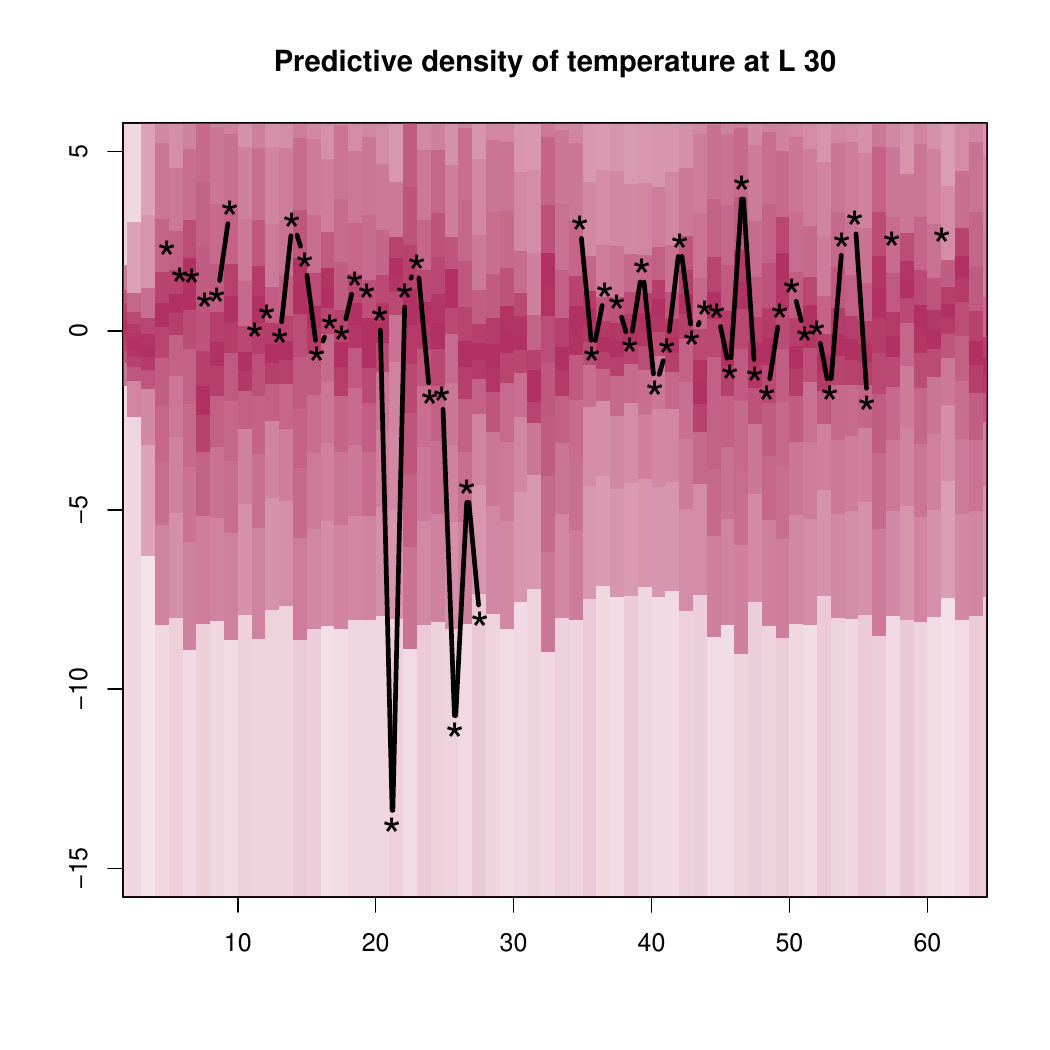}		
		\caption{Posterior predictive densities of the reconstructed detrended temperature data of Alaska and its surroundings at 4 locations. Higher the intensity of the color, higher is the probability density. The black stars represent the available true temperature values (detrended). Except for a few points, majority of the true values at each location fall within the high probability density regions.}		
		\label{fig:posterior predictive density of temp time series of 4 locations Alaska temp data}
\end{figure}

%The complete time series for the four locations, which were indicated as blue dots in Figure~\ref{fig:Alaska}, are reconstructed using our model. The Bayesian predictive densities at each time point for these locations are shown in Figure~\ref{fig:posterior predictive density of temp time series of 4 locations Alaska temp data} using probability plot. Higher the intensity of the color, higher is the density. The available true detrended temperature values at these locations are plotted and depicted by black stars in Figure~\ref{fig:posterior predictive density of temp time series of 4 locations Alaska temp data}. Except a very few points at these locations, the detrended true values lie well within the high density regions. Overall, the performance of our model on this particular data set is encouraging.
The complete time series for the four locations, represented as blue dots in Figure~\ref{fig:Alaska}, are reconstructed using our model. The Bayesian predictive densities at each time point for these locations are visualized in Figure~\ref{fig:posterior predictive density of temp time series of 4 locations Alaska temp data} using probability plots, where higher color intensity indicates higher density.  
The available true detrended temperature values at these locations are overlaid as black stars in Figure~\ref{fig:posterior predictive density of temp time series of 4 locations Alaska temp data}. With only a few exceptions, the true detrended values fall well within the high-density regions. Overall, our model demonstrates strong performance on this dataset.
%At one location ($L_{30}$), three values lie away from the high density region. However, from the pattern of the data values, it seems that these values are outlying in comparison to the rest of the values. Other than these, all the available true detrended temperatures lie within high density regions (except one value at $L_{28}$). In a nutshell, we can claim that our model performs well in analyzing the temperature data of Alaska and its surroundings.
%---------------------------------
\subsection{Sea surface temperature data}
\label{sea temp}
%--------------------------------
%Another temperature data is analysed in this subsection for showing our model performance on real life data. 
%The sea surface temperature data is obtained from \url{http://iridl.ldeo.columbia.edu/SOURCES/.CAC/}. For the purpose of illustration, we have taken spatio-temporal observations of the first 40 locations and the first 100 time points from the complete data set. The locations of the chosen data set varies from 56$^{\circ}$W to 110$^{\circ}$E and 19$^{\circ}$S to 25$^{\circ}$N. The locations are given in Table \ref{locations of sea temp data}. Hundred monthly average temperatures are taken starting from the month of January 1970. Among these 40 locations and 100 time points, observations corresponding to 30 locations and 99 time points are taken as the learning set. The observations corresponding to the last 10 locations and the last time point are kept aside for the prediction purpose (see Appendix \ref{temporal prediction} for the algorithm of obtaining observations from predictive densities). Only for the sake of convenience, the time series at each location is referenced to its mean, in the sense that the mean of the time series at each location is subtracted from the original data. 
The sea surface temperature data is taken from \url{http://iridl.ldeo.columbia.edu/SOURCES/.CAC/}. For illustration, we use spatio-temporal observations from the first 40 locations and the first 100 time points. These locations range from 56$^{\circ}$W to 110$^{\circ}$E and 19$^{\circ}$S to 25$^{\circ}$N, as listed in Table \ref{locations of sea temp data}. The dataset consists of 100 monthly average temperatures starting from January 1970. Among these, observations from the first 30 locations and 99 time points form the learning set, while the last 10 locations and the final time point are reserved for prediction. %(see Appendix \ref{temporal prediction} for the predictive density algorithm). 
For computational convenience, each location’s time series is mean-referenced by subtracting its mean from the original data.
\begin{table}[!ht]
\begin{adjustbox}{width=\columnwidth,center}
\begin{tabular}{|c|rr|c|rr|c|rr|c|rr|}													
\hline																				
Sl. No. 	& Lat.	& Long.	&Sl. No. 	&Lat.	&Long.	&Sl. No. 	&Lat.	&Long. &Sl. No. 	&Lat.	&Long. \\
\hline																		
1	& 19S	& 108E	& 11	& 1N 	& 84E	& 21	& 13N	& 30E	& \textbf{31}	& 25N	& 4E	\\
2	& 19N	& 20W 	&12	& 19S 	& 14E	& 22	& 7S 	& 36E	& \textbf{32}	& 7N	& 52E 	\\
3	& 19S 	& 42E 	&13	& 13N 	& 48E	& 23	& 29N	& 104E	& \textbf{33}	& 11N	& 18E	\\
4	& 17N 	& 58E	&14	& 27N 	& 18W	& 24	& 15S 	& 24W	& \textbf{34}	& 5N	& 16W	\\
5	& 7N	& 2W	&15	& 5S 	& 110E 	& 25	& 1N	& 72E	& \textbf{35}	& 7N	& 56W	\\
6	& 11S	& 76E	& 16	& 17S 	& 46W	& 26	& 1S 	& 46W	& \textbf{36}	& 19S	& 106E	\\
7	& 3S	& 2E	& 17	& 7S 	& 10W	& 27	& 13S	& 64E	& \textbf{37}	& 1S	& 20W	\\
8	& 25S	& 28W	& 18	& 23S 	& 14E	& 28	& 9N	& 12E	& \textbf{38}	& 17S	& 28W 	\\
9	& 25N	& 100E	& 19	& 21S 	& 36W	& 29	& 15N	& 24E	& \textbf{39}	& 19S	& 0	\\
10	& 7N	& 26W	& 20	& 1S 	& 28W	& 30	& 25S	& 52E	& \textbf{40}	& 29N	& 48W	\\				
\hline																			\end{tabular}
\end{adjustbox}	
	\caption{Latitude and Longitude (in degrees) of 40 locations for the sea surface temperature data. The locations corresponding to the serial numbers, 
		indicated in bold, are used for complete time series prediction. These bold serial numbers are indicated as $L_{31}, \ldots, L_{40}$ for future reference.}
	\label{locations of sea temp data}
\end{table}

Similar to the Alaska temperature data analysis, we denote the learning data set by $\mathbb D = \{\bi{y}_1, \ldots, \bi{y}_{99}\}$, where $\bi{y}_{t}$ is a 30 dimensional 
spatial observation at time point $t$. %Here also we have made both temporal and spatial prediction. 
At these 30 locations, we obtain the posterior predictive densities for the 100th time point (see Appendix \ref{temporal prediction} for temporal predictive density) and find out the high probability density regions for the entire time series of the locations corresponding to the serial numbers indicated in bold within Table \ref{locations of sea temp data}. 
These locations are referred to as $L_{31}, \ldots, L_{40}$ hereafter. To compute the posterior predictive densities of the entire time series for these 10 spatial locations, we follow the similar path as in the Alaska temperature data analysis. We skip the details here to avoid repetitive description. The detailed algorithm can be found in Appendix \ref{spatial prediction}. %the dataset $\mathbb{D}$ is first augmented with initial values for the 99 time points of these locations, as done in Alaska temperature data analysis. 
%The values are then updated by sampling from the corresponding conditional distributions given $\mathbb{D}$. Specifically, we begin with $\mathbb{D}^* = {\bi{y}{1}^*, \ldots, \bi{y}{99}^}$, where $\bi{y}_{t}^{} = [\bi{y}t^T : \bi{z}{t}^T]^T$ is a 40-dimensional vector, with $\bi{z}t$ representing the initial estimates for locations $L{31}, \ldots, L_{40}$, for $t \in {1, \ldots, 99}$. The model parameters, including latent variables, are then updated given $\mathbb{D}^$, followed by updating the last 10 values of $\bi{y}^_{t}$ based on the estimated parameters and latent values. This process continues throughout the MCMC run (see Appendix \ref{spatial prediction} for the detailed spatial prediction algorithm).
Notably, this sea surface temperature data originates from a non-stationary (both weakly and strongly), non-Gaussian spatio-temporal process with lagged correlations tending to zero (see \cite{bhattacharya2021bayesian} for further details).
%
%\begin{figure}[!ht]
%\begin{subfigure}{.5\textwidth}
%\centering
%\includegraphics[trim={0.25cm 0cm 0cm 0cm},width=0.65\textwidth,height=8cm, clip]{Sea_temp_locations_plot.jpg}
%\end{subfigure}
%\caption{Locations of sea surface temperature data after the transformation given in equations~\ref{Lambert}. The red points represent the locations of the observations in the learning set. The blue solid dots are the locations of the observations in the test set.}
%\label{fig: sea temp data locations}
%\end{figure}
%%%%%%%%%%%%%%%%%%%%%%%%%%%%%%%%%%%%%%%%%%%%%%%%%%%%%%
\subsubsection{Prior choices for sea surface temperature data}
%%%%%%%%%%%%%%%%%%%%%%%%%%%%%%%%%%%%%%%%%%%%%%%%%%%%%%%
\label{choice of prior param for sea temp data}
%------------------------------------------------------
We completely specify the priors as follows:
%Following cross-validation technique to obtain the hyper parameters, we completely specify the priors as follows:	
\begin{align*}
& \alpha_* \sim N(0,\sqrt{500}), \beta_* \sim N(0,\sqrt{300}), [\sigma^2]  \propto IG(45000,2/2), [\sigma^2_{\theta}]  \propto IG (1000,780/2),\\
&  [\sigma^2_{p}]  \propto IG (100,100/2), [\eta_{1}^*] \propto N (-3,1), [\eta_{2}^*] \propto N (-3,1).
\end{align*}
The hyperparameters are determined by cross-validation, similar to the previous analyses. 
%As before, we fixed $\eta_3$ at its MLE; here the value is 5.8879. The shape parameters for the prior distribution of the variance parameters are kept to be large to ensure that the posterior variances of the variance parameters remain finite, as done in other studies. We performed a total of 2,50,000 MCMC iterations with 1,00,000 burn-in. In addition, for a better mixing, a thinning of width 5 has been done. It took around 12 hours and 53 minutes on our desktop computer.
%As before, we fixed $\eta_3$ at its MLE, which in this case is 5.8879. To ensure that the posterior variances of the variance parameters remain finite, we set the shape parameters for their prior distributions to be large, as done in other cases.  A total of 250,000 MCMC iterations were performed, with the first 100,000 iterations discarded as burn-in. Additionally, to improve mixing, a thinning of width 5 was applied. The entire computation took approximately 12 hours and 53 minutes on our desktop computer.
As before, we fixed $\eta_3$ at its MLE, which in this case is 5.8879, and to ensure finite posterior variances for the variance parameters, we set large shape parameters for their prior distributions.

A total of 250,000 MCMC iterations were performed, discarding the first 100,000 as burn-in. To improve mixing, a thinning interval of 5 was applied. The entire computation took approximately 12 hours and 53 minutes on our desktop computer.
%---------------------------------
\subsubsection{Controlling $M_s$}
\label{Ms control}
%----------------------------------
%Since the locations are distributed very widely over the space, the value of $M_s$ becomes too large for this data for a given $s$. Now $M_s$ appears in the denominator of the variance covariance matrix of the predictive densities of the time series at a particular location (see equation (\ref{eq4})). Therefore, the variability becomes too less. So, to control the variability we modify the definition of $M_s$ as $M_{s} = \exp(c\,\max\{||s^2-u^2||^2: u\in S\})$, where $c$ is a positive small constant. Note that it does not hamper the properties of $M_s$ and hence all the theoretical properties of the processes remain unchanged. The constant $c$ controls the spatial variability which can be thought of as a distance scaling factor. The choice of $c$ was also done by cross-validation. It was found that $c=0.25$ works reasonably well for the sea surface temperature data. 
Since the locations are widely distributed in space, the value of $M_s$ becomes excessively large for a given $s$ in this dataset. As $M_s$ appears in the denominator of the variance-covariance matrix of the predictive densities of the time series at a given location (see Equation~(\ref{eq4})), this leads to an unrealistically low variability. To address this, we redefine $M_s$ as $M_{s} = \exp\left(c\max\left\{||s^2 - u^2||^2 : u \in S\right\}\right),$ where $c$ is a small positive constant. This modification preserves all theoretical properties of $M_s$, ensuring the process characteristics remain unchanged.  %and, consequently, the properties of the process remain unchanged. 
%The constant $c$ acts as a spatial variability control parameter and can be interpreted as a distance scaling factor. 
The constant $c$ serves as a spatial variability control parameter, acting as a distance scaling factor. 
%The optimal choice of $c$ was determined via cross-validation, and it was found that $c = 0.25$ provided reasonable performance for the sea surface temperature data.
The optimal choice of $c$ was determined via cross-validation, with $c = 0.25$ yielding reasonable performance for the sea surface temperature data.
%----------------------------
\subsubsection{Results of the sea surface temperature data}
\label{results of sea temp data}
%----------------------------
Figure~\ref{fig:trace plot of sea temp data} of Appendix~\ref{Sea temp data: trace plot}, representing the MCMC trace plots of the parameters, exhibits no evidence of non-convergence of our MCMC algorithm. 
%The posterior predictive color density plots for the latent variables are shown in Figure \ref{fig:posterior density of latent variables of sea temp data} of Appendix \ref{Sea temp data: latent predictive density}.

\begin{figure}[!h]
	\centering
	\includegraphics[trim={0cm 0cm 0cm 0cm},width=0.75\textwidth,height=12cm,keepaspectratio,clip]{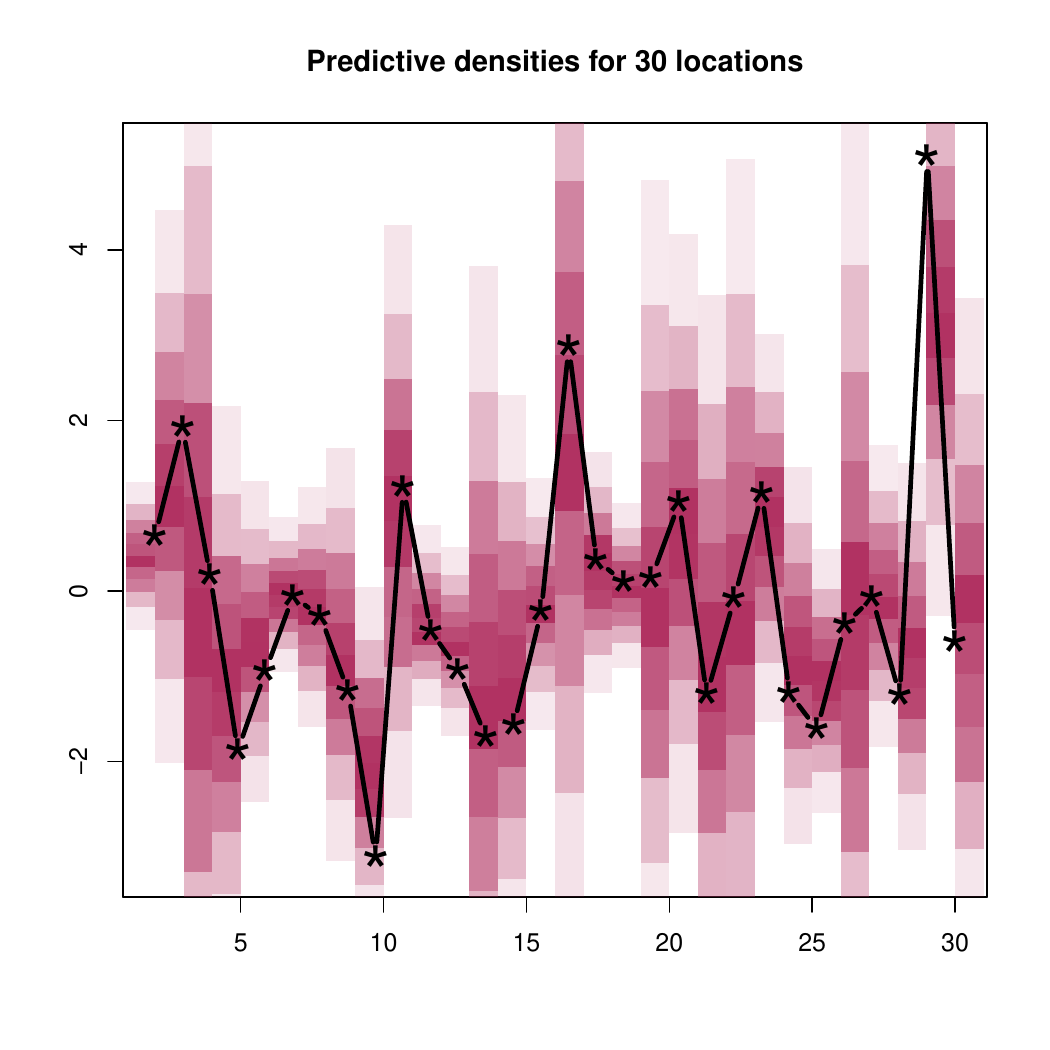}
	\caption{The highest probability density regions of the predictive densities and the true values of sea surface temperature data for the 100th month at 30 locations are depicted in this figure. The x-axis of of the figure indicates the locations, and at each location, color-coded densities are plotted in the vertical axis. The intensity of the color is proportional to the density. The true values are indicated by the black stars. Majority of the true values (average referenced) fall in the high density regions.}
	\label{fig: hpd predictive density for 30 locations Sea temp data}
\end{figure}

%Next, we provide the posterior predictive densities at the 100th month for each of the 30 locations. The algorithm for obtaining the temporal predictive densities is given in the Appendix \ref{temporal prediction}. The highest predictive density regions are given for 30 locations and are displayed in Figure \ref{fig: hpd predictive density for 30 locations Sea temp data}. All the true values fall in the high probability density region. Further, the individual predictive density plots are shown in Figure \ref{fig: predictive density for 30 locations sea temp data} of Appendix \ref{Trace plt and posterior plts for Sea temp data}. As the plots indicate, the posterior predictive densities for the future time point correctly contains the true values for all of these locations. The results encourage us to use the proposed model for the future time point prediction  for a non-stationary spatio-temporal data. 
Next, we provide the posterior predictive densities at the 100th month for each of the 30 locations, following the algorithm of temporal predictive densities given in the Appendix \ref{temporal prediction}. The highest predictive density regions for these 30 locations are shown in Figure~\ref{fig: hpd predictive density for 30 locations Sea temp data}, where most of the true values fall in the high probability density regions. %Further, the individual predictive density plots are displayed in Figure~\ref{fig: predictive density for 30 locations sea temp data} of Appendix~\ref{Trace plt and posterior plts for Sea temp data}. 
The individual predictive densities along with the 95\% predictive intervals indicate the posterior predictive densities correctly contain the true values at all locations (plots are not shown for brevity). The average length of the 95\% predictive intervals is 8.3161 (standard deviation = 4.8172). %These results support the effectiveness of our proposed model for future time point predictions in non-stationary spatio-temporal data. 

\begin{figure}[!ht]
%	\begin{subfigure}{.5\textwidth}
	\centering
	\includegraphics[trim={0.25cm 0cm 0cm 0cm},width=0.45\textwidth,height=6.5cm, keepaspectratio, clip]{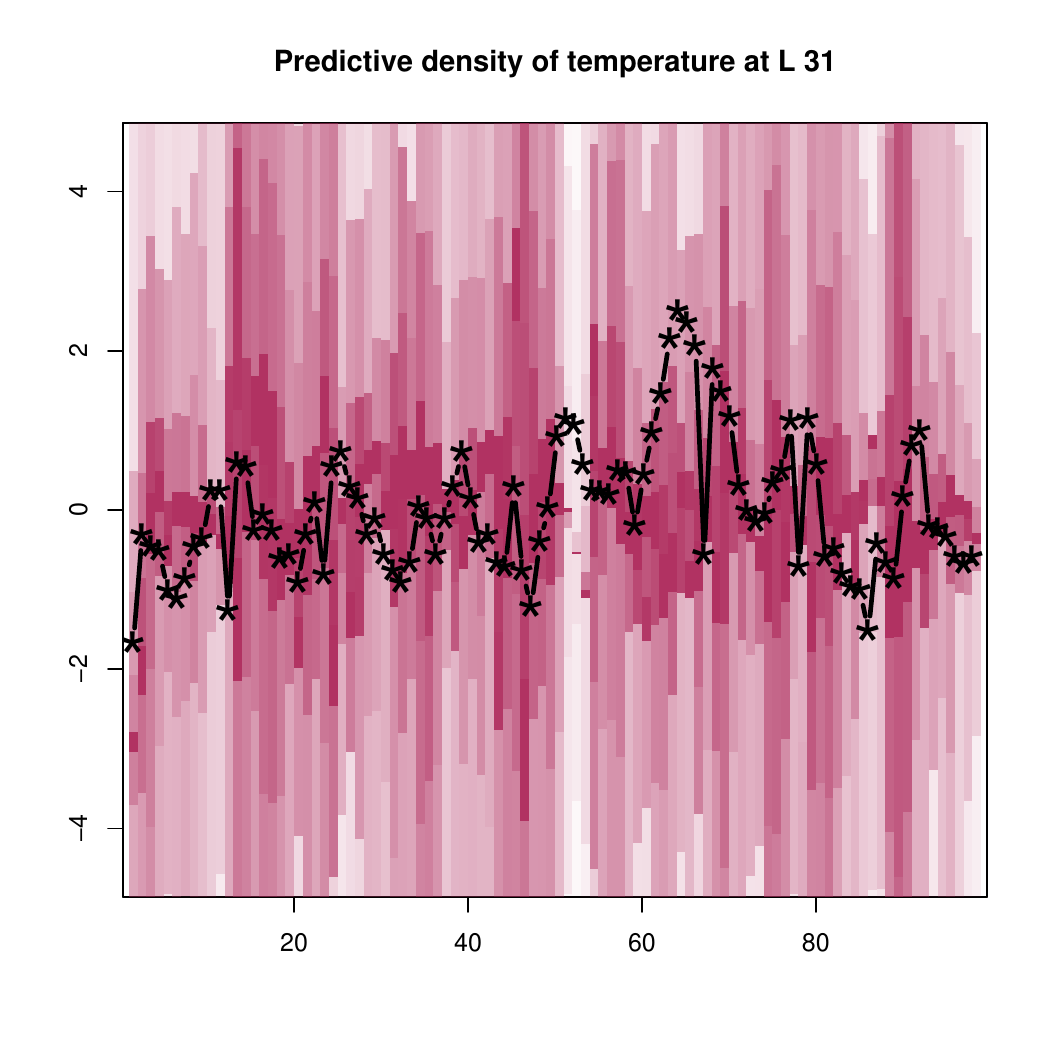}
	\includegraphics[trim={0.25cm 0cm 0cm 0cm},width=0.45\textwidth,height=6.5cm, keepaspectratio, clip]{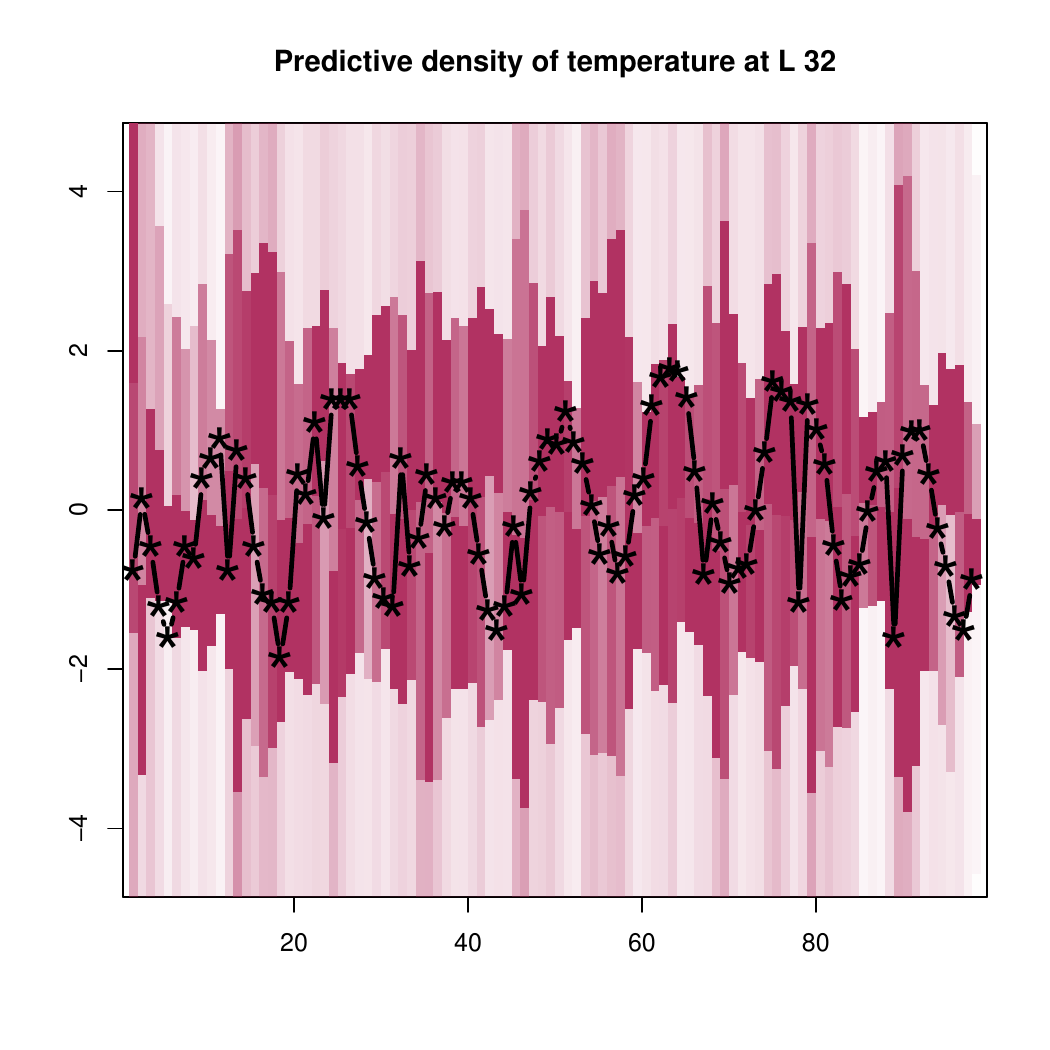}
	\includegraphics[trim={0.25cm 0cm 0cm 0cm},width=0.45\textwidth,height=6.5cm,keepaspectratio, clip]{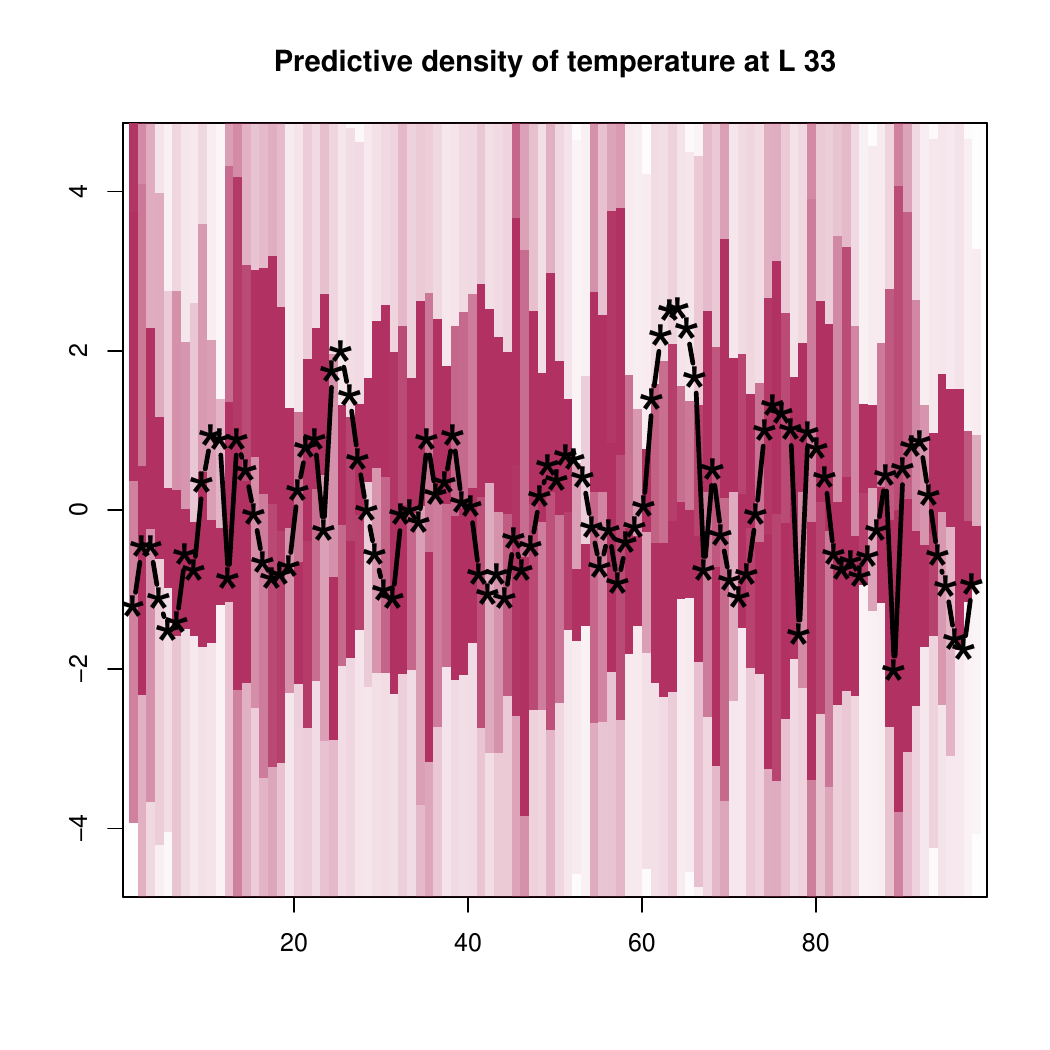}
	\includegraphics[trim={0.25cm 0cm 0cm 0cm},width=0.45\textwidth,height=6.5cm,keepaspectratio, clip]{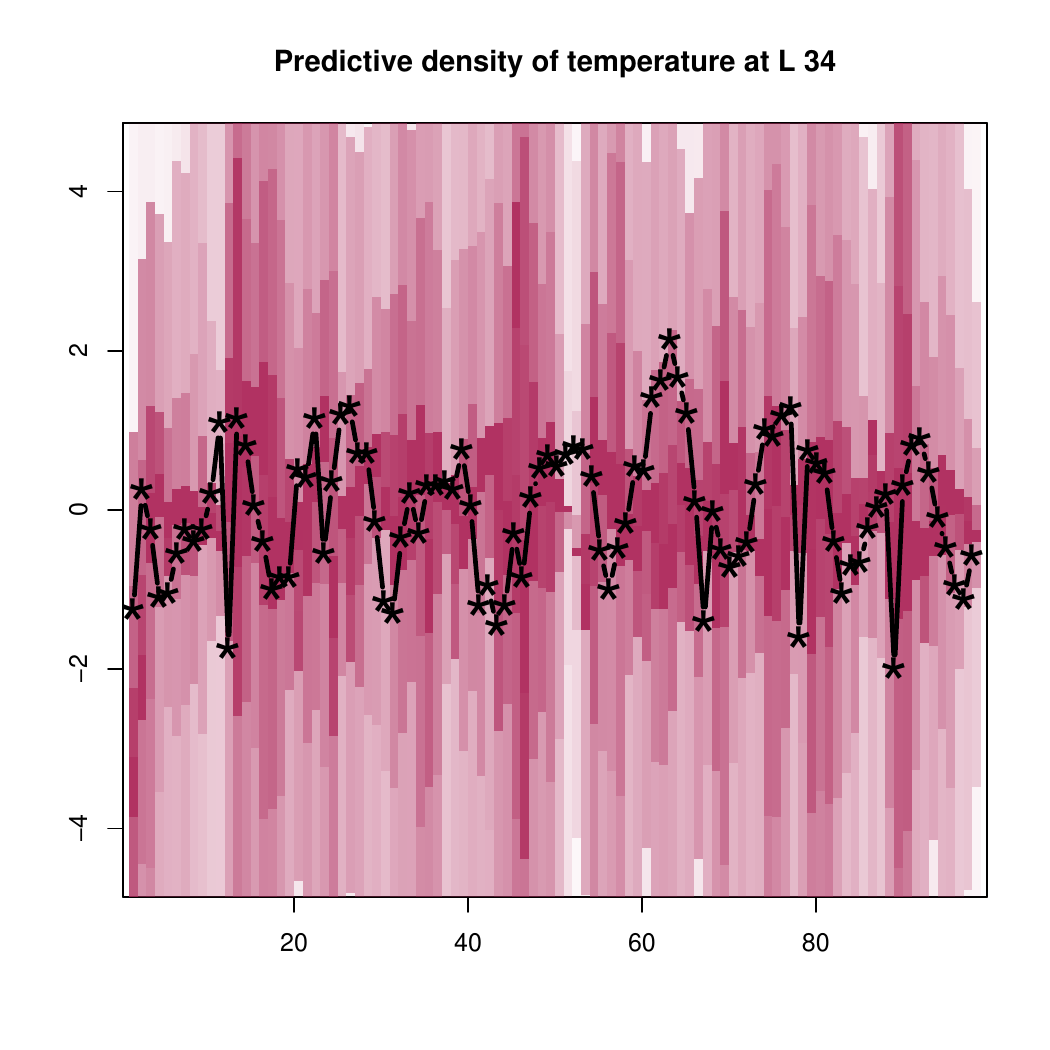}
	\includegraphics[trim={0.25cm 0cm 0cm 0cm},width=0.45\textwidth,height=6.5cm,keepaspectratio, clip]{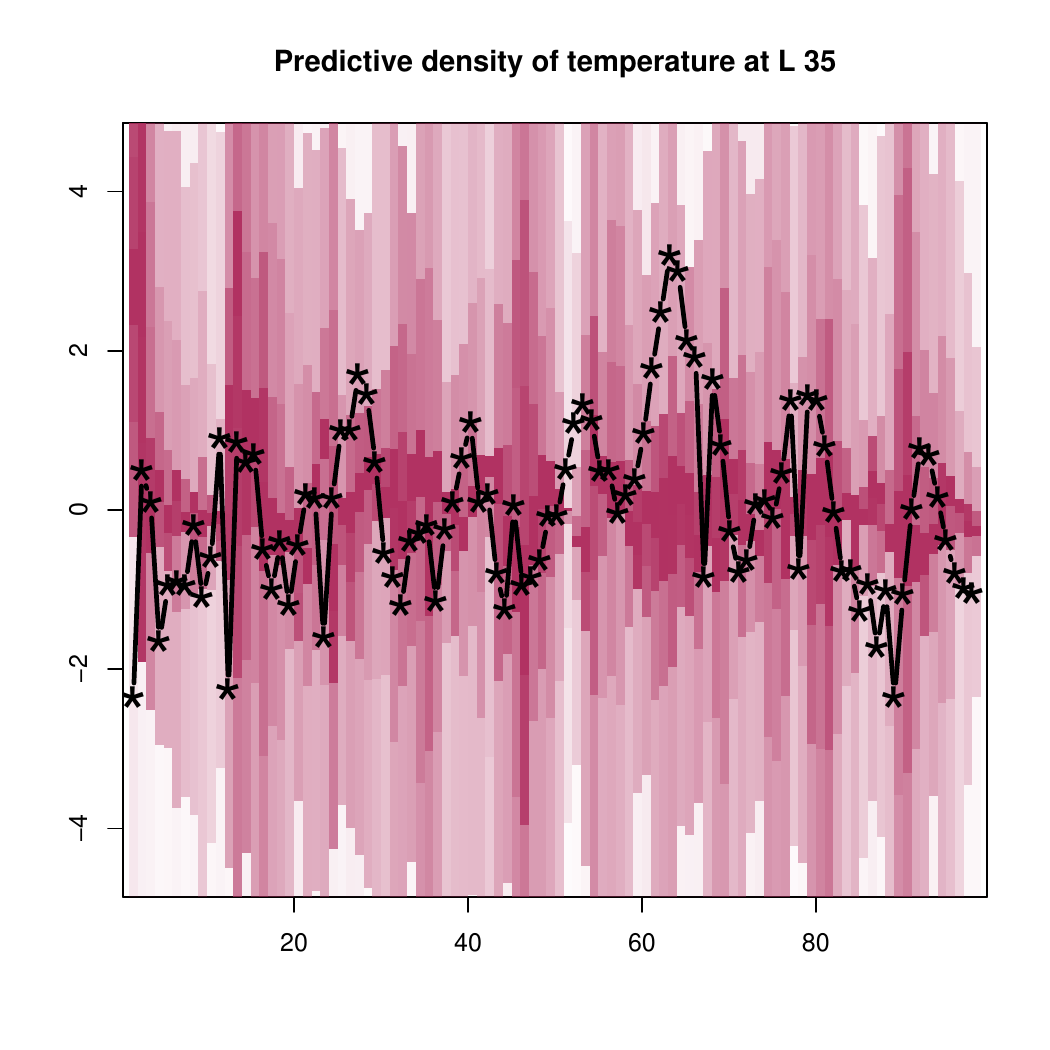}	
	\includegraphics[trim={0.25cm 0cm 0cm 0cm},width=0.45\textwidth,height=6.5cm,keepaspectratio, clip]{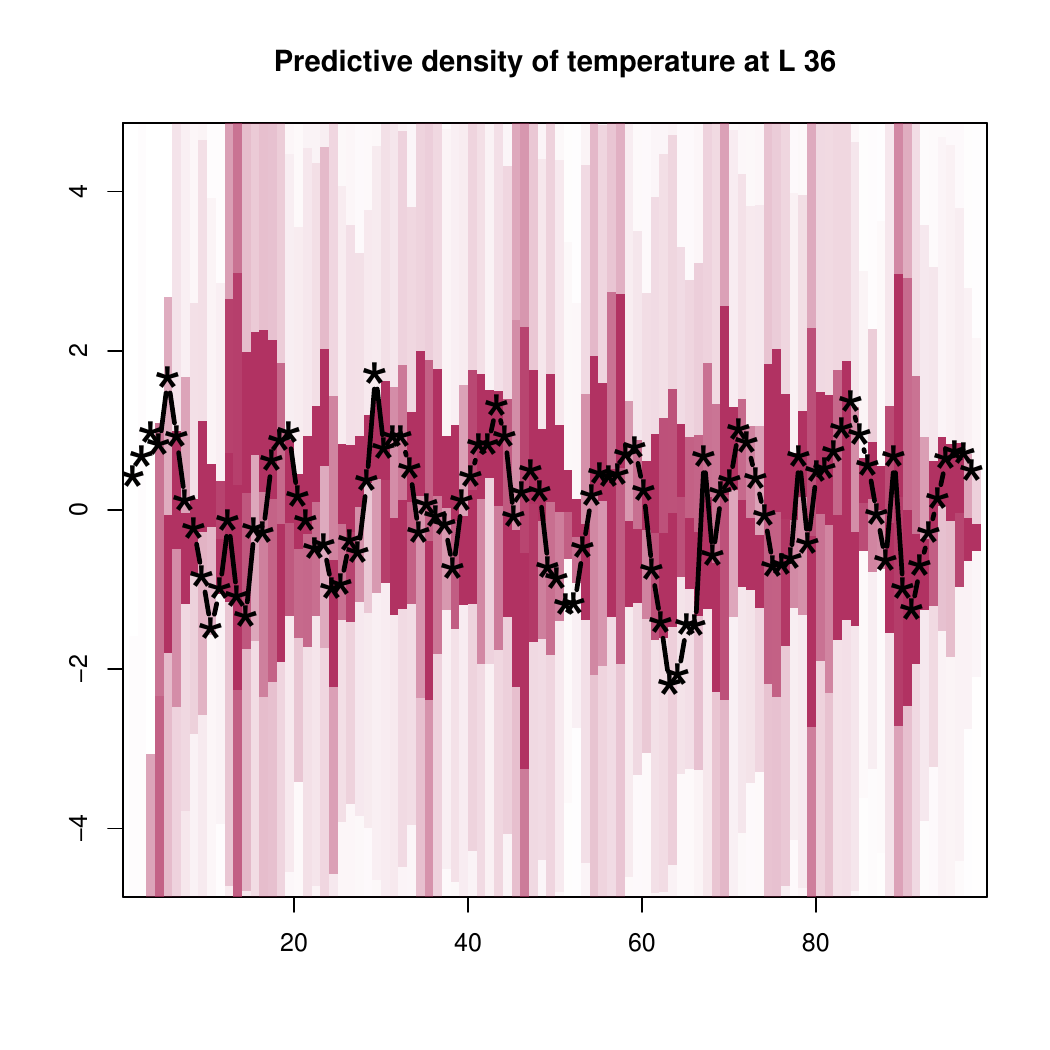}
	\caption{Posterior predictive densities of the reconstructed time series of sea surface temperature data (average referenced) at 6 locations. 
		Higher the intensity of the color, higher is the probability density. The black stars represent the true temperature values (average referenced). 
		Most of the true values fall within the high density regions.}		
	\label{fig:posterior predictive density of temp time series of 6 locations sea temp data}
\end{figure}

\begin{figure}[!h]
    \centering
    \includegraphics[trim={0.25cm 0cm 0cm 0cm},width=0.45\textwidth,height=6.5cm,keepaspectratio, clip]{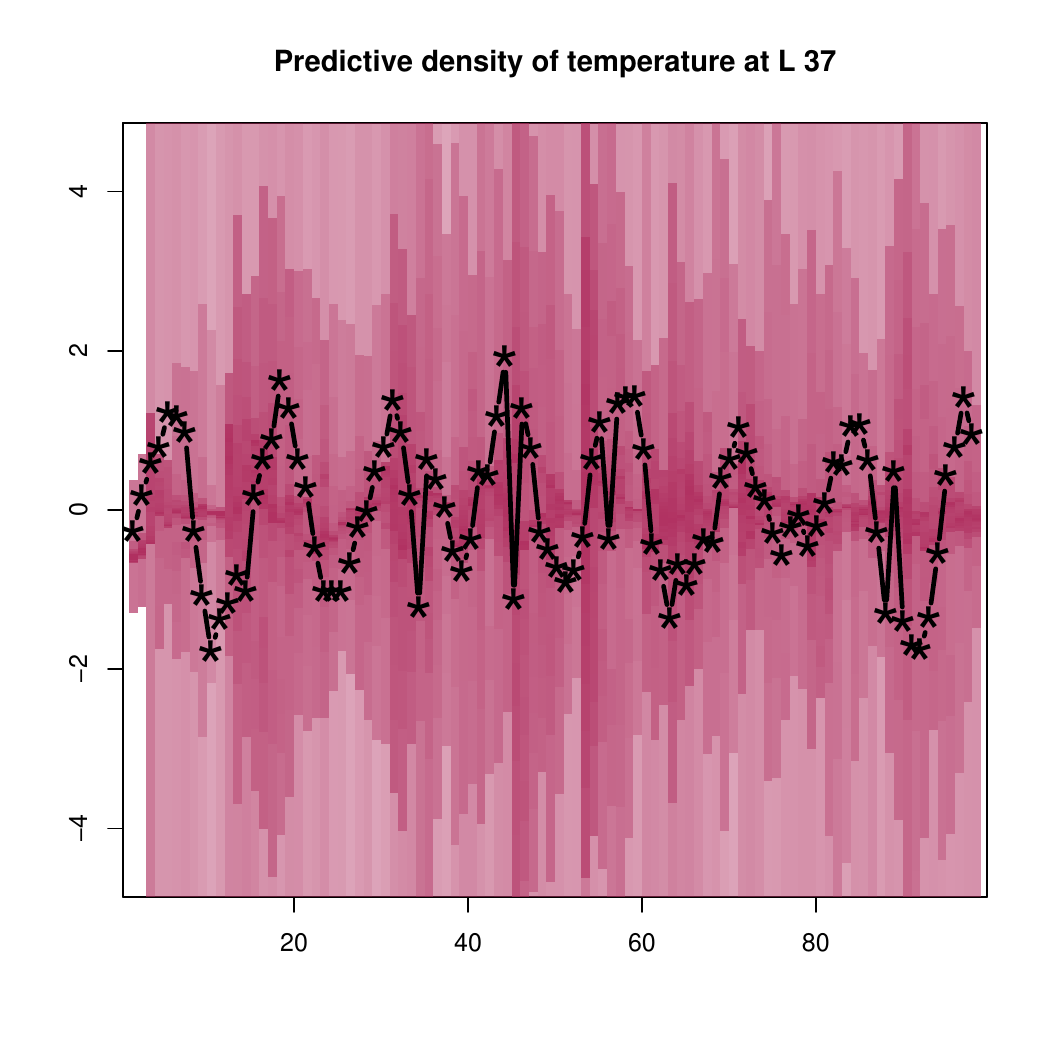}
	\includegraphics[trim={0.25cm 0cm 0cm 0cm},width=0.45\textwidth,height=6.5cm,keepaspectratio, clip]{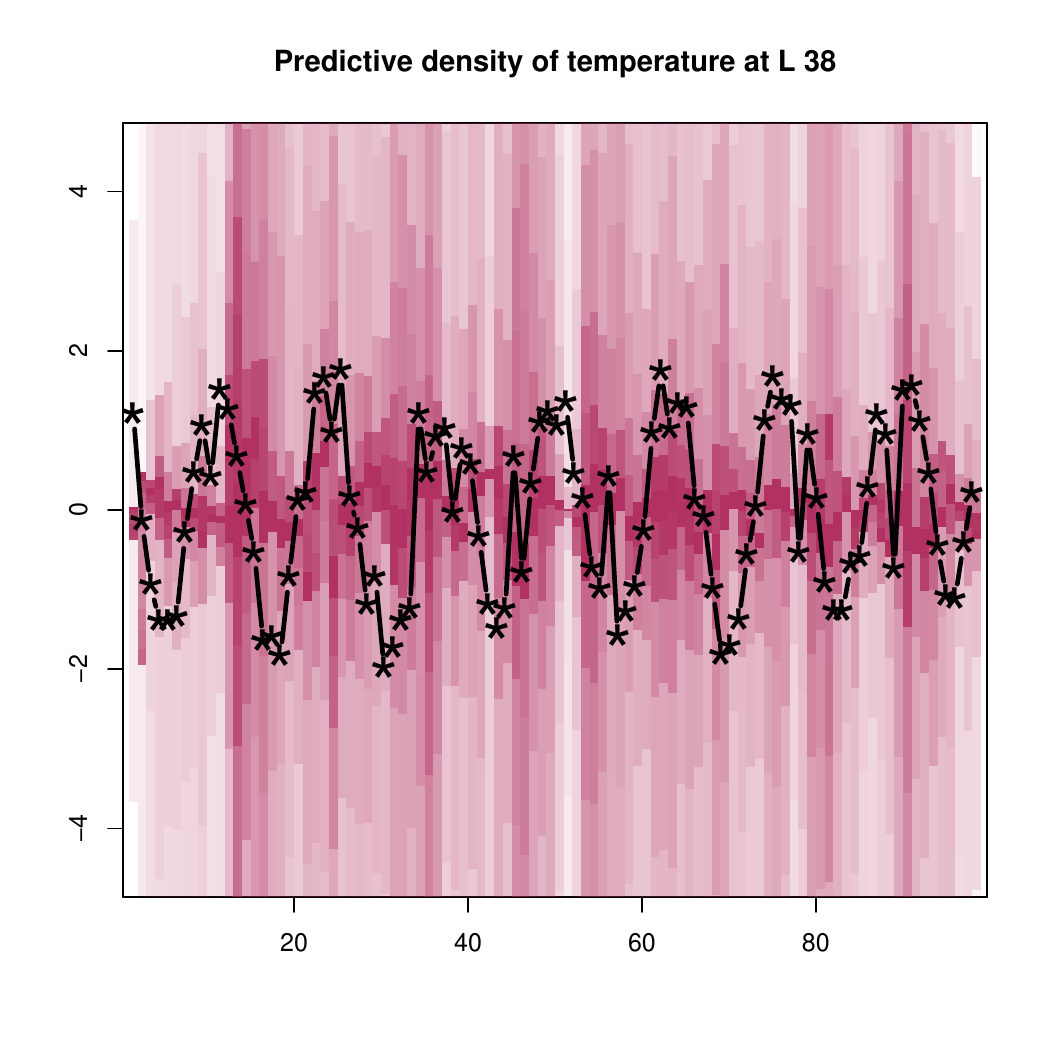}
	\includegraphics[trim={0.25cm 0cm 0cm 0cm},width=0.45\textwidth,height=6.5cm,keepaspectratio, clip]{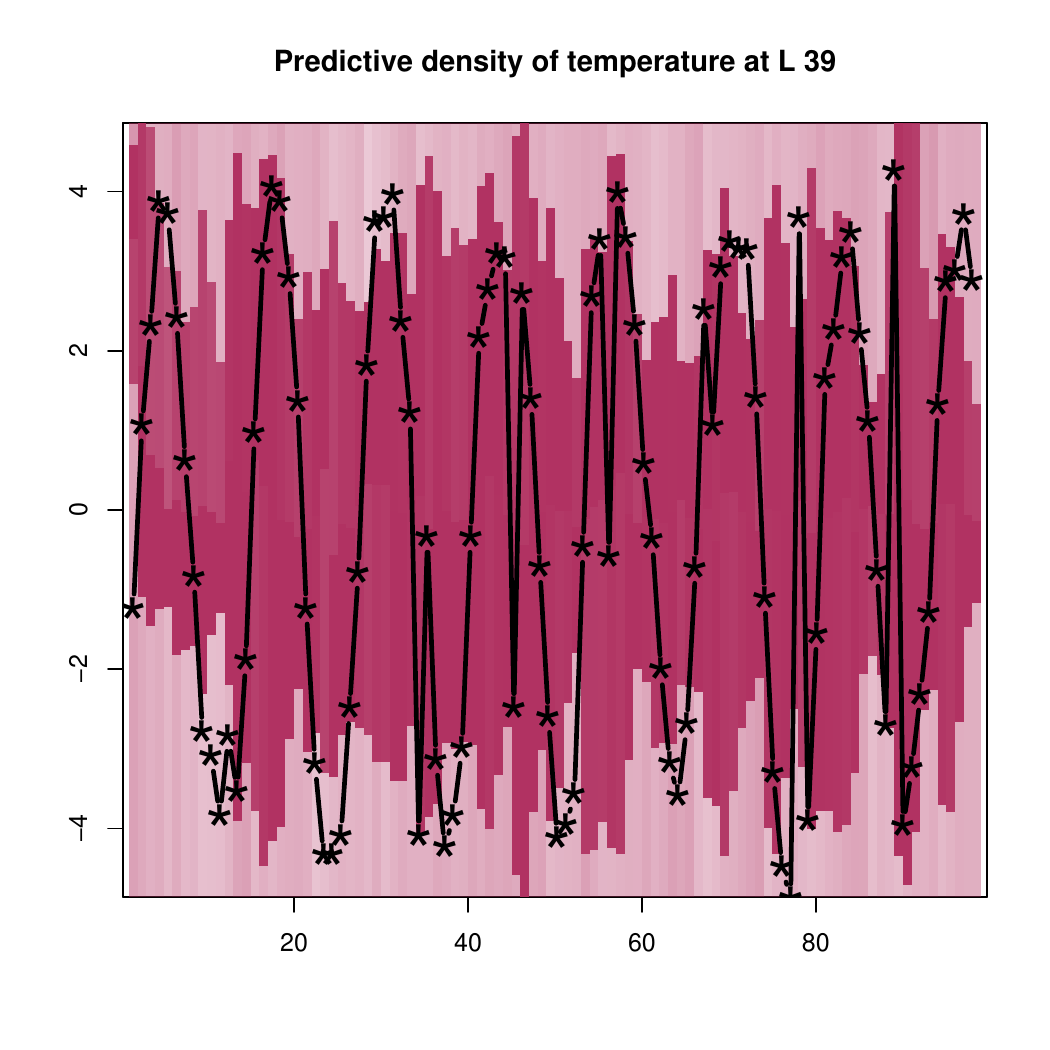}
	\includegraphics[trim={0.25cm 0cm 0cm 0cm},width=0.45\textwidth,height=6.5cm,keepaspectratio, clip]{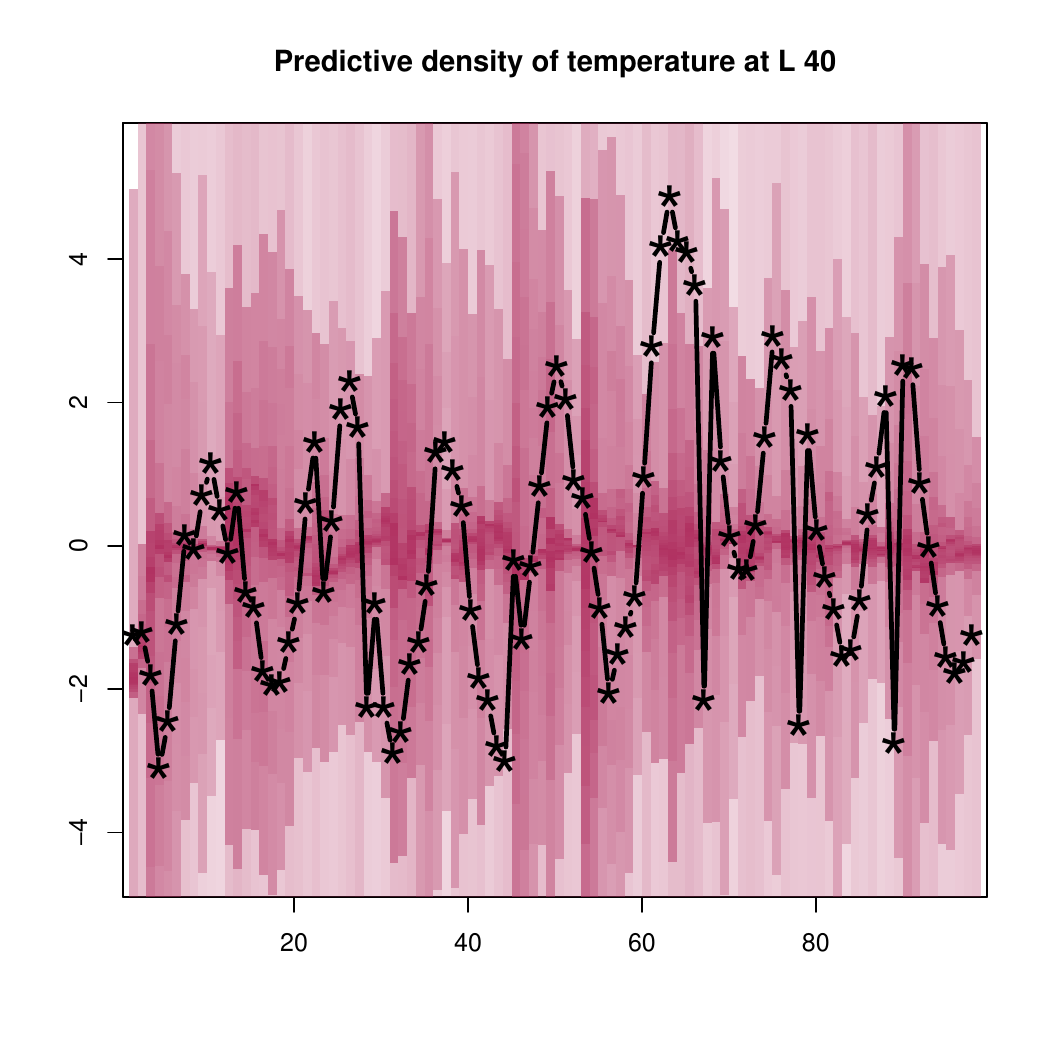}	
    \caption{Posterior predictive densities of the reconstructed time series of sea surface temperature data (average referenced) at another 4 locations. 
		Higher the intensity of the color, higher is the probability density. The black stars represent the true temperature values (average referenced). 
		Most of the true values fall within the high density regions.}
    \label{fig:posterior predictive density of temp time series of 4 locations sea temp data}
\end{figure}

%Another important aspect of the spatio-temporal modeling is to predict the complete time series at the given locations. 
%Similar to the other real data analyses, here also we predict at various spatial locations.As mentioned in Section~\ref{sea temp}, we kept aside the complete temporal observations for the ten locations for evaluating the performance of the proposed model. We obtained the posterior predictive densities for each of the 99 time points for a given location (see \ref{spatial prediction} for the algorithm). The highest density regions of the predictive densities are depicted in Figure~\ref{fig:posterior predictive density of temp time series of 10 locations sea temp data}. Clearly, most of the true values, indicated by black stars in Figure~\ref{fig:posterior predictive density of temp time series of 10 locations sea temp data}, fall within the high probability density regions. 
%Similar to the other real data analyses, we also perform predictions at various spatial locations in this study. 
Similar to previous real data analyses, we perform spatial predictions in this study.
As outlined in Section~\ref{sea temp}, we set aside the complete temporal observations for ten locations to evaluate our model's performance. For each of these locations, we obtained the posterior predictive densities for all 99 time points (see Appendix~\ref{spatial prediction} for the algorithm). The highest density regions of the predictive densities are depicted in Figures~\ref{fig:posterior predictive density of temp time series of 6 locations sea temp data} and \ref{fig:posterior predictive density of temp time series of 4 locations sea temp data}. As evident from the figure, most of the true values fall in the high probability density regions, further validating the effectiveness of our model.
%As one can see, except for the two locations, which are represented by $L_{36}$ and $L_{39}$, most of the true values, indicated by black stars in Figure~\ref{fig:posterior predictive density of temp time series of 10 locations sea temp data}, fall well within the high probability density regions. Next, we give a plausible explanation for the poor performance at the locations $L_{36}$ and $L_{39}$.

\section{Summary and conclusion}
\label{sec:conclusion}
Although Hamiltonian equations are well-known in physics, their role in statistics has primarily been limited to Hamiltonian Monte Carlo for approximate posterior simulations. However, given their success in phase-space modeling, it is natural to anticipate their potential in spatio-temporal statistics if properly utilized. This key insight motivated us to develop a novel spatio-temporal model based on the leap-frog algorithm applied to a suitably modified set of Hamiltonian equations, where stochasticity is induced through appropriate Gaussian processes.  

Our marginal stochastic process is \emph{nonparametric, non-Gaussian, non-stationary, and non-separable}, with a well-structured dynamic temporal component and continuous-time treatment. Additionally, the lagged correlations between observations decay to zero as the space-time lag approaches infinity, making our process more realistic compared to existing spatio-temporal non-stationary models. Simulation studies demonstrate that our model consistently outperforms the widely used latent process spatio-temporal modeling approach based on non-stationary Gaussian processes. Furthermore, the model's flexibility and applicability are confirmed by its strong performance on two real datasets. The continuity and smoothness properties further enhance the elegance of our proposed process.  

{In all simulations and real data analyses, we assume evenly spaced time points. However, our model can be readily adapted to unevenly spaced time points with minimal modifications. We propose such a modification in Section \ref{Modification for irregularly spaced data} of Appendix. We shall elaborate with simulation and real data analyses in a later work.} 
%Specifically, unevenly spaced observations can be handled by treating certain time points as missing values, which can be imputed during the MCMC process using predictive densities. For clarity, suppose data are observed at time points $t_1$, $t_1+\delta$, $t_1+3\delta$, and $t_1+5\delta$, where $\delta > 0$. The analysis would then consider the sequence $t_1, t_1+\delta, t_1+2\delta, t_1+3\delta, t_1+4\delta, t_1+5\delta$, treating $t_1+2\delta$ and $t_1+4\delta$ as missing values, which are simulated during MCMC iterations using the predictive densities.  

In this study, we have not focused on analyzing particularly large datasets. In future work, we plan to extend our approach to massive datasets using basis function representations and a transdimensional MCMC technique known as TTMCMC, introduced by \cite{das2019transdimensional}.
%=====================================
\section*{Acknowledgment}
%======================================
The authors are thankful to Dr. Moumita Das and Dr. Suman Guha for their valuable remarks, which significantly improved the presentation of the manuscript. %The authors also thank ChatGPT which is solely used for editing purpose. 
During the preparation of this work the authors used ChatGPT in order to improve the English writing solely. After using this tool, the authors reviewed and edited the content as needed and take full responsibility for the content of the publication.

\appendix
%==========================================
\section{Derivation of equations \ref{observation equation} and \ref{latent equation}}
\label{Modified leap-frog}
%------------------------------------------
Here we show how the equations \ref{observation equation} and \ref{latent equation} are obtained using Leap-frog algorithm on modified Hamiltonian equation \ref{eq: proposed stochastic differential equations}. From equation \ref{eq: proposed stochastic differential equations} we have $\frac{dy(s,t)}{dt} = \beta^*y(s,t) + M_s^{-1}x(s,t)$ and applying Leap-frog (\cite{young2014leapfrog}) on $y(s,\cdot)$, we get
\begin{align}
	\label{eq2: modified eq for theta}
	y(s,t+\delta t) &= y(s,t)+ \beta^* y(s,t)\delta t + \delta t M_s^{-1} x(s,t+\delta t/2) \notag \\
	&= y(s,t)\left(1+ \beta^* \delta t\right) +  \delta t M_s^{-1} x(s,t+\delta t/2) \notag \\
	& = \beta y(s,t) +  \delta t M_s^{-1} x(s,t+\delta t/2),
\end{align}
where $\beta =\left(1+ \beta^* \delta t\right).$ We set $|\beta|<1$, which implies $-2<\beta^* \delta t<0$. 
%As we will see, this restriction ensures that lagged correlations between observations decay to zero as the space-time lag approaches infinity. 
From equation \ref{eq: proposed stochastic differential equations}, we obtain $\frac{dx(s,t)}{dt} = \alpha^* x(s,t) - \nabla V(y(s,t))$, and a numerical approximation for $x(s,\cdot)$ is given as:
	\begin{align}
		\label{eq1: modified eq for p}
		x\left(s,t+\frac{\delta t}{2}\right) &= x(s,t)+ \alpha^* x(s,t)\frac{\delta t }{2} - \nabla V(y(s,t))\frac{\delta t}{2} \notag \\
		&= x(s,t)\left(1+ \alpha^*\frac{\delta t}{2}\right) - \nabla V(y(s,t))\frac{\delta t}{2} \notag \\
		& = \alpha x(s,t) - \nabla V(y(s,t))\frac{\delta t}{2},
	\end{align}
	where $\alpha = \left(1+ \alpha^*\delta t/2\right)$. We set $|\alpha|<1$ implying $ -2<\alpha^* \delta t/2 < 0$. 
%	Restricting $\alpha$ on $(-1,1)$ led to good MCMC mixing in our Bayesian applications. 
    Replacing equation (\ref{eq1: modified eq for p}) into equation (\ref{eq2: modified eq for theta}), we obtain equation \ref{observation equation} of Section \ref{Modified HM and leap frog} as 
	\begin{align*}
		%\label{eq4: Leap-frog for theta}
		\theta(t+\delta t) &= \beta \theta(t) + \delta t M^{-1} \left(\alpha p(t) - \nabla V(\theta(t))\frac{\delta t}{2}\right).
	\end{align*}	
	Again, invoking leap-frog algorithm for $x(s,\cdot)$ \citep{young2014leapfrog}, we get \ref{latent equation} of Section \ref{Modified HM and leap frog} as
	\allowdisplaybreaks
	\begin{align*}
		%\label{eq3: Leap-frog form of p}
		x(s,t+\delta t) & = x\left(s,t+\frac{\delta t}{2}\right) + \alpha^* x\left(s,t+\frac{\delta t}{2}\right) \frac{\delta t}{2} - \nabla V\left(y(s,{t+\delta t})\right) \frac{\delta t}{2} \notag \\ 
		& = x\left(s,t+\frac{\delta t}{2}\right)\left(1+ \alpha^* \frac{\delta t}{2}\right) - \nabla V\left(y(s,{t+\delta t})\right) \frac{\delta t}{2}  \notag \\
		& = \alpha x\left(s,t+\frac{\delta t}{2}\right) - \nabla V\left(y(s,{t+\delta t})\right) \frac{\delta t}{2} \notag \\
		& = \alpha \left(\alpha x(s,t) - \nabla V(y(s,t))\frac{\delta t}{2}\right) - \nabla V\left(y(s,{t+\delta t})\right) \frac{\delta t}{2} \text{ (using equation \ref{eq1: modified eq for p})}\notag \\
		& = \alpha^2 x(s,t) - \frac{\delta t}{2} \left\{\alpha \nabla V(y(s,t)) + \nabla V(y(s,{t+\delta t})) \right\}. 
	\end{align*}
	%where the fourth equality follows from equation (\ref{eq1: modified eq for p}).
	%Equations (\ref{eq4: Leap-frog for theta}) and(\ref{eq3: Leap-frog form of p}) constitute the modified leap-frog equations and are the key ingredients of our proposed spatio-temporal process.
   % Equations (\ref{eq4: Leap-frog for theta}) and(\ref{eq3: Leap-frog form of p}) form the modified leap-frog equations, serving as the core framework of our proposed spatio-temporal process.
%========================================
\section{Correlation analysis of GQN model}
\label{Correlation analysis of GQN model}
\begin{figure}[!h]
\includegraphics[width=0.45\textwidth,height=6.5cm]{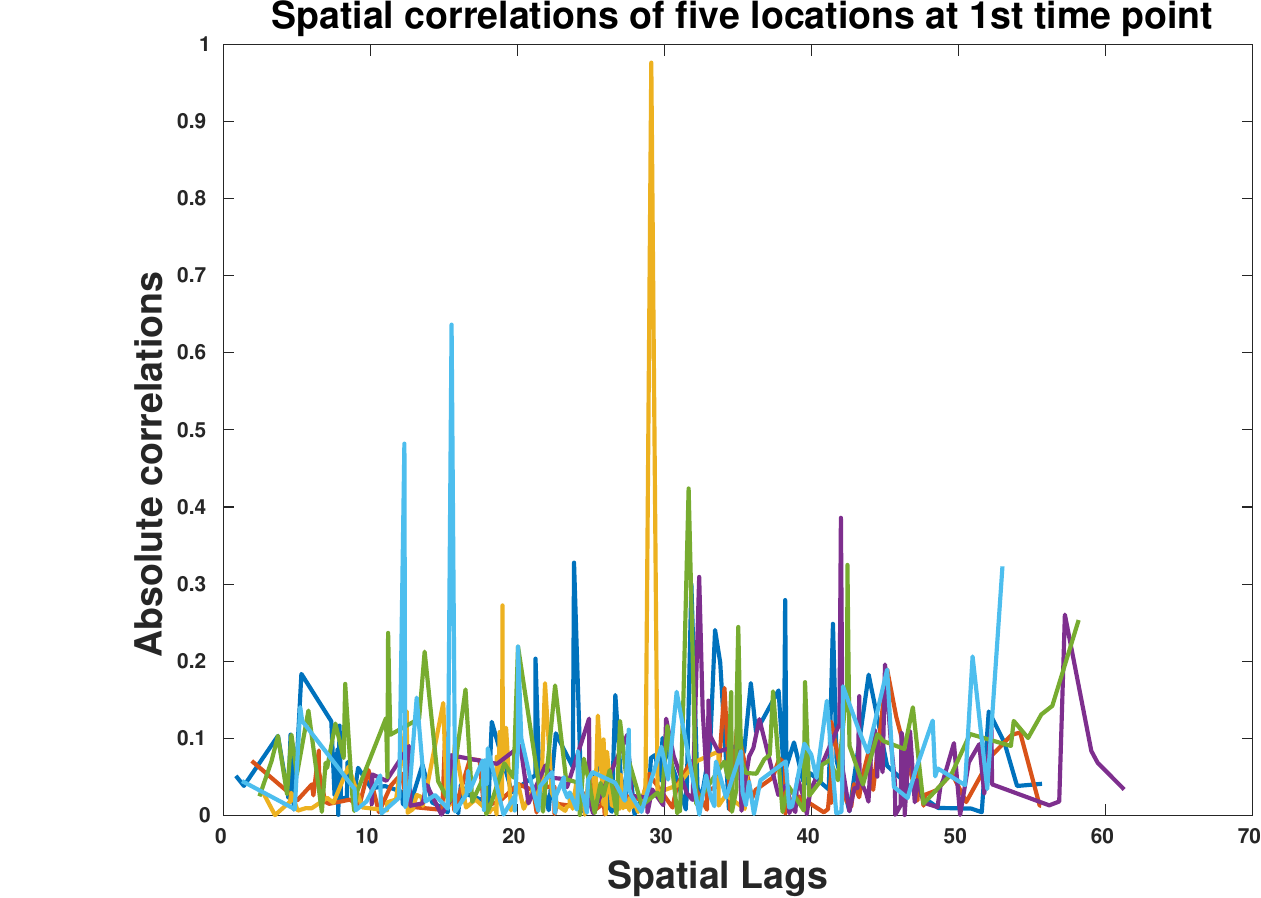}
\includegraphics[width=0.45\textwidth,height=6.5cm]{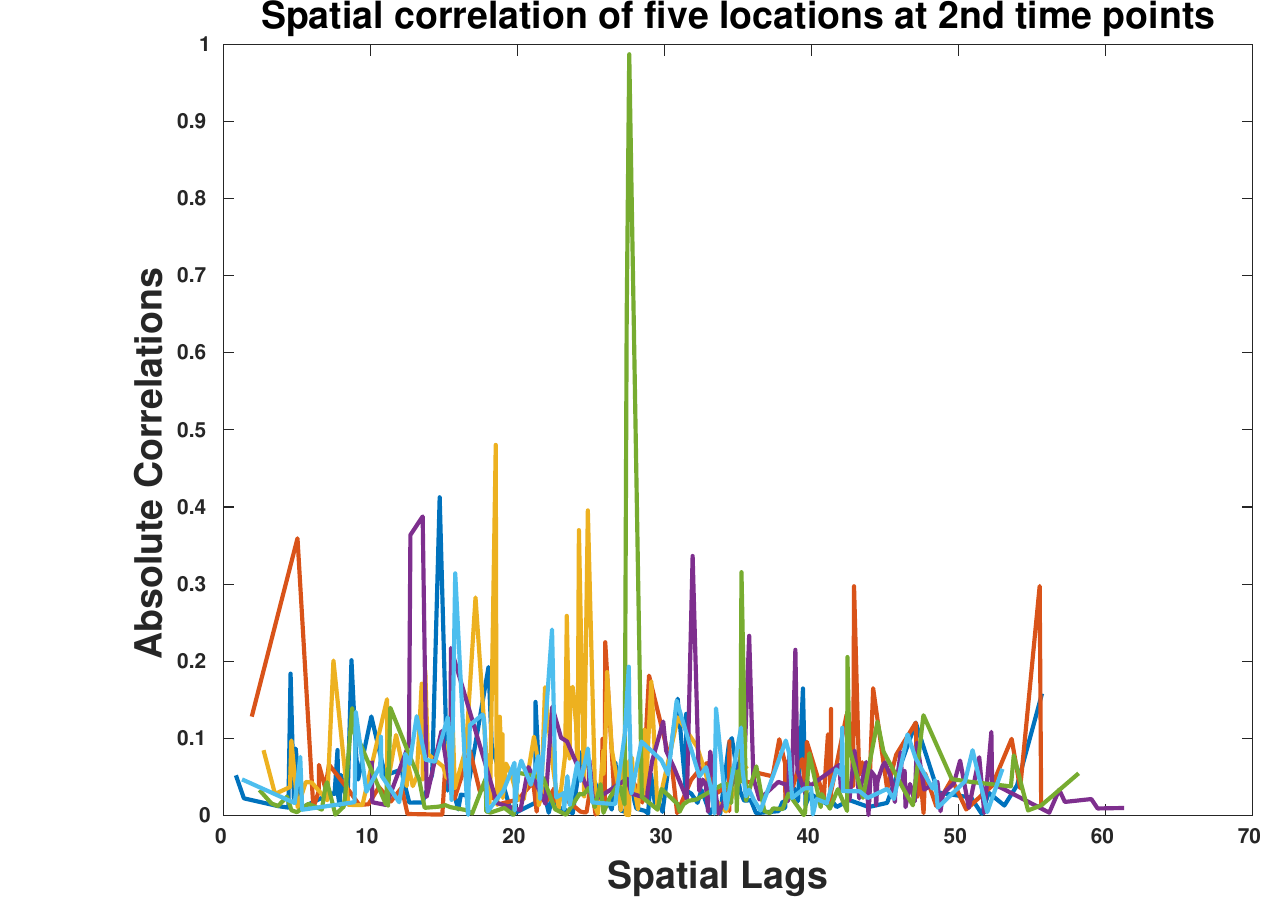}
\includegraphics[width=0.45\textwidth,height=6.5cm]{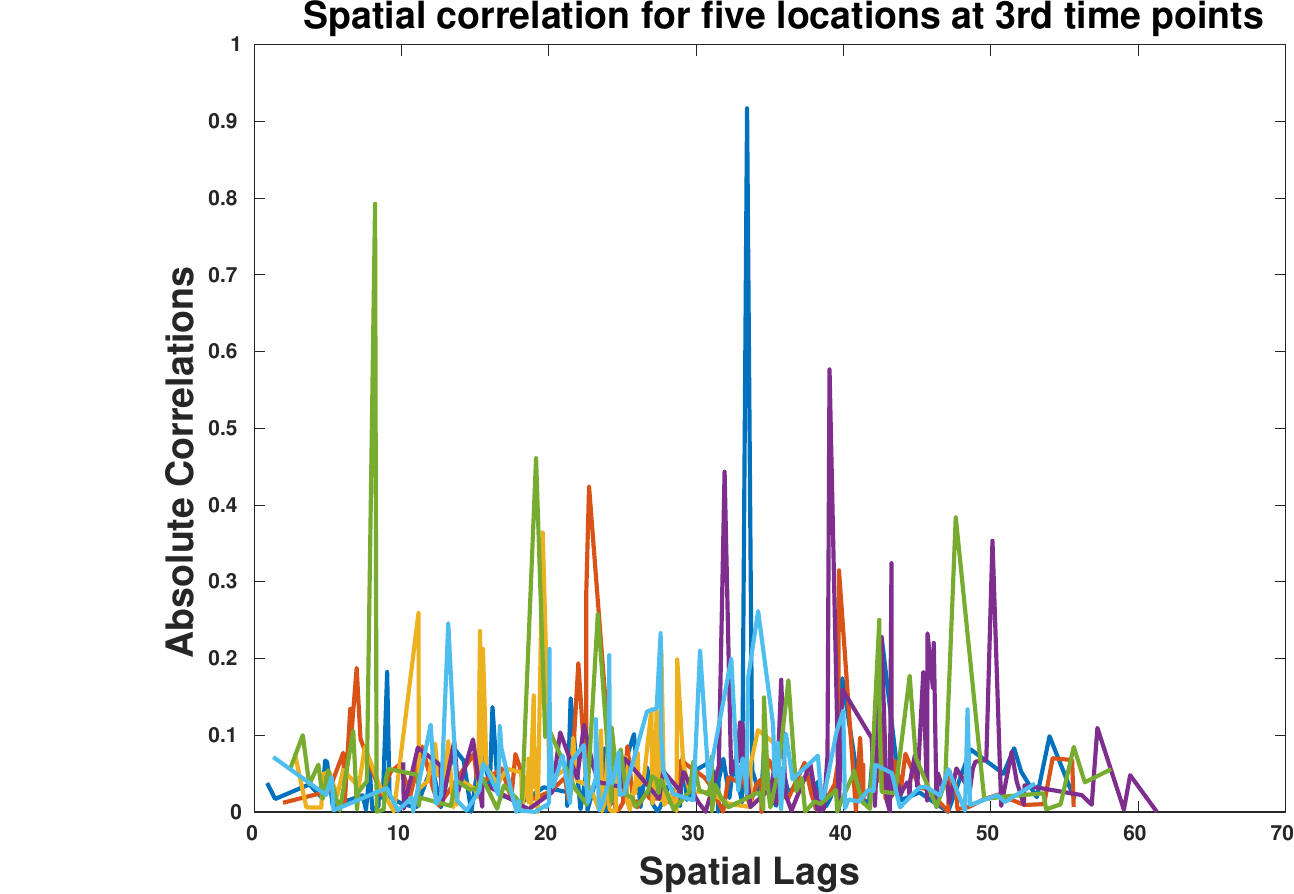}
\includegraphics[width=0.45\textwidth,height=6.5cm]{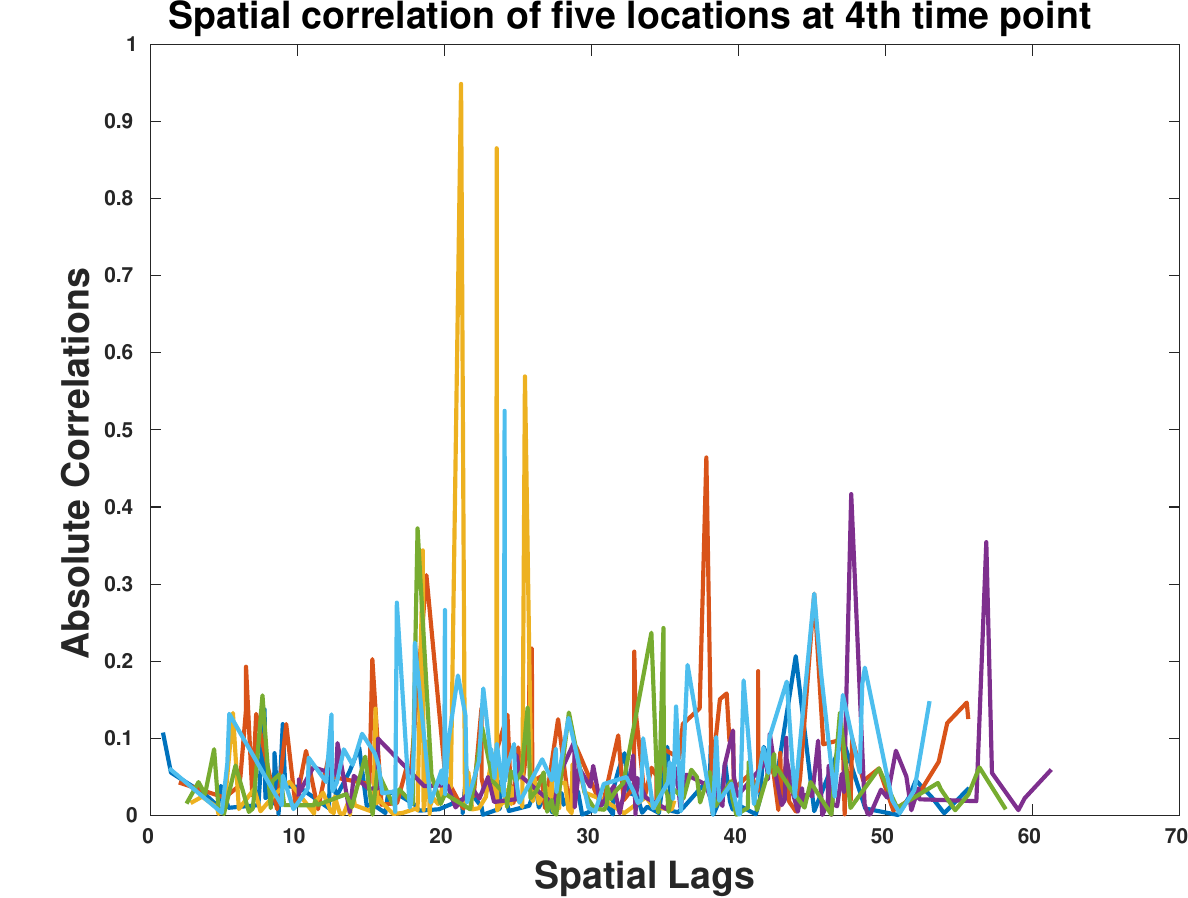}
\centering
\includegraphics[width=0.45\textwidth,height=6.5cm]{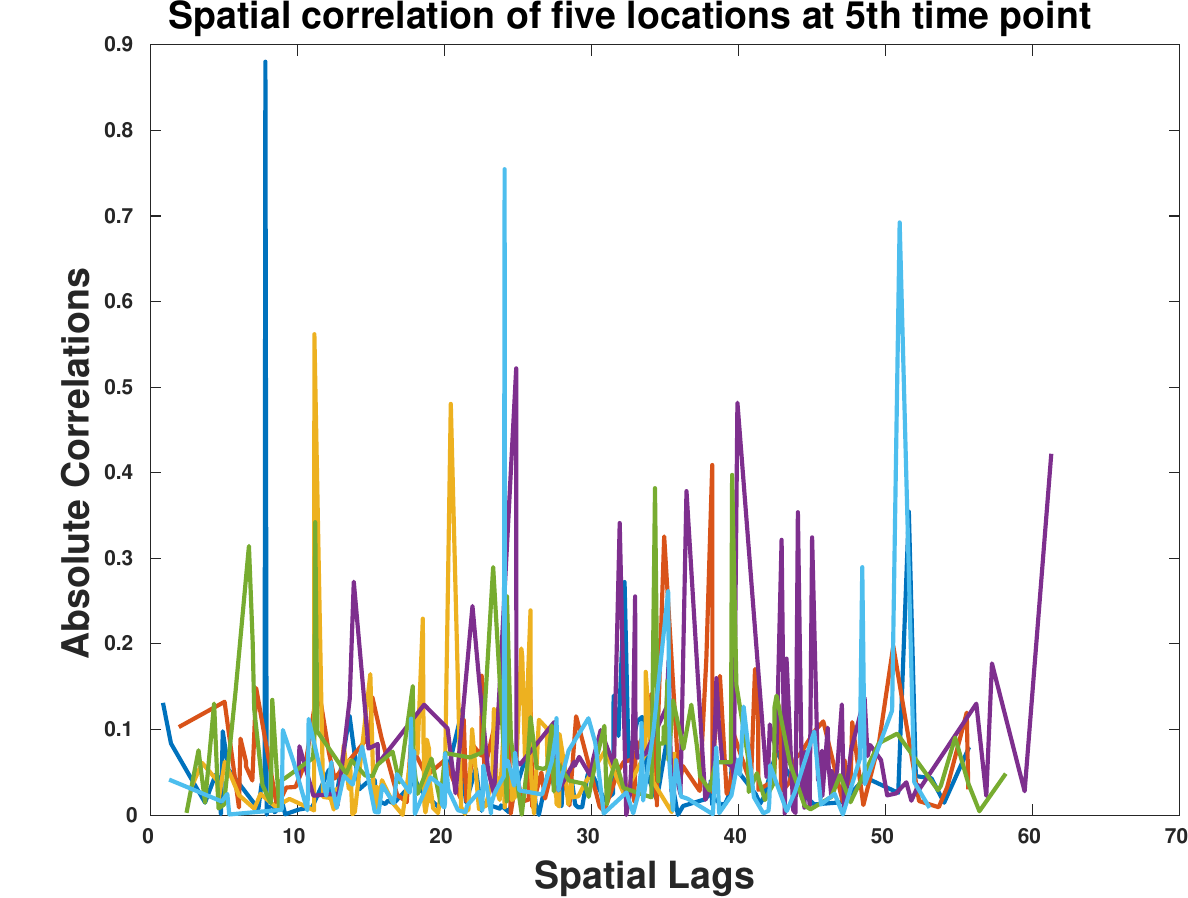}
    \caption{Spatial correlation plots at 5 times points using observations simulated from the GQN model are depicted. x-axis and y-axis denote the spatial lags and the correlations, respectively. The five different colors indicate 5 locations in each plot.}
    \label{Spatial correlation plot for GQN model}
\end{figure}

\begin{figure}[!h]
\includegraphics[width=0.45\textwidth,height=6.5cm]{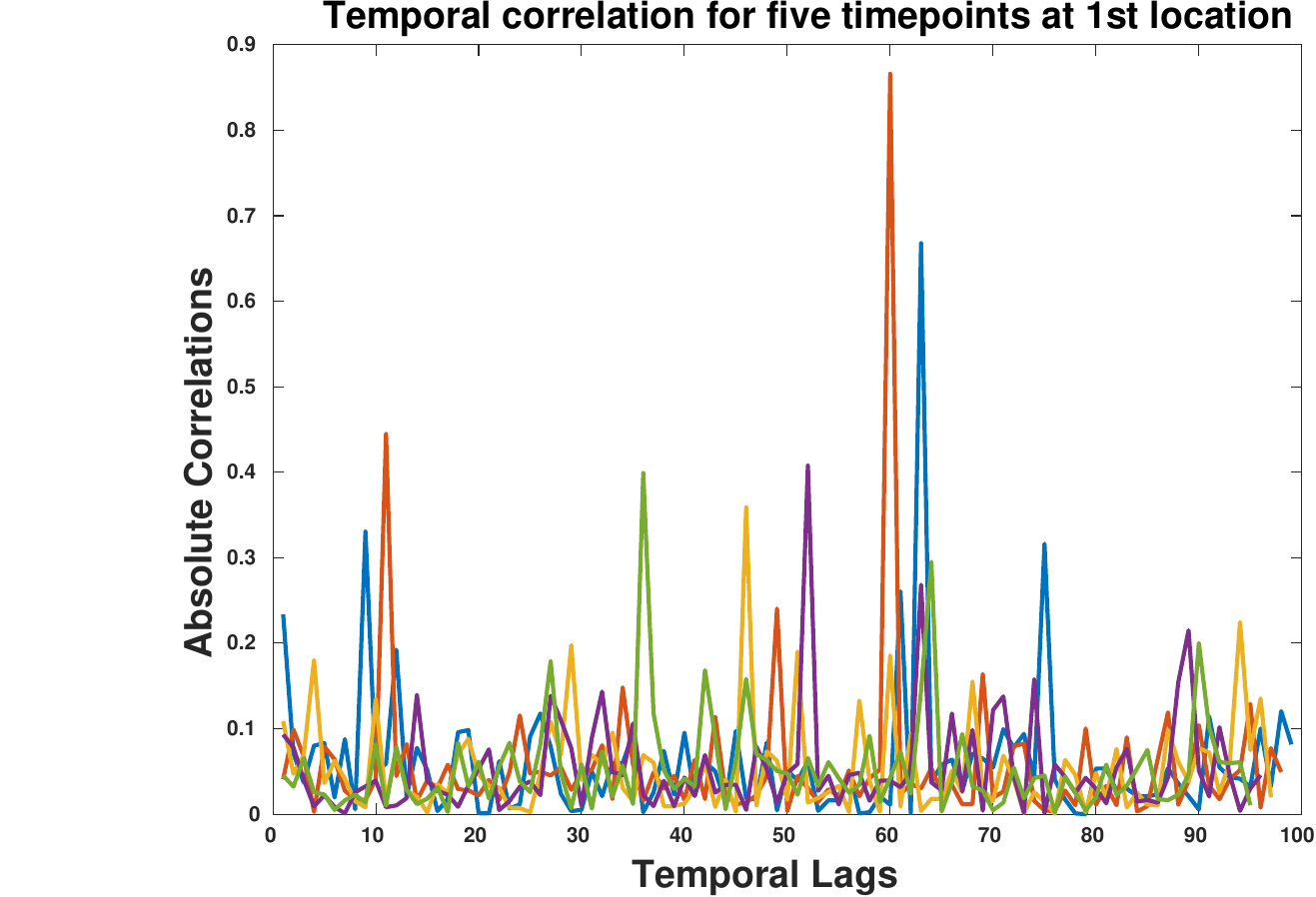}
\includegraphics[width=0.45\textwidth,height=6.5cm]{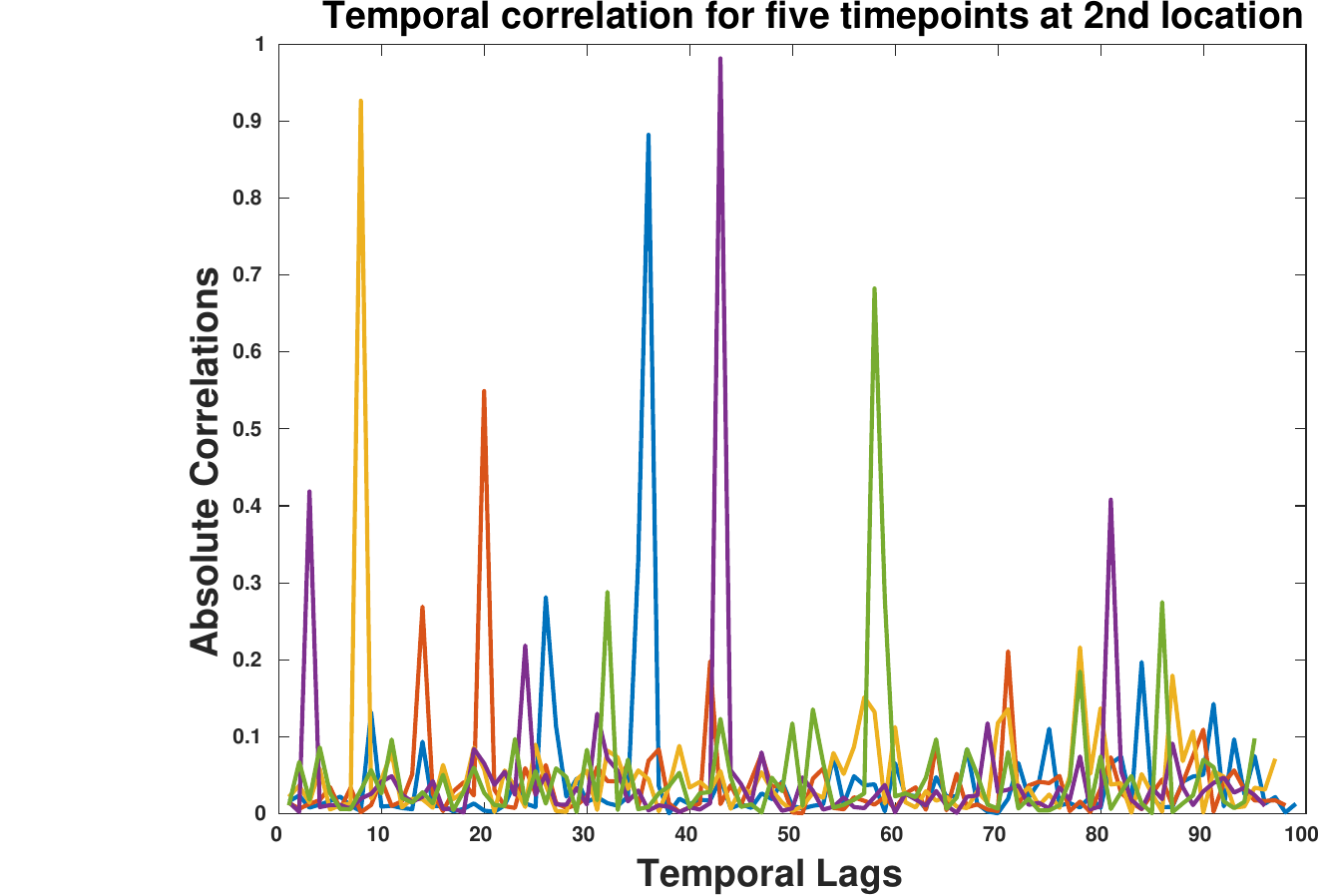}
\includegraphics[width=0.45\textwidth,height=6.5cm]{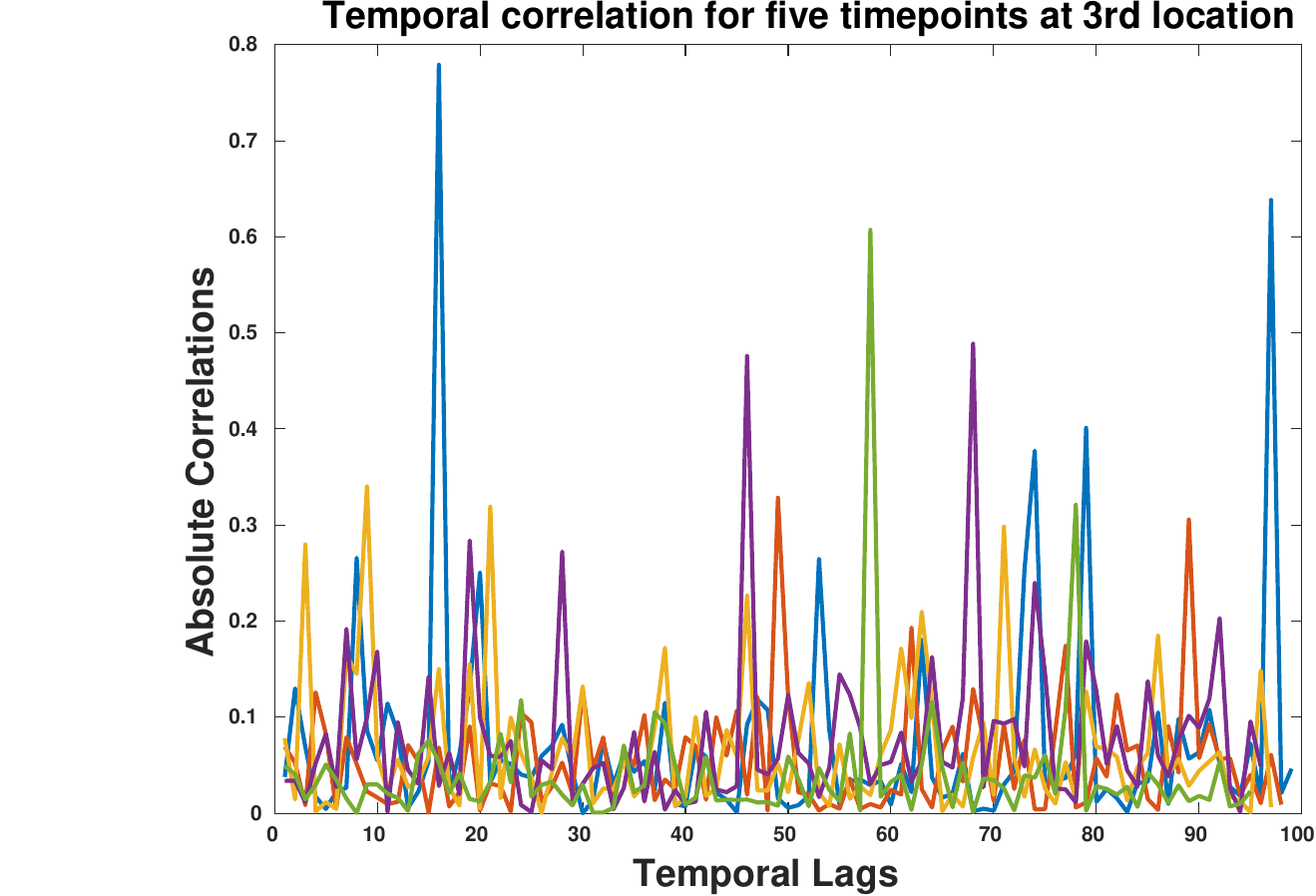}
\includegraphics[width=0.45\textwidth,height=6.5cm]{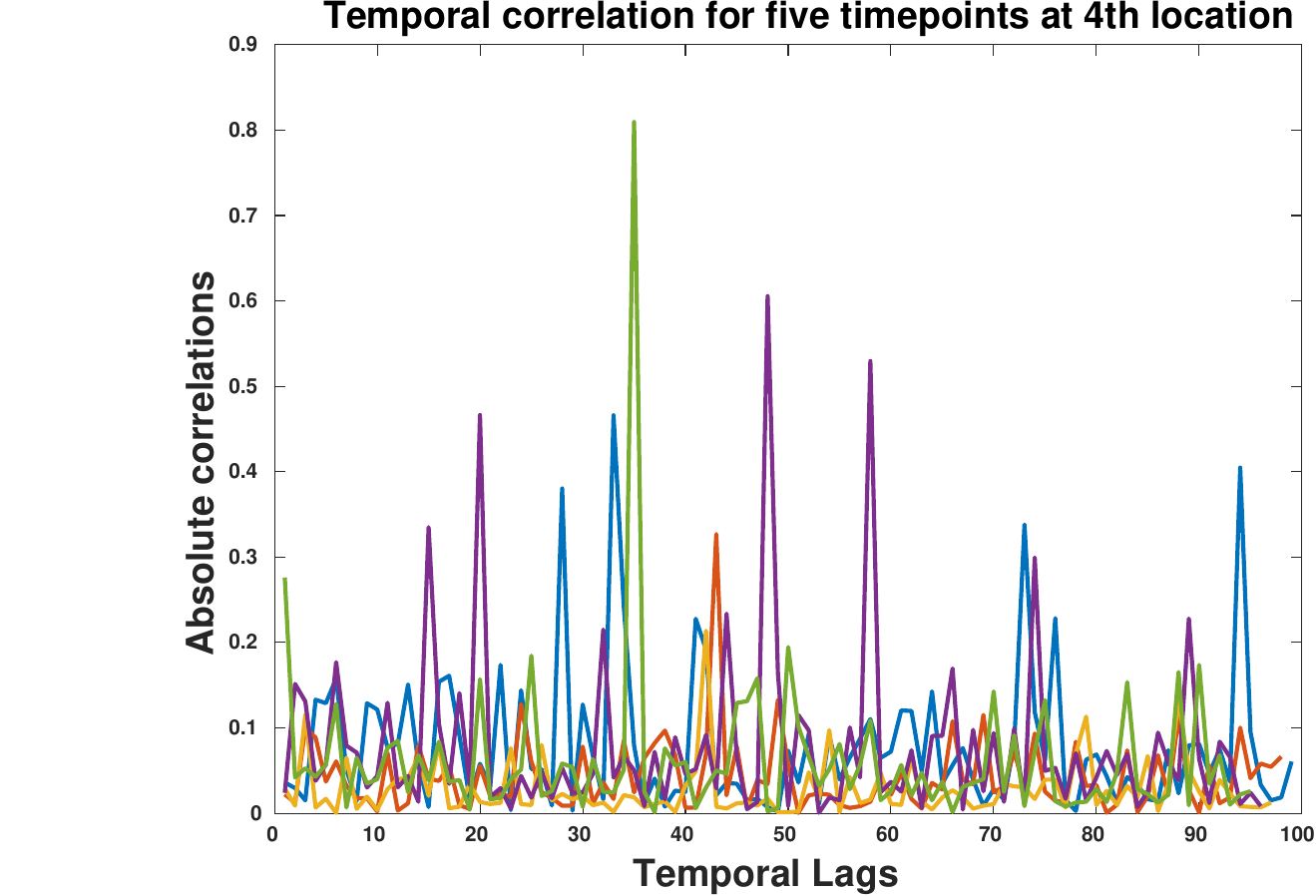}
\centering
\includegraphics[width=0.45\textwidth,height=6.5cm]{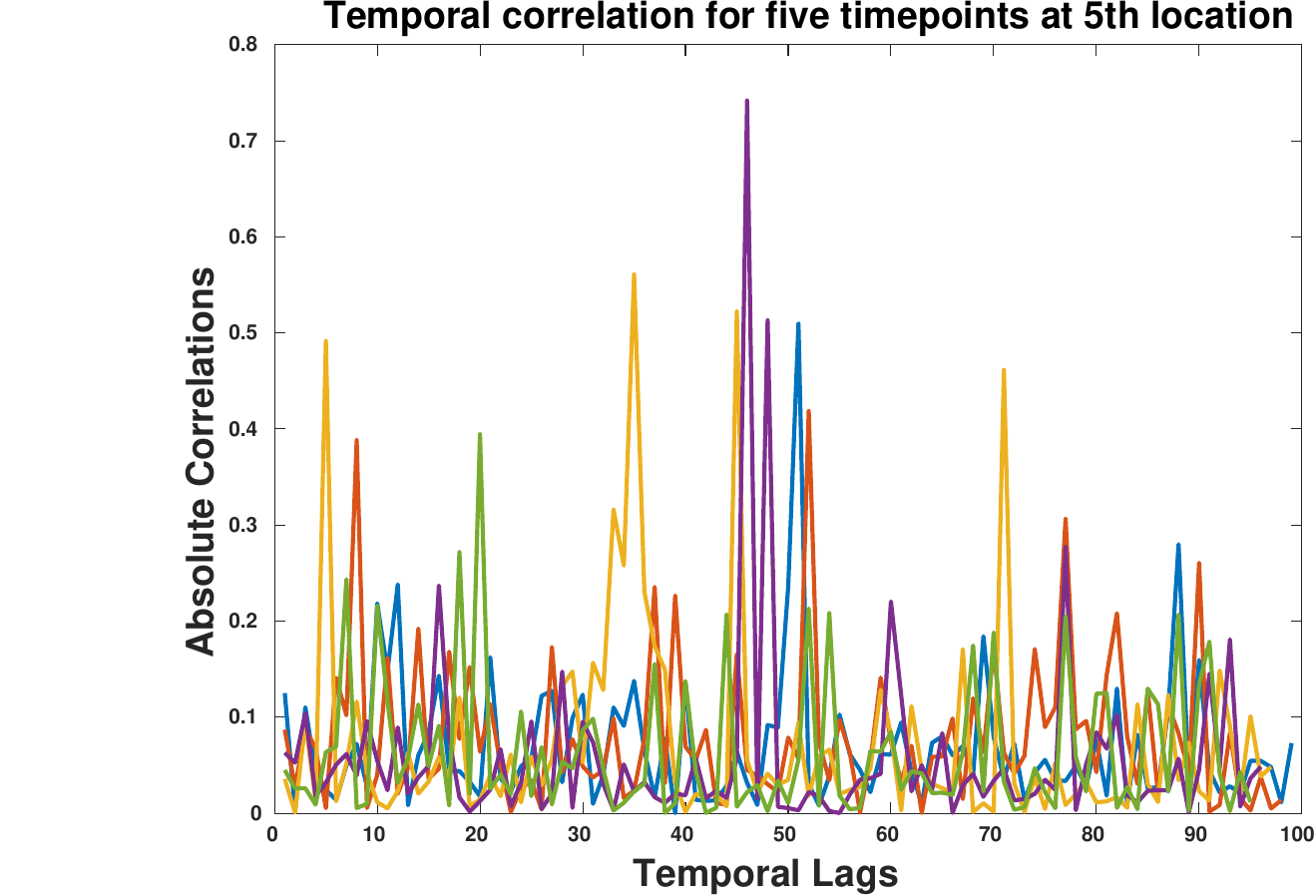}
    \caption{Temporal correlation plots at 5 locations using the observations simulated from GQN model are depicted. x-axis and y-axis denote the temporal lags and the correlations, respectively. The five different colors denote 5 time points in each plots.}
    \label{Temporal correlation plot for GQN model}
\end{figure}

%--------------------------------------------
We performed two simulation studies on the GQN model, taking large temporal and spatial lags, which shows no evidence of correlation going to zero. To obtain the lagged spatial correlations, we simulated 100 spatio-temporal observations from the GQN model independently with five time points and 100 locations; to obtain lagged temporal correlations, we simulated spatio-temporal observations from the same model with 100 time points and 5 locations. We provide five lagged spatial correlation plots corresponding to five time points, each containing five locations (\ref{Spatial correlation plot for GQN model}). Similarly, we provide five lagged temporal correlation plots corresponding to five locations containing five different time points (\ref{Temporal correlation plot for GQN model}).
%---------------------------------------
\begin{comment}
\begin{figure}[!h]
    \centering
    \begin{subfigure}{0.33\textwidth}
        \includegraphics[width=\linewidth,height=3cm]{GQN_spatial_lagged_corr_time1.jpg}
    \end{subfigure}
        \begin{subfigure}{0.33\textwidth}
        \includegraphics[width=\linewidth,height=3cm]{GQN_spatial_lagged_corr_time2.jpg}
    \end{subfigure}
        \begin{subfigure}{0.33\textwidth}
        \includegraphics[width=\linewidth,height=3cm]{GQN_spatial_lagged_corr_time3.jpg}
    \end{subfigure}
        \begin{subfigure}{0.33\textwidth}
        \includegraphics[width=\linewidth,height=3cm]{GQN_spatial_lagged_corr_time4.jpg}
    \end{subfigure}
        \begin{subfigure}{0.33\textwidth}
        \includegraphics[width=\linewidth,height=3cm]{GQN_spatial_lagged_corr_time5.jpg}
    \end{subfigure}
    \caption{Spatial correlation plots at 5 time points using observations simulated from GQN model are depicted. x-axis and y-axis denote the spatial lags and the correlations, respectively.}
    \label{Spatial correlation plot for GQN model}
\end{figure}
\end{comment}
%------------------------------------------------------

%
\begin{comment}
\begin{figure}[!h]
\includegraphics[width=\textwidth]{Temporal_corr_GQN-crop.pdf}
    \caption{Temporal correlation plots at 5 locations using the observations simulated from GQN model are depicted. x-axis and y-axis denote the temporal lags and the correlations, respectively. The five different colors denote 5 time points in each plots.}
    \label{Temporal correlation plot for GQN model}
\end{figure}
\end{comment}
%--------------------------------------------
\section{Spatial correlation plots}
\label{sample correlation plots}
\begin{figure}[!ht]
	%\centering
		\begin{subfigure}{0.48\textwidth}
			\includegraphics[width=\linewidth,height=7.5cm]{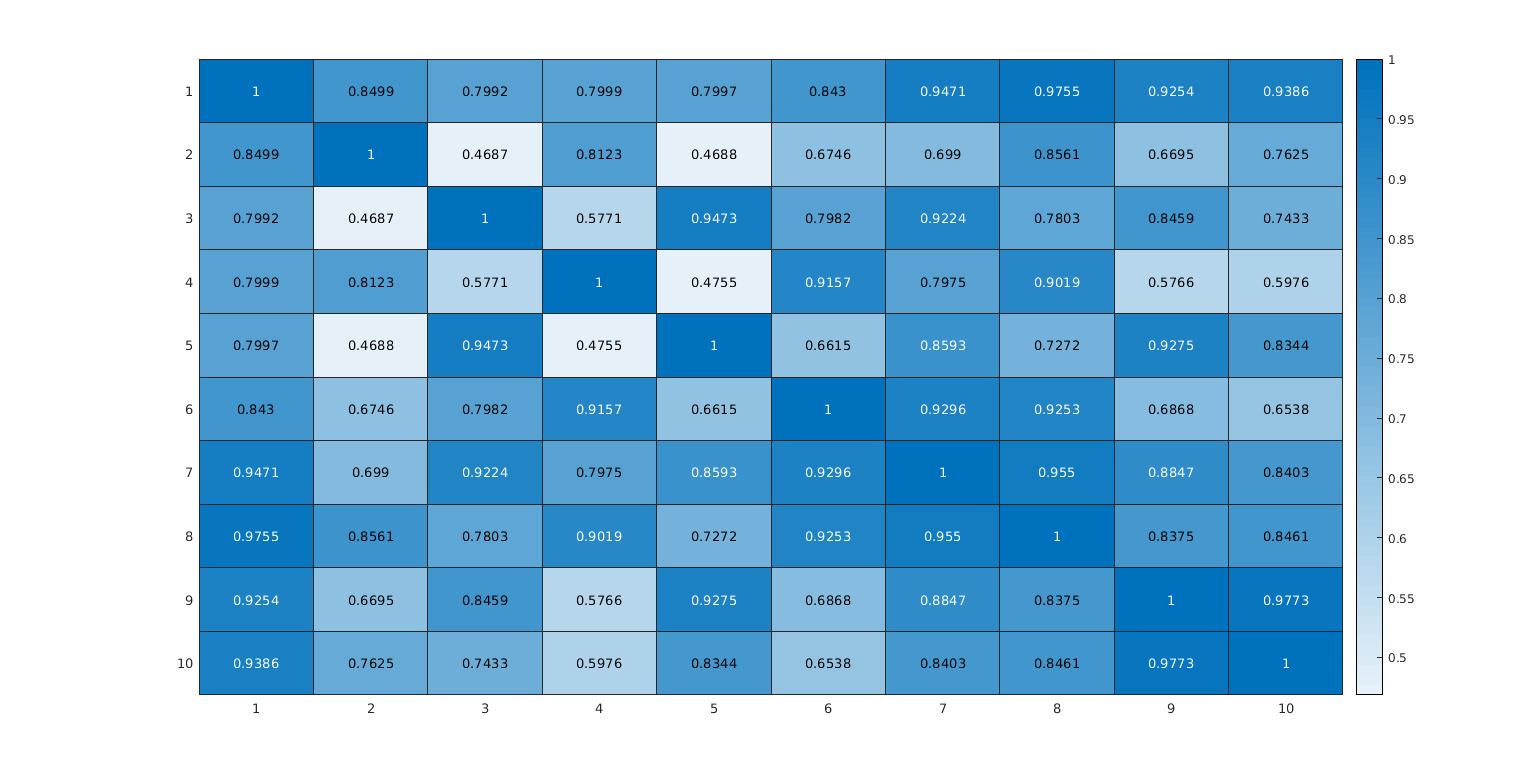}
			\caption{ }
		\end{subfigure}
		\begin{subfigure}{0.5\textwidth}
			\includegraphics[width=\linewidth,height=7.5cm]{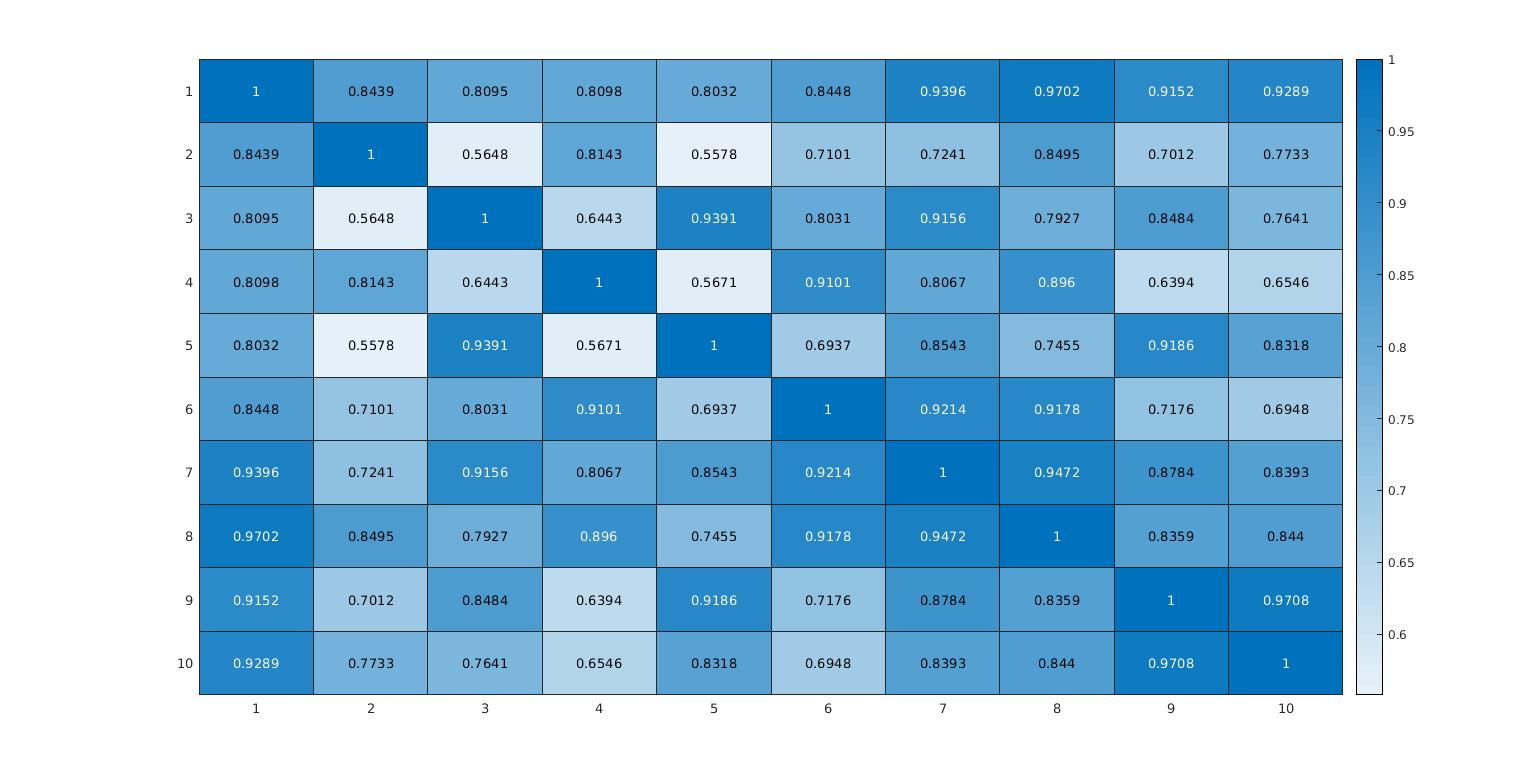}
			\caption{ }
		\end{subfigure}
			\begin{subfigure}{0.48\textwidth}
			\includegraphics[width=\linewidth,height=7.5cm]{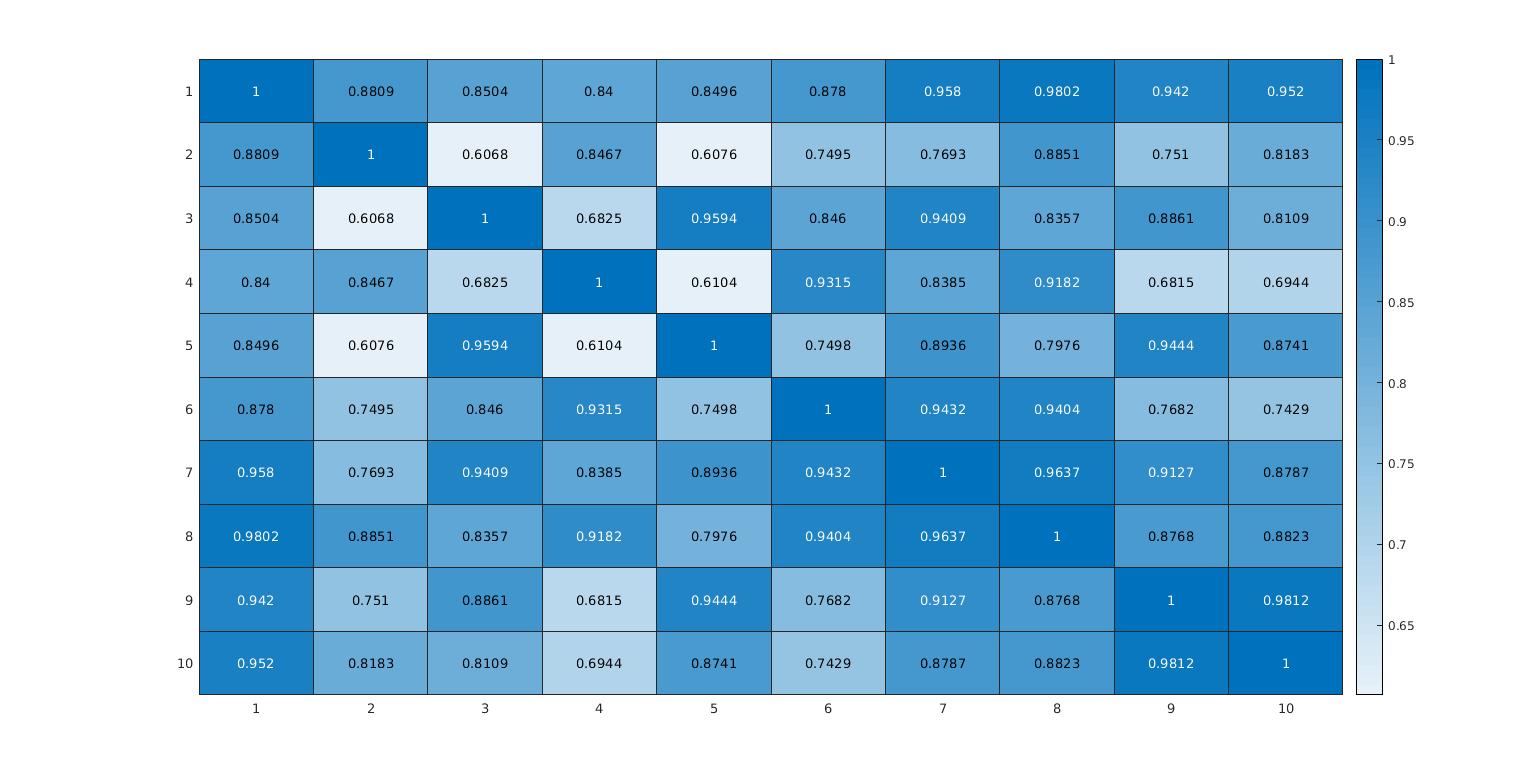}
			\caption{}
		\end{subfigure}
		\begin{subfigure}{0.5\textwidth}
			\includegraphics[width=\linewidth,height=7.5cm]{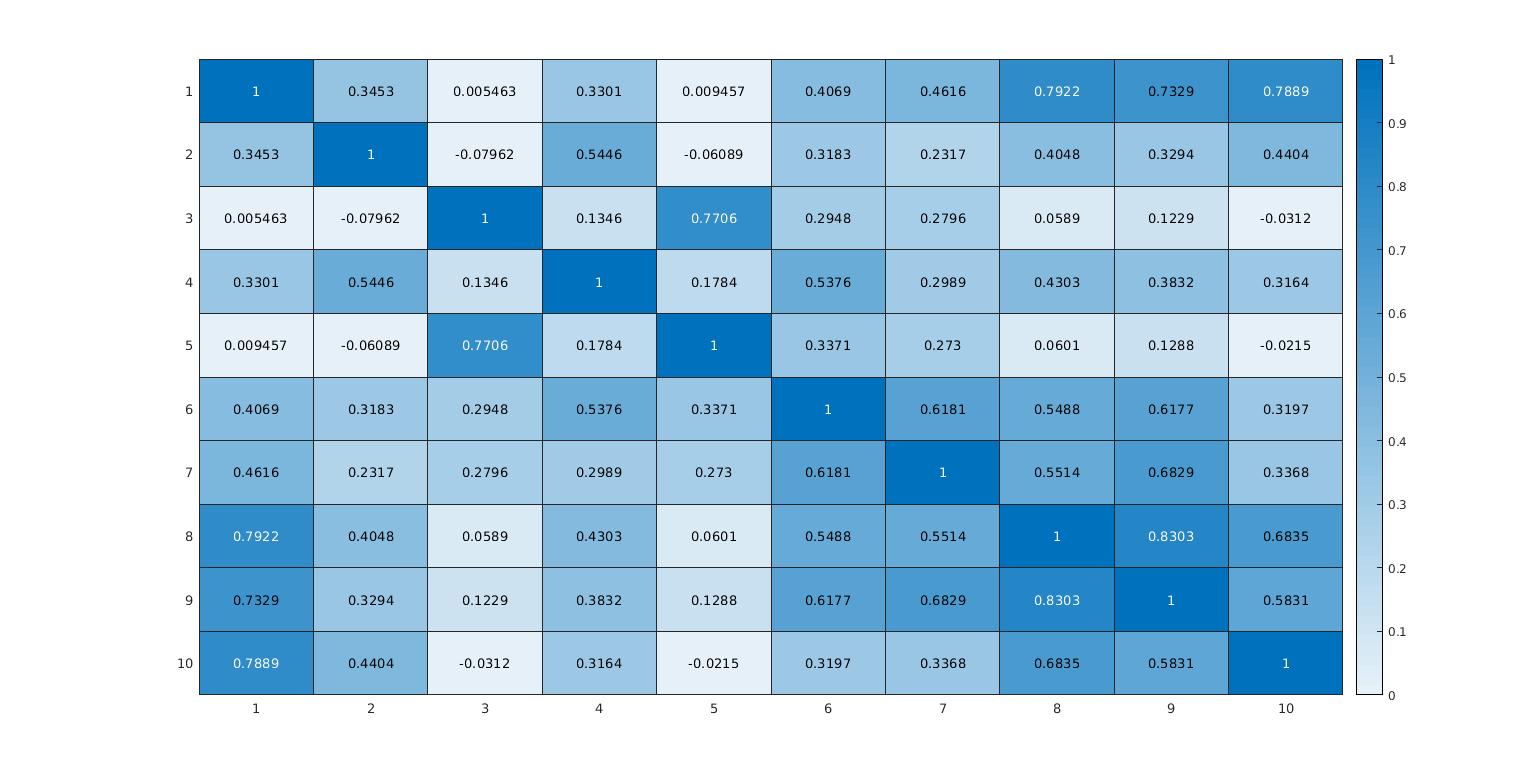}
			\caption{}
		\end{subfigure}
		\caption{The figure depicts examples of spatial correlation matrices for four different models. The x-axis and the y-axis represent the indices of the locations. 
		Higher color intensity represents larger correlation. Plot (a) shows spatial correlations obtained from a squared exponential covariance kernel with variance 
		and decay parameter 1, abbreviated as SE(1,1). Plots (b) and (c) depict spatial correlations for Mat\'{e}rn(3/2) and Mat\'{e}rn(5/2) with range parameter 1. 
		Finally plot (d) shows spatial correlations for the proposed model, with $\eta_1 = \eta_2 =\eta_3 =1$, $\sigma^2 = \sigma^2_{\theta} = \sigma_{p}^2 = 1$, $\alpha= \beta = 0.9$. 
        Our model can have negative entries in a correlation matrix whereas the correlation matrices for the other models are always positive.}
		\label{Fig:spatial correlation matrices}
	\end{figure}

%===========================================
We compared the spatial correlations of our model with that of squared exponential and Mat\'{e}rn spatial correlation using simulations. 
	{We simulated spatio-temporal observations 1000 times from our model taking $\eta_1 = \eta_2 =\eta_3 =1$, $\sigma^2 = \sigma^2_{\theta} = \sigma_{p}^2 = 1$, $\alpha= \beta = 0.9$. The number of locations and time points are taken to be 10 and 4, respectively. Based on 1000 simulations we calculated the sample spatial correlation matrix. Similarly, we simulated spatio-temporal observations from a Gaussian process with zero mean and squared exponential kernel with variance and decaying parameter 1. The corresponding sample spatial correlation matrix is calculated based on the 1000 repetitions. The same exercise has been performed by replacing the squared exponential kernel with two Mat\'{e}rn covariance kernels, namely, Mat\'{e}rn(3/2) and Mat\'{e}rn(5/2), respectively. The range parameter for the Mat\'{e}rn covariance kernel is taken as 1 for both the cases.}	
		{The sample spatial correlations are provided in Figure~\ref{Fig:spatial correlation matrices}.}
        %Spatial correlation matrix for SE(1,1).
        % Spatial correlation matrix for Mat\'{e}rn(3/2).
        % Spatial correlation matrix for Mat\'{e}rn(5/2).
        % Spatial correlation matrix for our model.
    \clearpage
%=========================================
\section{Proofs of our results}
\label{theorem proofs}
%==========================================
This appendix contains proofs of all the relevant results presented in Section \ref{properties of the processes}. 
\begin{lem}
	%\label{lemma3: differentiability of Ms}
	$M_s$ is infinitely differentiable in $s\in S$. 
\end{lem}
Proof: Since the exponential function is infinitely differentiable, it is sufficient to show that $\max\{||s^2-u^2||^2: u\in S\}$ is infinitely differentiable 
with respect to $s$. 
We first prove the result in one dimension and then will generalize to higher dimension. 
Let $S$ be a compact set in $\R$. We consider different cases as follows.

Case 1: Let $S = [a,b]$, where $0 < a < b$. Then $\max\{|s^2-u^2|^2: u\in [a,b]\} = s^4 + \max(a^4,b^4) - 2s^2 a^2$, which is an infinitely smooth function of $s$. 

Case2: Let $S = [a,b]$, where $a < b < 0$. Then $\max\{|s^2-u^2|^2: u\in [a,b]\} = s^4 + \max(a^4,b^4) - 2s^2b^2$, which is also infinitely differentiable function of $s$. 

Case 3: Let $S = [-a,b]$, where $a,b>0$. Then 
\begin{align}
	\label{eqn on M_s}
	\max\{|s^2-u^2|^2: u\in [-a,b]\} = %\begin{cases}
	s^4 + \max(a^4,b^4) - 2s^2\min(a^2,b^2)
\end{align}
%	s^2 + \max(a^2,b^2) - 2sb & \text{ if } s<0 \\
%\max(a^4,b^4) & \text{ if } s=0. 
%\end{cases}
Clearly, $\max\{|s^2-u^2|^2: u\in [-a,b]\}$ is infinitely  differentiable function of $s$, for any $s\in [-a,b]$. 

Now suppose that $S = [-a_1,b_1]\times [-a_2, b_2]$, where $a_i, b_i$ are positive for each $i =1, 2$. Let $s\in S$. Define $f_{s}: S\rightarrow \R$, such that $f_{s}(u) = ||s^2 - u^2||^2 = (s_1^2-u_1^2)^2 + (s_2^2-u_2^2)^2$. Define $f_{1s}(u_1) = (s_1^2-u_1^2)^2$ and $f_{2s}(u_2) = (s_2^2-u_2^2)^2$. Therefore, $f_{s}(u) = f_{1s}(u_1) + f_{2s}(u_2)$ and $\max\limits_{u\in S} f_{s}(u) = \max\limits_{u\in S}[f_{1s}(u_1) + f_{2s}(u_2)] = \max\limits_{u_1\in [-a_1,b_1]} f_{1s}(u_1) + \max\limits_{u\in [-a_2,b_2]} f_{2s}(u_2)$. The equality follows as $u = (u_1,u_2)\in S,$ with $u_1 \in [-a_1,b_1]$ and $u_2\in [-a_2,b_2]$. We have already proved that (Case 3, equation (\ref{eqn on M_s})) $\max\limits_{u_i\in [-a_i,b_i]} f_{is}(u_i) = s_i^4 + \max(a_i^4,b_i^4) - 2s_i^2 \min(a_i^2,b_i^2)$ for $i=1,2$. Hence $\max\limits_{u\in S} f_{s}(u) = \sum\limits_{i=1}^2 s_i^4 + \max(a_i^4,b_i^4) - 2s_i^2 \min(a_i^2,b_i^2) $, which is infinitely differentiable with respect to $s_1$ and $s_2$. 

If $S = [a_1,b_1]\times [a_2,b_2]$ where (i) $0<a_1<b_1$ and $0<a_2<b_2$ or (ii) $a_1<b_1<0$ and $a_2<b_2<0$ or (iii) $0<a_1<b_1$ and $a_2<b_2<0$ or (iv) $a_1<b_1<0$ and $0<a_2<b_2$, the proof goes in a similar manner as above except that now we have to use Case 1 and Case 2 for one dimension, instead of Case 3. \hfill $\square$

In the proofs of the theorems and lemmas that follow, we denote $Y(s,t)$ by $\theta_s(t)$ and $X(s,t)$ by $p_s(t)$, for any $s\in S$, and $t\in [0,\infty)$. 
\begin{thm}
	%		\label{lem: covariance} 
	Under assumptions A1 to A3, cov$\left(Y(s,h\delta t), Y(s',h'\delta t)\right)$ converges to 0 as $||s-s'||\rightarrow \infty$ and/or $|h-h'|\rightarrow \infty$.  %	
\end{thm}
Proof: %Let us denote $Y(s,\cdot)$ by $\theta_s(\cdot)$, $Y(s',\cdot)$ by $\theta_{s'}(\cdot)$ and $X(s,t)$ by $p(s,t)$, for any $s\in S$, and $t\in [0,\infty)$. 
First we shall show that 
$$\mbox{cov}\left(\theta_{s}(h\delta t),\theta_{s'}(h'\delta t)\Big | p\right)\rightarrow 0, $$ as $||s-s'||\rightarrow \infty$ and/or $|h-h'|\rightarrow \infty$. %Here by $\left(\cdot\Big p\right)$, we mean the complete latent process is given. 
Without loss of generality let us assume that $h>h'$. Now
\allowdisplaybreaks
\begin{align}
	\label{eq1 of covariance proof}
	& \mbox{cov}\left(\theta_{s}(h\delta t), \theta_{s'}(h'\delta t)\Big |p\right)\notag \\
	& \quad = \mbox{cov}\left(\beta \theta_{s}((h-1)\delta t) + \frac{\delta t}{M_{s}}\left\{\alpha p_{s}((h-1)\delta t) - \frac{\delta t}{2}\frac{\partial}{\partial \theta_{s}}V\left(\theta_{s}((h-1)\delta t)\right)\right\},\theta_{s'}(h'\delta t)\Big | p\right) \notag \\
	& \quad = \mbox{cov} (\beta \theta_{s}((h-1)\delta t), \theta_{s'}(h'\delta t) \Big | p) - \frac{1}{2}\frac{(\delta t)^2}{M_s}\mbox{cov}\left(\frac{\partial}{\partial \theta_{s}}V(\theta_{s}((h-1)\delta t)),\theta_{s'}(h'\delta t) \Big | p \right) \notag \\
	& \quad = \cdots \notag \\
	& \quad = \beta^{h-h'}\mbox{cov}\left(\theta_{s}(h'\delta t),\theta_{s'}(h'\delta t)|p\right) - \frac{1}{2}\frac{(\delta t)^2}{M_s}\sum_{k=1}^{h-h'}\beta^{k-1}\mbox{cov}\left[\frac{\partial}{\partial \theta_{s}}V\left(\theta_{s}((h-k)\delta t)\right),\theta_{s'}(h'\delta t)|p\right].
\end{align}
Since, for any $1\leq \ell\leq h-h'$, 

\allowdisplaybreaks
\begin{align*}
	%\label{eq2}
	&\mbox{cov}\left(\frac{\partial}{\partial \theta_{s}}V\left(\theta_{s}((h-\ell)\delta t)\right),\theta_{s'}(h'\delta t)\Big | p\right) \notag \\
	& \quad = \mbox{cov}\left(\frac{\partial}{\partial \theta_{s}}V\left(\theta_{s}((h-\ell)\delta t)\right), \beta \theta_{s'}((h'-1)\delta t) + \frac{\delta t}{M_{s'}}\left\{\alpha p_{s}((h'-1)\delta t) - \frac{\delta t}{2}\frac{\partial}{\partial \theta_{s'}}V\left(\theta_{s}((h'-1)\delta t)\right)\right\}\Big | p \right) \notag \\
	& \quad = \beta \mbox{cov}\left(\frac{\partial}{\partial \theta_{s}}V\left(\theta_{s}((h-\ell)\delta t)\right),\theta_{s'}((h'-1)\delta t) \Big | p\right) - \frac{(\delta t)^2}{2M_{s'}}\mbox{cov}\left(\frac{\partial}{\partial \theta_{s}}V\left(\theta_{s}((h-\ell)\delta t)\right),\frac{\partial}{\partial \theta_{s'}} \right. \notag \\
	& \qquad \left. V\left(\theta_{s'}((h-1)\delta t)\right)\Big| p\right) \notag \\ 
	& \quad = \cdots \notag \\
	& \quad = \beta^{h'}\mbox{cov}\left(\frac{\partial}{\partial \theta_{s}}V\left(\theta_{s}((h-\ell)\delta t)\right),\theta_{s'}(0) \Big | p\right) - \frac{(\delta t)^2}{2M_{s'}} \sum_{k=1}^{h'} \beta^{k-1}\mbox{cov}\left(\frac{\partial}{\partial \theta_{s}}V\left(\theta_{s}((h-\ell)\delta t)\right), \frac{\partial}{\partial \theta_{s'}} \right. \notag \\
	& \qquad \left. V\left(\theta_{s'}((h-k)\delta t)\right)\Big | p \right),
\end{align*}
we have, 
\begin{align}
	\label{eq2 of covariance proof}
	&\left|\mbox{cov}\left(\frac{\partial}{\partial \theta_{s}}V\left(\theta_{s}((h-\ell)\delta t)\right),\theta_{s'}(h'\delta t)\Big | p\right)\right| \notag \\
	& \quad \leq \left| \beta^{h'}\mbox{cov}\left(\frac{\partial}{\partial \theta_{s}}V\left(\theta_{s}((h-\ell)\delta t)\right),\theta_{s'}(0) \Big | p\right)\right| + \frac{(\delta t)^2}{2M_{s'}} \sum_{k=1}^{h'} \bigg| \beta^{k-1}\mbox{cov}\bigg[\frac{\partial}{\partial \theta_{s}}V\left(\theta_{s}((h-\ell)\delta t)\right), \frac{\partial}{\partial \theta_{s'}} \notag \\
	& \qquad  V\left(\theta_{s'}((h-k)\delta t)\right) \bigg] \bigg| \notag \\
	& \quad \leq |\beta|^{h'} \sigma_{\theta} \sigma + \frac{1-|\beta|^{h'}}{1-|\beta|} \frac{(\delta t)^2}{2M_{s'}} \sigma^2 = \epsilon, \mbox{say},
\end{align}
where $\sigma^2$ and $\sigma_{\theta}^2$ are the variance terms of the processes $\frac{\partial}{\partial \theta_{s}}V\left(\theta_{s}(h-1)\delta t\right)$ and $\theta_{s}(0)$, respectively (see A2 and A3). 
Therefore, from Equation (\ref{eq1 of covariance proof}) we obtain
\allowdisplaybreaks
\begin{align}
	\label{eq3 of covariance proof}
	&\Big |\mbox{cov}\left(\theta_{s}(h\delta t), \theta_{s'}(h'\delta t)\Big |p\right) - \beta^{h-h'}\mbox{cov}\left(\theta_{s}(h'\delta t),\theta_{s'}(h'\delta t)|p\right)\Big | \notag \\
	& \quad \leq \frac{1}{2}\frac{(\delta t)^2}{M_s}\sum_{k=1}^{h-h'}|\beta|^{k-1}\Big |\mbox{cov}\left[\frac{\partial}{\partial \theta_{s}}V\left(\theta_{s}((h-k)\delta t)\right),\theta_{s'}(h'\delta t)|p\right] \Big | \notag \\
	& \quad \leq \frac{1}{2}\frac{(\delta t)^2}{M_s} \frac{1-|\beta|^{h-h'}}{1-|\beta|} \epsilon,
\end{align}
using equation (\ref{eq2 of covariance proof}). Now from the first term of the right hand side of equation (\ref{eq1 of covariance proof}), we obtain
\begin{align*}
	%\label{eq4}
	& \mbox{cov}\left(\theta_{s}(h'\delta t),\theta_{s'}(h'\delta t)\Big|p\right) \notag \\
	& \quad = \mbox{cov}\left(\beta \theta_{s}((h'-1)\delta t) + \frac{\delta t}{M_{s}}\left\{\alpha p_{s}((h'-1)\delta t) - \frac{\delta t}{2}\frac{\partial}{\partial \theta_{s}}V\left(\theta_{s}((h'-1)\delta t)\right)\right\}, \right. \notag \\
	& \qquad \left. \beta \theta_{s'}((h'-1)\delta t) + \frac{\delta t}{M_{s'}}\left\{\alpha p_{s}((h'-1)\delta t) - \frac{\delta t}{2}\frac{\partial}{\partial \theta_{s'}}V\left(\theta_{s}((h'-1)\delta t)\right)\right\}\Big| p
	\right) \notag \\ 
	& \quad = \beta^2 \mbox{cov}\left(\theta_{s}((h'-1)\delta t), \theta_{s'}((h'-1)\delta t)\Big | p \right) - \beta\frac{1}{2}\frac{(\delta t)^2}{M_{s'}} \mbox{cov}\left(\theta_{s}((h'-1)\delta t),\frac{\partial}{\partial \theta_{s'}}V\left(\theta_{s'}((h'-1)\delta t)\right)\big| p\right) \notag \\
	& \qquad - \beta\frac{1}{2}\frac{(\delta t)^2}{M_{s}} \mbox{cov}\left(\theta_{s'}((h'-1)\delta t),\frac{\partial}{\partial \theta_{s}}V\left(\theta_{s}((h'-1)\delta t)\right)\big| p\right) + \frac{1}{4}\frac{(\delta t)^4}{M_{s}M_{s'}}\mbox{cov}\left(\frac{\partial}{\partial \theta_{s}}V\left(\theta_{s}((h'-1)\delta t)\right),\right. \notag \\
	& \qquad \left. \frac{\partial}{\partial \theta_{s'}}V\left(\theta_{s'}((h'-1)\delta t)\right)\Big | p \right) \notag \\
	& \quad = \cdots \notag \\
	& \quad = \beta^{2h'} \mbox{cov}\left(\theta_{s}(0),\theta_{s'}(0) \Big | p\right) -\frac{(\delta t)^2}{2M_{s'}}\sum_{k=1}^{h'} \beta^{2k-1} \mbox{cov}\left(\theta_{s}((h'-k)\delta t),\frac{\partial}{\partial \theta_{s'}}V\left(\theta_{s'}((h'-k)\delta t)\right)\big| p\right) \notag \\
	& \quad - \frac{(\delta t)^2}{2M_{s}}\sum_{k=1}^{h'} \beta^{2k-1} \mbox{cov}\left(\theta_{s'}((h'-k)\delta t),\frac{\partial}{\partial \theta_{s}}V\left(\theta_{s}((h'-k)\delta t)\right)\big| p\right) + \frac{1}{4}\frac{(\delta t)^4}{M_{s}M_{s'}} \sum_{k=1}^{h'} \beta^{2(k-1)} \times \notag \\
	& \qquad ~~ \mbox{cov}\left(\frac{\partial}{\partial \theta_{s}}V\left(\theta_{s}((h'-k)\delta t)\right), \frac{\partial}{\partial \theta_{s'}}V\left(\theta_{s'}((h'-k)\delta t)\right)\Big | p \right),
\end{align*}
which in turn implies 
\allowdisplaybreaks
\begin{align}
	\label{eq4 of covariance proof}
	& \Big| \mbox{cov}\left(\theta_{s}(h'\delta t),\theta_{s'}(h'\delta t)\Big|p\right) - \beta^{2h'} \mbox{cov}\left(\theta_{s}(0),\theta_{s'}(0) \Big | p\right)\Big| \notag \\
	& \leq \frac{(\delta t)^2}{2M_{s'}} \sum_{k=1}^{h'} |\beta|^{2k-1} \Big| \mbox{cov}\left(\theta_{s}((h'-k)\delta t),\frac{\partial}{\partial \theta_{s'}}V\left(\theta_{s'}((h'-k)\delta t)\right)\big| p\right)\Big| + \frac{(\delta t)^2}{2M_{s}} \sum_{k=1}^{h'}  |\beta|^{2k-1} \times \notag \\
	& \quad \Big| \mbox{cov}\left(\theta_{s'}((h'-k)\delta t),\frac{\partial}{\partial \theta_{s}}V\left(\theta_{s}((h'-k)\delta t)\right)\big| p\right)\Big| + \frac{1}{4}\frac{(\delta t)^4}{M_{s}M_{s'}} \sum_{k=1}^{h'} |\beta|^{2(k-1)} \Big| \mbox{cov}\left(\frac{\partial}{\partial \theta_{s}}V\left(\theta_{s}((h'-k)\delta t)\right),\right. \notag \\
	& \qquad \left. \frac{\partial}{\partial \theta_{s'}}V\left(\theta_{s'}((h'-k)\delta t)\right)\Big | p \right)\Big| \leq \frac{(\delta t)^2}{2M_{s'}} \sum_{k=1}^{h'} |\beta|^{2k-1} \epsilon + \frac{(\delta t)^2}{2M_{s}} \sum_{k=1}^{h'}  |\beta|^{2k-1} \epsilon + \frac{1}{4}\frac{(\delta t)^4}{M_{s}M_{s'}} \sum_{k=1}^{h'} |\beta|^{2(k-1)} \sigma^2 \notag \\
	%& \quad  \notag \\
	& \quad \leq \frac{(\delta t)^2}{2M_{s'}} |\beta| \frac{1-|\beta|^{2h'}}{1-|\beta|^2} \epsilon + \frac{(\delta t)^2}{2M_{s}} |\beta| \frac{1-|\beta|^{2h'}}{1-|\beta|^2} \epsilon + \frac{1}{4}\frac{(\delta t)^4}{M_{s}M_{s'}} \frac{1-|\beta|^{2h'}}{1-|\beta|^2} \sigma^2.
\end{align}
From equations (\ref{eq3 of covariance proof}) and (\ref{eq4 of covariance proof}), we get 
\begin{align}
	\label{eq5 of covariance proof}
	&\left| \mbox{Cov}\left[\theta_{s}(h\delta t),\theta_{s'}(h'\delta t)\bigg|p\right] - \beta^{(h-h')+2h'}\mbox{cov}\left(\theta_{s}(0),\theta_{s'}(0) \bigg| p\right) \right| \notag \\
	& \quad \leq \left| \mbox{Cov}\left[\theta_{s}(h\delta t),\theta_{s'}(h'\delta t)\bigg|p\right] - \beta^{h-h'} \mbox{Cov}\left[\theta_{s}(h'\delta t),\theta_{s'}(h'\delta t)\bigg|p\right]\right| + |\beta|^{h-h'}\bigg| \mbox{Cov}\left[\theta_{s}(h'\delta t),\theta_{s'}(h'\delta t)\bigg|p\right] - \notag \\
	& \qquad \beta^{2h'} \mbox{cov}\left(\theta_{s}(0),\theta_{s'}(0) \Big | p\right)\bigg| \notag \\
	& \quad \leq \frac{1}{2}\frac{(\delta t)^2}{M_s} \frac{1-|\beta|^{h-h'}}{1-|\beta|} \epsilon + |\beta|^{h-h'} \left\{ \frac{(\delta t)^2}{2M_{s'}} |\beta| \frac{1-|\beta|^{2h'}}{1-|\beta|^2} \epsilon + \frac{(\delta t)^2}{2M_{s}} |\beta| \frac{1-|\beta|^{2h'}}{1-|\beta|^2} \epsilon + \frac{1}{4}\frac{(\delta t)^4}{M_{s}M_{s'}} \frac{1-|\beta|^{2h'}}{1-|\beta|^2} \sigma^2 \right\},
\end{align}
which goes to 0 as $|h-h'|\rightarrow \infty$ and/or $||s-s'||\rightarrow \infty$, under assumptions A1-A3 in conjunction with Remark \ref{Remark on M_s goes to infinity}. Writing $\bigg|\mbox{Cov}\left[\theta_{s}(h\delta t),\theta_{s'}(h'\delta t)\bigg\vert p\right]\bigg|$ as $\Big |\mbox{Cov}\left[\theta_{s}(h\delta t),\theta_{s'}(h'\delta t)\bigg\vert p\right] - \beta^{(h-h')+2h'}\mbox{Cov}\left[\theta_{s}(0),\theta_{s'}(0)\bigg\vert p\right] +  \beta^{(h-h')+2h'}\mbox{Cov}\left[\theta_{s}(0),\theta_{s'}(0)\bigg\vert p\right]\Big |$ and noting that $\Big|\beta^{(h-h')+2h'}$
\newline 
$\mbox{Cov}\left[\theta_{s}(0),\theta_{s'}(0)\bigg\vert p\right]\Big|$ is $|\beta|^{(h-h')} |\beta|^{2h'}\sigma_{\theta}^2\exp{\{-\eta_2 ||s_1-s_2||^2\}}$ (by assumption A2), we have 
$$\mbox{cov}\left(\theta_{s}(h\delta t), \theta_{s'}(h'\delta t)\Big |p\right)\rightarrow 0,$$ as $||s-s'||\rightarrow \infty$ and/or $|h-h'|\rightarrow \infty.$
Now the dominated convergence theorem indicates that the unconditional covariance	cov $\left(Y(s,h\delta t), Y(s',h'\delta t)\right)\rightarrow 0$ as $||s-s'||\rightarrow \infty$ and $|h-h'|\rightarrow \infty$. \hfill $\square$
\begin{thm}
	%		\label{lemma1: almost sure continuity}
	If the assumptions A1-A3 hold true, then 
	$Y(s,h\delta t)$ and $X(s,h\delta t)$ are continuous in $s$, for all $h\geq 1$, with probability 1. 
\end{thm}
Proof: %Let us denote $Y(s,\cdot)$ by $\theta_s(\cdot)$ and $X(s,\cdot)$ by $p_s(\cdot)$. 
Note that, for $h\geq 1$, 
\begin{align}
	\label{eqn theta}
	\theta_{s}(h \delta t) &= \beta \theta_{s}((h-1)\delta t) + \frac{\delta t}{M_{s}} \left\{\alpha p_{s}((h-1)\delta t)-\frac{\delta t}{2} \frac{\partial}{\partial \theta_s} V\left(\theta_{s} ((h-1)\delta t)\right) \right\} \mbox{ and }\\
	\label{eqn p}
	p_{s}(h\delta t) &= \alpha^2 p_{s}((h-1)\delta t) - \frac{\delta t}{2} \left\{\alpha \frac{\partial}{\partial \theta_s} V\left(\theta_{s}((h-1)\delta t)\right) + \frac{\partial}{\partial \theta_s} V\left(\theta_{s}(h\delta t)\right) \right\}.
\end{align}
Putting $h=1$ in equations (\ref{eqn theta}) and (\ref{eqn p}), we get 
\begin{align*}
	\theta_{s}(\delta t) &= \beta \theta_{s}(0) + \frac{\delta t}{M_{s}} \left\{\alpha p_{s}(0)-\frac{\delta t}{2} \frac{\partial}{\partial \theta_s} V\left(\theta_{s} (0)\right) \right\} \mbox{ and }\\
	p_{s}(\delta t) &= \alpha^2 p_{s}(0) - \frac{\delta t}{2} \left\{\alpha \frac{\partial}{\partial \theta_s} V\left(\theta_{s}(0)\right) + \frac{\partial}{\partial \theta_s} V\left(\theta_{s}(\delta t)\right) \right\}.
\end{align*}
Now $p_{s}(0)$ and $\theta_{s}(0)$ are Gaussian processes with continuous sample paths with probability 1 under assumptions A1 and A2 (also see Remark \ref{R1}), and Lemma~\ref{lemma3: differentiability of Ms} shows that $M_{s}$ is continuous in $s$. Furthermore, assumption A3 implies that the derivative of $V(\cdot)$ is also a Gaussian process with continuous sample paths (see Remark \ref{R2}). Since composition of two continuous function is also a continuous function, therefore, $\frac{\partial}{\partial \theta_s}V(\theta_{s}(0))$ is also continuous in $s$ with probability 1. This implies $\theta_{s}(\delta t)$ is continuous in $s$ with probability 1 as it is a linear combination of three almost sure continuous functions in $s$. This immediately implies that $p_{s}(\delta t)$ is also continuous in $s$ with probability 1.  
\par 
Assume that $\theta_{s}(h \delta t)$ and $p_{s}(h\delta)$ are continuous is $s$ with probability 1, for $h=k+1$. We will show that $\theta_{s}(h \delta t)$ and $p_{s}(h\delta)$ are almost surely continuous in $s$ for $h=k+2$. Now
\begin{align*}
	\theta_{s}((k+2) \delta t) &= \beta \theta_{s}((k+1)\delta t) + \frac{\delta t}{M_{s}} \left\{p_{s}((k+1)\delta t)-\frac{\delta t}{2} \frac{\partial}{\partial \theta_s} V\left(\theta_{s} ((k+1)\delta t)\right) \right\} \mbox{ and }\\
	p_{s}((k+2)\delta t) &= \alpha^2 p_{s}((k+1)\delta t) - \frac{\delta t}{2} \left\{\alpha \frac{\partial}{\partial \theta_s} V\left(\theta_{s}((k+1)\delta t)\right) + \frac{\partial}{\partial \theta_s} V\left(\theta_{s}((k+2)\delta t)\right) \right\}.
\end{align*}
Since $\theta_{s}((k+1)\delta t)$ and $p_{s}((k+1)\delta t)$ are assumed to be continuous in $s$ with probability 1, derivative of a Gaussian process is also a Gaussian process, composition of two continuous functions is also a continuous function, and linear combinations of continuous functions is a continuous function, we have $\theta_{s}((k+2)\delta t)$ is continuous is $s$ with probability 1. %Using a very similar logic on $p_{s}((k+2)\delta t)$, it is immediate to 
Similar arguments show that $p_{s}((k+2)\delta t)$ is also continuous in $s$ with probability 1. Hence using the argument of induction, we claim that $\theta_{s}(h\delta t)$ and $p_{s}(h\delta t)$ are continuous in $s$ for any $h\geq 1$, with probability 1. \hfill $\square$
\begin{thm}
	%		\label{lemma2:mean square continuity}
	Under the assumptions A1-A3, 
	$Y(s,h\delta t)$ and $X(s,h\delta t)$ are continuous in $s$ in the mean square sense, for all $h\geq 1$.
\end{thm}
Proof: %Let us denote $Y(s,\cdot)$ by $\theta_s(\cdot)$ and $X(s,\cdot)$ by $p_s(\cdot)$. 
We follow similar steps as in Lemma \ref{lemma1: almost sure continuity}. From equation (\ref{eqn theta}) we have
\begin{align*}
	\theta_{s}(\delta t) &= \beta \theta_{s}(0) + \frac{\delta t}{M_{s}} \left\{\alpha p_{s}(0)-\frac{\delta t}{2} \frac{\partial}{\partial \theta_s} V\left(\theta_{s} (0)\right) \right\}.
\end{align*}
Under assumptions A1 and A2, $p_{s}(0)$ and $\theta_{s}(0)$ are continuous in $s$ in the mean square sense, and by Lemma~\ref{lemma3: differentiability of Ms}, $M_{s}$ is continuous in $s$. Now since the random function $V(\cdot)$ is twice differentiable under assumption A3 (see also Remark \ref{R2}), the partial derivative process of $V$ is Lipschitz and hence the composition function $\frac{\partial}{\partial \theta_s} V(\theta_{s}(0))$ is also mean square continuous in $s$ (see page 416 of \cite{Banerjee2015}). Now using the fact that the linear combination of mean square continuous processes is also mean-square continuous, we have $\theta_{s}(\delta t)$ is mean square continuous in $s$. This, in turn, implies that, $\frac{\partial}{\partial \theta_s} V(\theta_{s}(\delta t))$ is also mean square continuous in $s$ using the same argument as above. Therefore, 
$$p_{s}(\delta t) = \alpha^2 p_{s}(0)-\frac{\delta t}{2} \left\{\alpha \frac{\partial}{\partial \theta_{s}}V(\theta_{s}(0))+\frac{\partial}{\partial \theta_s} V(\theta_{s}(\delta t)) \right\}$$ is mean square continuous in $s$. 
\par
Now applying the similar argument of induction as in Lemma \ref{lemma1: almost sure continuity}, we have the desired result. \hfill $\square$
\begin{thm}
	%	\label{lemma4: almost sure differentiability}
	Under the assumptions A1-A3, %$M_{s} = \exp(\max\{||s-u||^2: u\in S\})$, 
	$Y(s,h\delta t)$ and $X(s,h\delta t)$ have differentiable sample paths with respect to $s$, almost surely. 
\end{thm}
Proof: %Let us denote $Y(s,\cdot)$ by $\theta_s(\cdot)$ and $X(s,\cdot)$ by $p_s(\cdot)$.
The proof goes in the same line as that of Lemma \ref{lemma1: almost sure continuity}. We start with $\theta_{s}(\delta t)$ which is a linear combination of $\theta_{s}(0)$, $p_{s}(0)$ and $\frac{\partial}{\partial \theta_s} V(\theta_{s}(0))$. Note that $\theta_{s}(0)$, $p_{s}(0)$ have differentiable sample paths in $s$ by assumptions A1 and A2 (see Remark \ref{R1}). Now using the fact that composition of two differentiable functions is also differentiable (see the Lemma 3.4 of \ctn{Suman2017}), $\frac{\partial}{\partial \theta_s} V(\theta_{s}(0))$ has differentiable sample paths in $s$ (refer to assumption A3 and Remark \ref{R2}). Moreover, since $M_{s}$ is differentiable (Lemma~\ref{lemma3: differentiability of Ms}) and since the linear combination of differentiable functions is a differentiable function, $\theta_{s}(\delta t)$ has differentiable sample paths in $s$. The existence of differentiable sample paths of $\theta_{s}(\delta t)$ implies that $p_{s}(\delta t)$ will have differentiable sample paths in $s$. The rest of the proof 
is similar to that of Lemma \ref{lemma1: almost sure continuity}. \hfill $\square$
\begin{lem}
	%		\label{lemma5: mean square differentiability of composition of GP}
	Let $f:\mathbb{R}\rightarrow \mathbb{R}$ be a zero mean Gaussian random function with covariance function $c_{f}(x_1,x_2)$, $x_1, x_2\in \R$, which is four times continuously differentiable. Let $\left\{Z(s):s\in S\right\}$ be a random process with the following properties 
	\begin{enumerate}
		\item $E(Z(s)) = 0$,
		\item The covariance function $c_{Z}(s_1,s_2)$, $s_1, s_2 \in S$, where $S$ is a compact subspace of $\mathbb{R}^2$, is four times continuously differentiable, and
		\item $\frac{\partial Z(s)}{\partial s_i}$ has finite fourth moment. 
	\end{enumerate}
	%Assume the covariance functions are such that the second order derivative of $f$ is Gaussian and the second order derivative of $Z$ exists in mean square sense. In particular, we
	%Assume that the covariance functions of the processes are four-time continuously differentiable. 
	Then the process $\{g(s):s\in S\}$, where $g(s)=f(Z(s))$, is mean square differentiable in $s$.     
\end{lem}
Proof: To show that $\{g(s):s\in S\}$ is mean square differentiable in $s$ we have to show that, for any $p\in S$, there exists a function $L_{s}(p)$, linear in $p$, such that $$g(s+p) = g(s) + L_{s}(p) + R(s,p),$$ where $$\frac{R(s,p)}{||p||}\stackrel{L_{2}}{\longrightarrow} 0.$$
Let $s_o\in S$ be any point in $S$. Using multivariate Taylor series expansion we have 
\begin{align*}
	g(s_0+p) = g(s_0)+p^{T}\nabla g(s_0) + R(s_0,p),
\end{align*}
where $\nabla g(s_0) = \left(\frac{\partial f(Z(s))}{\partial s_1}, \frac{\partial f(Z(s))}{\partial s_2}\right)^T,$ with $\frac{\partial f(Z(s))}{\partial s_i} = \frac{df(Z(s))}{d(Z(s))}\frac{\partial Z(s)}{\partial s_i}$, for $i=1,2$. Therefore, $L_{s_0}(p) = p^{T}\nabla g(s_0)$, a linear function in $p$. To complete the proof we note that, from multivariate Taylor series expansion, $|R(s_0,p)|\leq M^* ||p||^2$, where $$M^* = \max\left\{ \Big|\frac{\partial^2 f(Z(s))}{\partial s_1^2} \Big|, \Big|\frac{\partial^2 f(Z(s))}{\partial s_1 \partial s_2}\Big|, \Big|\frac{\partial^2 f(Z(s))}{\partial s_2 \partial s_1}\Big|, \Big|\frac{\partial^2 f(Z(s))}{\partial s_2^2} \Big| \right\}, $$ 
with 
$$\frac{\partial^2 f(Z(s))}{\partial s_i^2} =
\frac{d^2 f(Z(s))}{d((Z(s)))^2}\left(\frac{\partial Z(s)}{\partial s_i}\right)^2 + 
\frac{d f(Z(s))}{dZ(s)} \frac{\partial^2 Z(s)}{\partial s_i^2}, \mbox{ for } i=1,2, $$ and 
$$ \frac{\partial^2 f(Z(s))}{\partial s_1 \partial s_2} = \frac{\partial^2 f(Z(s))}{\partial s_2 \partial s_1} = \frac{d^2f(Z(s))}{d(Z(s))^2}\frac{\partial Z(s)}{\partial s_1}\frac{\partial Z(s)}{\partial s_2} + \frac{df(Z(s))}{d(Z(s))} \frac{\partial^2 Z(s)}{\partial s_1 \partial s_2}.$$
Since we have assumed that $Z(\cdot)$ and $f(\cdot)$ have covariance functions which are four times continuously differentiable, $Z(\cdot)$ and $f(\cdot)$ are twice differentiable (in the mean square sense) and hence the above terms involving first and second derivatives of $f(\cdot)$ and $Z(\cdot)$ are well-defined.
%Since we assume that $f$ and $Z$ both are twice mean square differentiable so, each of the partial derivative processes is bounded in $L^2$. Therefore, $\frac{R(s_0,p)}{||p||}\stackrel{L_{2}}{\longrightarrow} 0$. Since $s_0$ is any point in $S$ which completes the proof. \hfill $\square$
\par 
We will now show that the second moment of $\frac{\partial^2 f(Z(s))}{\partial s_i^2}$, for $i=1, 2$, and $\frac{\partial^2 f(Z(s))}{\partial s_1 \partial s_2}$  are finite. To prove the above fact, we first show that 
$\mbox{var}\left(\frac{d^2 f(Z(s))}{d((Z(s)))^2}\right)
< \infty$. Note that 
\begin{align}
	\label{eq 7}
	\mbox{var}\left(\frac{d^2 f(Z(s))}{d((Z(s)))^2}\right) = \mbox{var}\left(E\left\{\frac{d^2 f(Z(s))}{d((Z(s)))^2}\Bigg\vert Z(s)\right\}\right)
	+ E\left(\mbox{var}\left\{\frac{d^2 f(Z(s))}{d((Z(s)))^2}\Bigg\vert Z(s)\right\}\right).
\end{align}
Moreover, since $f''(x)$ is a Gaussian function with 0 mean and constant variance, we have

$E\left\{\frac{d^2 f(Z(s))}{d((Z(s)))^2}\Bigg\vert Z(s)\right\} = 0$ and $\mbox{var}\left\{\frac{d^2 f(Z(s))}{d((Z(s)))^2}\Bigg\vert Z(s)\right\}  = \mbox{constant} < \infty$. Thus
\begin{align}
	\label{eq8}
	\mbox{var}\left(E\left\{\frac{d^2 f(Z(s))}{d((Z(s)))^2}\Bigg\vert Z(s)\right\}\right) = 0 \mbox{ and }
\end{align} 
\begin{align}
	\label{eq9}
	E\left(\mbox{var}\left\{\frac{d^2 f(Z(s))}{d((Z(s)))^2}\Bigg\vert Z(s)\right\}\right) = \mbox{constant}.
\end{align} 
Therefore, combining equations (\ref{eq8}) and (\ref{eq9}), and using equation (\ref{eq 7}) we see that 
$\mbox{var}\left(\frac{d^2 f(Z(s))}{d((Z(s)))^2}\right) <\infty$. 
Similar argument shows that $\mbox{var}\left(\frac{d f(Z(s))}{dZ(s)}\right) < \infty$. Now to show that $E\left(\frac{\partial^2 f(Z(s))}{\partial s_i^2}\right)^2 < \infty$, we use $(a+b)^2\leq 2(a^2 + b^2)$ and have
\begin{align}
	\label{eq10}
	E\left(\frac{\partial^2 f(Z(s))}{\partial s_i^2}\right)^2 
	& \leq 2\left\{E\left(\left[\frac{d^2 f(Z(s))}{d((Z(s)))^2}\right]^2 
	\left[\frac{\partial Z(s)}{\partial s_i}\right]^4\right) + E\left(\left[\frac{d f(Z(s))}{dZ(s)}\right]^2 \left[\frac{\partial^2 Z(s)}{\partial s_i^2}\right]^2\right)\right\} \notag \\ 
	& = 2 E\left(E\left(\left[\frac{d^2 f(Z(s))}{d((Z(s)))^2}\right]^2\Bigg\vert Z(s)\right) \left[\frac{\partial Z(s)}{\partial s_i}\right]^4\right) \notag \\
	& \quad + 2 E\left(E\left(\left[\frac{d f(Z(s))}{dZ(s)}\right]^2\Bigg \vert Z(s)\right) \left[\frac{\partial^2 Z(s)}{\partial s_i^2}\right]^2\right).
\end{align}
Again using the fact that $f'(x)$ and $f''(x)$ are Gaussian with 0 mean and constant variance, we have
$E\left(\left[\frac{d^2 f(Z(s))}{d((Z(s)))^2}\right]^2\Bigg\vert Z(s)\right) =\mbox{constant}$ and 
$E\left(\left[\frac{d f(Z(s))}{dZ(s)}\right]^2\Bigg \vert Z(s)\right) = \mbox{constant}$. 

%Further since we have the $Z(s)$ to be Gaussian and its first two order derivatives are also Gaussian with zero-mean and constant variances so, we have 
Further, since the covariance function of $Z(s)$ is assumed to be four times continuously differentiable (2nd assumption of the lemma), the first two derivatives of $Z(s)$, $\frac{\partial Z(s)}{\partial s_i}, i=1, 2$ and $\frac{\partial^2 Z(s)}{\partial s_i^2}, i=1, 2$, $\frac{\partial^2 Z(s)}{\partial s_1^2\partial s_2^2}$ will also exist in the mean square sense, with zero means. Thus,  $E\left[\frac{\partial^2 Z(s)}{\partial s_1^2}\right]^2 = \mbox{constant}$. Also, by the 3rd assumption of the Llmma, the fourth moment of $\frac{\partial Z(s)}{\partial s_i}$, for $i=1,2$, are finite. That is, 
$E\left[\frac{\partial Z(s)}{\partial s_i}\right]^4= \mbox{constant}$. Therefore, from equation (\ref{eq10}) we see that $E\left(\frac{\partial^2 f(Z(s))}{\partial s_1^2}\right)^2 < M_1 < \infty$, for some $M_1 \in \mathbb{R}.$

Next we show that $E\left(\frac{\partial^2 f(Z(s))}{\partial s_1 \partial s_2}\right)^2 < \infty$. Using the same inequality, $(a+b)^2 \leq 2(a^2 + b^2)$, we have 
\begin{align}
	\label{eq11}
	E\left(\frac{\partial^2 f(Z(s))}{\partial s_1 \partial s_2}\right)^2 & \leq 2\left[ E\left\{\frac{d^2f(Z(s))}{d(Z(s))^2}\frac{\partial Z(s)}{\partial s_1}\frac{\partial Z(s)}{\partial s_2}\right\}^2 + E\left\{\frac{df(Z(s))}{d(Z(s))} \frac{\partial^2 Z(s)}{\partial s_1 \partial s_2}\right\}^2\right] \notag \\
	& = 2 E\left[E\left(\left[\frac{d^2 f(Z(s))}{d((Z(s)))^2}\right]^2\Bigg\vert Z(s)\right) \left(\frac{\partial Z(s)}{\partial s_1}\right)^2\left(\frac{\partial Z(s)}{\partial s_2}\right)^2\right] \notag \\
	& \quad + 2E\left[E\left\{\left(\frac{df(Z(s))}{d(Z(s))}\right)^2\Bigg \vert Z(s)\right\} \left(\frac{\partial^2 Z(s)}{\partial s_1 \partial s_2}\right)^2 \right].
\end{align}
Since $f'(x)$ and $f''(x)$ are Gaussian functions with mean 0 and constant variance, therefore, 

$E\left(\left[\frac{d^2 f(Z(s))}{d((Z(s)))^2}\right]^2\Bigg\vert Z(s)\right)$ and 
$E\left(\left(\frac{df(Z(s))}{d(Z(s))}\right)^2\Bigg\vert Z(s)\right)$ are constants. 
Moreover, since by our assumption, the fourth moment of $\frac{\partial Z(s)}{\partial s_i}$, for $i=1,2$, are finite, we have
%the derivatives of the process $Z(s)$ are Gaussian processes with 0 mean and constant variances, we have 
$E\left[\left(\frac{\partial Z(s)}{\partial s_1}\right)^2\left(\frac{\partial Z(s)}{\partial s_2}\right)^2\right] \leq \sqrt{E\left(\frac{\partial Z(s)}{\partial s_1}\right)^4 E\left(\frac{\partial Z(s)}{\partial s_2}\right)^4}$ $=  \mbox{constant}$. Since the covariance function of $Z(s)$ is four times continuously differentiable, $E\left(\frac{\partial^2 Z(s)}{\partial s_1 \partial s_2}\right)^2 = \mbox{constant}$. Thus we have, from 
equation (\ref{eq11}), $E\left(\frac{\partial^2 f(Z(s))}{\partial s_1 \partial s_2}\right)^2 < M_2,$ for some $M_2 \in R$. Hence each term in 
$M^*= \max\left\{ \Big|\frac{\partial^2 f(Z(s))}{\partial s_1^2} \Big|, \Big|\frac{\partial^2 f(Z(s))}{\partial s_1 \partial s_2}\Big|, \Big|\frac{\partial^2 f(Z(s))}{\partial s_2 \partial s_1}\Big|, \Big|\frac{\partial^2 f(Z(s))}{\partial s_2^2} \Big| \right\}$
has bounded second moment. 

Next we have to show that $E(M^*)^2<\infty$. Denote $M^* = \max\{A,B,C,D\}$, where $A = \Big|\frac{\partial^2 f(Z(s))}{\partial s_1^2} \Big|$, $B = \Big|\frac{\partial^2 f(Z(s))}{\partial s_1 \partial s_2}\Big|$, $C = \Big|\frac{\partial^2 f(Z(s))}{\partial s_2 \partial s_1}\Big|$ and $D = \Big|\frac{\partial^2 f(Z(s))}{\partial s_2^2} \Big|$. Note that it is sufficient to show that $X= \max\{A,B\}$ has finite second moment, because then with the same argument $Y=\max\{C,D\}$ will have finite second moment, and finally $Z= \max\{X,Y\}~ (=\max\{A,B,C,D\})$ will have finite second moment. \par
Now $X = \max\{A,B\} = \frac{A + B + |A-B|}{2}\leq \frac{A + B |A|+|B|}{2}.$ Therefore, 
\begin{align*}
	EX^2 & \leq \frac{1}{4} 2\left[E\left\{(A+B)^2 + (|A|+|B|)^2\right\}\right] \notag \\
	& \leq E\{A^2 + B^2 + |A|^2 + |B|^2\} = 2 E(A^2 + B^2). 
\end{align*}
Since $E A^2< \infty$ and $E B^2< \infty$, therefore, $E X^2 < \infty$. Now exactly the same arguments imply that $E Y^2 < \infty$ and $E Z^2 < \infty$. 
Hence, $\frac{R(s_0,p)}{||p||}\stackrel{L_{2}}{\longrightarrow} 0$. Since $s_0$ is any point in $S$, the proof is complete. \hfill $\square$
\begin{thm}
	%		\label{Thm: Mean-square diff}
	Let the assumptions A1-A3 hold true, with the covariance functions of all the assumed Gaussian processes being square exponentials. Then $Y(s,h\delta t)$ and $X(s,h\delta t)$ are mean square differentiable in $s$, for every $h\geq 1$. 
\end{thm}

Proof: %Let us denote $Y(s,\cdot)$ by $\theta_s(\cdot)$ and $X(s,\cdot)$ by $p_s(\cdot)$. 
The proof will follow the similar argument of induction as done in Lemma \ref{lemma1: almost sure continuity}. For $h=1$, we have
$$\theta_{s}(\delta t) = \beta \theta_{s}(0) + \frac{\delta t}{M_{s}} \left\{\alpha p_{s}(0)-\frac{1}{2} \delta t\frac{\partial}{\partial \theta_s} V\left(\theta_{s} (0)\right) \right\}.$$ 
Since, under the assumption A2 and the assumption of the theorem, $\theta_{s} (0)$ is a centered Gaussian process with squared exponential covariance function, the fourth moment of $\frac{\partial \delta_s(0)}{\partial s_i}$, $i=1,2$ are finite, and by assumption A3, $V(\cdot)$ is a zero-mean Gaussian random function with squared exponential covariance. Therefore, using Lemma \ref{lemma5: mean square differentiability of composition of GP}, $\frac{\partial}{\partial \theta_s} V\left(\theta_{s} (0)\right)$ is mean square differentiable. Under the assumptions A1 and A2, and using the fact that the covariance functions are assumed to squared exponential, $p_{s}(0)$ and $\theta_s(0)$ are mean squared differentiable. Using the fact that $M_s$ is differentiable in $s$ (see Lemma \ref{lemma3: differentiability of Ms}), $\theta_{s}(\delta t)$, being a linear combination of mean square differentiable functions, is also mean square differentiable.  

%Note that $\theta_{s}(\delta t)$, being a linear combinations of Gaussian process and Gaussian function, is also a Gaussian process. 
Next we show that $E(\theta_{s}(\delta t)) = 0$ and $\mbox{cov}(\theta_{s_1}(\delta t), \theta_{s_2}(\delta t))$ is 4-times differentiable. Denoting the first derivative of $V$ by $V'$, we obtain 

$$E(\theta_{s}(\delta t)) = \beta E(\theta_s(0)) + \frac{\delta t}{M_s} \left\{\alpha E(p_s(0)) - \frac{1}{2} \delta t E(V'(\theta_{s}(0)))\right\} = 0, $$ where we have used the fact that $E(V'(\theta_{s}(0))) = E[E(V'(\theta_{s}(0)))\vert \theta_{s}(0)] = E(0) = 0$.

Therefore, $\mbox{cov}(\theta_{s_1}(\delta t), \theta_{s_2}(\delta t)) = E\left(\theta_{s_1}(\delta t) \theta_{s_2}(\delta t) \right)$. Since the processes are assumed to  be independent, we have 
\begin{align}
	\label{eq14: cov of theta_s}
	E\left(\theta_{s_1}(\delta t) \theta_{s_2}(\delta t)\right) = \beta^2 \sigma_{\theta}^2 e^{-\eta_2||s_1 - s_2||^2} + \frac{(\delta t)^2\alpha^2}{M_{s_1} M_{s_2}} \sigma^2_{p}e^{-\eta_1||s_1-s_2||^2} + \frac{(\delta t)^4}{4 M_{s_1} M_{s_2}} E\left[V'(\theta_{s_1}(0))V'(\theta_{s_2}(0))\right].
\end{align}
Now according to our assumption, $\mbox{cov}(V(x_1),V(x_2)) = \sigma^2 e^{-\eta_3||x_1 - x_2||^2}$, that is, the covariance function of $V(\cdot)$ can be written as $\kappa(h) = \sigma^2 e^{-\eta_3 h^2}$, where $h= ||x_1 - x_2||$. The second derivative of $\kappa(h)$ is given by $-2\sigma^2 \eta_3 e^{-\eta_3 h^2}(1- 2\eta_3 h^2)$. Hence the covariance function of $V' (\cdot)$ will be $\mbox{cov}(V'(x_1),V'(x_2)) = 2\sigma^2 \eta_3 e^{-\eta_3 h^2}(1- 2\eta_3 h^2)$ (see \ctn{Stein1999}, page 21). Therefore, the last term of 
the right hand side of equation (\ref{eq14: cov of theta_s}) can be computed as 
\allowdisplaybreaks
\begin{align}
	\label{eq15}
	E\left[V'(\theta_{s_1}(0))V'(\theta_{s_2}(0))\right] & = EE\left[V'(\theta_{s_1}(0))V'(\theta_{s_2}(0)) \big \vert \theta_{s_1}(0) \theta_{s_2}(0) \right] \notag \\
	& = E\left[2\sigma^2 \eta_3 e^{-\eta_3 (\theta_{s_1}(0) - \theta_{s_2}(0))^2} (1-2\eta_3 (\theta_{s_1}(0) - \theta_{s_2}(0))^2) \right] \notag \\
	& = 2\sigma^2 \eta_3 \left[E e^{-\eta_3 (\theta_{s_1}(0) - \theta_{s_2}(0))^2} - 2\eta_3 E\left\{(\theta_{s_1}(0) - \theta_{s_2}(0))^2 e^{-\eta_3 (\theta_{s_1}(0) - \theta_{s_2}(0))^2}\right\} \right].	
\end{align}
Since $(\theta_{s_1}(0), \theta_{s_2}(0))^T \sim N_{2} \left(\bi{0}, \begin{pmatrix}
	\sigma^2_{\theta} & \sigma^2_{\theta} e^{-\eta_2 ||s_1 - s_2||^2} \\
	\sigma^2_{\theta} e^{-\eta_2 ||s_1 - s_2||^2} & \sigma^2_{\theta}
\end{pmatrix}\right),$ therefore, $\theta_{s_1}(0)- \theta_{s_2}(0)\sim N(0 , 2\sigma^2_{\theta} - 2\sigma^2_{\theta} e^{-\eta_2 ||s_1 - s_2||^2})$, which in turn implies that $$ \frac{(\theta_{s_1}(0)- \theta_{s_2}(0))^2}{\nu} \sim \chi^2_{(1)}, $$ where $\nu = 2\sigma^2_{\theta} - 2\sigma^2_{\theta} e^{-\eta_2 ||s_1 - s_2||^2}.$ Using the fact that the moment generating function (mgf) of a $\chi^2_{(1)}$ random variable is $(1-2t)^{-1/2}$, we have 

\begin{align}
	\label{eq16}
	E \left[e^{-\eta_3 (\theta_{s_1}(0) - \theta_{s_2}(0))^2}\right] = \left(1+ 4\eta_3 \sigma^2_{\theta} \left(1-e^{\eta_2||s_1 - s_2||^2}\right)\right)^{-1/2}. 
\end{align} 
Differentiating equation (\ref{eq16}) with respect to $\eta_3$ and cancelling the minus sign from both sides yield
\begin{align}
	\label{eq17}
	E \left[(\theta_{s_1}(0) - \theta_{s_2}(0))^2 e^{-\eta_3 (\theta_{s_1}(0) - \theta_{s_2}(0))^2}\right] = 2\sigma_{\theta}^2 (1-e^{\eta_2||s_1 - s_2||^2}) \left(1+ 4\eta_3 \sigma^2_{\theta} (1-e^{\eta_2||s_1 - s_2||^2})\right)^{-3/2}. 
\end{align} 
Combining equations (\ref{eq15}), (\ref{eq16}), (\ref{eq17}), we get \begin{align}
	\label{eq18}
	E\left[V'(\theta_{s_1}(0))V'(\theta_{s_2}(0))\right] = 2\sigma^2 \eta_3 \left[ \frac{1}{\left(1+ 4\eta_3 \sigma^2_{\theta} (1-e^{\eta_2||s_1 - s_2||^2})\right)^{1/2}} -  \frac{4\eta_3\sigma^2_{\theta} (1-e^{\eta_2||s_1 - s_2||^2})}{\left(1+ 4\eta_3 \sigma^2_{\theta} (1-e^{\eta_2||s_1 - s_2||^2})\right)^{3/2}}\right].    
\end{align}
Then inserting the value of $E\left[V'(\theta_{s_1}(0))V'(\theta_{s_2}(0))\right]$ from equation (\ref{eq18}) in equation (\ref{eq14: cov of theta_s}) we obtain 
\begin{align*}
	E\left(\theta_{s_1}(\delta t) \theta_{s_2}(\delta t)\right) & = \beta^2 \sigma_{\theta}^2 e^{-\eta_2||s_1 - s_2||^2} + \frac{(\delta t)^2\alpha^2}{M_{s_1} M_{s_2}} \sigma^2_{p}e^{-\eta_1||s_1-s_2||^2} + \frac{(\delta t)^4}{4 M_{s_1} M_{s_2}} 2\sigma^2 \eta_3 \\ 
	& \qquad  \left[ \frac{1}{\left(1+ 4\eta_3 \sigma^2_{\theta} (1-e^{\eta_2||s_1 - s_2||^2})\right)^{1/2}} -  \frac{4\eta_3\sigma^2_{\theta} (1-e^{\eta_2||s_1 - s_2||^2})}{\left(1+ 4\eta_3 \sigma^2_{\theta} (1-e^{\eta_2||s_1 - s_2||^2})\right)^{3/2}}\right].
	%	& \qquad \left.  \\
\end{align*}
Clearly, the covariance function of $\theta_{s}(\delta t)$ is four times differentiable, provided $M_s$ is also four times differentiable.
To apply Lemma \ref{lemma5: mean square differentiability of composition of GP} on $V'(\theta_s(\delta t))$ we have to show that the fourth moment of $\frac{\partial}{\partial s_1}\theta_s(\delta t)$ finitely exists. 
From $$\theta_{s}(\delta t) = \beta \theta_{s}(0) + \frac{\delta t}{M_{s}} \left\{\alpha p_{s}(0)-\frac{\delta t}{2} \frac{\partial}{\partial \theta_s} V\left(\theta_{s} (0)\right) \right\} $$ we obtain
\begin{align}
	\label{eq19}
	\frac{\partial \theta_s(\delta t)}{\partial s_1} & = \beta \frac{\partial \theta_s(0)}{\partial s_1} + \frac{\delta t}{M_s}\left\{\frac{\partial p_{s}(0)}{\partial s_1} - \frac{\delta t}{2} \frac{\partial}{\partial s_1} V'(\theta_s(0))  \right\} - \frac{\delta t}{M_s^2} \frac{\partial M_s}{\partial s_1} \left\{p_{s}(0)-\frac{1}{2} \delta t\frac{\partial}{\partial \theta_s} V\left(\theta_{s} (0)\right) \right\} \notag \\ 
	& = \beta \frac{\partial \theta_s(0)}{\partial s_1} + \frac{\delta t}{M_s}\left\{ \frac{\partial p_{s}(0)}{\partial s_1} - \frac{\delta t}{2} V''(\theta_s(0)) \frac{\partial \theta_s(0)}{\partial s_1}\right\} - \frac{\delta t}{M_s^2} \frac{\partial M_s}{\partial s_1} \left\{p_{s}(0)-\frac{1}{2} \delta t\frac{\partial}{\partial \theta_s} V\left(\theta_{s} (0)\right) \right\}.
\end{align}

Now the fourth moment of $\frac{\partial \theta_s(\delta t)}{\partial s_1} $ will be finite if individually each term of equation (\ref{eq19}) has finite fourth moment, because 
\begin{align*}
	\left(\frac{\partial \theta_s(\delta t)}{\partial s_1}\right)^4 & \leq \kappa \left[\beta^4 \left(\frac{\partial \theta_s(0)}{\partial s_1}\right)^4 + \left(\frac{\delta t}{M_s}\right)^4\left\{ \left(\frac{\partial p_{s}(0)}{\partial s_1}\right)^4 + \left(\frac{\delta t}{2}\right)^4 \left(V''(\theta_s(0)) \frac{\partial \theta_s(0)}{\partial s_1}\right)^4\right\} +  \right. \notag \\
	& \qquad \left. \left(\frac{\delta t}{M_s^2} \frac{\partial M_s}{\partial s_1}\right)^4 \left\{\left(p_{s}(0)\right)^4+\left(\frac{1}{2} \delta t\right)^4\left(\frac{\partial}{\partial \theta_s} V\left(\theta_{s} (0)\right)\right)^4 \right\}\right],
\end{align*}
where $\kappa$ is a suitable constant. 
Note that due to assumptions A1-A2 and squared exponential covariance functions, $\frac{\partial \theta_s(0)}{\partial s_1}$, $\frac{\partial p_{s}(0)}{\partial s_1}$, and $p_{s}(0)$ are Gaussian processes with 0 mean and constant variances. Hence they have finite fourth moments. Therefore, we just need to show that $V''(\theta_s(0)) \frac{\partial \theta_s(0)}{\partial s_1}$ and $\frac{\partial}{\partial \theta_s} V\left(\theta_{s}(0)\right)$ have finite 4th moments. Observe that 
\begin{align}
	\label{eq20}
	E\left[V''(\theta_s(0)) \frac{\partial \theta_s(0)}{\partial s_1}\right]^4 & = E E\left\{ \left[V''(\theta_s(0)) \frac{\partial \theta_s(0)}{\partial s_1}\right]^4\bigg\vert \theta_s(0) \right\} \notag \\
	& = E \left[E\left\{ \left[V''(\theta_s(0))\right]^4\vert \theta_s(0) \right\} \left(\frac{\partial \theta_s(0)}{\partial s_1}\right)^4\right]
\end{align}
Now since $V''(\theta_s(0))$, given $\theta_s(0)$, is Gaussian with mean 0 and constant variance, 

\noindent $E\left\{ \left[V''(\theta_s(0))\right]^4\bigg\vert \theta_s(0) \right\}$ is constant (independent of $\theta_s(0)$), say $\kappa_1$. Hence from equation (\ref{eq20}) we obtain 
\begin{align}
	\label{eq21}
	E\left[V''(\theta_s(0)) \frac{\partial \theta_s(0)}{\partial s_1}\right]^4 = \kappa_1 E\left(\frac{\partial \theta_s(0)}{\partial s_1}\right)^4.
\end{align} Now note that $\frac{\partial \theta_s(0)}{\partial s_1}$ is also Gaussian with mean 0 and constant variance, so that $E\left(\frac{\partial \theta_s(0)}{\partial s_1}\right)^4$ is also constant. Thus, combining equations (\ref{eq20}) and (\ref{eq21}) we have \begin{align}
	\label{eq22}
	E\left[V''(\theta_s(0)) \frac{\partial \theta_s(0)}{\partial s_1}\right]^4 <\infty.
\end{align}

Next to show that $\frac{\partial}{\partial \theta_s} V\left(\theta_{s}(0)\right)$ has finite 4th moment, we notice that 
$$E(V'(\theta_s(0)))^4 = E\left[E(V'(\theta_s(0)))^4\vert \theta_s(0)\right] = \mbox{constant} = \kappa_2, \mbox{say}.$$
The last equality follows because $V'(\theta_s(0))$, given $\theta_s(0)$, is a Gaussian process with 0 mean and constant variance. 

Therefore, the fourth moment of $\frac{\partial}{\partial s_1}\theta_s(\delta t)$ exists finitely. So, by Lemma \ref{lemma5: mean square differentiability of composition of GP}, $V'(\theta_s(\delta t))$ is mean square differentiable, and hence under assumptions A1, A2, 
%Moreover, note that $\theta_{s}(\delta t)$ is a centered Gaussian process with covariance function 4 time differentiable. \textbf{\Large{Etar complete logic ta dite hobe, ekhankar gap ta fill korbo pore.}}
for $h=1$, $$ p_{s}(\delta t) = \alpha^2 p_{s}(0)-\frac{\delta t}{2} \left\{\alpha \frac{\partial}{\partial \theta_{s}}V(\theta_{s}(0))+\frac{\partial}{\partial \theta_s} V(\theta_{s}(\delta t)) \right\},$$ a linear combination of mean square differentiable function, is also mean square differentiable.    Before applying the steps of induction, we show that $\frac{\partial p_{s}(\delta t)}{\partial s_i}$, $i=1,2$, have finite 4th moment as it has to be used for the next step of induction. Note that, for $i=1,2$,  
\begin{align*}
	\frac{\partial p_{s}(\delta t)}{\partial s_i} = \alpha^2  \frac{\partial p_{s}(0)}{\partial s_i} -\frac{\delta t}{2} \left\{\alpha V''(\theta_s(0))\frac{\partial \theta_s(0)}{\partial s_i} + V''(\theta_s(\delta t))\frac{\partial \theta_s(\delta t)}{\partial s_i}\right\}.
\end{align*}
From equation (\ref{eq22}), $E\left(V''(\theta_s(0))\frac{\partial \theta_s(0)}{\partial s_i}\right)^4< \infty.$ We have already shown that $\frac{\partial \theta_s(\delta t)}{\partial s_i}$ has finite fourth moment, for $i=1,2$. Thus, each term in the expression of $\frac{\partial p_{s}(\delta t)}{\partial s_i}$ has finite fourth moment. Hence $E\left(\frac{\partial p_{s}(\delta t)}{\partial s_i}\right)^4< \infty$.

%Using the Lemma \ref{lemma5: mean square differentiability of composition of GP} once more, we have $\frac{\partial}{\partial \theta_s} V(\theta_{s}(\delta t))$ is mean square differentiable. 
%Using the assumptions A1, A2 and using the assumed squared exponential covariance function, $p_{s}(\delta t)$, a linear combination of mean square differentiable function, is also mean square differentiable. 
Thereafter, using the argument of induction as in the proof of Lemma \ref{lemma1: almost sure continuity}, we have the desired result. \hfill $\square$
%
% Note: in this sample, the section number is hard-coded in. Following
% proper LaTeX conventions, it should properly be coded as a reference:
%
%In this appendix we prove the following theorem from
%Section~\ref{sec:textree-generalization}:
%
%
%%%%%%%%%%%%%%%%%%%%%%%%%%%%%%%%%%%%%%%%%%%%%
\section{Calculation of joint conditional density of the observed data}	
\label{joint density of observed data}
%%%%%%%%%%%%%%%%%%%%%%%%%%%%%%%%%%%%%%%%%%%%%%%
Here we will find the data model, that is, the conditional distribution of Data given Latent, $\bi{y}_0$, $\bi{x}_0$ and the parameter $\bi{\theta}$. %We will do that in steps. 

First, we will find the conditional distribution of $\bi{y}_1$ given $\bi{y}_0$, $\bi{x}_0$ and the parameter $\bi{\theta}$. We have, for $i=1,2,\ldots,n$, 
$$y(s_i,1) = \beta\, y(s_i,0)+ \frac{\alpha x(s_i,0)}{M_{s_i}} - \frac{1}{2}\frac{V'(y(s_i,0))}{M_{s_i}},$$ and $$\frac{1}{2}\left[\frac{V'(y(s_1,0))}{M_{s_1}}, \ldots, \frac{V'(y(s_n,0))}{M_{s_n}}\right]' \sim N_{n}\left(\bi{0},\frac{\sigma^2}{4}\Sigma_{0}\right),$$ where $\Sigma_{0}$ is the $n\times n$ covariance matrix with $(i,j)$th element $\frac{2\eta_3 e^{-\eta_3 h_{ij}^2(0)}\left(1-2\eta_3 h_{ij}^2(0)\right)}{M_{s_i}M_{s_j}}.$ Therefore, 	
\begin{align}
	\label{eq2}
	\left[\bi{y}_1\bigg\vert \bi{y}_0; \bi{x}_0; \bi{\theta}\right] \sim N_{n}\left(\bi{\mu}_0,\frac{\sigma^2}{4}\Sigma_{0}\right). 
\end{align}
%	where $\bi{\mu}_0$ is an $n-$dimensional vector with $i$th element being $\beta\, y(s_i,0)+ \frac{x(s_i,0)}{M_{s_i}}$. 
Similarly, 
\begin{align}
	\label{eq3}
	&\bigg[\bi{y}_2\bigg\vert \bi{y}_1; \bi{x}_1; \bi{y}_0; \bi{x}_0; \bi{\theta}
	\bigg]\sim N_{n}\left(\bi{\mu}_1,\frac{\sigma^2}{4}\Sigma_1 \right),
\end{align}
where %$\bi{\mu}_1$ is $n-$dimension with $i$th element being $\beta\, y(s_i,1)+ \frac{x(s_i,1)}{M_{s_i}}$ and 
the $(i,j)$th element of the $n\times n$ covariance matrix $\Sigma_1$ is $\frac{2\eta_3 e^{-\eta_3 h_{ij}^2(1)}\left(1-2\eta_3 h_{ij}^2(1)\right)}{M_{s_i}M_{s_j}}.$ 
Following the same argument, we can write down the likelihood as
\begin{align*}
	%\label{eq4}
	L  &= [\mathbb{D}\big\vert \bi{x}_0;\ldots; \bi{x}_{T-1};\bi{y}_0;\bi{\theta}] \notag \\
	& \propto [\bi{y}_1\big\vert \bi{y}_0;\bi{x}_0; \bi{\theta}] \ldots \left[\bi{y}_T\big\vert \bi{y}_{T-1}, \ldots, \bi{y}_0; \bi{x}_{T-1}, \ldots, \bi{x}_0; \bi{\theta}\right] \notag \\
	& \propto \frac{(\sigma^2)^{-nT/2}}{\prod\limits_{t=1}^{T}|\Sigma_{t-1}|^{1/2}}\, e^{-\frac{2}{\sigma^2}\sum\limits_{t=1}^{T}\left(\bi{y}_{t} - \bi{\mu}_{t-1}\right)^T\Sigma_{t-1}^{-1}\left(\bi{y}_{t} - \bi{\mu}_{t-1}\right)},
\end{align*} 
where, for $j=1, 2, \ldots, T$, the
%	 $\bi{y_{js}} = (y(s_1,j), y(s_2,j), \ldots, y(s_n,j))'$, $i$th element of $\bi{\mu}_{j-1}$ is $\beta\, y(s_i,j-1)+ \frac{x(s_i,j-1)}{M_{s_i}}$ and 
$(k,\ell)$th element of $\Sigma_{j-1}$ is $$\frac{2\eta_3 e^{-\eta_3 h_{k\ell}^2(j-1)}\left(1-2\eta_3 h_{k\ell }^2(j-1)\right)}{M_{s_k}M_{s_\ell}}.$$ %with $h_{k\ell}(j-1) = |y(s_{k},j-1)-y(s_{\ell},j-1)|.$
%%%%%%%%%%%%%%%%%%%%%%%%%%%%%%%%%%%%%%%%%%%%%%%%%%%
\section{Calculation of the conditional joint density of latent data}
\label{joint density of latent data}
%%%%%%%%%%%%%%%%%%%%%%%%%%%%%%%%%%%%%%%%%%%%%%%%%%%
Here we will derive the conditional distribution of  
%latent variables $ X(s_1,1), X(s_2,1), \ldots, X(s_n,1), \ldots, X(s_1,T),X(s_2,T) \ldots, X(s_n,T)$
%$$\mbox{Latent} = \left\{X(s_1,1), X(s_2,1), \ldots, X(s_n,1); X(s_1,2), X(s_2,2), \ldots, X(s_n,2); \ldots; X(s_1,T),X(s_2,T) \ldots, X(s_n,T)\right\}$$
%given data, $\bi{y}_0$, $\bi{x}_0$ and the parameter $\bi{\theta}$. According to the notation used in \ref{complete likelihood}, here we will find the conditional distribution of
 $\mathbb{L}$ given $\mathbb{D}$, $\bi{y}_0$, $\bi{x}_0$ and the parameter $\bi{\theta}$.
%Like in the last subsection here also, the above mentioned conditional distribution will be found in steps. 

First, we shall find the conditional distribution of 
$\bi{x}_1$ given  $\bi{x}_0$, $\bi{y}_0$, and $\bi{y}_1.$ For $i=1,\ldots,n$, we have $$ x(s_i,1) = \alpha^2 x(s_i,0) - \frac{1}{2}\left\{\alpha V'(y(s_i,0)) + V'(y(s_i,1))\right\}.$$
%Let $i,k\in \{1,2,\ldots,n\}$ and $j\in \{0,1\}$. We use the following notation: 
%\begin{align*}
%V'(y(s_i,j)) & = W_{ij}, \\
%\ell_{ik}(0,1) & = |y(s_i,0) - y(s_k,1)|, \\
%h_{ik}(j) & =  |y(s_i,j) - y(s_k,j)|\mbox{ and } \\
%\bi{W} & = \begin{pmatrix}
	%	\bi{W}_0 \\
	%	\bi{W}_1
	%\end{pmatrix},
	%\end{align*}
	%where $\bi{W}_j$ is an $n\times 1$ vector whose $i$th element is $W_{ij}$. 
	Since $V'(\cdot)$ is a random Gaussian function with zero mean and covariance function given in equation~(\ref{eq1}) and $\mathbb{W}_0 = (\bi{W}_0'~ \bi{W}_1')'$ is a $2n\times 1$ vector, $\mathbb{W}_0 \sim N_{2n}(\bi{0},\sigma^2 \Sigma),$ where $\Sigma$ is the $2n \times 2n$ covariance matrix partitioned as 
	$$ \begin{pmatrix}
		\alpha^2 \Sigma_{00} & \alpha \Sigma_{01} \\
		\alpha \Sigma_{10} & \Sigma_{11}
	\end{pmatrix},$$ where the $(i,k)$th element of $\Sigma_{jj}$ is 
	$2\eta_3 e^{-\eta_3 h_{ik}^2(j)}\left(1-2\eta_3 h_{ik}^2(j)\right),$ for $j=0,1$, and the $(i,k)$th element of $\Sigma_{01} = \Sigma_{10}'$ is $2\eta_3 e^{-\eta_3 \ell_{ik}^2(0,1)}\left(1-2\eta_3 \ell_{ik}^2(0,1)\right).$ Therefore, $$\bi{W}_0 + \bi{W}_1 = [I_n \vdots I_n]\mathbb{W}_0 \sim N_n(\bi{0},\sigma^2\left(\alpha^2\Sigma_{00}+\alpha\Sigma_{01}+\alpha\Sigma_{10}+\Sigma_{11}\right)).$$
	Let $\Omega_1 = \alpha^2 \Sigma_{00}+\alpha \Sigma_{01}+\alpha\Sigma_{10}+\Sigma_{11}$.
	Then 
	\begin{align}
		\label{eq5}
		[\bi{x}_1\vert \bi{x}_0; \bi{y}_0;\bi{y}_1;\bi{\theta}] \sim N_n(\alpha^2 \bi{x}_0,\frac{\sigma^2}{4}\Omega_1).
	\end{align}
	%where $i$th element of $\bi{\eta}_0$ is $x(s_i,0)$, $i=1,\ldots, n$.
	Similarly, 
	\begin{align}
		\label{eq6}
		&[\bi{x}_2\vert \bi{x}_1;
		\bi{x}_0;\bi{y}_0; \bi{y}_1; \bi{y}_2; \bi{\theta}]\sim N_n(\alpha^2 \bi{x}_1,\frac{\sigma^2}{4}\Omega_2),
	\end{align}
	%where, for $i=1,\ldots,n$, $i$th element of $\bi{\eta}_1$ is $x(s_i,1)$. 
	where $\Omega_2 = \alpha^2 \Sigma_{11} +\alpha \Sigma_{12}+\alpha\Sigma_{21}+\Sigma_{22}$. For $j=1,2$, the $(i,k)$th element of $\Sigma_{jj}$ is $2\eta_3 e^{-\eta_3 h_{ik}^2(j)}\left(1-2\eta_3 h_{ik}^2(j)\right),$ and the $(i,k)$th element of $\Sigma_{12} = \Sigma_{21}'$ is $2\eta_3 e^{-\eta_3 \ell_{ik}^2(1,2)}\left(1-2\eta_3 \ell_{ik}^2(1,2)\right).$
	Now with the help of equations (\ref{eq5}) and (\ref{eq6}) we write down the conditional latent process model as 
	\begin{align*}
		%\label{eq7: Latent model}
		&\bigg[\mathbb{L}\bigg\vert \bi{y}_0;\ldots; \bi{y}_T;\bi{x}_0;\bi{\theta}\bigg] \notag \\
		& \propto [\bi{x}_1\vert \bi{x}_0; \bi{y}_0;\bi{y}_1;\bi{\theta}] \ldots [\bi{x}_T\vert \bi{x}_{T-1};\ldots;  \bi{x}_0; \bi{y}_0;\ldots; \bi{y}_T;\bi{\theta}] \notag \\
		%& \qquad  \notag \\ 
		& \propto \frac{(\sigma^2)^{-nT/2}}{\prod\limits_{t=1}^{T} |\Omega_t|^{1/2}} e^{-\frac{2}{\sigma^2}\sum\limits_{t=1}^T (\bi{x}_{t} - \alpha^2 \bi{x}_{t-1})^T\Omega_t^{-1}(\bi{x}_{t} - \alpha^2 \bi{x}_{t-1})}, 
	\end{align*}
	where, for $m\in\{1,2,\ldots, T\}$, 
	%$\bi{x}_{ms} = (x(s_1,m), x(s_2,m), \ldots, x(s_n,m))'$, 
	%$\bi{\eta}_{m-1,s} = (x(s_1,m-1), x(s_2,m-1), \ldots, x(s_n,m-1))'$,
	$\Omega_t = \alpha^2\Sigma_{t-1,t-1} +\alpha \Sigma_{t-1,t} + \alpha \Sigma_{t,t-1} + \Sigma_{t,t},$ where the $(i,k)$th element of $\Sigma_{jj}$, for $j=t-1,t$, is $2\eta_3 e^{-\eta_3 h_{ik}^2(j)}\left(1-2\eta_3 h_{ik}^2(j)\right),$ and the $(i,k)$th element of $\Sigma_{t-1,t} = \Sigma_{t,t-1}'$ is 
	$2\eta_3 e^{-\eta_3 \ell_{ik}^2(t-1,t)}$ $ 
	\left(1-2\eta_3 \ell_{ik}^2(t-1,t)\right).$
%%%%%%%%%%%%%%%%%%%%%%%%%%%%%%%%%%%%%%%%%%%%%%%%%%%%%%%%
\section{Calculation of full conditional distributions of the parameters and the latent variables, given the observed data}
\label{Appendix B: full conditional densities}
	%%%%%%%%%%%%%%%%%%%%%%%%%%%%%%%%%%%%%%%%%%%%%%%%%%
	\subsection{Full conditional distribution of $\beta_*$}
	%%%%%%%%%%%%%%%%%%%%%%%%%%%%%%%%%%%%%%%%%%%%%%%%%%%%
	Before finding the full conditional distribution of $\beta$ (hence $\beta_*$), we note that the only term that depends on $\beta$ in equation (\ref{eq9: complete joint}) is $\exp\left\{-\frac{2}{\sigma^2} \sum\limits_{t=1}^{T} \left[\bi{\mu}_{t-1}^T\Sigma_{t-1}^{-1}\bi{\mu}_{t-1} -2 \bi{y}_{t}^T\Sigma_{t-1}^{-1}\bi{\mu}_{t-1}\right]\right\}$. Further notice that $\bi{\mu}_{t} = \beta\bi{y}_{t-1} + \mbox{constant (with respect to $\beta$)}$. Therefore, the term which depends on $\beta$ (hence on $\beta^*$) simplifies to $e^{-\frac{2\beta^2}{\sigma^2} \sum\limits_{t=1}^{T} \bi{y}_{t-1}'\Sigma_{t-1}^{-1}\bi{y}_{t-1}}$
	$\times \, e^{\frac{4\beta}{\sigma^2} \sum\limits_{t=1}^{T}  \bi{y}_{t}'\Sigma_{t-1}^{-1}\bi{y}_{t-1}},$ where $\beta = -1+\frac{2e^{\beta_*}}{1+2e^{\beta_*}}.$ The full conditional density of $\beta_*$, therefore, is given by 
	%Let $c = \sum\limits_{m=1}^{T} \left[ \bi{y}_{T-m+1}'\Sigma_{T-m}^{-1}\bi{y}_{T-m}\right].$ Hence the full conditional distribution of $\beta$ is given by
	\allowdisplaybreaks
	\begin{align*}
		%	\label{eq1: fcd of beta}
		[\beta_*\vert \ldots] & \propto [\beta_*] e^{-\frac{2\beta^2}{\sigma^2} \sum\limits_{t=1}^{T} \bi{y}_{t-1}'\Sigma_{t-1}^{-1}\bi{y}_{t-1} + \frac{4\beta}{\sigma^2} \sum\limits_{t=1}^{T}  \bi{y}_{t}'\Sigma_{t-1}^{-1}\bi{y}_{t-1}} \notag \\
		& \propto e^{-\frac{{\beta_*}^2}{2\sigma_{\beta_*}^2}} e^{-\frac{2\beta^2}{\sigma^2} \sum\limits_{t=1}^{T} \bi{y}_{t-1}'\Sigma_{t-1}^{-1}\bi{y}_{t-1} + \frac{4\beta}{\sigma^2} \sum\limits_{t=1}^{T}  \bi{y}_{t}'\Sigma_{t-1}^{-1}\bi{y}_{t-1}} \notag \\ 
		& =\pi(\beta_*) g_1(\beta_*),
	\end{align*}
	where $\pi(\beta_*) = e^{-\frac{{\beta_*}^2}{2\sigma_{\beta_*}^2}}$ and $g_1(\beta_*) = e^{-\frac{2\beta^2}{\sigma^2} \sum\limits_{t=1}^{T} \bi{y}_{t-1}'\Sigma_{t-1}^{-1}\bi{y}_{t-1} + \frac{4\beta}{\sigma^2} \sum\limits_{t=1}^{T}  \bi{y}_{t}'\Sigma_{t-1}^{-1}\bi{y}_{t-1}}$.%, as mentioned in equation (\ref{eq1: fcd of beta}).
	%Hence the full conditional distribution of $\beta$ follows a normal distribution with mean $\beta_0 + \frac{4\sigma_{\beta}^2}{\sigma^2}c$ and variance $\sigma_{\beta}^2$.
	%The close form of the full conditional density for $\beta^*$ is not available and thus, it is updated using random walk MCMC technique. %will be generated using additive-multiplicative TMCMC (see \ctn{roy2020function}).
	
	%%%%%%%%%%%%%%%%%%%%%%%%%%%%%%%%%%%%%%%%%%%%%%%%%%%%%
	\subsection{Full conditional distribution of $\alpha_*$}
	%%%%%%%%%%%%%%%%%%%%%%%%%%%%%%%%%%%%%%%%%%%%%%%%%%%%%%%
	The full conditional density of $\alpha_*$ will be given by $[\alpha_*]\times [\mbox{Data,Latent}\vert \bi{\theta}]$. Now the term that depends on $\alpha$ (hence on $\alpha_*$) in $[\mbox{Data,Latent}\vert \bi{\theta}]$, that is, in equation (\ref{eq9: complete joint}), is given by $$g_2(\alpha_*) = \frac{1}{\prod\limits_{t=1}^T |\Omega_{t}|^{1/2}} e^{-\frac{2}{\sigma^2}\sum\limits_{t=1}^T \left[ (\bi{x}_t-\alpha^2\bi{x}_{t-1})^T\Omega_{t}^{-1}(\bi{x}_t-\alpha^2\bi{x}_{t-1})\right]} \times e^{-\frac{2\alpha}{\sigma^2}\sum\limits_{t=1}^T \left[\alpha \bi{x}_{t-1}^TD\Sigma_{t-1}^{-1}D\bi{x}_{t-1}-2\bi{y}_t^T\Sigma_{t-1}^{-1}D\bi{x}_{t-1}\right]},$$ where $D$ is the $n\times n$ diagonal matrix containing the diagonal elements $\frac{1}{M_{s_i}}, i=1, \ldots,n.$ In the above calculation we use $\bi{\mu}_{t} = \beta\bi{y}_{t} + \alpha D\bi{x}_{t}.$
	Thus the full conditional density of $\alpha_*$ is given by 
	\begin{align*}
		%\label{fcd of alpha*}
		[\alpha_*\vert \ldots] &\propto e^{-\frac{{\alpha_*}^2}{2\sigma_{\alpha_*}^2}} g_2(\alpha_*).
	\end{align*}
	%However, the close form is not available and hence will be updated using random walk MCMC technique, similar to $\beta^*$. 
	%%%%%%%%%%%%%%%%%%%%%%%%%%%%%%%%%%%%%%%%%%%%55
	\subsection{Full conditional distribution of $\sigma_{\theta}^2$}
	%%%%%%%%%%%%%%%%%%%%%%%%%%%%%%%%%%%%%%%%%%%%%%%%%%%%
	The only term that depends on $\sigma_{\theta}^2$ in equation (\ref{eq9: complete joint}) is $\left(\frac{1}{\sigma_{\theta}^2}\right)^{n/2}\exp\left\{-\frac{1}{2\sigma_{\theta}^2}\bi{y}_0'\Delta_0^{-1}\bi{y}_0 \right\}$, and therefore, the full conditional distribution of $\sigma_{\theta}^2$ is 
	\begin{align*}
		%\label{eq2: fcd of sigma_theta}
		[\sigma_{\theta}^2\vert \ldots ] & \propto [\sigma_{\theta}^2] \left(\frac{1}{\sigma_{\theta}^2}\right)^{n/2} \exp\left\{-\frac{1}{2\sigma_{\theta}^2}\bi{y}_0'\Delta_0^{-1}\bi{y}_0 \right\} \notag \\
		& \propto \left(\frac{1}{\sigma_{\theta}^2}\right)^{\alpha_{\theta}+n/2+1} \exp\left\{ -\frac{1}{\sigma_{\theta}^2}\left(\frac{\gamma_{\theta}+ \bi{y}_0'\Delta_0^{-1}\bi{y}_0}{2}\right)\right\}.
	\end{align*} 
	That is, the full conditional distribution of $\sigma_{\theta}^2$ is IG$\left(\alpha_{\theta}+n/2, \frac{\gamma_{\theta}+\bi{y}_0'\Delta_0^{-1}\bi{y}_0}{2}\right)$.
%	
	%%%%%%%%%%%%%%%%%%%%%%%%%%%%%%%%%%%%%%%%%%
	\subsection{Full conditional distribution of $\sigma_{p}^2$}
	%%%%%%%%%%%%%%%%%%%%%%%%%%%%%%%%%%%%%%%%%%%
	Since the only term that depends on $\sigma_{p}^2$ in equation (\ref{eq9: complete joint}) is $\left(\frac{1}{\sigma_{p}^2}\right)^{n/2}\exp\left\{-\frac{1}{2\sigma_p^2}\bi{x}_0'\Omega_0^{-1}\bi{x}_0 \right\}$ and we have assumed inverse gamma with parameters $\alpha_{p}$ and $\gamma_{p}$, 
	\begin{align*}
		%\label{eq3: fcd of sigma_p}
		[\sigma_p^2\vert \ldots ] & \propto \left(\frac{1}{\sigma_{p}^2}\right)^{\alpha_{p}+1} \left(\frac{1}{\sigma_{p}^2}\right)^{n/2}\exp\left\{ -\frac{1}{\sigma_{p}^2}\left(\frac{\gamma_{p}+ \bi{x}_0'\Omega_0^{-1}\bi{x}_0}{2}\right)\right\}.
	\end{align*}	
	This implies that the full conditional distribution of $\sigma_p^2$ is IG$\left(\alpha_{p}+n/2, \frac{\gamma_{p}+\bi{x}_0'\Delta_0^{-1}\bi{x}_0}{2}\right)$.
%	
	%%%%%%%%%%%%%%%%%%%%%%%%%%%%%%%%%%%%%
	\subsection{Full conditional distribution of $\sigma^2$}
	%%%%%%%%%%%%%%%%%%%%%%%%%%%%%%%%%%%%%%%%%%%%%%
	Note that $[\mbox{Data}\vert \mbox{Latent}, \bi{x}_0, \bi{y}_0, \bi{\theta}]$ and $[\mbox{Latent}\vert \mbox{Data}, \bi{x}_0, \bi{y}_0, \bi{\theta}]$ depend on $\sigma^2$. Therefore, the full conditional distribution of $\sigma^2$ can be achieved as follows:
	\begin{align*}
		%\label{eq: fcd of sigma}
		[\sigma^2\vert \ldots] &\propto [\sigma^2]\prod\limits_{t=1}^{T} [\bi{y}_t\vert \bi{y}_{t-1},\bi{x}_{t-1}, \bi{\theta}] [\bi{x}_t\vert \bi{y}_{t},\bi{y}_{t-1},\bi{x}_{t-1}, \bi{\theta}] \notag \\ 
		& \propto [\sigma^2] (\sigma^2)^{-Tn} \exp\left\{-\frac{2}{\sigma^2}\sum\limits_{t=1}^T \left[(\bi{y}_t - \bi{\mu}_{t})^T\Sigma_{t-1}^{-1} (\bi{y}_t - \bi{\mu}_{t}) + (\bi{x}_t - \alpha^2\bi{x}_{t-1})^T\Omega_{t}^{-1} (\bi{x}_t - \alpha^2\bi{x}_{t-1})\right] \right\} \notag \\ 
		& \propto \left(\frac{1}{\sigma^2}\right)^{\alpha_v+ Tn + 1} \exp\left\{-\frac{1}{\sigma^2} \left[\frac{\gamma_v}{2} + 2 \zeta  \right]\right\},
	\end{align*}
	where $\zeta = \sum\limits_{t=1}^T \left[(\bi{y}_t - \bi{\mu}_{t})^T\Sigma_{t-1}^{-1} (\bi{y}_t - \bi{\mu}_{t}) + (\bi{x}_t - \alpha^2\bi{x}_{t-1})^T\Omega_{t}^{-1} (\bi{x}_t - \alpha^2\bi{x}_{t-1})\right].$ Hence the full conditional distribution of $\sigma^2$ is inverse-Gamma with parameters $\alpha_v+Tm$ and $\gamma_v/2 + 2\zeta$.
%	
	%%%%%%%%%%%%%%%%%%%%%%%%%%%%%%%%%%%%%%%%%%%5
	\subsection{Full conditional distributions of $\eta_1^*$, $\eta_2^*,$ and $\eta_3^*$}
	%%%%%%%%%%%%%%%%%%%%%%%%%%%%%%%%%%%%%%%%%%%%%%
	%The full conditional distributions of $\eta_1$, $\eta_2,$ and $\eta_3$ do not have a closed form and hence, will be updated using TMCMC (\ctn{dutta2014markov}). The algorithm is provided in the next section. 
	We observe that only $[\bi{x}_0\vert \bi{\theta}]$ depends $\eta_1$ (hence on $\eta_1^*$) and $[\bi{Y}_0\vert \bi{\theta}]$ depends on $\eta_2$ (hence on $\eta_2^*$). Therefore, the full conditional densities of $\eta_1^*$ and $\eta_2^*$ are given by 
	\allowdisplaybreaks
	\begin{align}
		\label{fcd of eta1*}
		[\eta_1^*\vert \ldots] & \propto [\eta_1^*][\bi{x}_0\vert \bi{\theta}] \notag \\
		& \propto e^{-{\eta_1^*}^{2}/2} \frac{1}{|\Omega_0|^{1/2}} e^{-\frac{1}{2\sigma_p^2} \bi{x}_0^T\Omega_0^{-1}\bi{x}_0} \notag \\
		& = \pi(\eta_1^*)g_3(\eta_1^*)
	\end{align}
	and 
	\begin{align}
		\label{fcd of eta2*}
		[\eta_2^*\vert \ldots] & \propto [\eta_2^*][\bi{y}_0\vert \bi{\theta}] \notag \\
		& \propto e^{-{\eta_2^*}^{2}/2} \frac{1}{|\Delta_0|^{1/2}} e^{-\frac{1}{2\sigma_{\theta}^2} \bi{y}_0^T\Delta_0^{-1}\bi{y}_0} \notag \\
		& = \pi(\eta_2^*)g_4(\eta_2^*),
	\end{align}
	respectively, where $\eta_1 = e^{\eta_1^*}$, $\eta_2 = e^{\eta_2^*}$, $\pi(\eta_1^*) = e^{-{\eta_1^*}^{2}/2}$, $\pi(\eta_2^*) = e^{-{\eta_2^*}^{2}/2}$, $g_{3}(\eta_1^*) = \frac{1}{|\Omega_0|^{1/2}} e^{-\frac{1}{2\sigma_p^2} \bi{x}_0^T\Omega_0^{-1}\bi{x}_0}$ and $g_4 (\eta_2^*)= \frac{1}{|\Delta_0|^{1/2}} e^{-\frac{1}{2\sigma_{\theta}^2} \bi{y}_0^T\Delta_0^{-1}\bi{y}_0}$.
	
	On the other hand, the joint conditional distribution of (Data, Latent) given $\bi{x}_0, \bi{y}_0, \bi{\theta}$ depends on $\eta_3$ (hence on $\eta_3^*$). Thus, the full conditional distribution of $\eta_3^*$ is given by 
	\begin{align}
		\label{fcd of eta3^*}
		[\eta_3^*] &\propto [\eta_3^*]\prod\limits_{t=1}^{T} [\bi{y}_t\vert \bi{y}_{t-1},\bi{x}_{t-1}, \bi{\theta}] [\bi{x}_t\vert \bi{y}_{t},\bi{y}_{t-1},\bi{x}_{t-1}, \bi{\theta}] \notag \\
		& \propto e^{-{\eta_3^*}^{2}/2} \frac{1}{\prod\limits_{t=1}^{T}|\Sigma_{t-1}|^{1/2}|\Omega_t|^{1/2}} e^{-\frac{2}{\sigma^2}\sum\limits_{t=1}^T \left[(\bi{y}_t - \bi{\mu}_{t})^T\Sigma_{t-1}^{-1} (\bi{y}_t - \bi{\mu}_{t}) + (\bi{x}_t - \alpha^2\bi{x}_{t-1})^T\Omega_{t}^{-1} (\bi{x}_t - \alpha^2\bi{x}_{t-1})\right]} \notag \\
		& = \pi(\eta_3^*)g_5(\eta_3^*),
	\end{align}
	where $\eta_3 = e^{\eta_3^*}$, $\pi(\eta_3^*) = e^{-{\eta_3^*}^{2}/2}$ and \newline $g_5(\eta_3^*) = \frac{1}{\prod\limits_{t=1}^{T}|\Sigma_{t-1}|^{1/2}|\Omega_t|^{1/2}} e^{-\frac{2}{\sigma^2}\sum\limits_{t=1}^T \left[(\bi{y}_t - \bi{\mu}_{t})^T\Sigma_{t-1}^{-1} (\bi{y}_t - \bi{\mu}_{t}) + (\bi{x}_t - \alpha^2\bi{x}_{t-1})^T\Omega_{t}^{-1} (\bi{x}_t - \alpha^2\bi{x}_{t-1})\right]}$. 
%	
	%%%%%%%%%%%%%%%%%%%%%%%%%%%%%%%%%%%%%%%%%%%5
	\subsection{Full conditional distribution of $\bi{x}_0$}
	%%%%%%%%%%%%%%%%%%%%%%%%%%%%%%%%%%%%%%%%%%%%%%
	Using the fact that only $\bi{x}_1$ and $\bi{y}_1$ depend on $\bi{x}_0$ and writing $\bi{\mu}_0 = \beta\bi{y}_{0} +\alpha D\bi{x}_0$, we have
	\allowdisplaybreaks
	\begin{align}
		\label{eq4: fcd of x0}
		[\bi{x}_0\vert \ldots] & \propto [\bi{x}_0\vert \bi{\theta}] [\bi{x}_1\vert \bi{y}_1, \bi{y}_0, \bi{x}_0,\bi{\theta}][\bi{y}_1\vert \bi{y}_0, \bi{x}_0,\bi{\theta}] \notag \\
		& \propto e^{-\frac{1}{2\sigma^2_p}\bi{x}_0^T\Omega_0^{-1}\bi{x}_0} e^{-\frac{2}{\sigma^2}(\bi{x}_1-\alpha^2\bi{x}_0)^T\Omega_1^{-1}(\bi{x}_1-\alpha^2\bi{x}_0)} e^{-\frac{2}{\sigma^2}(\bi{y}_1-\beta\bi{y}_0-\alpha D\bi{x}_0)^T\Sigma_0^{-1}(\bi{y}_1-\beta\bi{y}_0-\alpha D\bi{x}_0)} \notag \\
		& \propto e^{-\frac{1}{2\sigma_p^2}\left[\bi{x}_0^T\Omega_0^{-1}\bi{x}_0 + \frac{4\sigma_p^2\alpha^4}{\sigma^2}\bi{x}_0^T\Omega_1^{-1}\bi{x}_0 + \frac{4\sigma_p^2 \alpha^2}{\sigma^2} \bi{x}_0^T D\Sigma_0^{-1}D\bi{x}_0 - \frac{8\sigma_p^2\alpha^2}{\sigma^2}\bi{x}_0^T \Omega_1^{-1}\bi{x}_1 - \frac{8\sigma_p^2 \alpha}{\sigma^2}\bi{x}_0^T D \Sigma_0^{-1}(\bi{y}_1-\beta \bi{y}_0)\right]} \notag \\
		& \propto e^{-\frac{1}{2\sigma_p^2}\left[\bi{x}_0^T A\bi{x}_0 - 2 \bi{x}_0^T B \bi{x}_1 - 2\bi{x}_0^T C(\bi{y}_1 - \beta \bi{y}_0) \right]} \notag \\
		& \propto e^{-\frac{1}{2\sigma_p^2} (\bi{x}_0 - A^{-1}B\bi{x}_1 - A^{-1}C(\bi{y}_1 - \beta \bi{y}_0))^T A(\bi{x}_0 - A^{-1}B\bi{x}_1 - A^{-1}C(\bi{y}_1 - \beta \bi{y}_0))},
	\end{align}
where % $D$ is an $n\times n$ diagonal matrix with $i$th diagonal entry as $\frac{1}{M_{s_i}}$, 
	$A = \Omega_0^{-1}+\frac{4\sigma_p^2\alpha^4}{\sigma^2} \Omega_1^{-1}+ \frac{4\sigma_p^2 \alpha^2}{\sigma^2} D\Sigma_{0}^{-1}D$, $B = \frac{4\sigma_p^2\alpha^2}{\sigma^2}\Omega_1^{-1}$, $C = \frac{4\sigma_p^2 \alpha}{\sigma^2}D\Sigma_0^{-1}$. We note here that $D$ (= diag$(1/M_{s_i})_{i=1}^{n}$) is a positive definite matrix as all the diagonal entries are strictly positive, $A$ being a sum of three positive definite matrices is also positive definite and hence invertible. Thus, from equation (\ref{eq4: fcd of x0}), we get $[\bi{x}_0\vert \ldots ]\sim N_{n}(A^{-1}(B\bi{x}_1+C(\bi{y}_1 - \beta \bi{y}_0)),\sigma_p^2A^{-1}).$ 
%
%%%%%%%%%%%%%%%%%%%%%%%%%%%%%%%%%%%%%%%%%%%
\section{Algorithm for temporal prediction}
\label{temporal prediction}
%%%%%%%%%%%%%%%%%%%%%%%%%%%%%%%%%%%%%%%%%%%
We provide the algorithms for simulating observations from the posterior predictive densities for single-time point and multiple-time point predictions. 
%%%%%%%%%%%%%%%%%%%%%%%%%%%%%%%%%%%%%%%%%%%
%%%%%%%%%%%%%%%%%%%%%%%%%%%%%%%
Let $\bi{Y}_{T+1} = (Y(s_1, T+1), \ldots, Y(s_n,T+1))^T$ and $\bi{x}_{T+1} = (x(s_1, T+1), \ldots, x(s_n,T+1))^T$ be the response and latent variable  at time point $T+1$. The aim is to simulate observations from the predictive density of $\bi{Y}_{T+L}$, for $L=1, 2, \ldots$. For $k=1, \ldots, N$, where $N$ is the number of iterations after the burn-in period $B$, do the following: 
\begin{enumerate}
    \item[S1.] simulate from $[\mathbb{L},\bi{\theta}\vert \mathbb{D}]$ using Gibbs, and Metropolis Hastings within Gibbs via the full conditionals as outlined in Section \ref{full conditional}. Let the simulated value at the $k$th iteration be denoted by $\bi{\theta}^{(k)}$ and $\mathbb{L}^{(k)}$. 
    \item[S2.] Given $\bi{\theta}^{(k)}$ and $\mathbb{L}^{(k)}$, simulate $n-$variate observation from normal distribution with the mean $\bi{\mu}_{T}$, where the $i$th element of $\bi{\mu}_{T}$ is as mentioned in Section \ref{complete likelihood} with $m$ replaced with $T$, $\beta$ replaced with $\beta^{(k)}$, $\alpha$ replaced with $\alpha^{(k)}$, and 
covariance matrix $\Sigma_{T}$. The elements of $\Sigma_{T}$ is the same as given in Section \ref{data model} after equation \ref{eq4}, with $\eta_3$, $\alpha$ and $j$ replaced with $\eta_3^{(k)}$, $\alpha^{(k)}$ and $T+1$, respectively. Let the simulated observations be denoted by $\bi{y}_{T+1}^{(k)}$. 
    \item[S4.] Then simulate $\bi{x}_{T+1}$ from the $n-$variate normal distribution with mean $\alpha^{(k)^2}\bi{x}_{T}$ and covariance matrix $\Omega_{T+1}$. 
  The elements of $\Omega_{T}$ is the same as described in Section \ref{process model} with $t$ replaced by $T+1$, $\alpha$ replaced by $\alpha^{(k)}$ and $\eta_3$ replaced by $\eta_3^{(k)}$. 
    \item[S5.] Update the data $\mathbb{D}$ to $\mathbb{D}_1^{(k)} = \mathbb{D}\cup \{\bi{y}_{T+1}^{(k)}\}$ and the latent $\mathbb{L}^{(k)}$ to $\mathbb{L}_1^{(k)} = \mathbb{L}\cup \{\bi{x}_{T+1}^{(k)}\}.$
    \item[S6.] Given $\mathbb{D}_1^{(k)},\mathbb{L}_1^{(k)}$ and $\bi{\theta}^{(k)}$, simulate $\bi{y}_{T+2}^{(k)}$ from the $n-$variate normal with mean $\bi{\mu}_{T+1}$ and covariance matrix $\Sigma_{T+1}$, where the $i$th element of $\bi{\mu}_{T+1}$ is as mentioned in Section \ref{complete likelihood} with $m$, $\beta$ and $\alpha$ replaced with $T+1$, $\beta^{(k)}$ and $\alpha^{(k)}$, respectively, and covariance matrix $\Sigma_{T+1}$. The elements of $\Sigma_{T+1}$ is the same as given in Section \ref{data model} after equation \ref{eq4}, with $\eta_3$, $\alpha$ and $j$ replaced with $\eta_3^{(k)}$, $\alpha^{(k)}$ and $T+2$, respectively. Let the simulated observations be denoted by $\bi{y}_{T+2}^{(k)}$.
    \item[S7.] Then simulate $\bi{x}_{T+2}$ from the $n-$variate normal distribution with mean $\alpha^{(k)^2}\bi{x}_{T+1}^{(k)}$ and covariance matrix $\Omega_{T+1}$. The elements of $\Omega_{T+1}$ are the same as described in Section \ref{process model} with $t$ replaced by $T+2$, $\alpha$ replaced by $\alpha^{(k)}$ and $\eta_3$ replaced by $\eta_3^{(k)}$. 
    \item[S8.] Continue the procedure to get a simulated value $\bi{y}_{T+L}^{(k)}$. 
\end{enumerate}
%------------------------------------------------
\section{Algorithm for prediction of time series at a spatial point}
\label{spatial prediction}
%=================================================
Let $s^*$ be a spatial location at which we want to obtain prediction for the time points $t_1, \ldots, t_T$. Therefore, it is required to obtain observations from the distribution given by $\left[y(s^*, t), t=1, \ldots, T, \vert \mathbb{D}\right]$. Let us denote $\left(y(s^*, t), t = 1, \ldots, T\right)$ by $\bi{y}^*$ and $\left(x(s^*, t), t=1\ldots,T \right)$ by $\bi{x}^*$. 
Notice that $\left(y(s_1,t), \ldots, y(s_n,t), y(s^*,t)\right)$ given $\mathbb{L}, \bi{x}^{*}, \bi{\theta}$ and $\left(y(s_1,t-1), \ldots, y(s_n,t-1), y(s^*,t-1)\right)$ is a $(n+1)-$variate normal with mean $\bi{\mu}^*_{t-1} = \left(\beta y(s_1,t-1)+\frac{\alpha x(s_1,t-1)}{M_{s_1}}, \ldots, \beta y(s_n,t-1)+\frac{\alpha x(s_n,t-1)}{M_{s_n}}, \beta y(s^*,t-1)+\frac{\alpha x(s^*,t-1)}{M_{s^*}}\right)^T$, and covariance matrix $\Sigma^*_{t-1}$ of order $n+1\times n+1$. The elements of $\Sigma^*_{t-1}$ are the same as given in Section \ref{data model}. Let $\Sigma^*_{t-1}$ be partitioned as $\Sigma^*_{t-1} = \begin{pmatrix}
\Sigma_{t-1} & \bi{S}_{1,t-1} \\
\bi{S}_{1,t-1}^T & s_{n+1,n+1}
\end{pmatrix}$, where $\Sigma_{t-1}$ is a $n\times n$ matrix, $\bi{S}_{1,t-1}$ is a $n\times 1$ vector. Therefore, $[y(s^*,t)\vert \ldots]$ follows a univariate normal with mean $\beta y(s^*,t-1)+\frac{\alpha x(s^*,t-1)}{M_{s^*}} - \bi{S}_{1,t-1}^T \Sigma_{t-1}^{-1}(\bi{y}_t - \bi{\mu}_t)$ and variance $s_{n+1,n+1} - \bi{S}_{1,t-1}^T \Sigma_{t-1}^{-1} \bi{S}_{1,t-1}$. We use this fact to simulate $y(s^*,t)$, for $t=1, \ldots, T$, sequentially.

Initialize $\bi{y}^*$ and $\bi{x}^*$ by $\bi{y}^{*(0)}$ and $\bi{x}^{*(0)}$, and all the other unknowns. For $k=1, \ldots, N+B$, do the following.
\begin{enumerate}
    \item[S1.] For $t=1, \ldots, T$,
    \item[S2.] Define $\mathbb{D}^{(k-1)} = [\mathbb{D},\bi{y}^{*(k-1)}]$ and $\mathbb{L}^{(k-1)} = [\mathbb{L}, \bi{x}^{*(k-1)}]$. 
    \item[S3.] Simulate $\bi{\theta}^{(k)}$ from $[\bi{\theta}|\mathbb{D}^{(k-1)}, \mathbb{L}^{(k-1)}]$ using Gibbs and MH within Gibbs via full conditionals as given in Section \ref{full conditional}. Let the simulated value be $\bi{\theta}^{(k)}$. 
    \item[S4.] Given, $\bi{\theta}^{(k)}$ and $\mathbb{D}^{(k-1)}$, simulate the latent variables from their full conditionals as given in Section \ref{full conditional}. Let the simulated value be $\mathbb{L}^{(k)}$.
    \item[S5.] Given $\bi{\theta}^{(k)}$ and $\mathbb{L}^{(k)}$, simulate $y(s^*,t)$ from $[y(s^*,t)\vert L^{(k)}, \bi{\theta}^{(k)}, \mathbb{D}]$, say $y(s^*,t)^{(k)}$. 
\end{enumerate}
Store $N$ observations after discarding $B$ observations as burn-in. 
%
%======================================================
\section{Modification for irregularly spaced data}
\label{Modification for irregularly spaced data}
%-===================================================
%Below I give a **conceptual extension, an explicit algorithm, and Bayesian inference code skeleton** for adapting the Hamiltonian spatio-temporal model in your paper to **irregularly spaced time points**, staying faithful to the assumptions and construction in equations (3)–(4) and Supplementary Algorithms H and I.

%I will **not re-derive the paper**, but instead show *how to modify exactly what is already there* so that irregular time steps are handled rigorously and cleanly.
%All statements are grounded in the model and likelihood derivations in Sections 2–5 and Appendix H/I of the paper.
Below we provide an idea how our model can be used in irregularly spaced time points. The detailed simulation and real data analysis will be taken up in a separate work. To deal with the irregularly spaced time points, we modify the equations (\ref{observation equation}) and (\ref{latent equation}) following the discretization of Ornstein-Uhlenbeck models.
%---

%## 1. Key observation: the model is already continuous-time in spirit

%Although equations (3)–(4) are written with a fixed step size (\delta t),

%[
%\begin{aligned}
%y(s,t+\delta t) &= \beta y(s,t)

%* \delta t, M_s^{-1}!\left(\alpha x(s,t) - \frac{\delta t}{2} V'(y(s,t))\right), \
%  x(s,t+\delta t) &= \alpha^2 x(s,t)
%  -\frac{\delta t}{2}{\alpha V'(y(s,t)) + V'(y(s,t+\delta t))},
%  \end{aligned}
%  ]

%the **underlying motivation is continuous Hamiltonian dynamics** (Section 2), discretized via leap-frog.
%Therefore:

%> **Irregular time steps are handled by letting (\delta t) vary with time.**

Let observed times be
$[0 = t_0 < t_1 < \cdots < t_T,\quad \Delta t_k = t_k - t_{k-1}]$.
%We replace $\delta t$ with $\Delta t_k$.
%
%---
%
%## 2. Irregular-time Hamiltonian update equations
%
For each spatial location (s) and time interval $[t_{k-1},t_k]$, the modified observation equation is given by
%### Observation equation
$$y(s,t_k) =\beta^{\Delta t_k} y(s,t_{k-1})
+
\Delta t_k, M_s^{-1}
\left(
\alpha x(s,t_{k-1})-
\frac{\Delta t_k}{2} V'(y(s,t_{k-1}))
\right)$$
%
%### Latent evolution
and the modified latent equation is provided as
$$x(s,t_k)
=
\alpha^{2\Delta t_k} x(s,t_{k-1})-
\frac{\Delta t_k}{2}
\Big[
\alpha V'(y(s,t_{k-1}))
+
V'(y(s,t_k))
\Big].$$
%## 6. Why this works (important theoretical point)
The theoretical properties that we have proved in Section \ref{process proposal} remain valid. These can be argued as follows: Leap-frog 
algorithm is valid for variable step sizes; 
we already assume that the model has continuous paths in time; and, finally, the correlation decays to 0 as the time lags increases to infinity because 
$\prod_k \beta^{\Delta t_k} \to 0
  \quad\text{as}\quad \sum_k \Delta t_k \to \infty.$ %Further, we already assumes that the model has continuous paths in time therefore, with the above modification the continuity in time does not change. 

%because the model already assumes continuous paths in time (Theorem 2.2–2.5).
%* Leap-frog is valid for **variable step sizes**.
%* Correlation decay still holds because

%Hence the **core theoretical guarantees of the paper remain intact**.

%### Why the powers?

%* In the regular grid, stability requires (|\alpha|<1), (|\beta|<1).
%* In irregular time, the correct continuous-time analogue is
%  [
%  \alpha^{\Delta t_k},\quad \beta^{\Delta t_k},
%  ]
%  exactly as in discretized OU / linear SDE models.
%* This preserves **correlation decay** (Theorem 2.1) and stability.

%---

%3. Likelihood with irregular time gaps
Moreover, the likelihoods in Section \ref{complete likelihood}
can be modified easily with changed covariance matrices. They are described below. 
%
%### Conditional data likelihood (generalization of eq. (6))
First we modify the conditional data likelihood, that is, for $k=1,\dots,T$,
\[
y_k \mid y_{k-1}, x_{k-1}, \Theta
\sim
\mathcal N_n\left(\mu_{k-1}, ;\sigma^2 \Sigma_{k-1}^{(\Delta t_k)}\right),
\]
where
\[
\mu_{k-1}(i)
=
\beta^{\Delta t_k} y(s_i,t_{k-1})
+
\frac{\Delta t_k}{M_{s_i}}
\left(
\alpha x(s_i,t_{k-1})
-\frac{\Delta t_k}{2} V'(y(s_i,t_{k-1}))
\right),
\]
and
\[
\Sigma_{k-1}^{(\Delta t_k)}(i,j)
=
\frac{\Delta t_k^2}{M_{s_i}M_{s_j}}
\operatorname{cov}\left(
V'(y(s_i,t_{k-1})),
V'(y(s_j,t_{k-1}))
\right).
\]
with 
\[
\operatorname{cov}(V'(x),V'(x'))
=
2\eta_3\sigma^2 e^{-\eta_3|x-x'|^2}(1-2\eta_3|x-x'|^2).
\]
as given in equation (\ref{eq1}). 
%
%### Latent likelihood (generalization of eq. (7))
Similarly, the likelihood of the latent variables can be written as 
\[
x_k \mid x_{k-1}, y_{k-1}, y_k
\sim
\mathcal N_n
\left(
\alpha^{2\Delta t_k} x_{k-1},;
\frac{\sigma^2 \Delta t_k^2}{4}\Omega_k
\right),
\]
where the definition of $\Omega_k$ remains the same as in Section \ref{process model}, with replacing the distances by
\[
|y(s_i,t_{k-1}) - y(s_j,t_k)|.
\]
\clearpage
%===================================================
\section{Supplementary plots for 3 component mixture of GPs}
\label{Trace plt and posterior plts for 3 comp mixture GP}
%=================================================
In this appendix, we provide the trace plots and the posterior densities of the complete time series of the latent variables for the simulated data used in Section \ref{mixture of 3-comp GPs}. 
%------------------------------------------
\subsection{Trace plots}
\label{3 comp mixture GP: trace plots}
%------------------------------------------
\begin{figure}[!h]
    \includegraphics[width=0.45\textwidth,height=4.5cm]{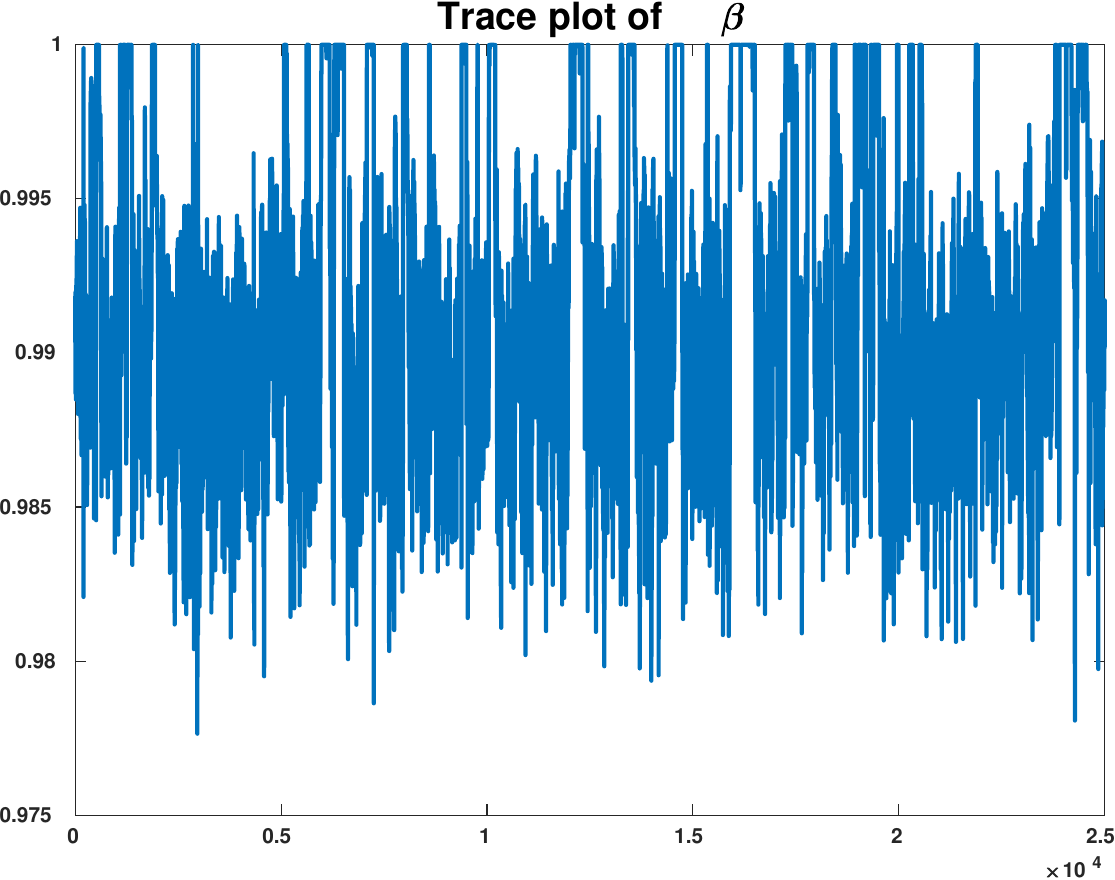}
    \includegraphics[width=0.45\textwidth,height=4.5cm]{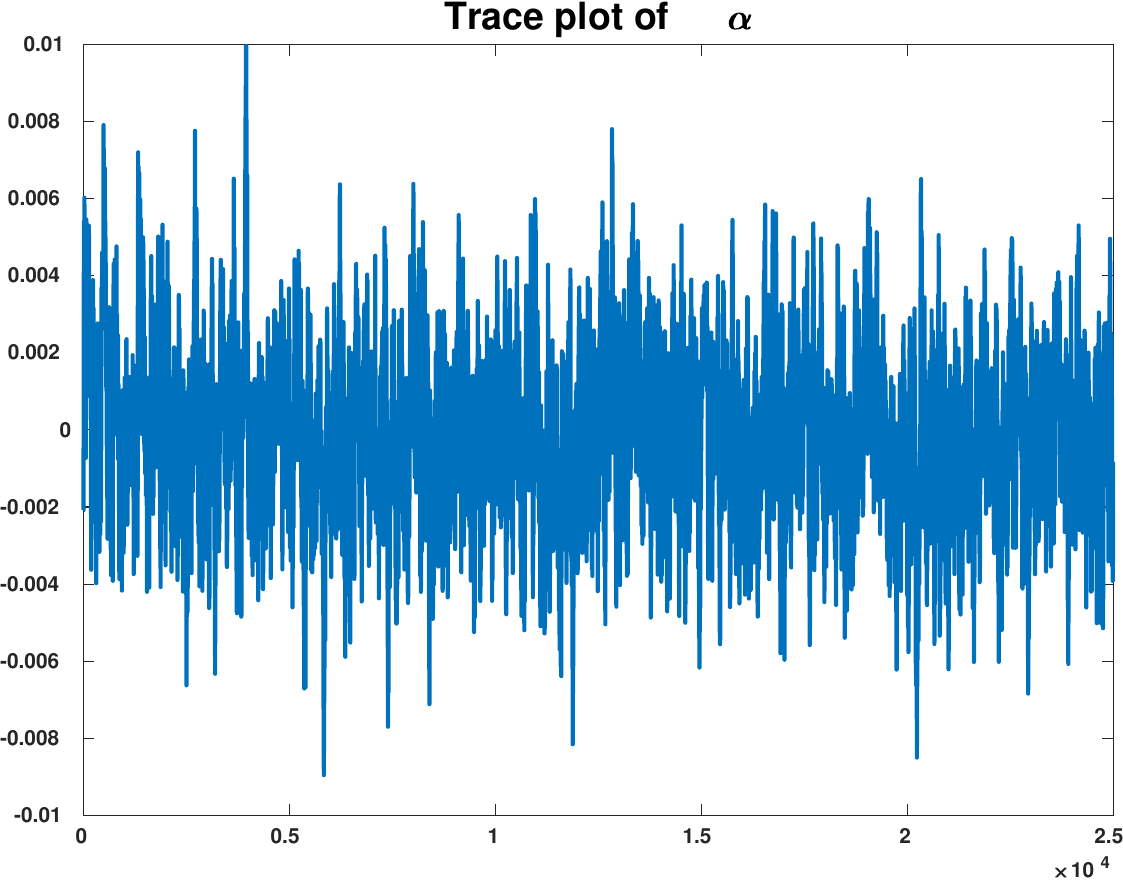}
    \includegraphics[width=0.45\textwidth,height=4.5cm]{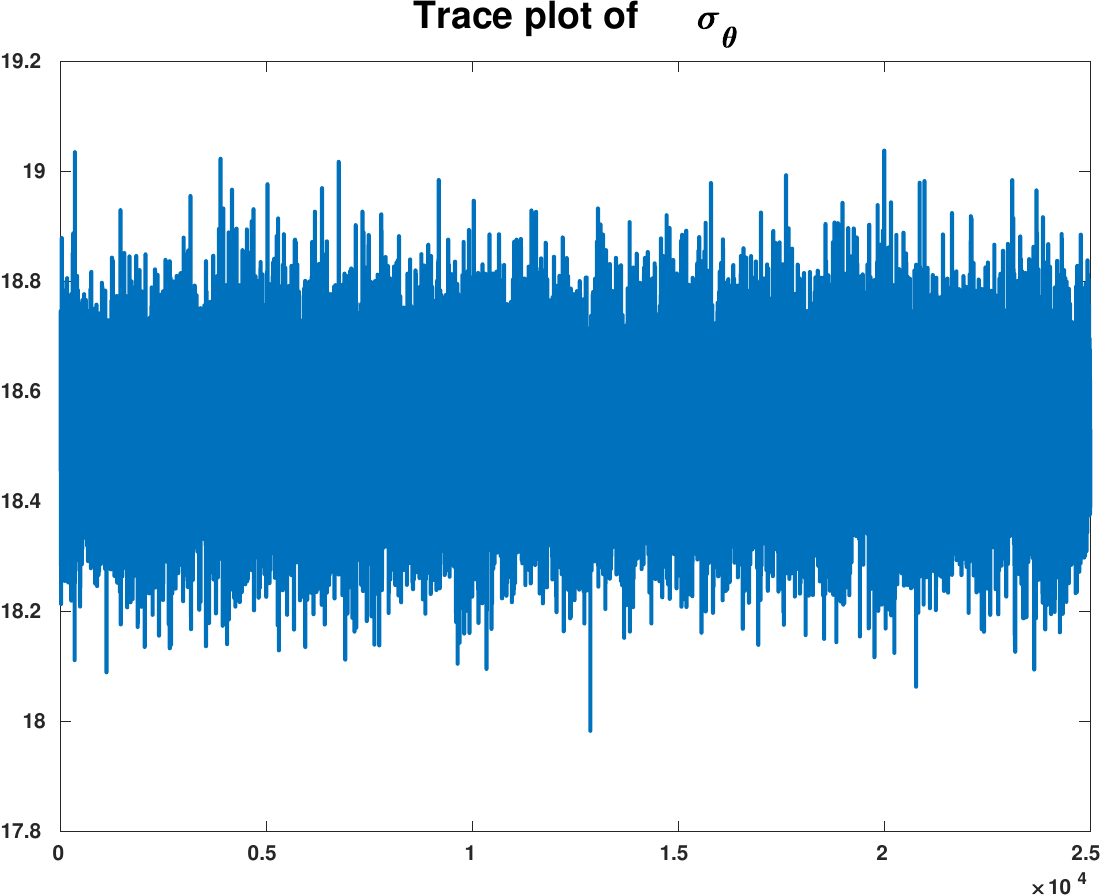}
    \includegraphics[width=0.45\linewidth,height=4.5cm]{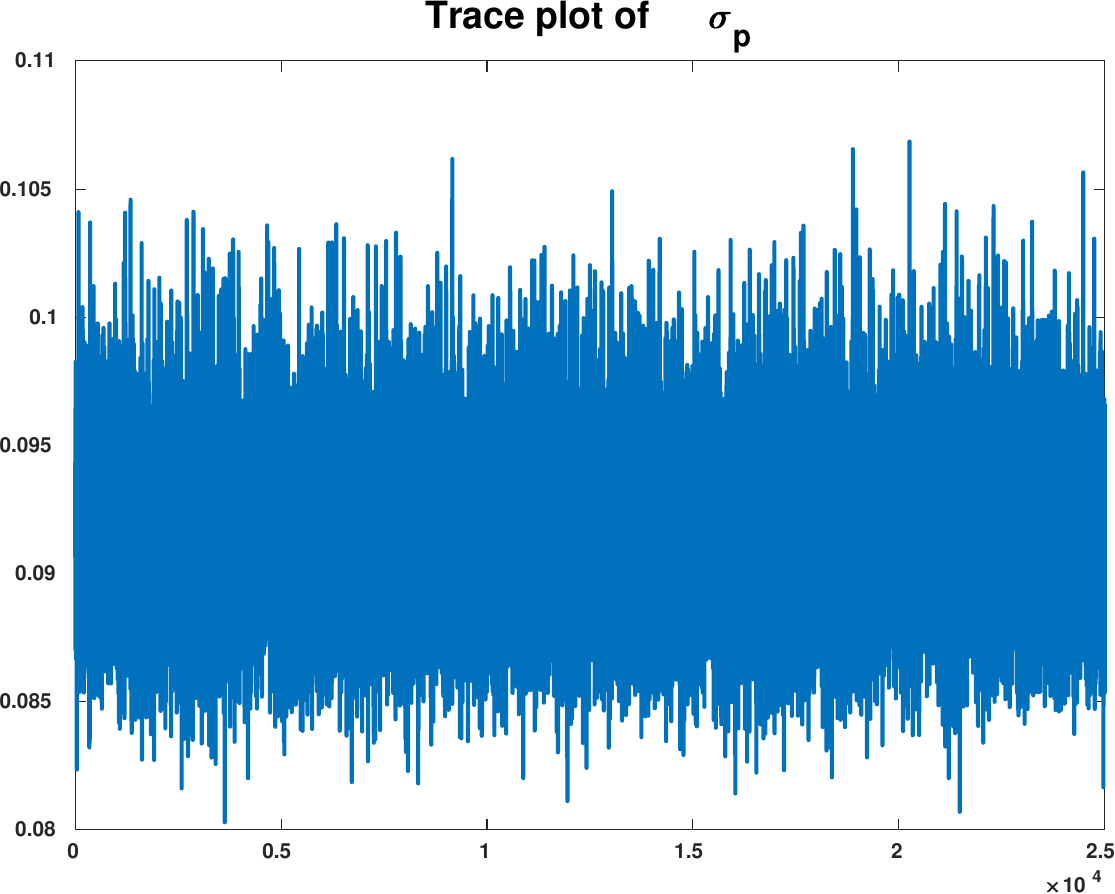}
    \includegraphics[width=0.45\textwidth,height=4.5cm]{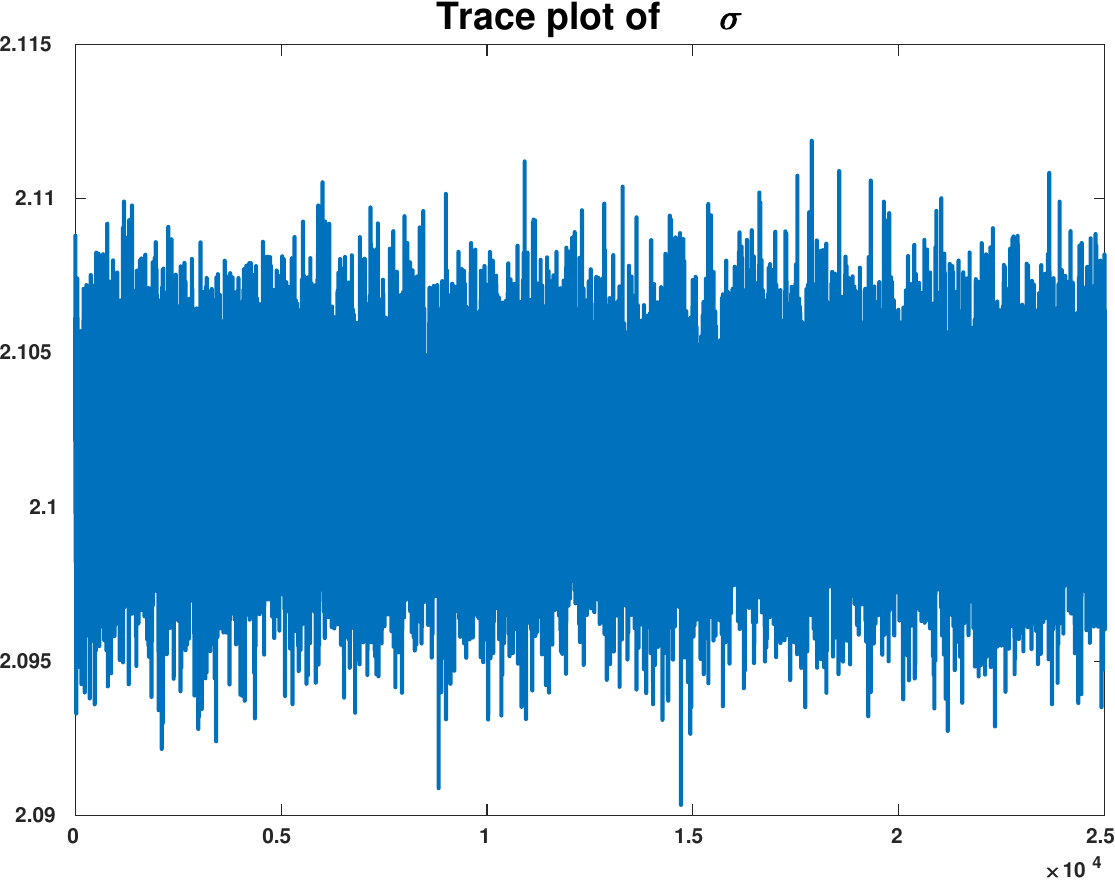}
    \includegraphics[width=0.45\textwidth,height=4.5cm]{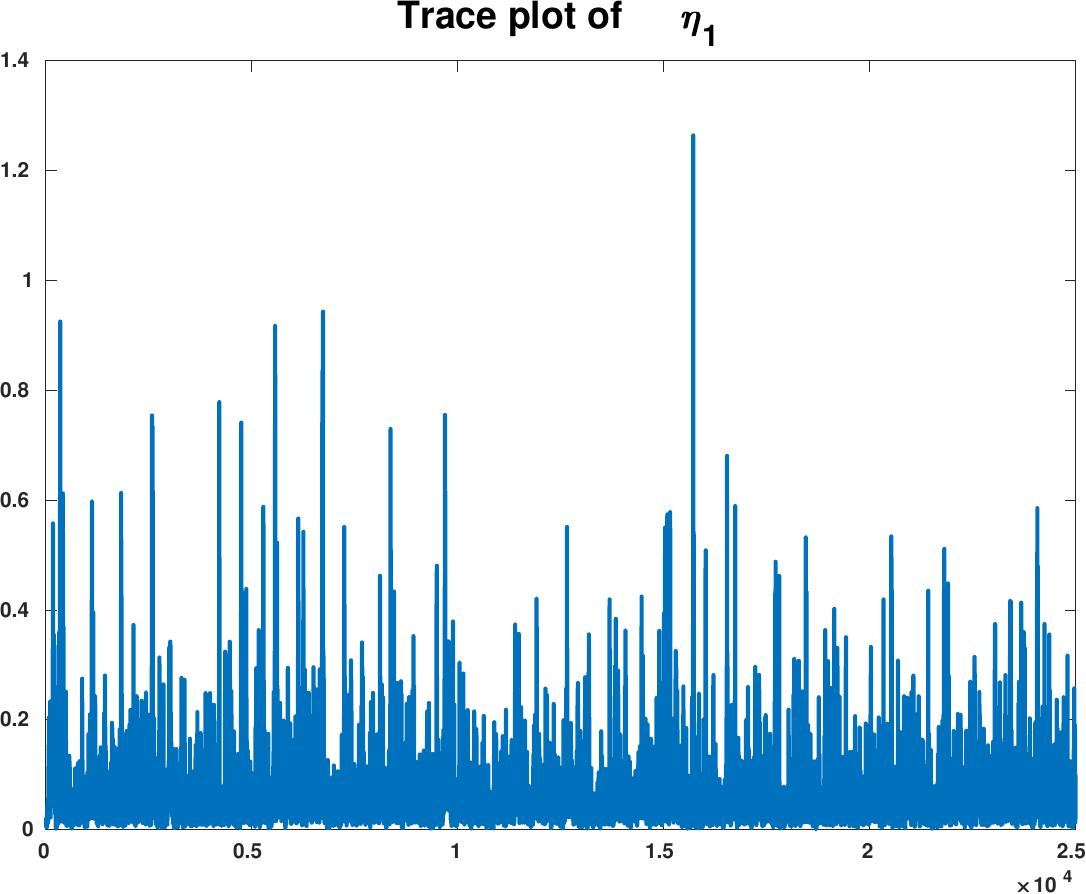}
    \centering
    \includegraphics[width=0.45\textwidth,height=4.5cm]{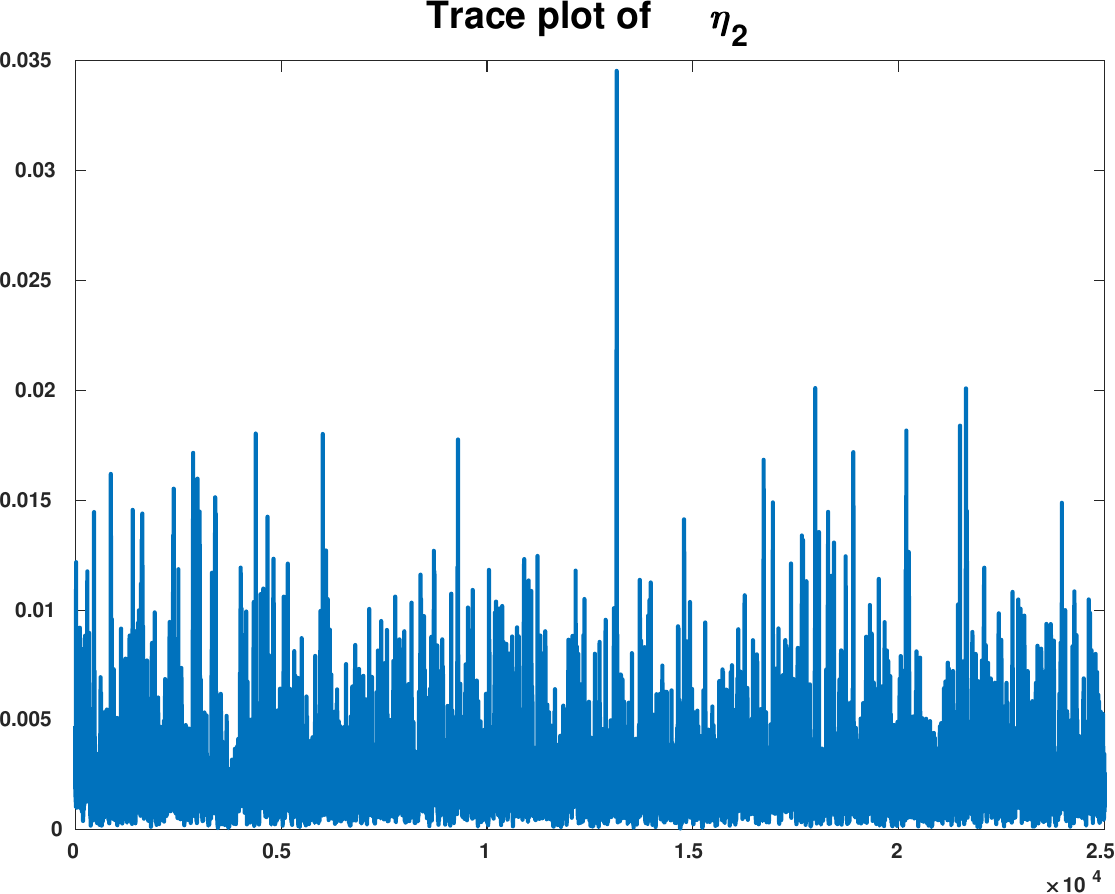}
    \caption{Trace plots for the parameters for data simulated from 3-component GP mixtures.}
	\label{fig:trace plot 3 comp GP mixtures}
\end{figure}
\begin{comment}
\begin{figure}[!ht]
	%\begin{subfigure}{.5\textwidth}
	\centering
	\includegraphics[width=0.8\textwidth,height=4cm,]{3_comp_GP_mixtures_Trace_plots-crop.pdf}
	\caption{Trace plots for the parameters for data simulated from 3-component GP mixtures.}
	\label{fig:trace plot 3 comp GP mixtures}
\end{figure}
\end{comment}
%===================================
\subsection{Posterior densities of the complete time series of latent variables}
\label{3 comp mixture GP: posterior predictive densities of latent variables}
%====================================
\begin{figure}[!h]
    \includegraphics[width=0.5\textwidth,height=6.5cm]{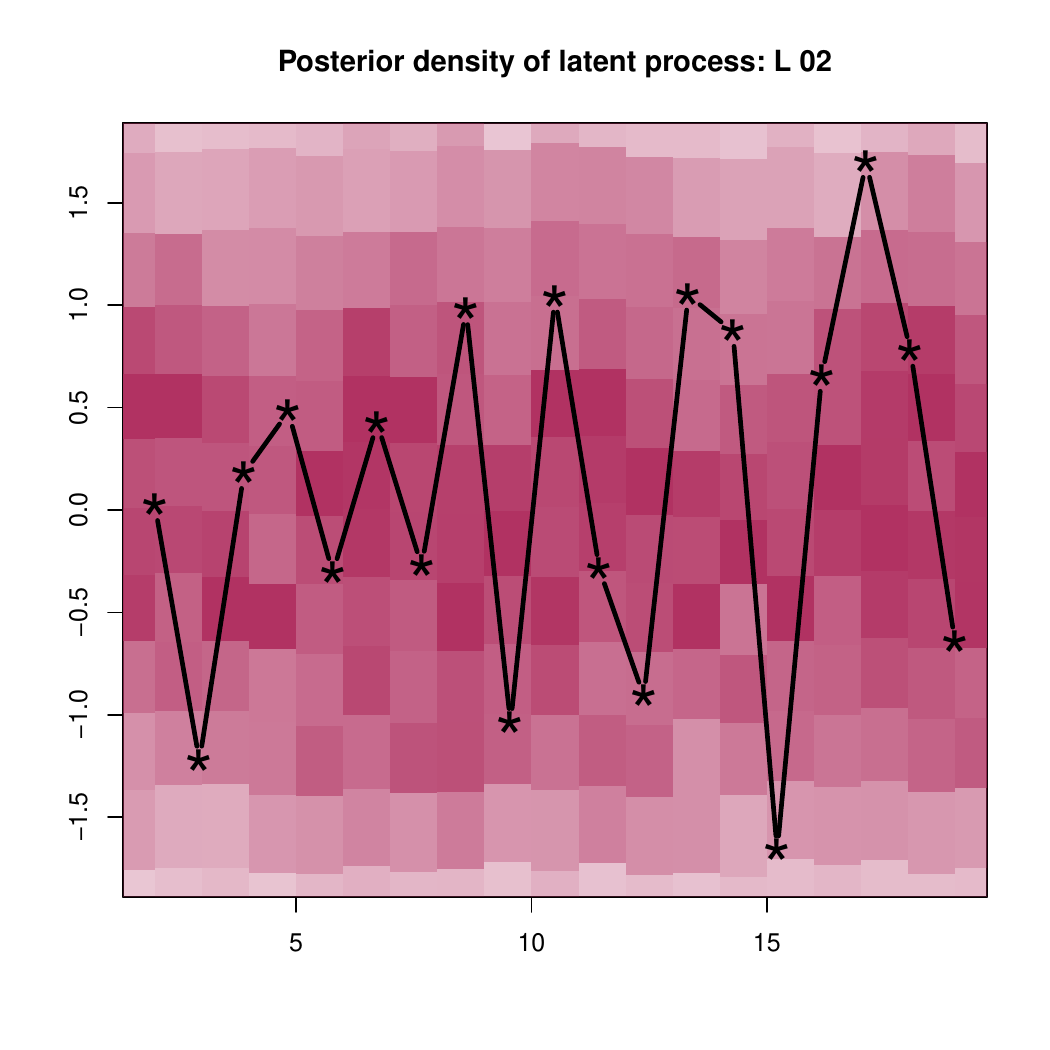}
    \includegraphics[width=0.5\textwidth,height=6.5cm]{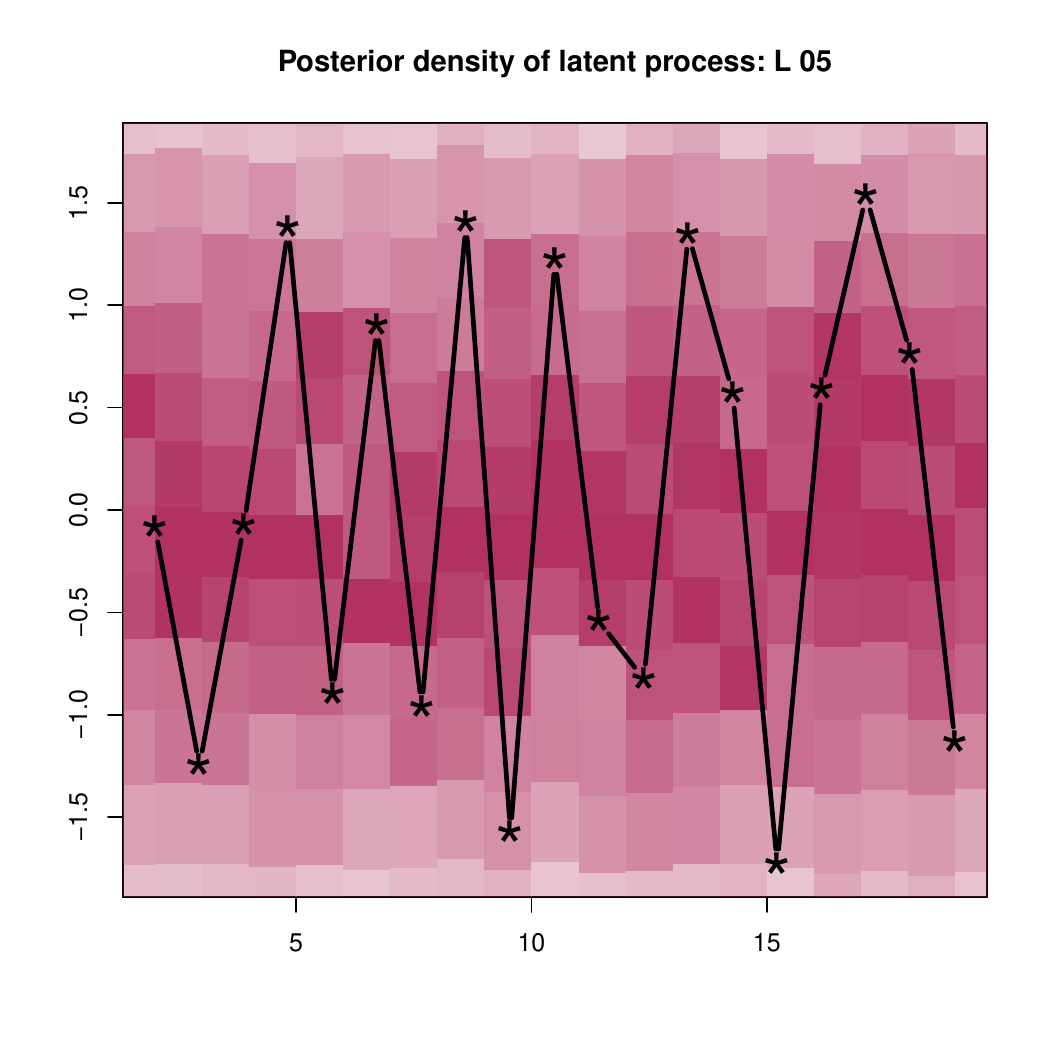}
    \includegraphics[width=0.5\textwidth,height=6.5cm]{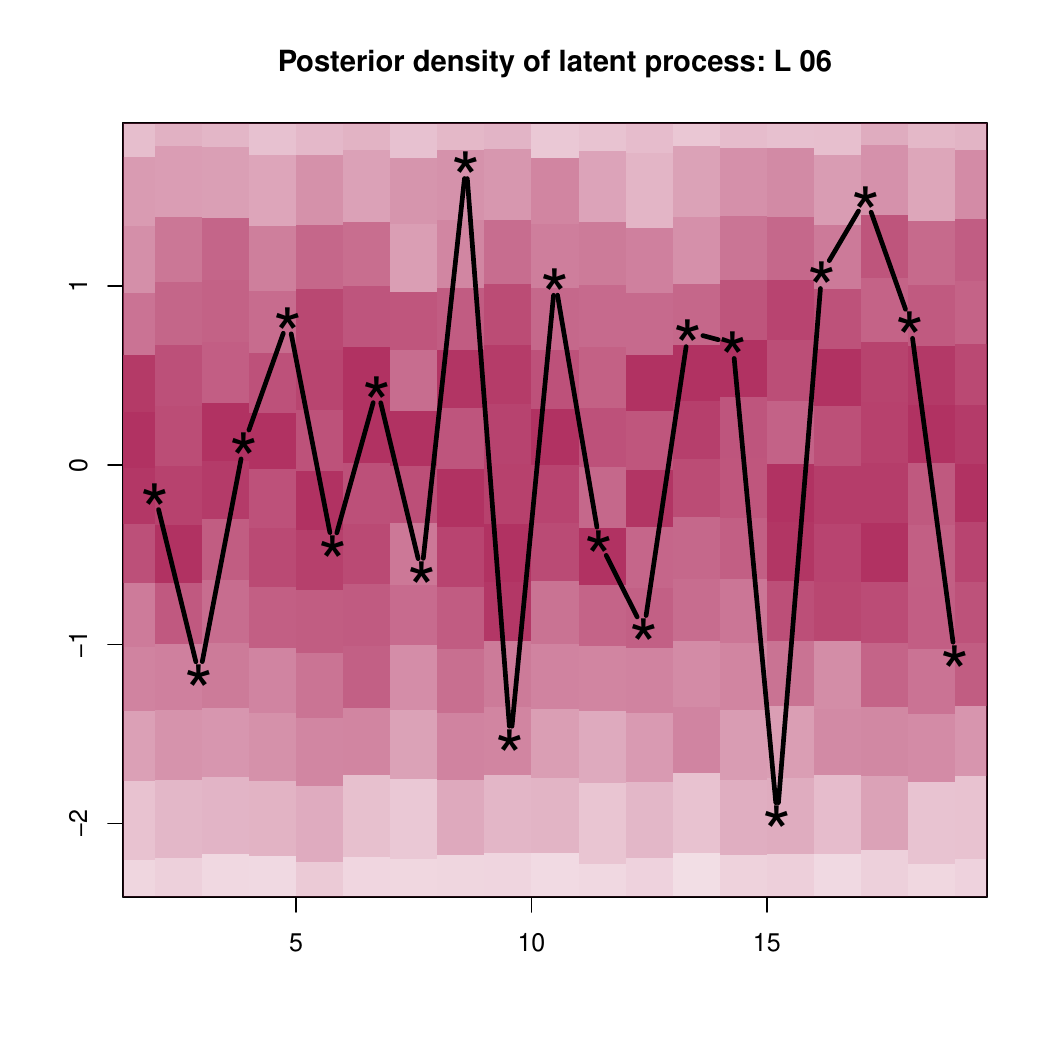}
    \includegraphics[width=0.5\textwidth,height=6.5cm]{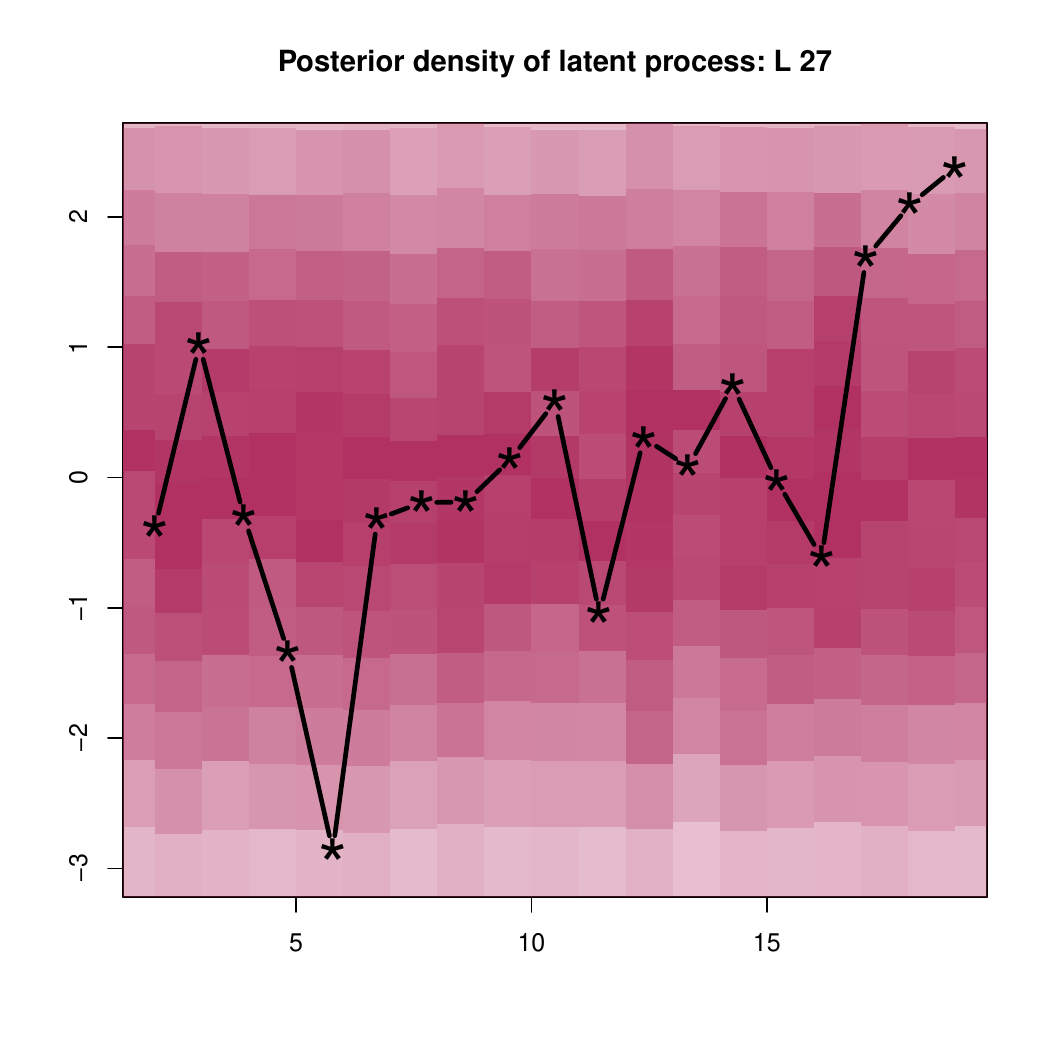}
    \includegraphics[width=0.5\textwidth,height=6.5cm]{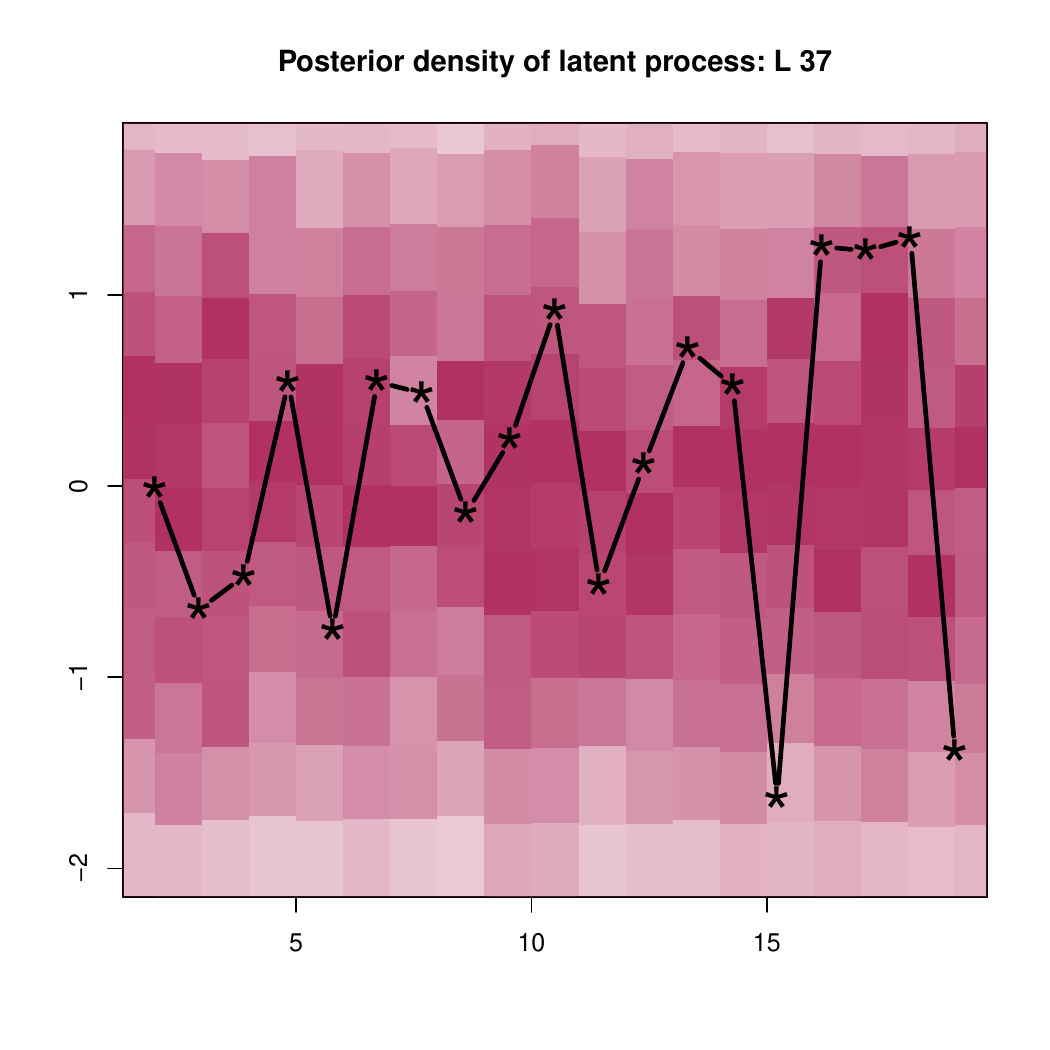}
    \includegraphics[width=0.5\textwidth,height=6.5cm]{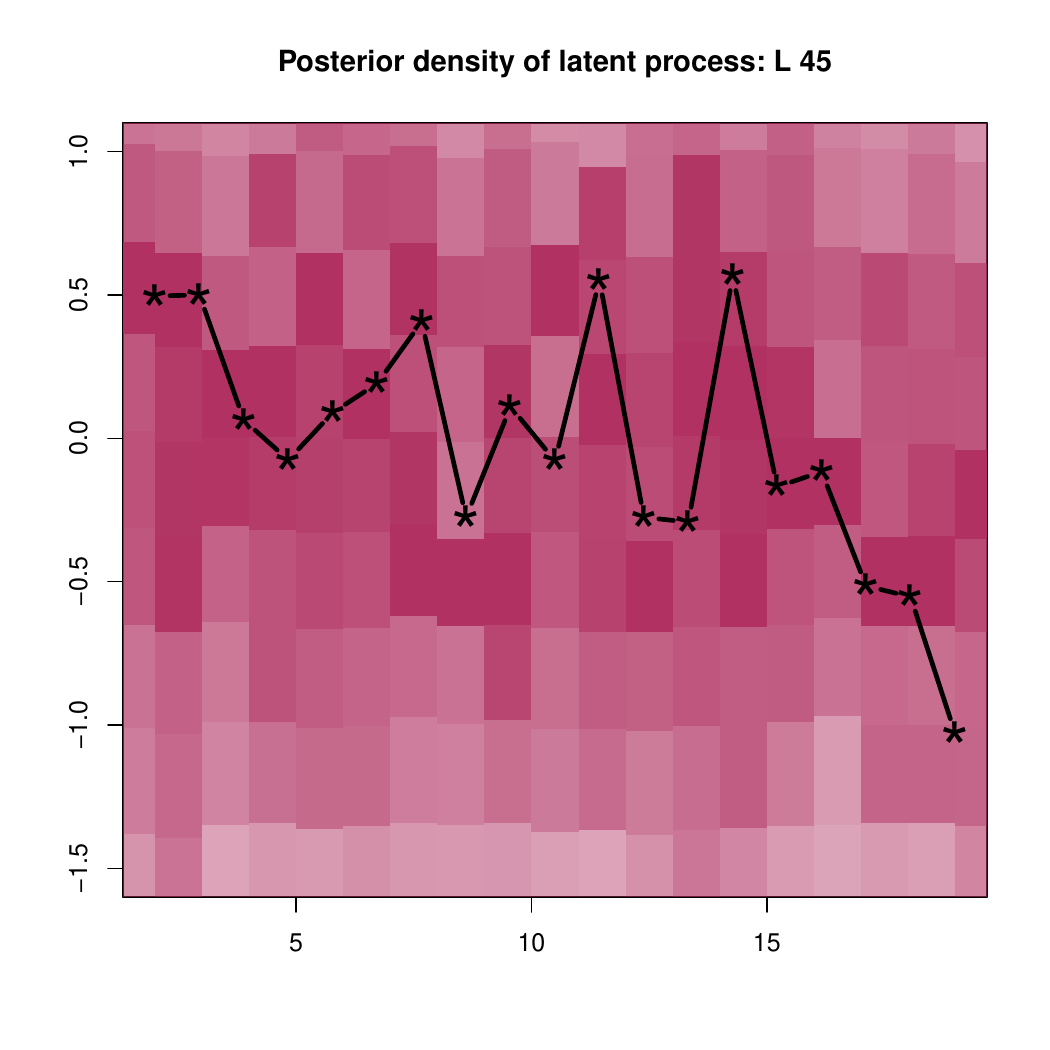}
    \caption{Posterior densities of latent variables for six locations for data simulated from 3-component mixture of GPs. Higher the intensity of the color, higher is the probability density. The black line represents the true values of the latent variables. $L_i$ denote the locations.}
    \label{fig:posterior density of latent variables 3-comp GP mixtures: first 25 locations}
\end{figure}
\clearpage
\section{Supplementary plots for mixture of GQNs}
\label{Trace plt and posterior plts for 2 comp mixture GQN}
%=================================================
Trace plots of the parameters and the posterior densities of the complete time series of the latent variables for two component mixture of GQNs are depicted below. The details of the data are given in Section \ref{Mixture_GQN}. 
%---------------------------------------------
\subsection{Trace plots}
\label{Mixture of GQNs: trace plots}
%------------------------------------------
\begin{figure}[!h]
    \includegraphics[width=0.45\textwidth,height=4.5cm]{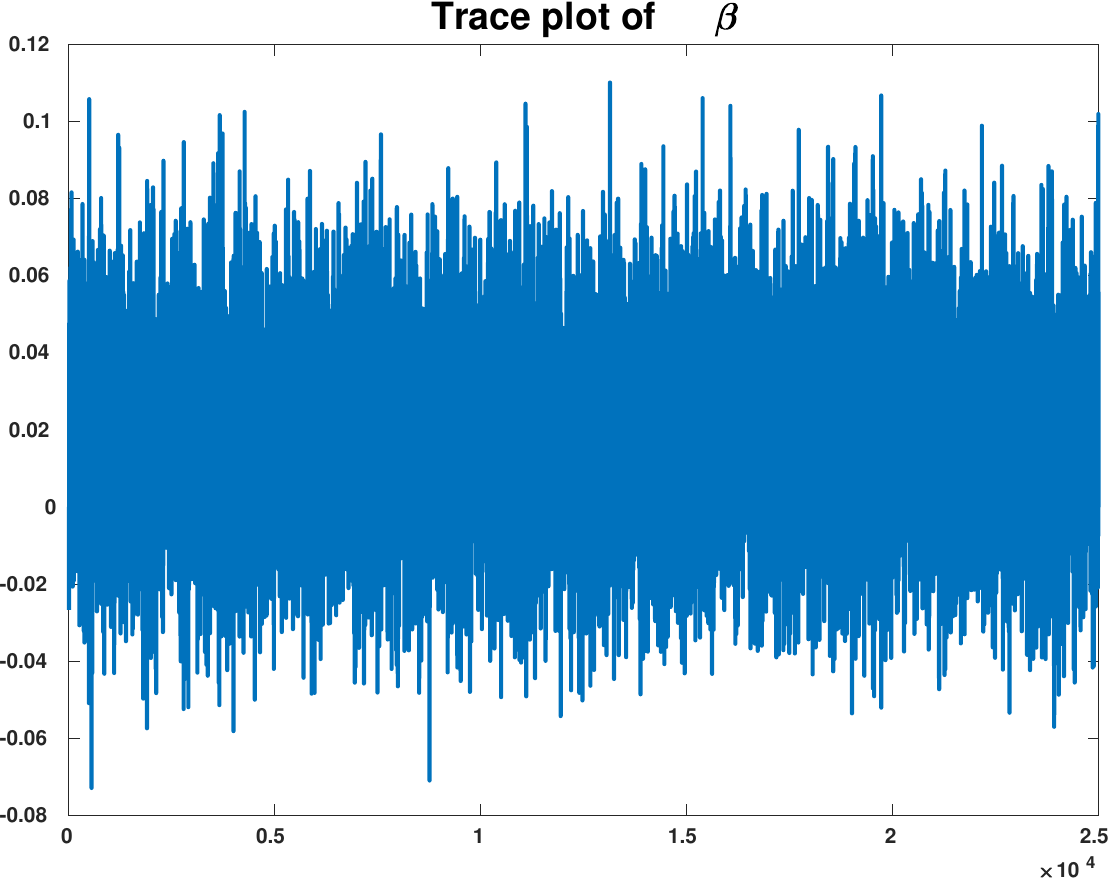}
    \includegraphics[width=0.45\textwidth,height=4.5cm]{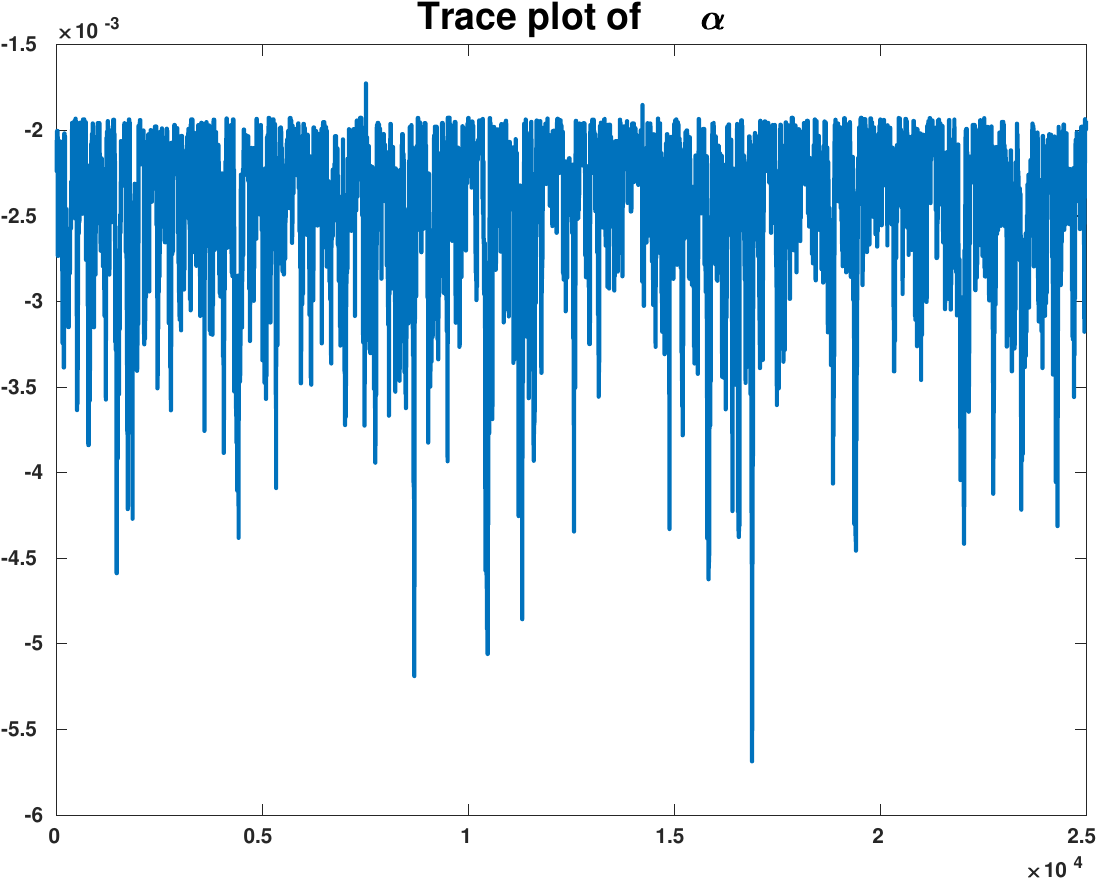}
    \includegraphics[width=0.45\textwidth,height=4.5cm]{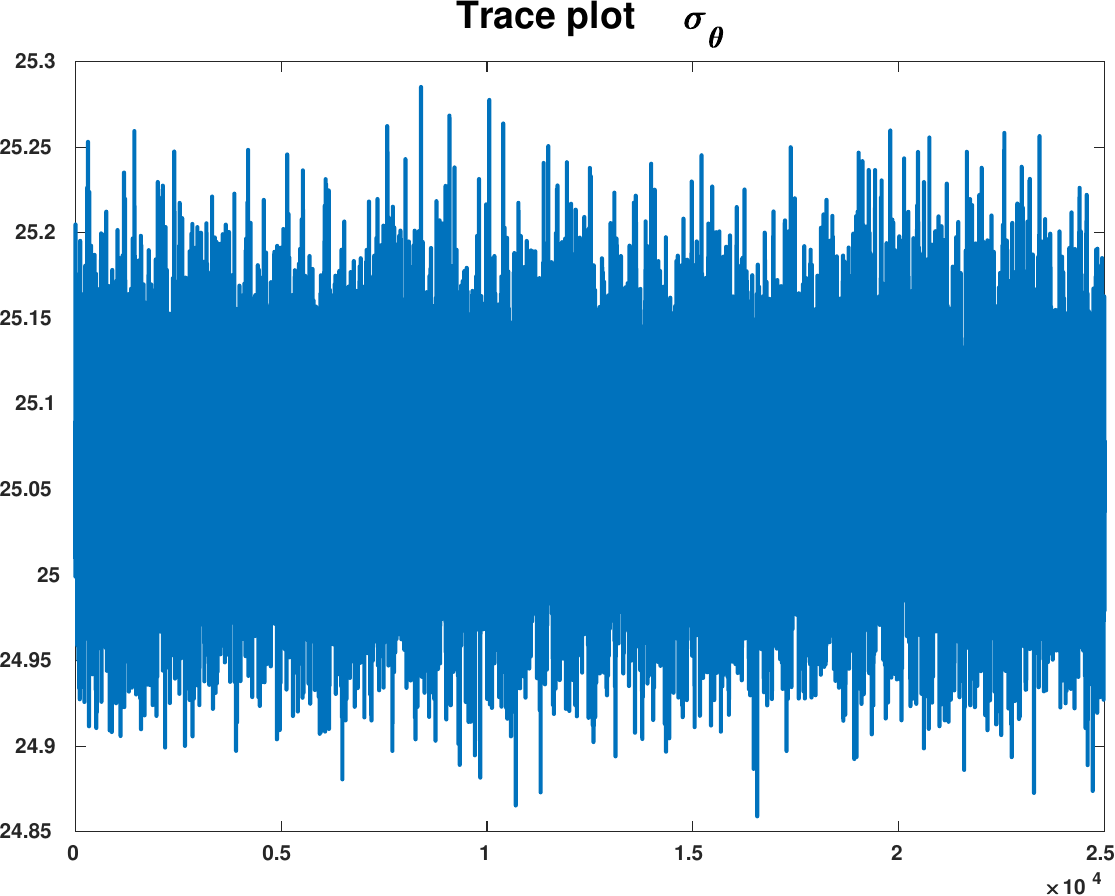}
    \includegraphics[width=0.45\linewidth,height=4.5cm]{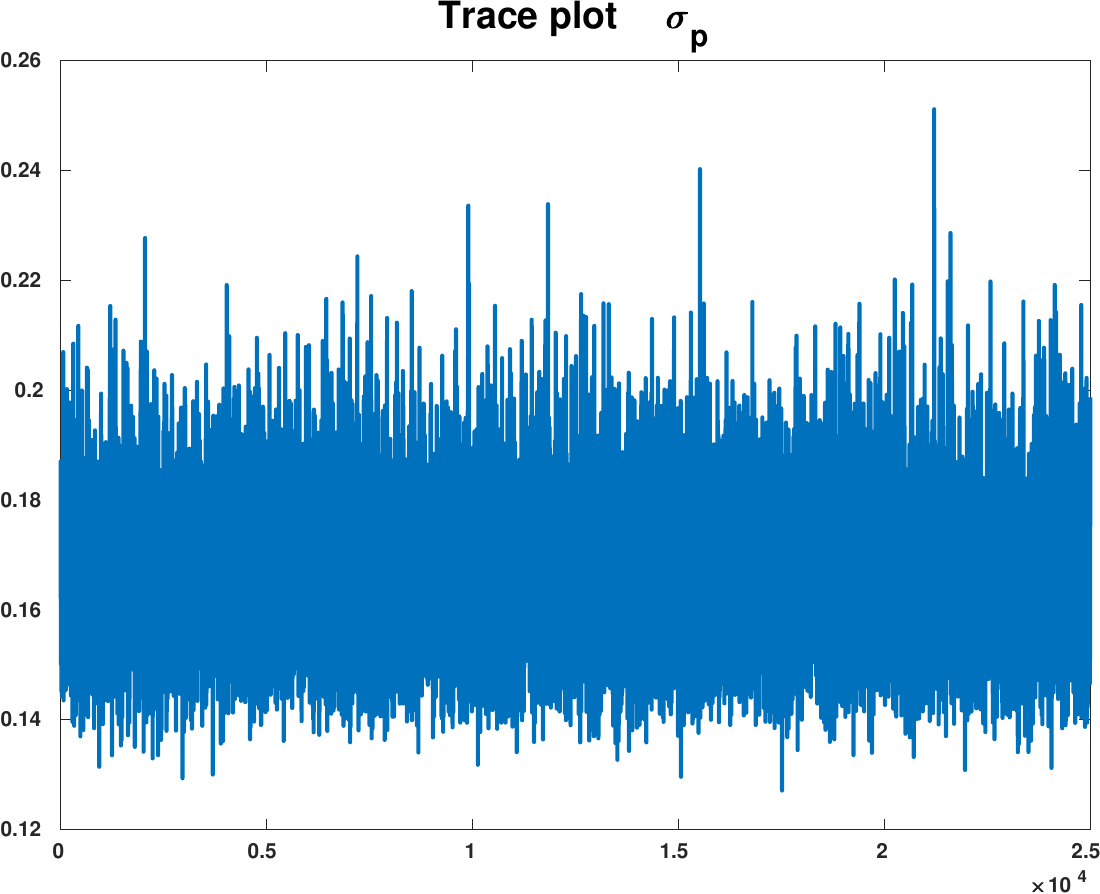}
    \includegraphics[width=0.45\textwidth,height=4.5cm]{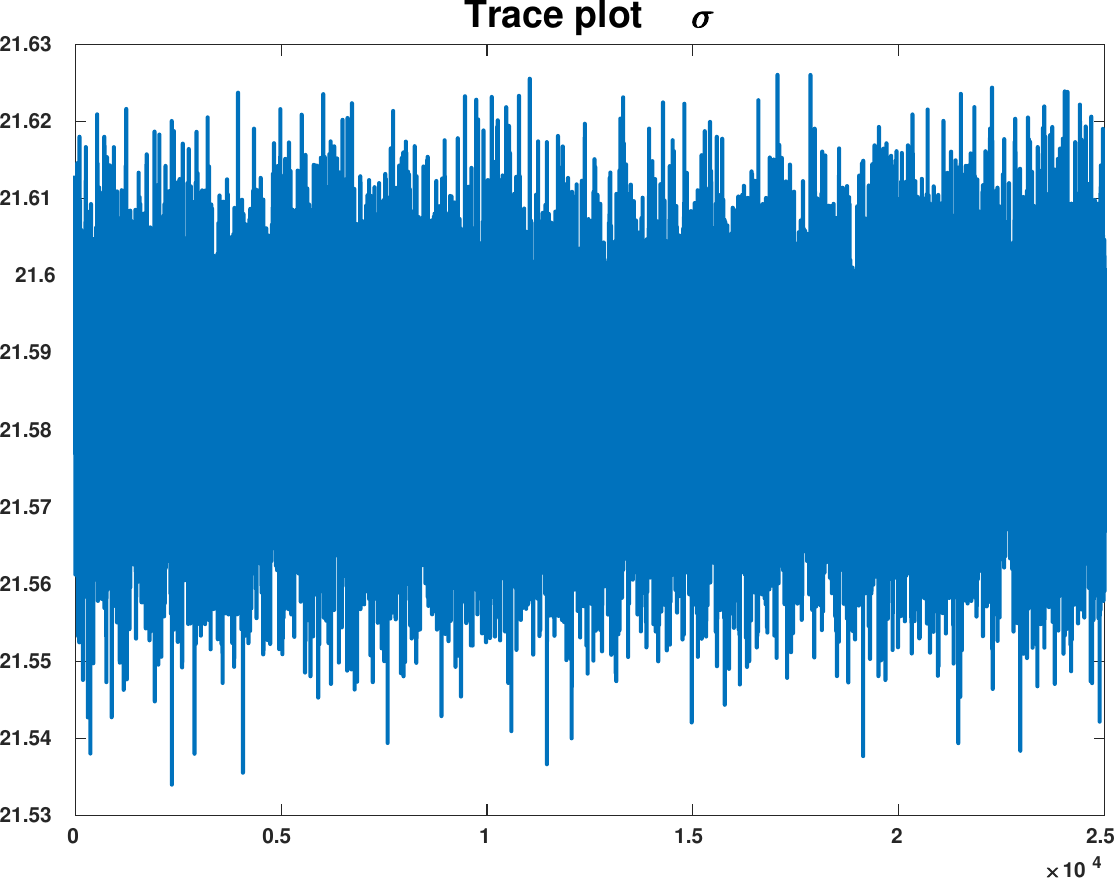}
    \includegraphics[width=0.45\textwidth,height=4.5cm]{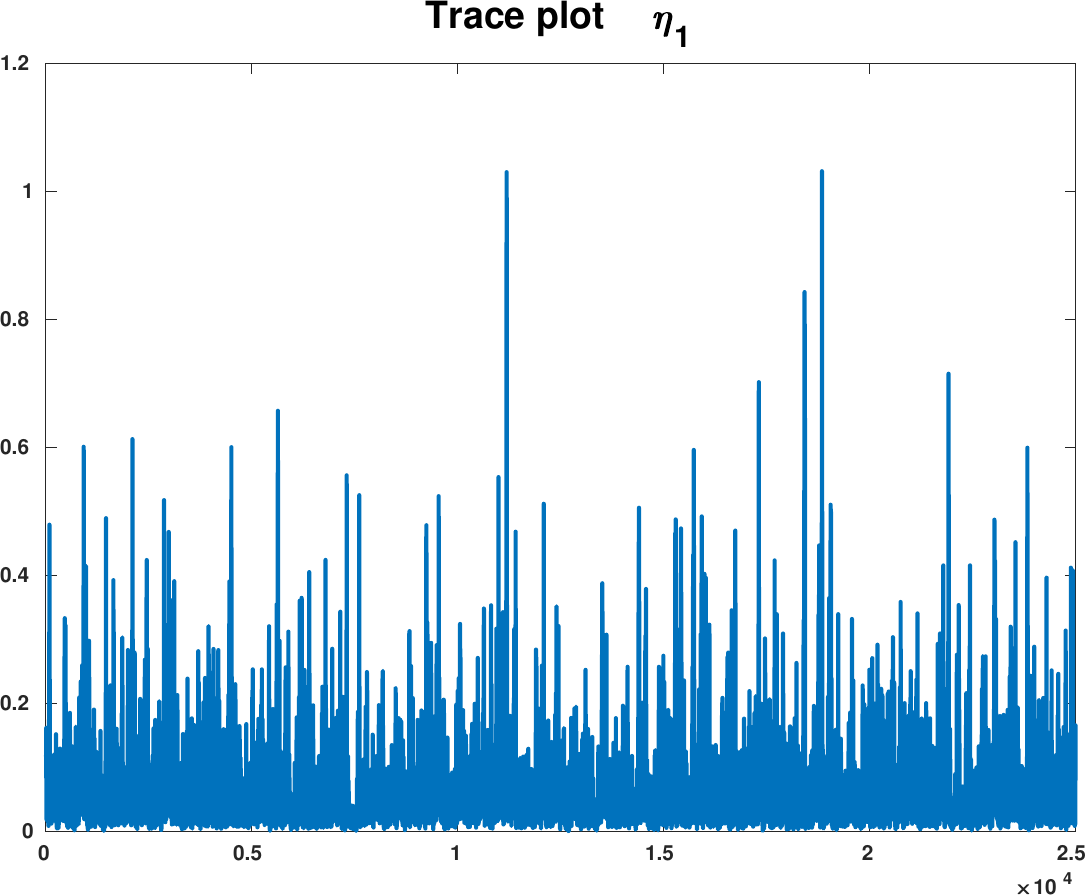}
    \centering
    \includegraphics[width=0.45\textwidth,height=4.5cm]{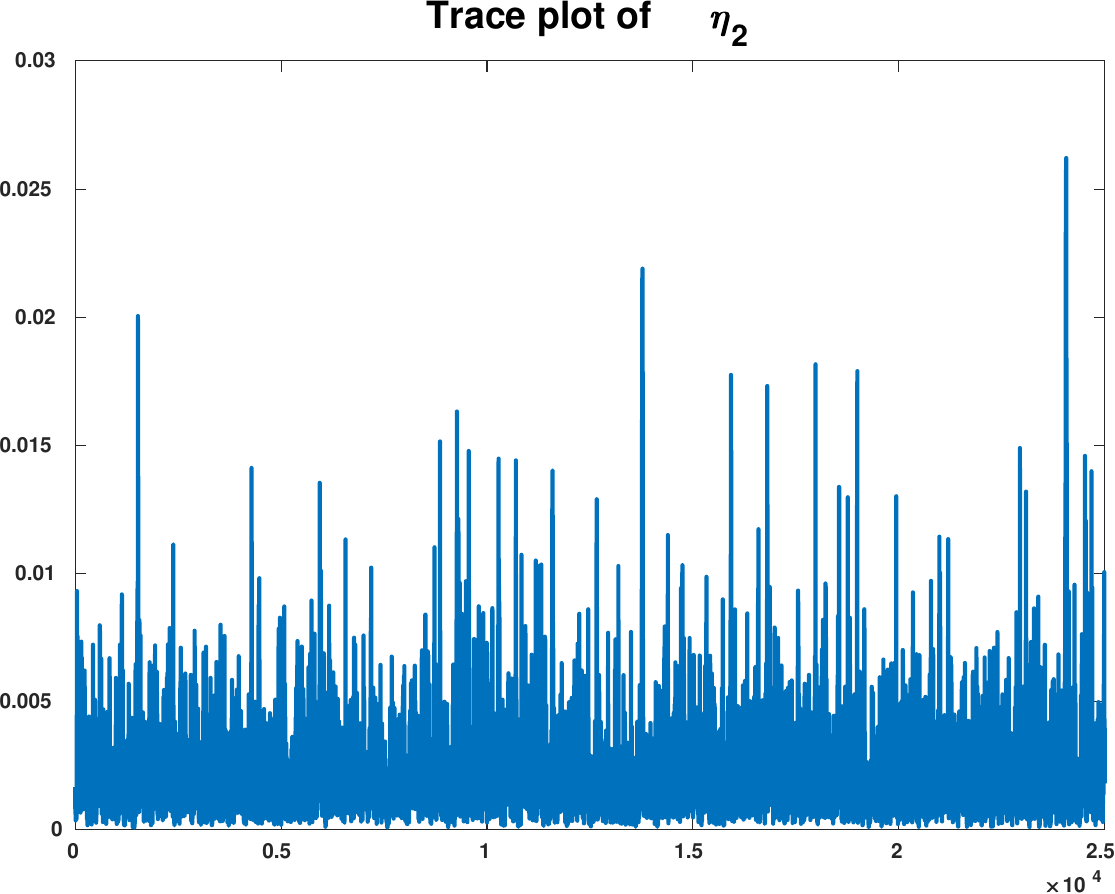}
    \caption{Trace plots for the parameters for data simulated from mixture of GQNs.}
	\label{fig:trace plot mixture GQNs}
\end{figure}
\subsection{Posterior densities of the complete time series of the latent variables}
\label{mixture of GQNs: posterior predictive densities of latent variables}
\begin{figure}[!h]
    \includegraphics[width=0.5\textwidth,height=6.5cm]{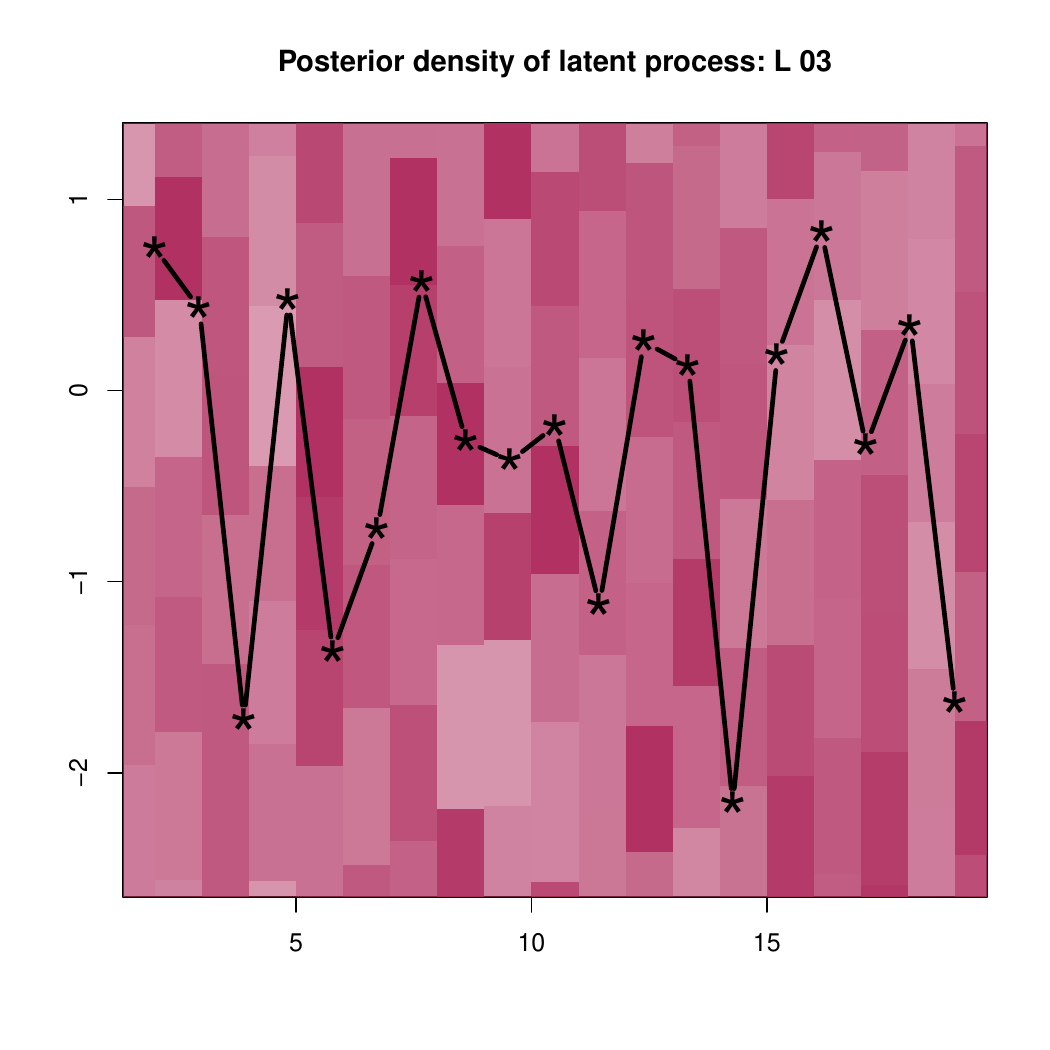}
    \includegraphics[width=0.5\textwidth,height=6.5cm]{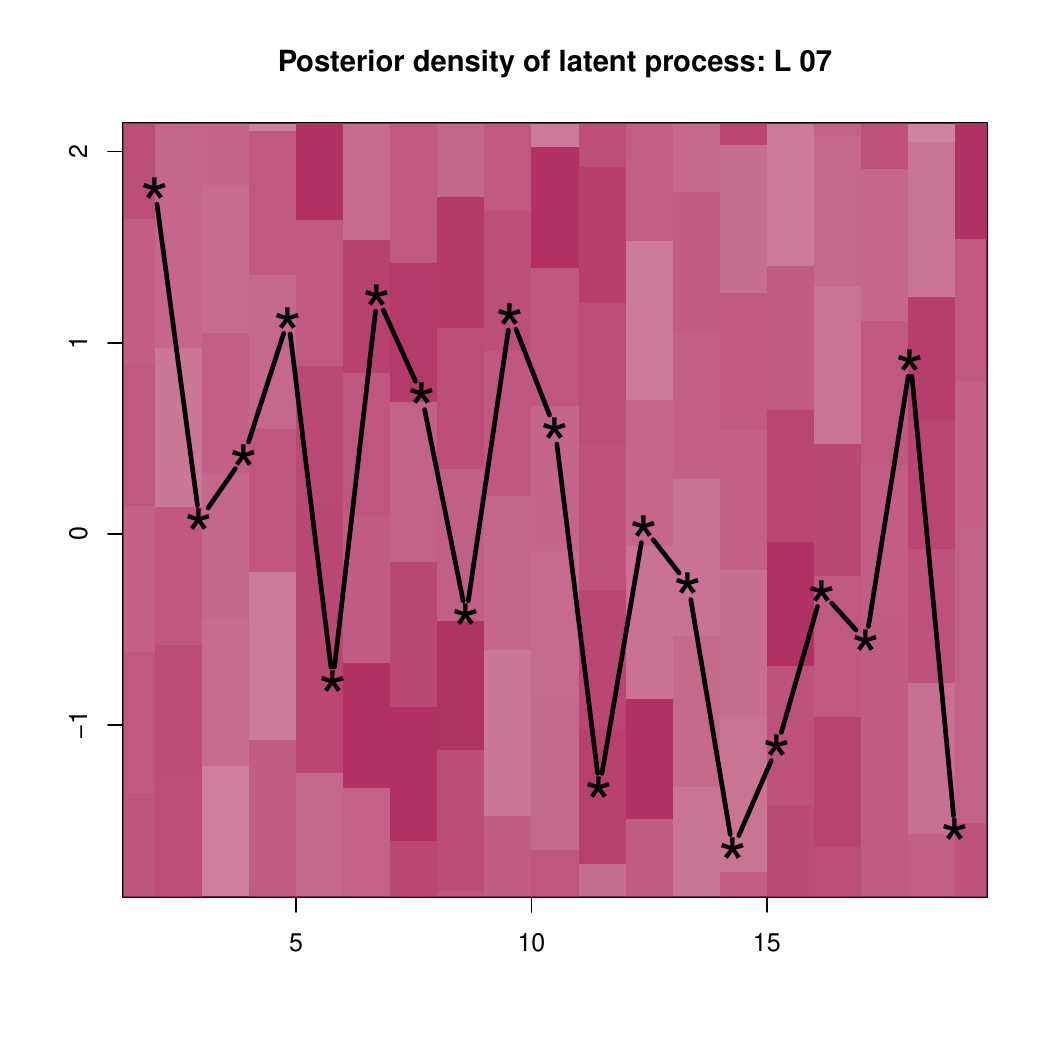}
    \includegraphics[width=0.5\textwidth,height=6.5cm]{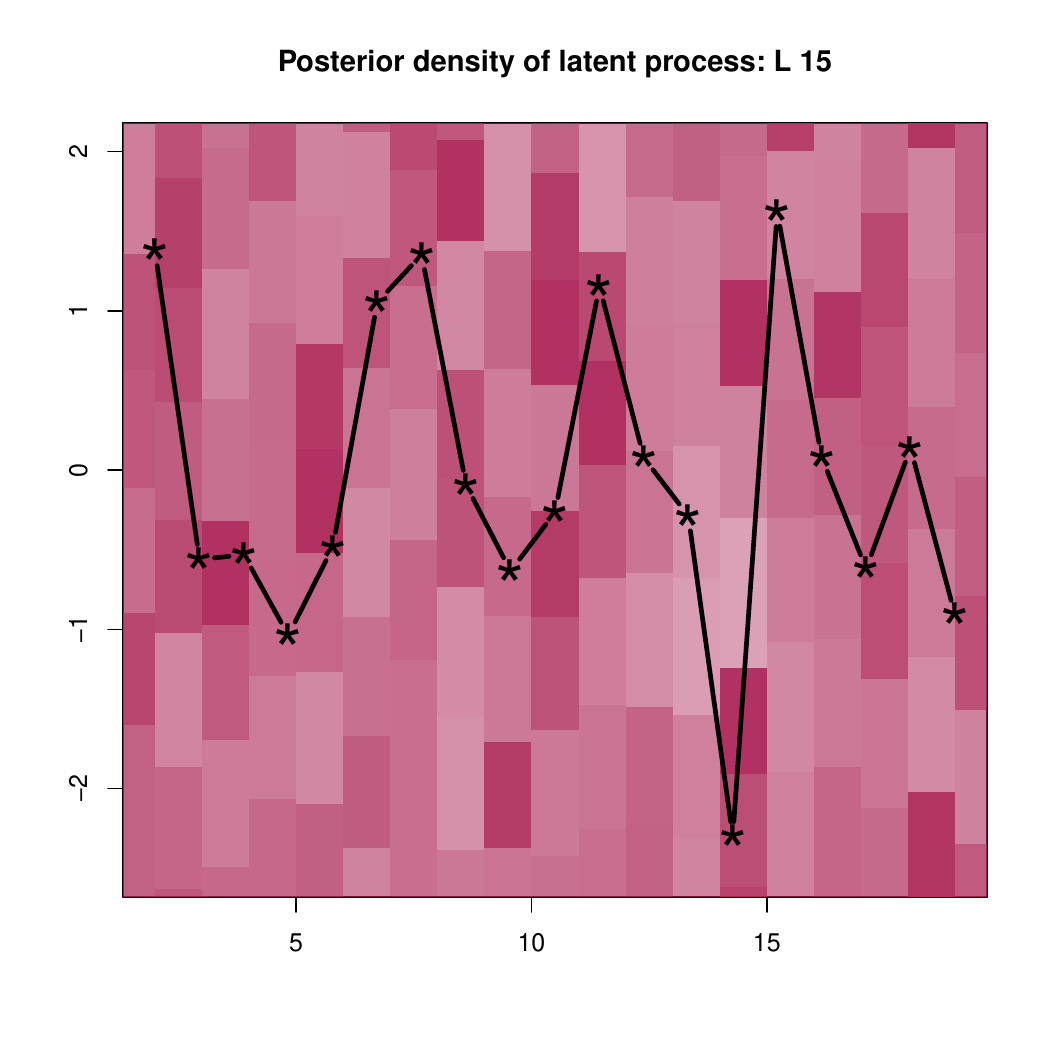}
    \includegraphics[width=0.5\textwidth,height=6.5cm]{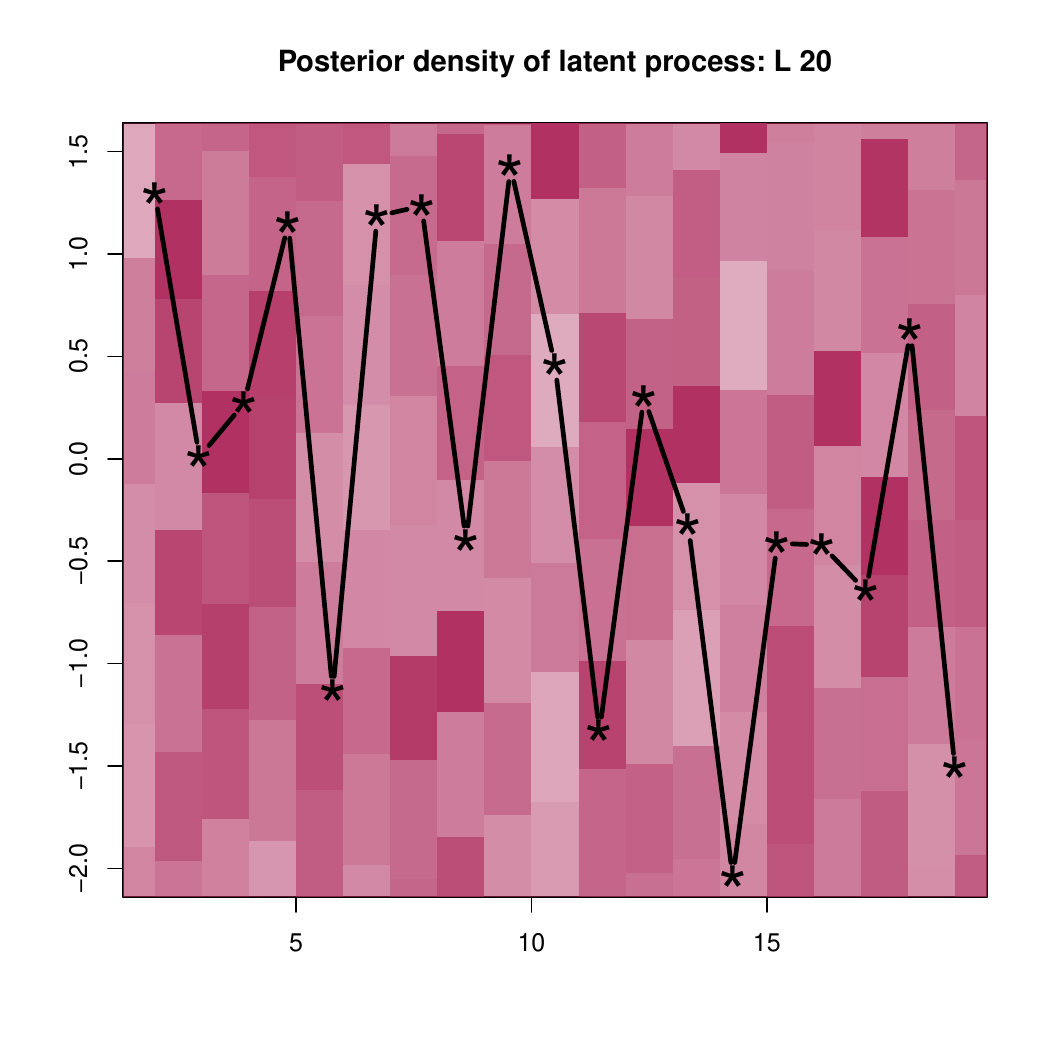}
    \includegraphics[width=0.5\textwidth,height=6.5cm]{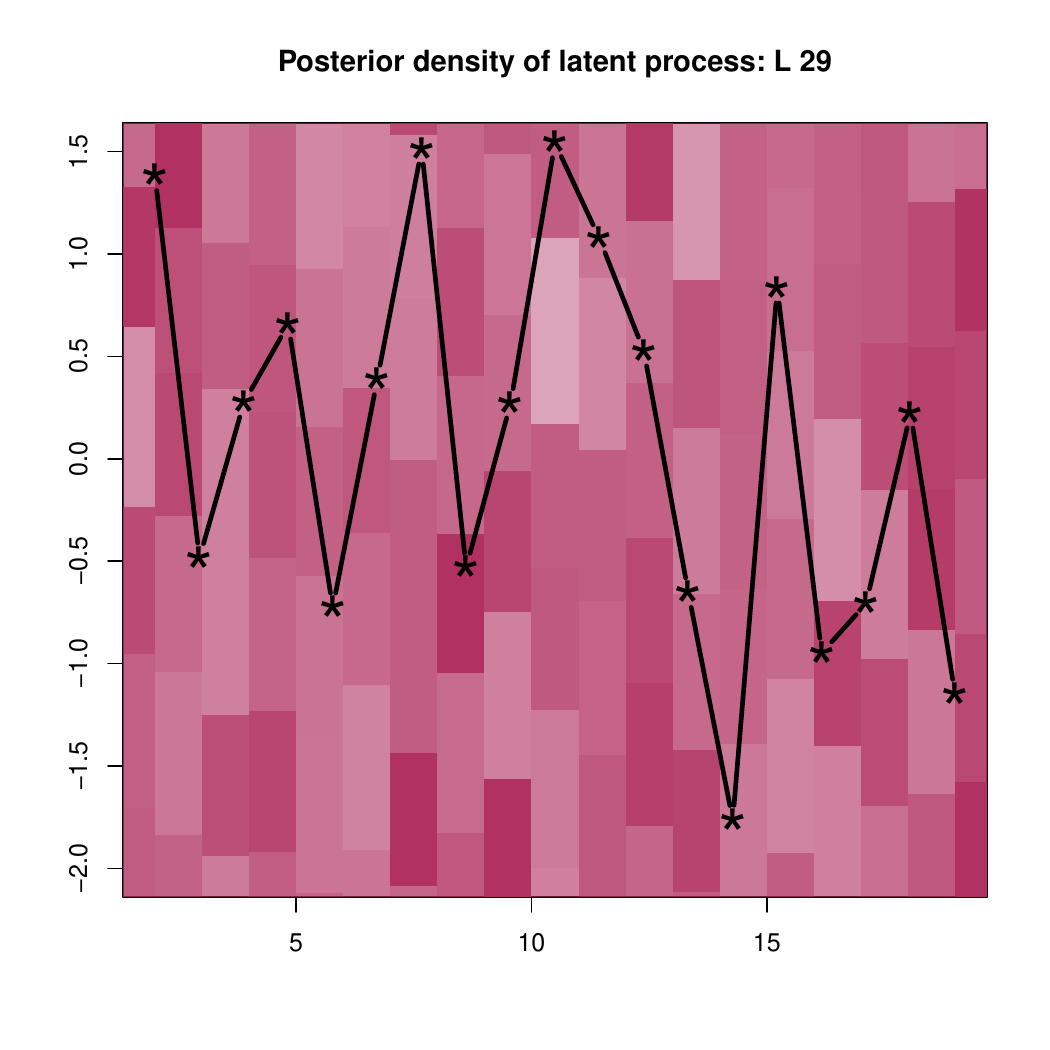}
    \includegraphics[width=0.5\textwidth,height=6.5cm]{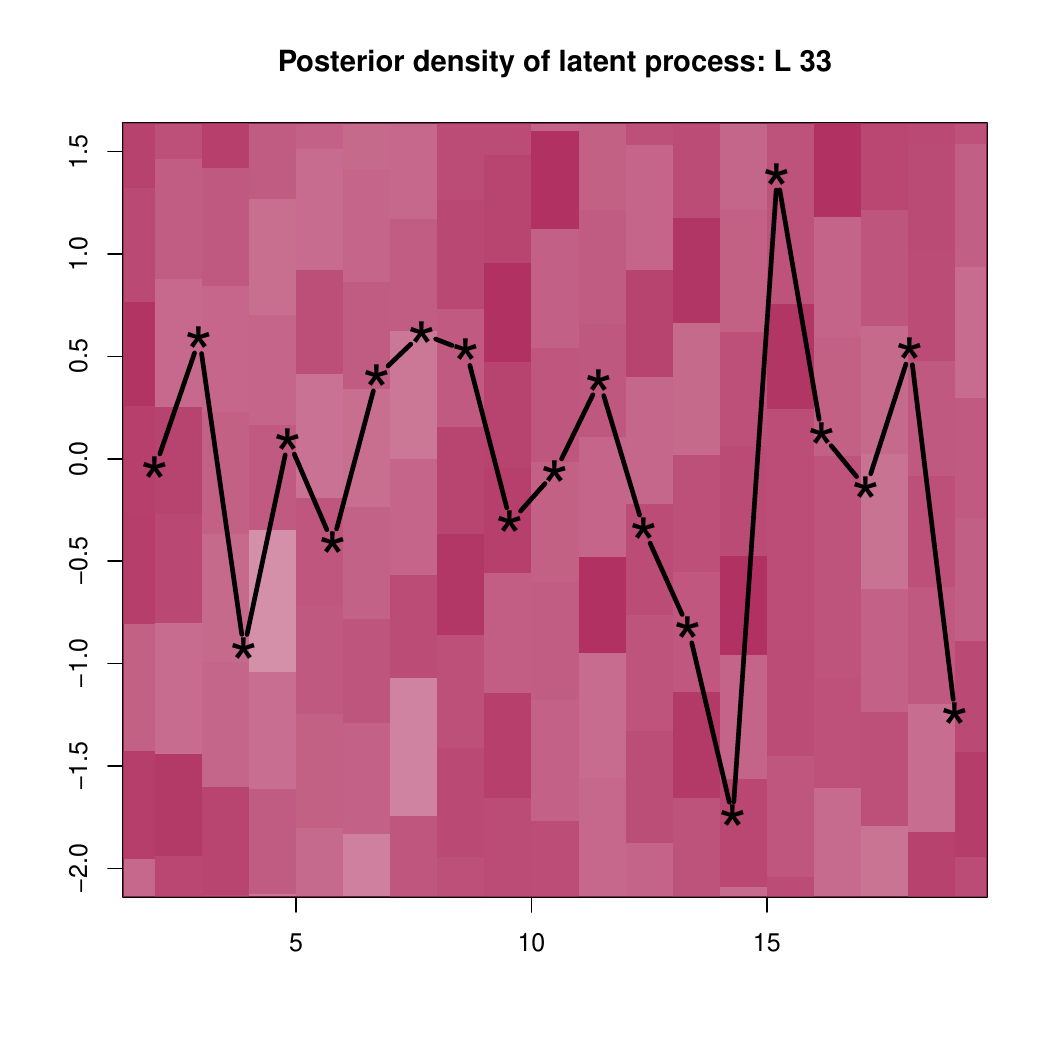}
	\caption{Posterior densities of latent variables for six locations for data simulated from mixture of two GQNs. Higher the intensity of the color, higher is the probability density. The black line represents the true values of the latent variables. $L_i$ denote the locations.}		
		\label{fig:posterior density of latent variables mixture of GQNs: first 25 locations}
\end{figure}
\clearpage
%%%%%%%%%%%%%%%%%%%%%%%%%%%%%%%%%%
\section{Supplementary plots for Alaska temperature data}
\label{Trace plt and posterior plts for Alaska tmp data}
%=========================================
\subsection{Trace plots}
\label{Alaska: trace plot}
%-----------------------------------------
\begin{figure}[!h]
    \includegraphics[width=0.45\textwidth,height=4.85cm]{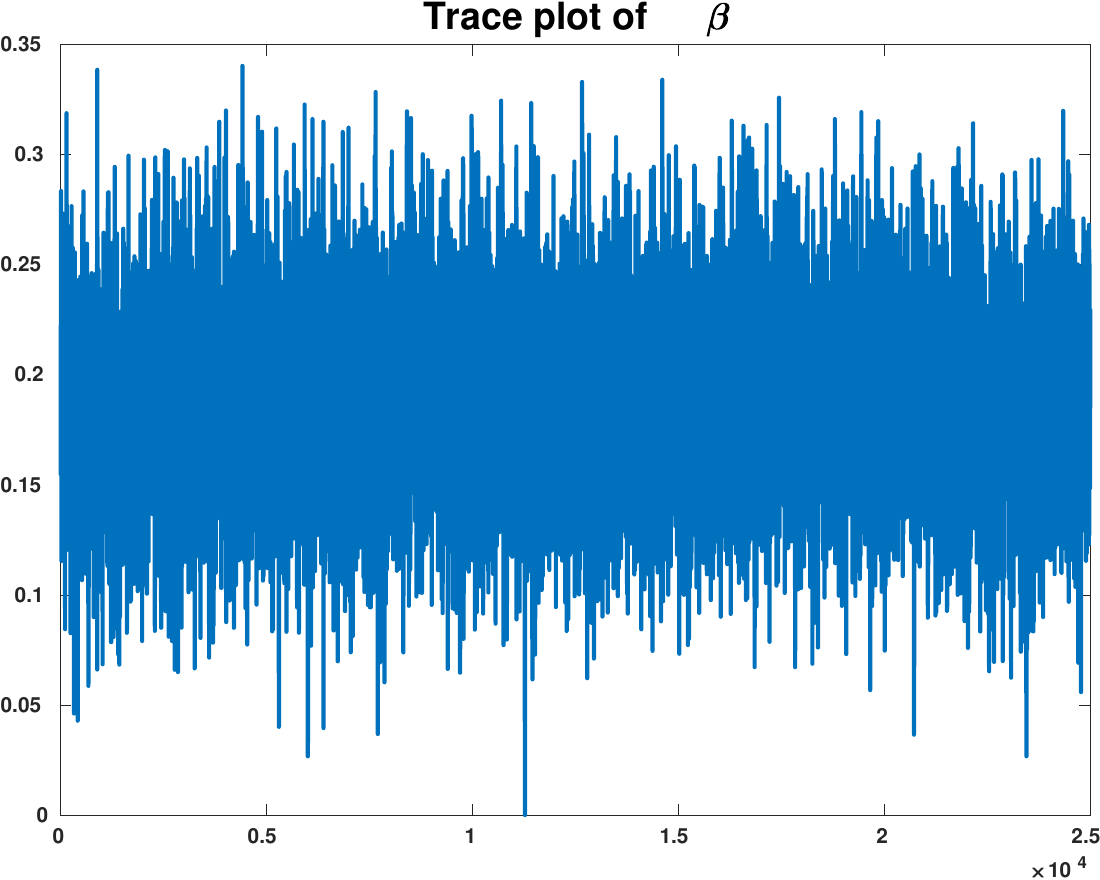}
    \includegraphics[width=0.45\textwidth,height=4.85cm]{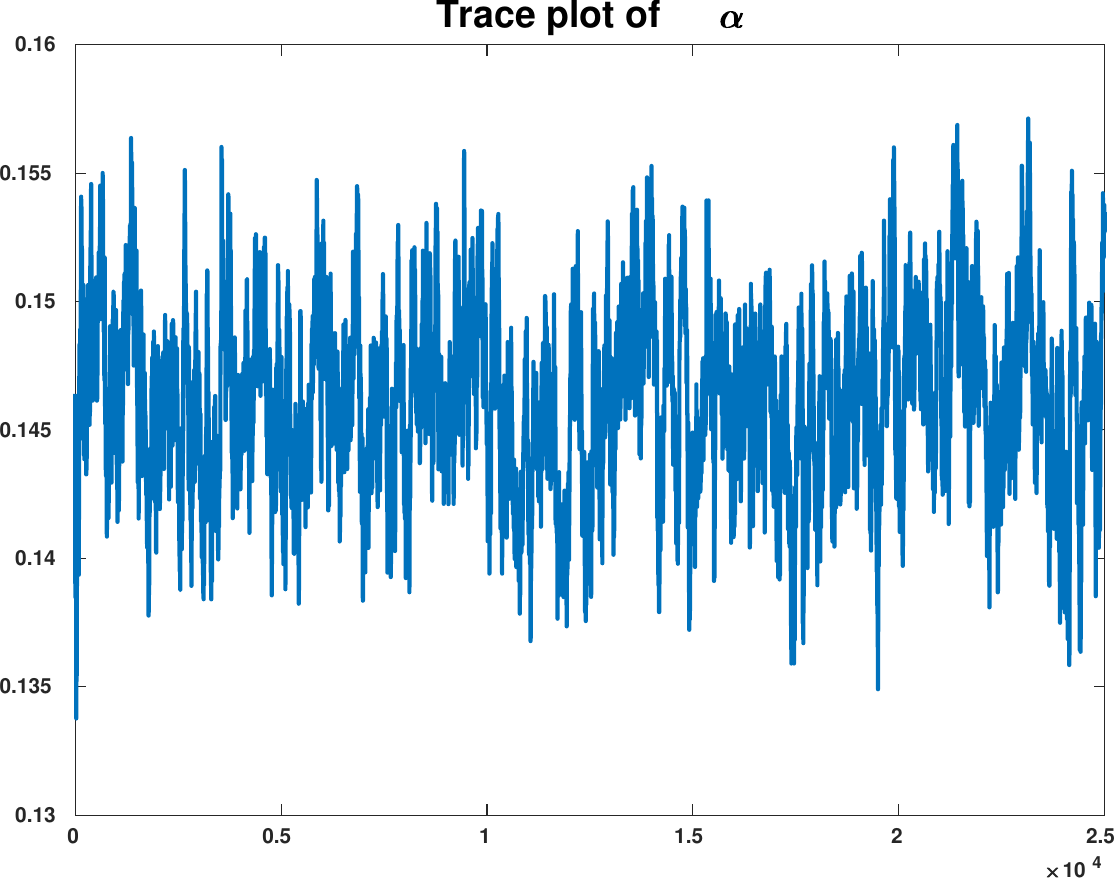}
    \includegraphics[width=0.45\textwidth,height=4.85cm]{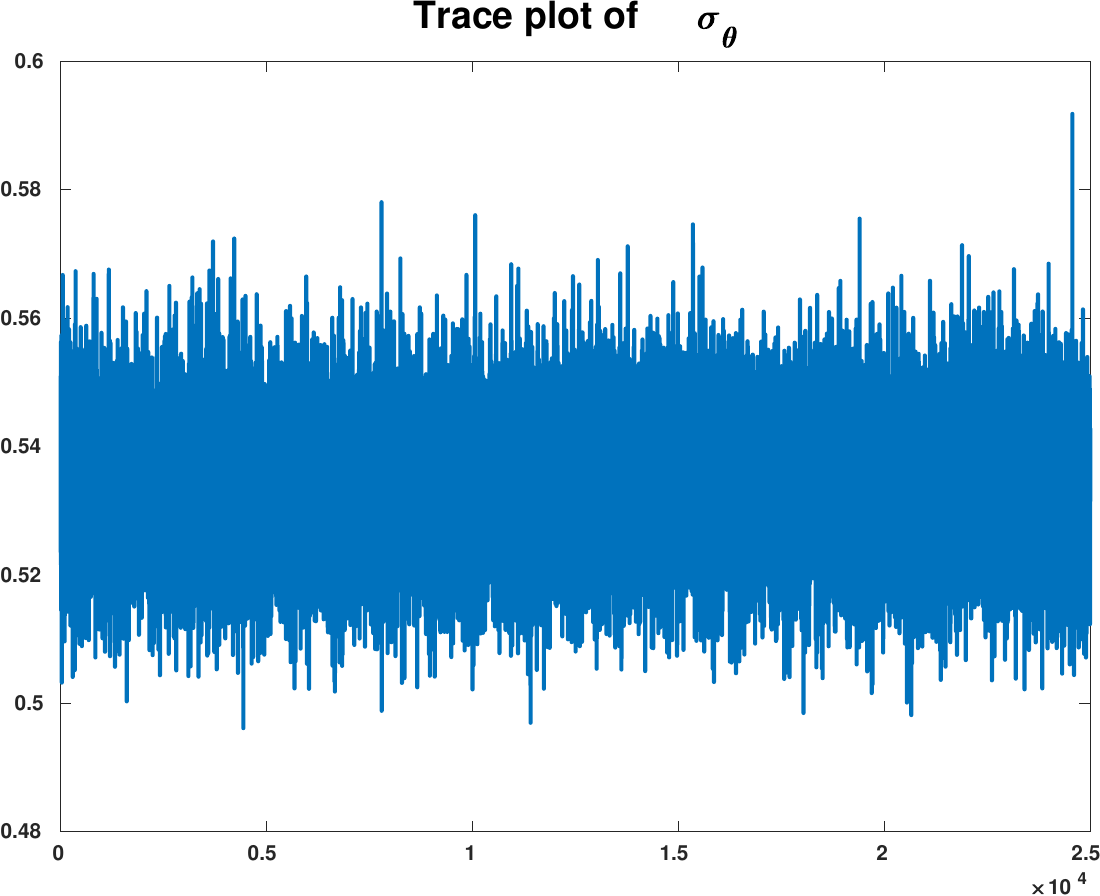}
    \includegraphics[width=0.45\linewidth,height=4.85cm]{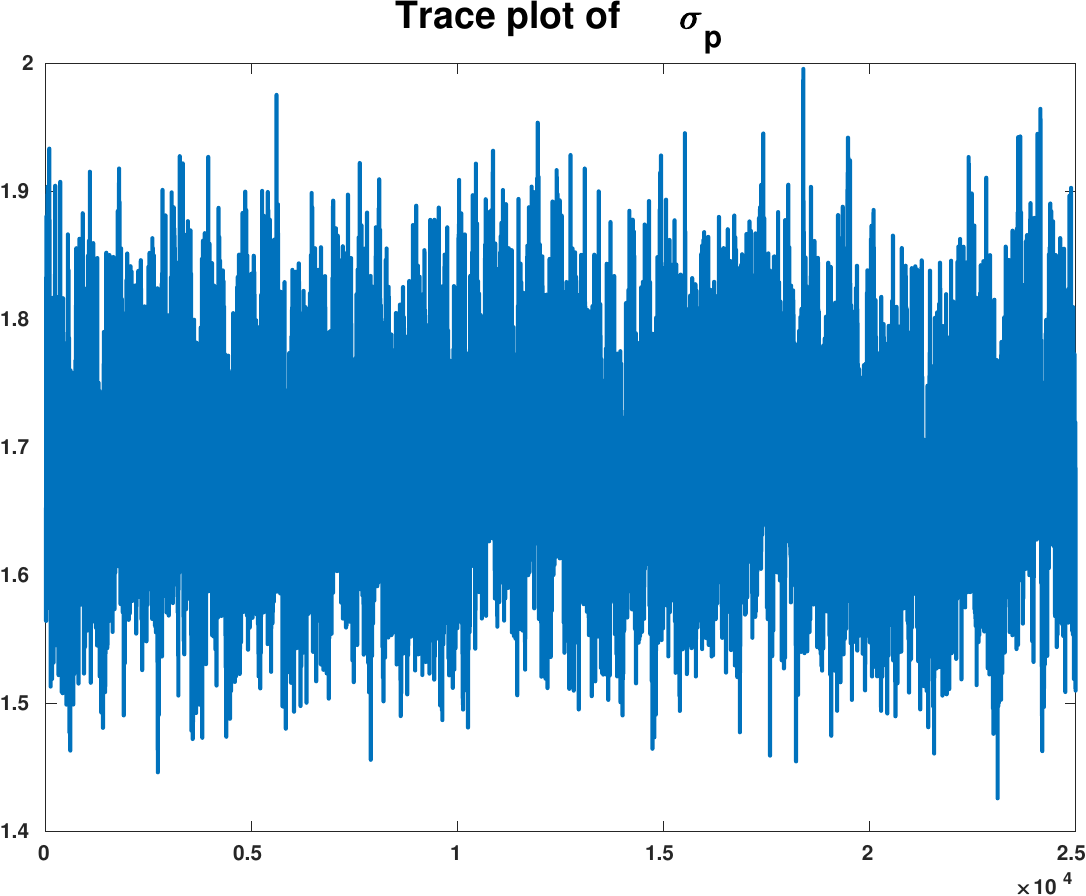}
    \includegraphics[width=0.45\textwidth,height=4.85cm]{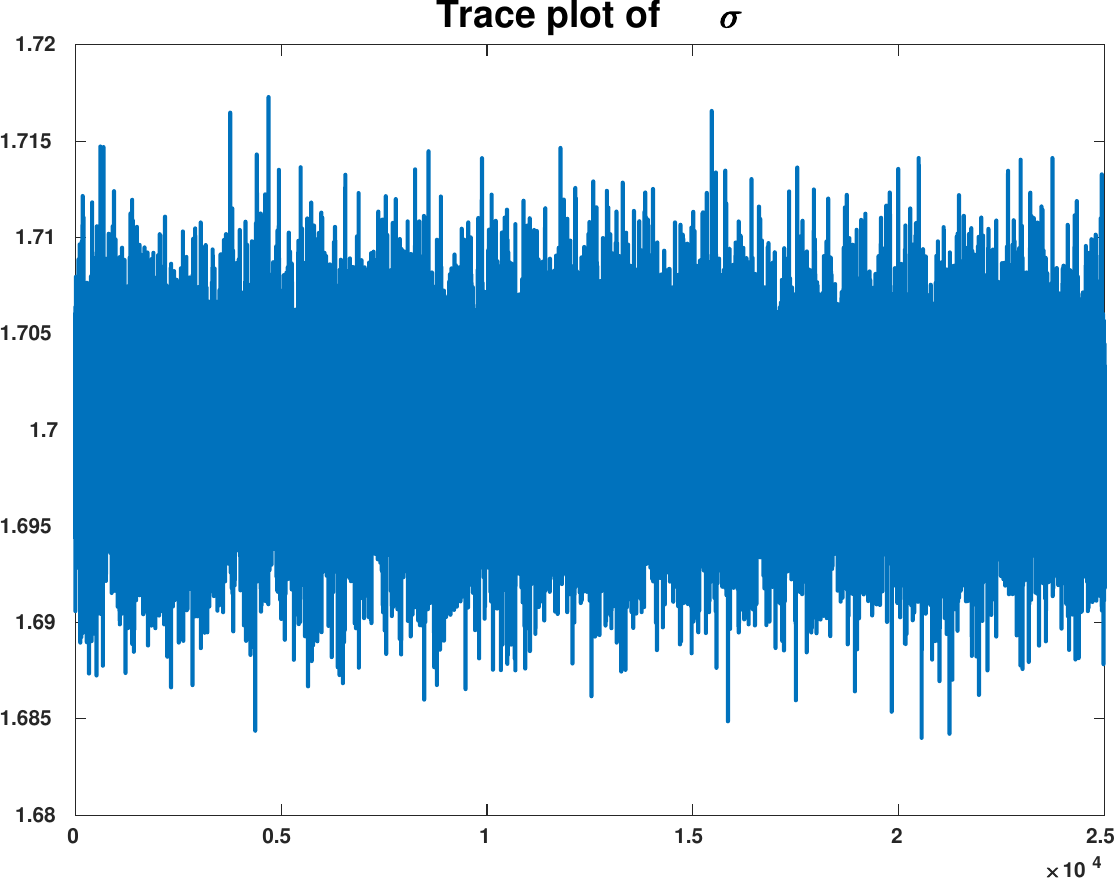}
    \includegraphics[width=0.45\textwidth,height=4.85cm]{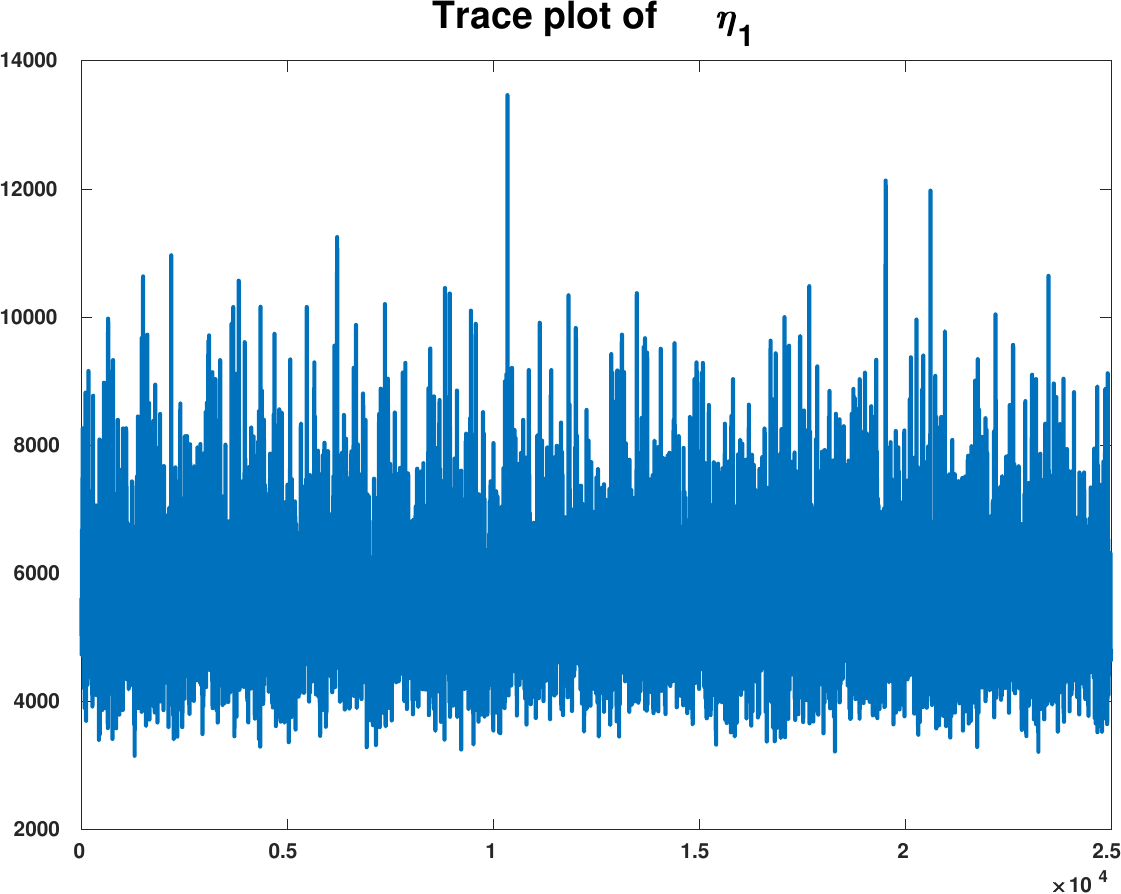}
    \centering
    \includegraphics[width=0.45\textwidth,height=4.85cm]{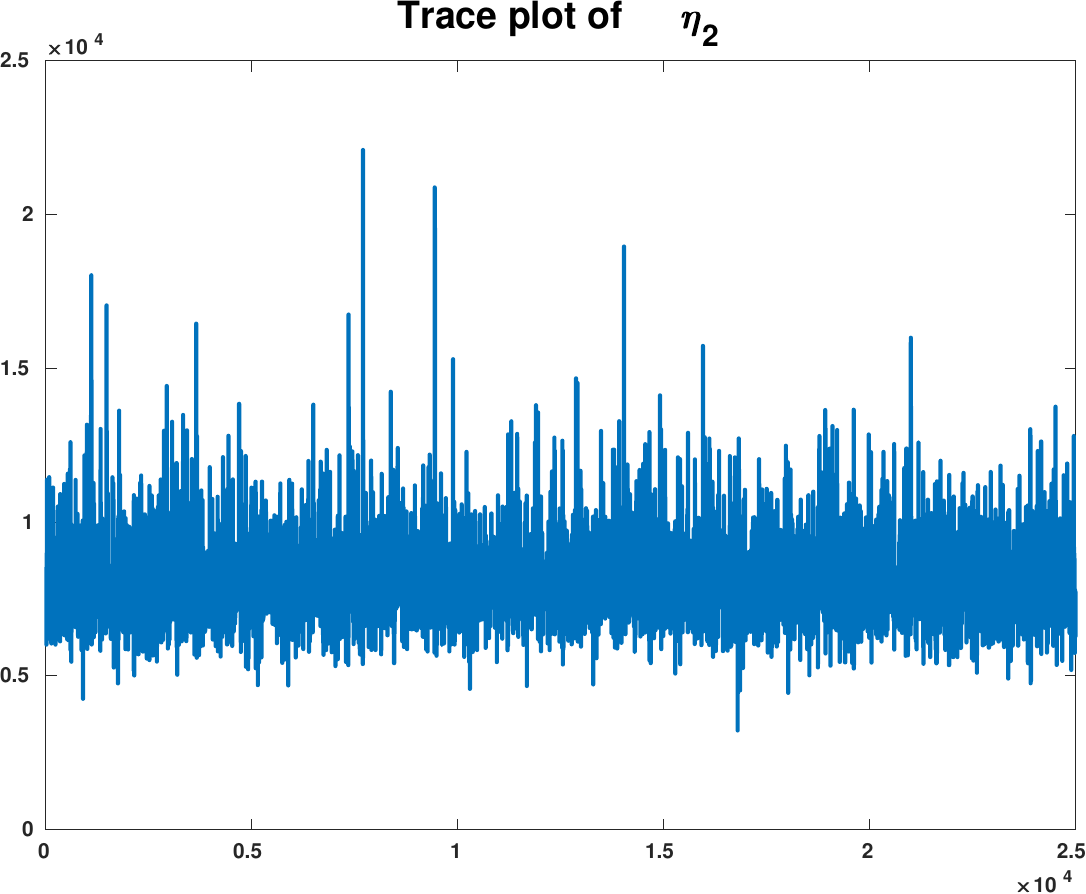}
\caption{Trace plot for the parameters except for $\eta_3$ corresponding to the Alaska temperature data.}
\label{fig:trace plot of Alaska temp data}
\end{figure}
\clearpage
\section{Supplementary plots for sea temperature data}
\label{Trace plt and posterior plts for Sea temp data}
%==============================
%Corresponding the analysis of Sea temperature data (subsection \ref{sea temp}), the trace plots of the parameters and the posterior densities of the complete time series of the latent variables are displayed in this appendix. 
\subsection{Trace plots}
\label{Sea temp data: trace plot}
\begin{figure}[!h]
    \includegraphics[width=0.45\textwidth,height=4.85cm]{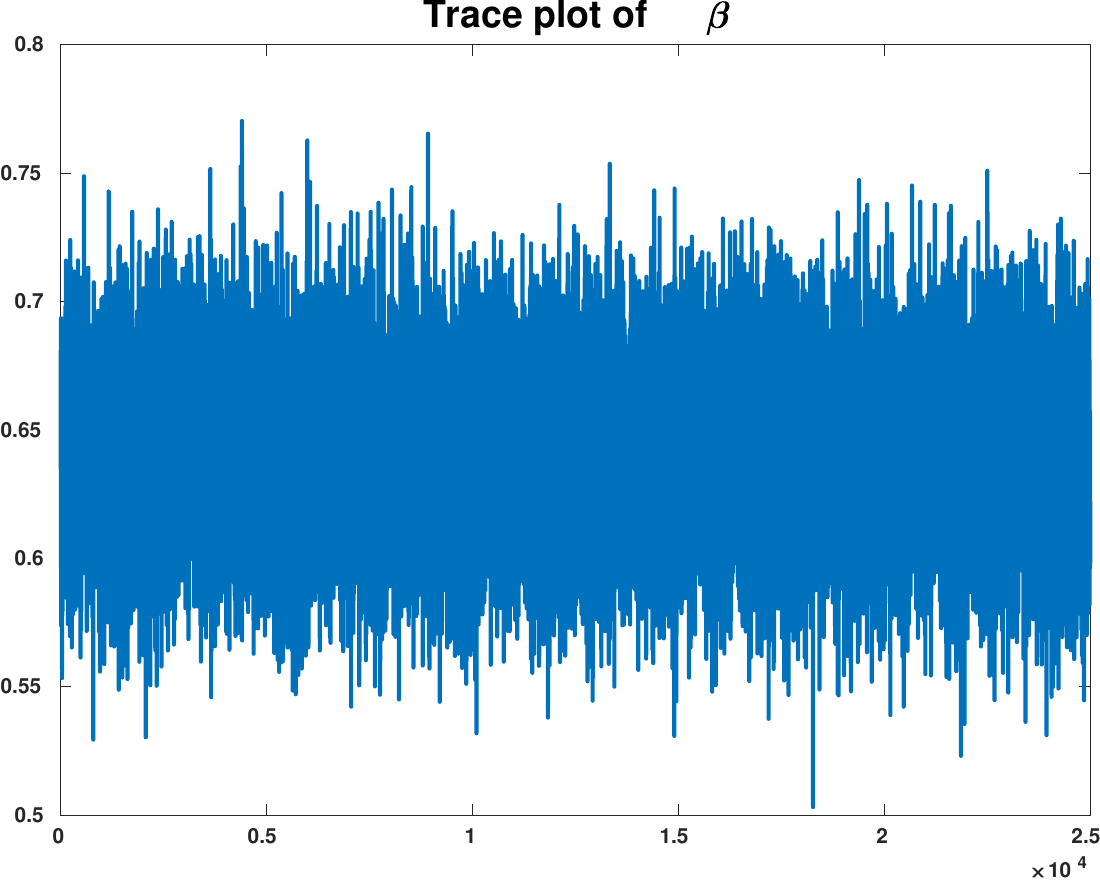}
    \includegraphics[width=0.45\textwidth,height=4.85cm]{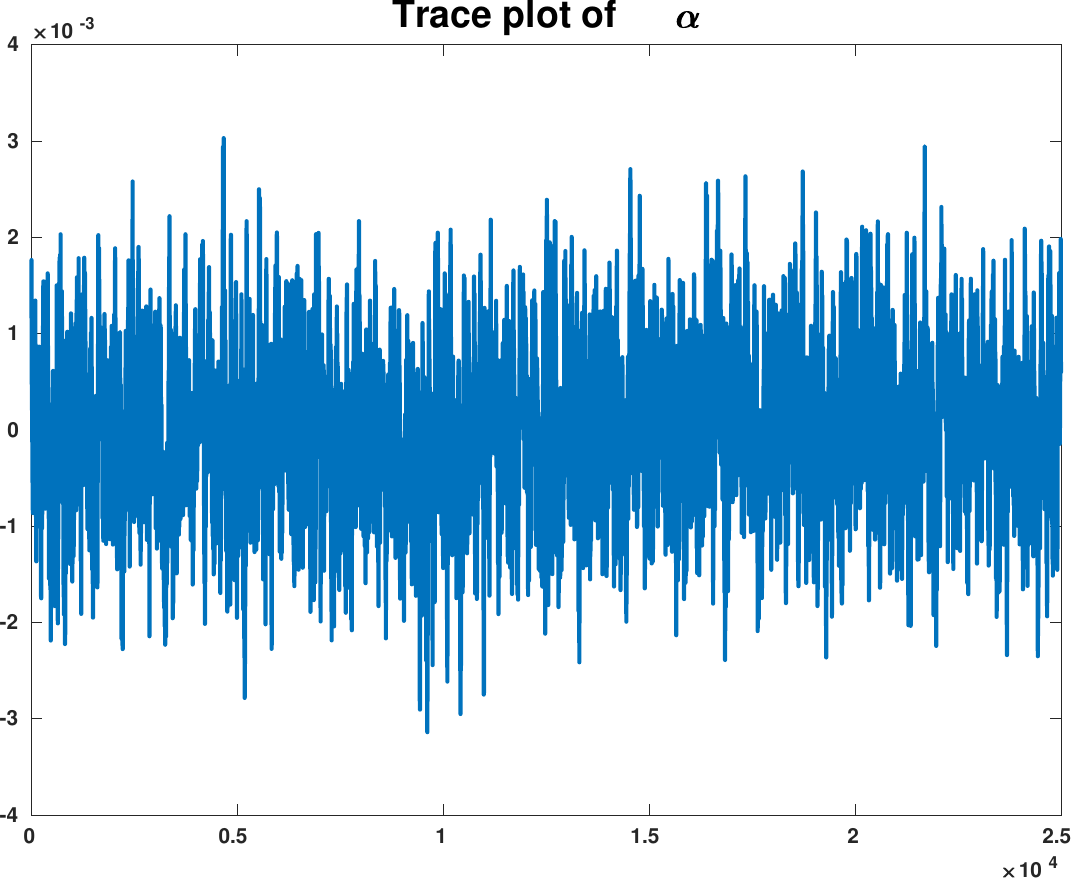}
    \includegraphics[width=0.45\textwidth,height=4.85cm]{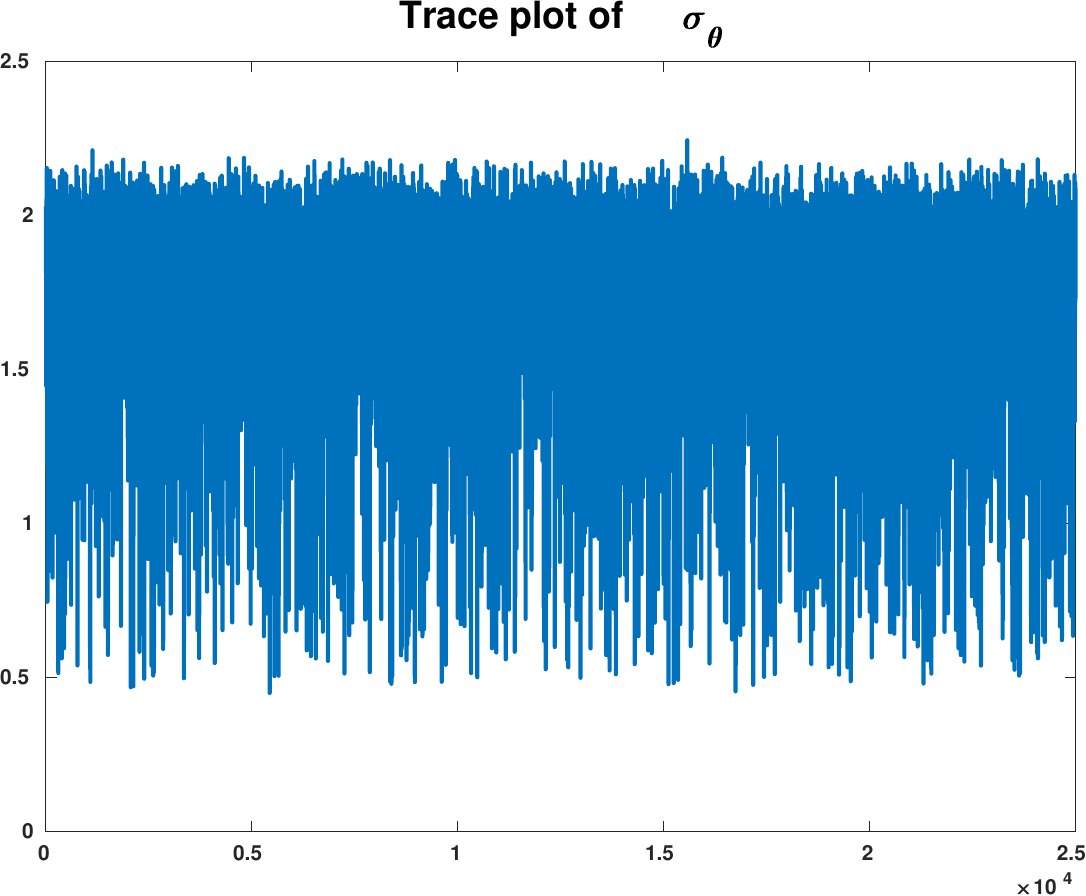}
    \includegraphics[width=0.45\linewidth,height=4.85cm]{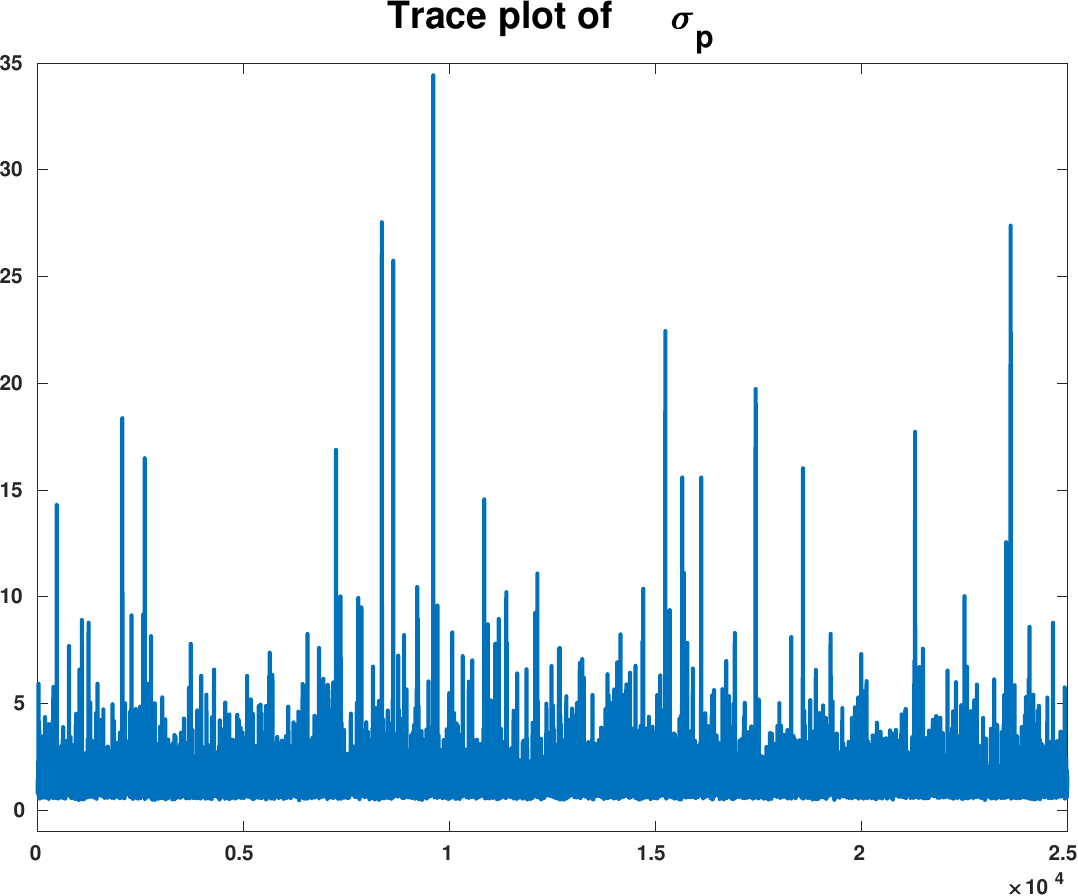}
    \includegraphics[width=0.45\textwidth,height=4.85cm]{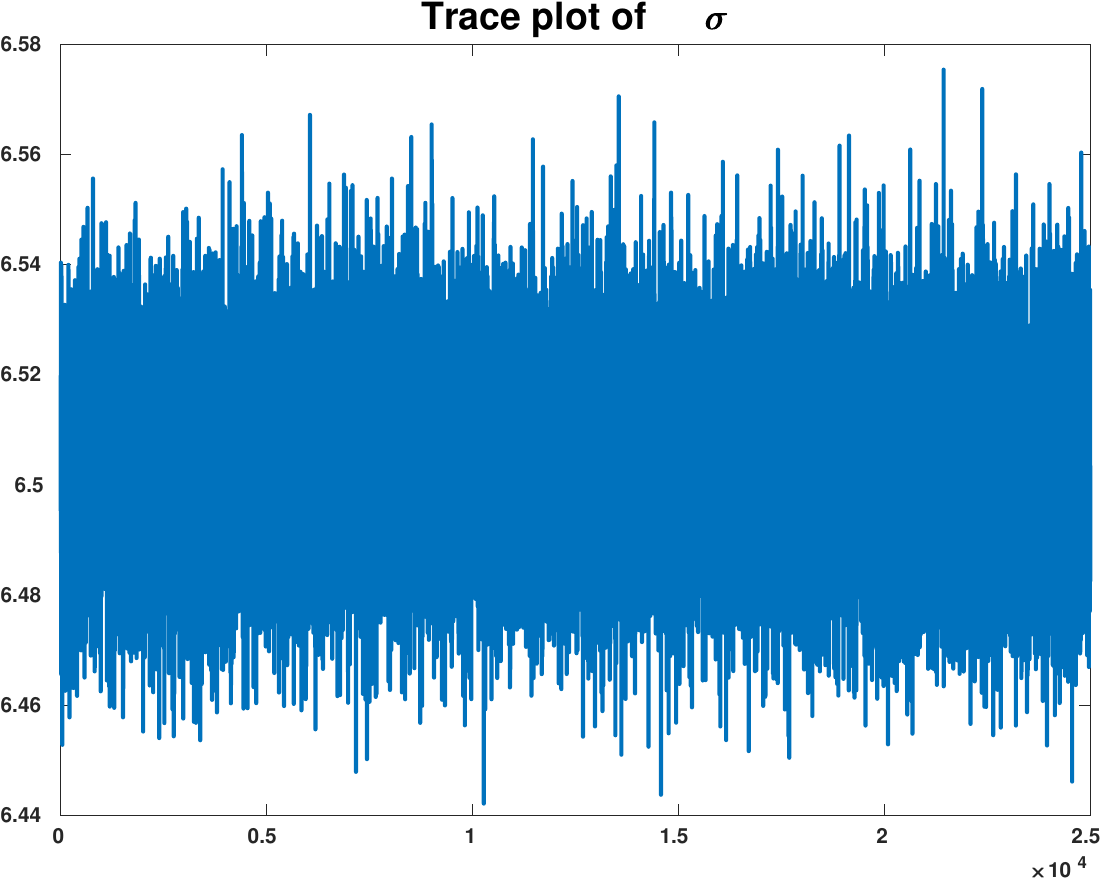}
    \includegraphics[width=0.45\textwidth,height=4.85cm]{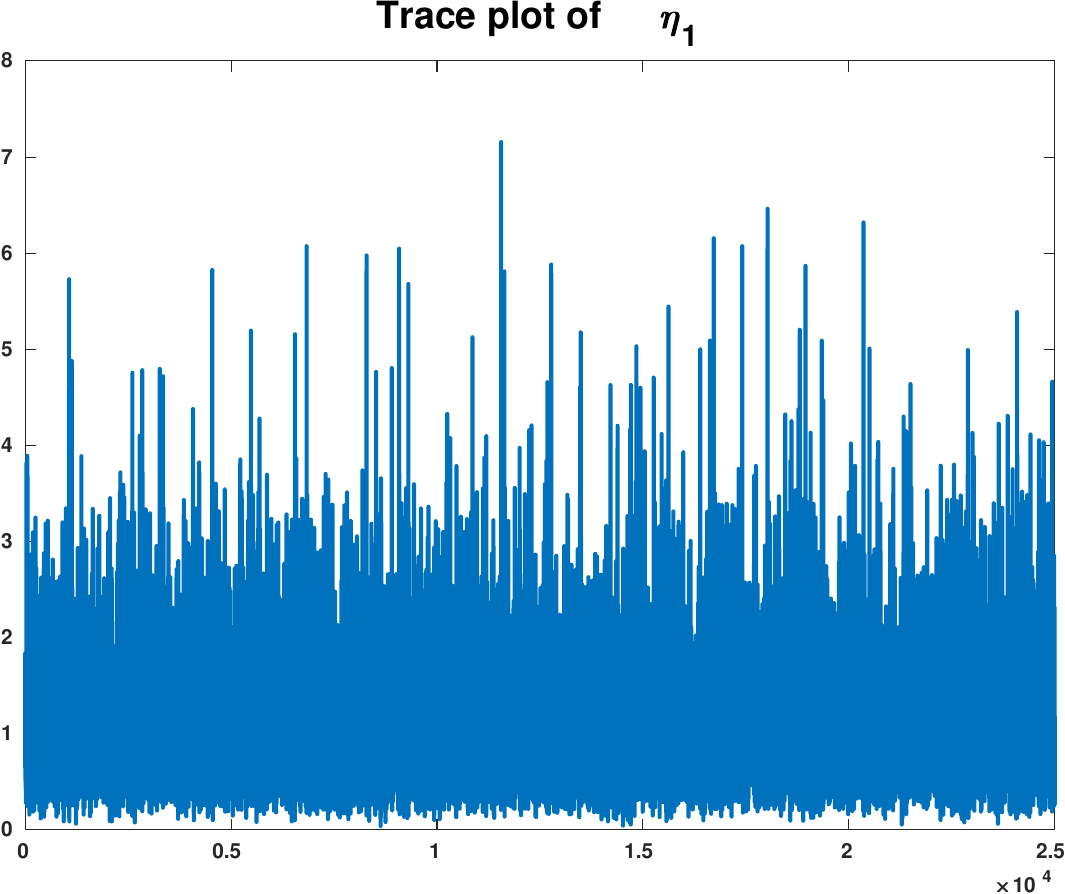}
    \centering
    \includegraphics[width=0.45\textwidth,height=4.85cm]{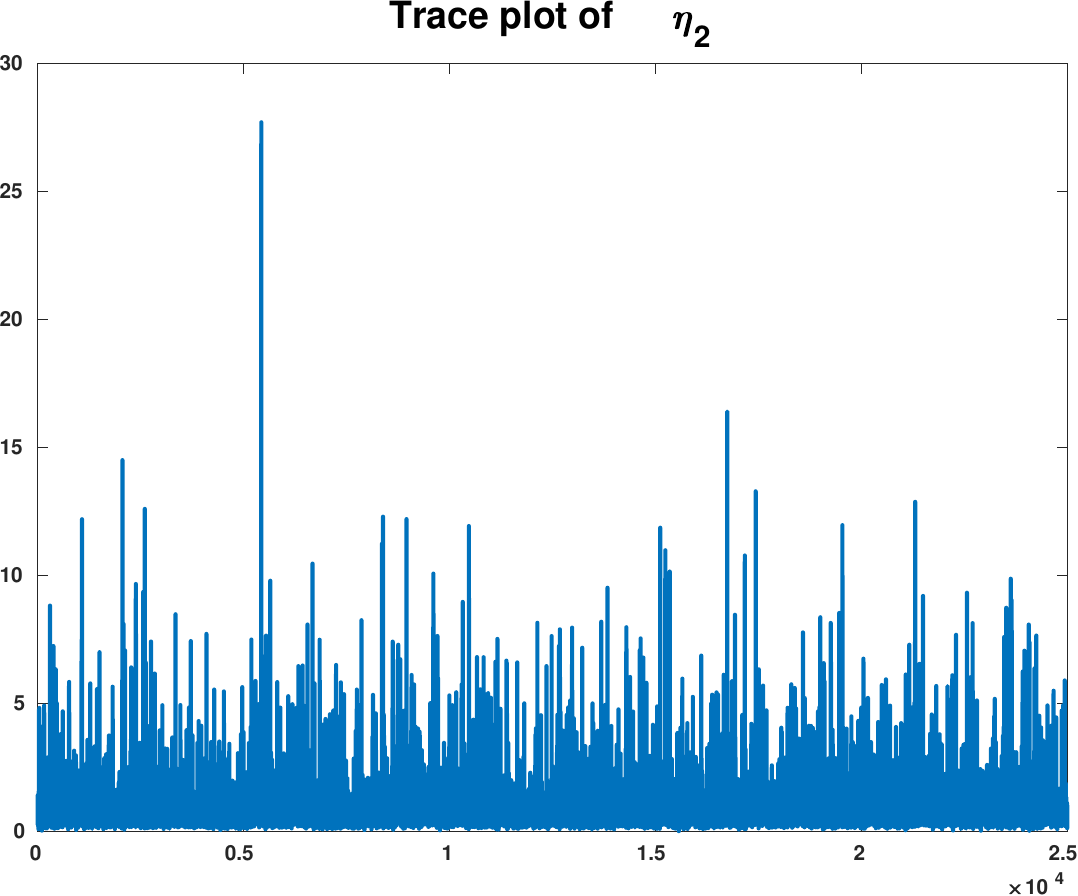}
\caption{Trace plots of the parameters except for $\eta_3$ for the sea surface temperature data.}
\label{fig:trace plot of sea temp data}
\end{figure}
\begin{comment}
\label{trace plot}
\begin{figure}[!ht]
\centering
\includegraphics[width=0.8\textwidth,height=4cm]{Sea_temp_trace_plots.jpg}
\caption{Trace plots of the parameters except for $\eta_3$ for the sea surface temperature data.}
\label{fig:trace plot of sea temp data}
\end{figure}
\end{comment}
%
\clearpage
\clearpage
\section{Stationarity, convergence of lagged correlations to zero and non-Gaussianity of the detrended Alaska data process}
\label{appendix:alaska}
This appendix demonstrates that the data set analyzed in Section \ref{Alaska} arises from a non-stationary and non-Gaussian process. Further, the analysis vindicates that the sample lagged correlation tends to 0 as the spatial and/or temporal lag tends to infinity. 
%============================================
\subsection{Stationarity of the detrended Alaska data process}
\label{subsec:stationarity_alaska}
%===========================================
To check if the data arrived from a stationary or non-stationary process, we resorted to the recursive Bayesian theory and methods developed by \ctn{roy2020bayesian}. 
In a nutshell, their key idea is to consider the Kolmogorov-Smirnov distance
between distributions of data associated with local and global space-times. 
Associated with the $j$-th local space-time region is an unknown probability $p_j$ of the event 
that the underlying process is stationary when the observed data corresponds to the $j$-th local region and the Kolmogorov-Smirnov distance falls below $c_j$,
where $c_j$ is any non-negative sequence tending to zero as $j$ tends to infinity. With suitable priors for $p_j$, \ctn{roy2020bayesian} constructed recursive posterior
distributions for $p_j$ and proved that the underlying process is stationary if and only if for sufficiently large number of observations
in the $j$-th region, the posterior of $p_j$ converges to one as $j\rightarrow\infty$.
Nonstationarity is the case if and only if the posterior of $p_j$ converges to zero as $j\rightarrow\infty$. 

In our implementation of the ideas of \ctn{roy2020bayesian}, we set the $j$-th local region to be the entire time series for the spatial location $\bi{s}_j$, for $j=1,\ldots,29$.
Thus, the size of each local region is $65$. 
%The number of regions, $2520$, is also large enough for the Bayesian
%recursive theories to be applicable. 
We choose $c_j$ to be of the same nonparametric, dynamic and adaptive form as detailed in \ctn{roy2020bayesian}. 
The dynamic form requires an initial value for the sequence. 
In practice, the choice of the initial value usually has significant effect on the convergence of the posteriors of $p_j$,
and so, the choice must be carefully made. However, in our case, for all initial values that we experimented with, lying between 0.05 and 1,
the recursive Bayesian procedure led to the conclusion of stationarity of the underlying spatio-temporal process. 

\begin{figure}
\centering
\includegraphics[width=0.75\textwidth,height=10cm,keepaspectratio]{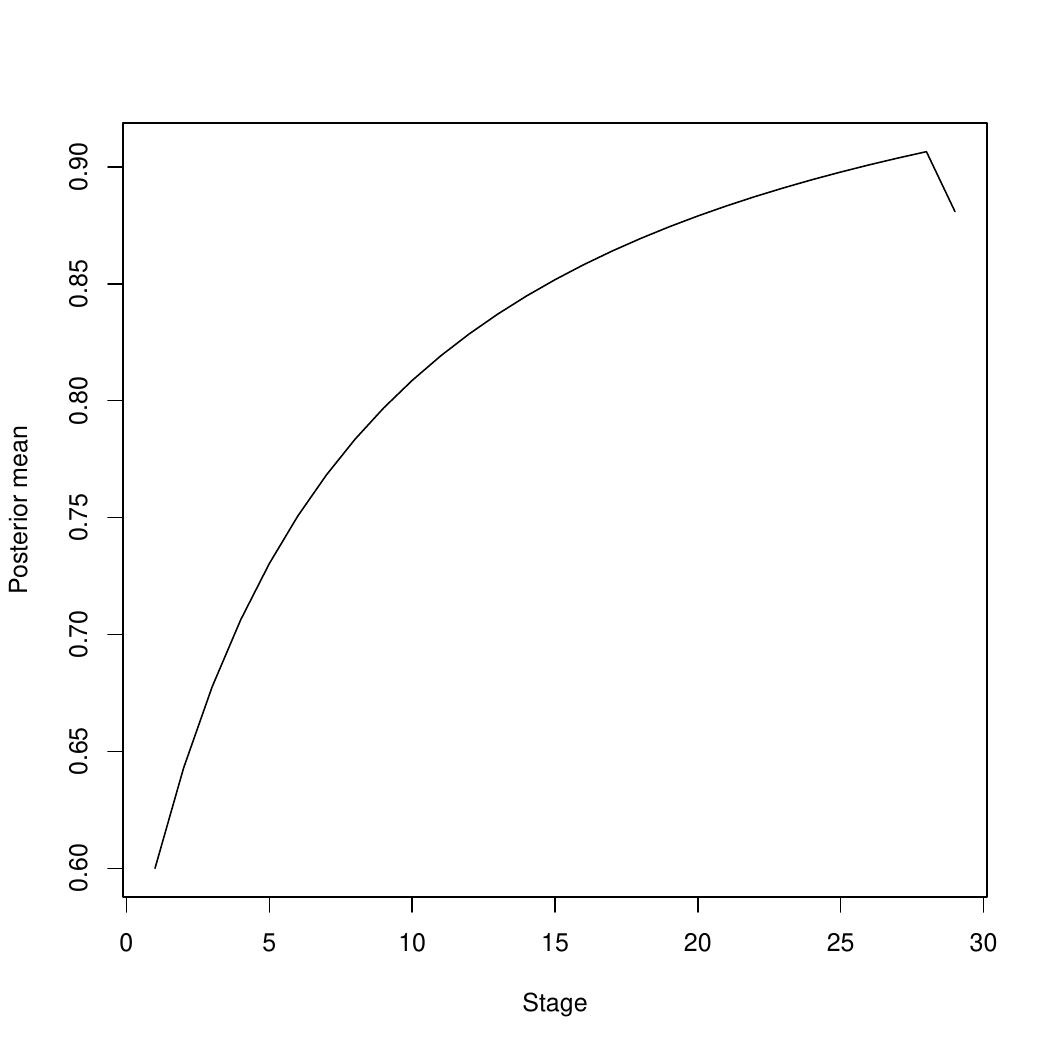}
\caption{Alaska data analysis: detection of strict stationarity.}
\label{fig:stationarity_alaska}
\end{figure}

We implemented the idea with our parallelised C code on $29$ parallel processors of a VMWare of Indian Statistical Institute; the time taken is less than a second.
For the initial value $0.05$, 
Figure \ref{fig:stationarity_alaska}
displays the means of the posteriors of $p_j$; $j=1,\ldots,29$, showing convergence to 1. The respective posterior variances are negligibly small
and hence not shown.  
Thus, the detrended spatio-temporal process that generated the Alaska data, can be safely regarded
as stationary.

%Figure \ref{fig:cov_nonstatonarity_realdata2} shows the results of our investigation of covariance stationarity. Panels (b), (c), (e) and (f) show
%convergence of the posterior means of $p_j$ in this context to zero for different partitioned intervals of $\|\bh\|$ associated with sufficient data such that 
%the covariances are well-defined. The posterior variances are again negligibly small as before.
%Thus, covariance nonstationarity of the underlying spatio-temporal process is also clearly indicated. In these cases, we chose the initial values of $c_j$ to be $0.05$.
%The time taken for our parallelised C code implementation on our dual-core laptop is only $5$ seconds for each $\|\bh\|$.
%Hence, the sea surface temperature phenomenon is not even weakly stationary.

\subsection{Convergence of lagged spatio-temporal correlations to zero for the Alaska data}
\label{subsec:zero_correlation}
\begin{figure}
\centering
\includegraphics[width=0.65\textwidth,height=10cm,keepaspectratio]{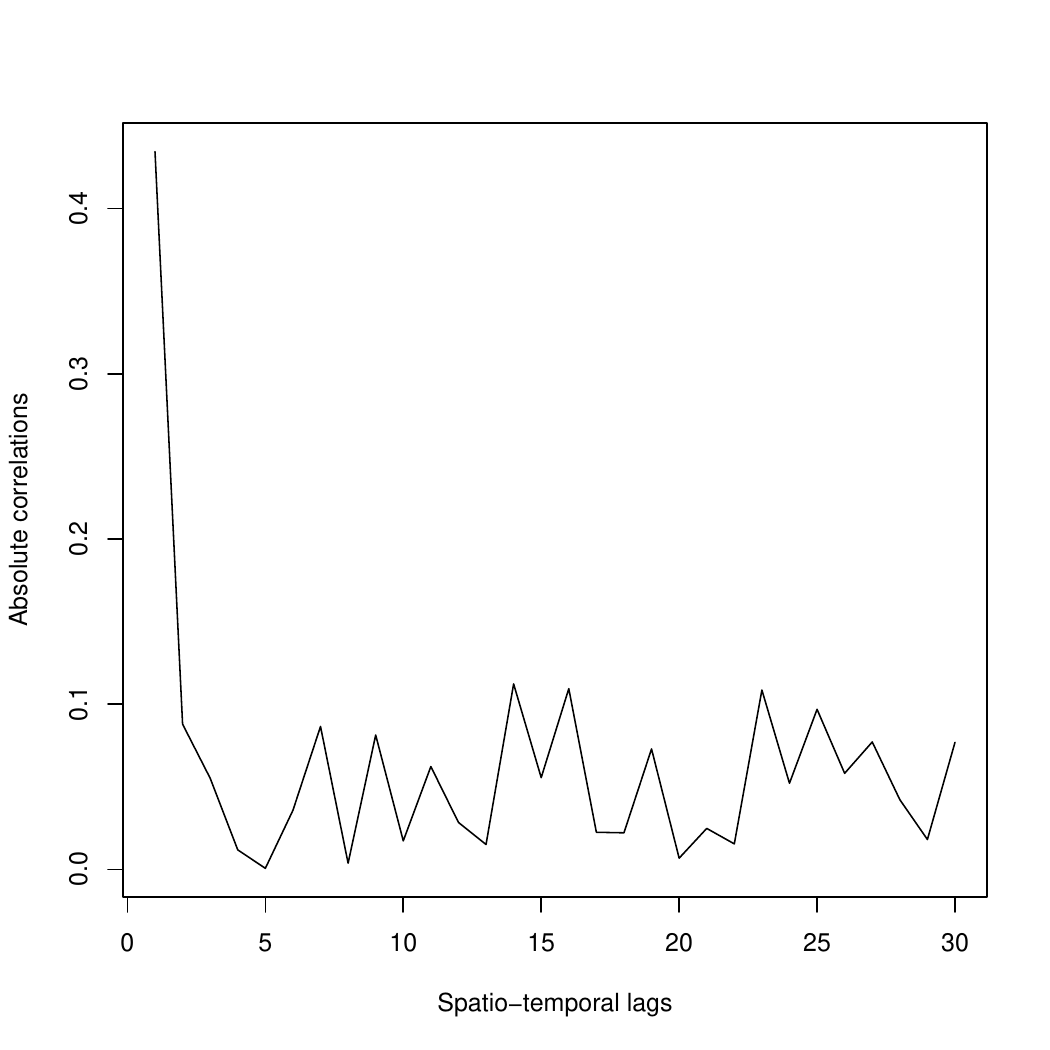}
\caption{Alaska data analysis: lagged spatio-temporal correlations converging to zero.}
\label{fig:corr_zero_alaska}
\end{figure}

Recall that one major purpose of our Hamiltonian spatio-temporal model is to emulate the property 
of most real datasets that the lagged spatio-temporal correlations tend to zero as the spatio-temporal lag tends to infinity, irrespective of stationarity or nonstationarity. 
Here we compute the lagged correlations on $30$ parallel processors on our VMWare, each processor computing the correlation for a partitioned interval of lag $\|\bi{h}\|$
such that the interval is associated with sufficient data, making the correlation well-defined.
The time taken for this exercise are a few seconds. 
Figure \ref{fig:corr_zero_alaska} demonstrates convergence of the lagged spatio-temporal correlations to zero; with larger amount of data such demonstration
would have been more convincing.
%
%This is a rather small dataset compared to the entire one, but this is necessary
%for computational feasibility, and moreover, larger datasets of sizes $20000$ and $30000$ did not yield significantly different results   
%
\subsection{Non-Gaussianity of the Alaska data}
\label{subsec:non_gaussian}
Simple quantile-quantile plots (not shown for brevity) revealed that the distributions of the time series data at the spatial locations, distributions of the spatial data
at the time points, and the overall distribution of the entire data set, are far from normal. Thus, traditional Gaussian process based models of the 
underlying spatio-temporal process are ruled out. Since the temporal distributions at the spatial locations and the spatial distributions at different time points are
also much different, it does not appear feasible to consider parametric stochastic process models for the data. These seem to make the importance of our nonparametric 
Hamiltonian process more pronounced.

%\clearpage
\bibliography{sample}

\end{document}